\documentclass[tradiabstract,longauth]{aa}

\usepackage{txfonts}
\usepackage{graphicx}
\usepackage{float}
\usepackage{natbib}
\usepackage[english]{babel}
\usepackage{color}
\usepackage{longtable}
\usepackage{rotating}

\begin{document}

\title{The ALPINE-ALMA [CII] survey: Data processing, catalogs, and statistical source properties}

\author{M.~B\'ethermin\inst{1} \and Y.~Fudamoto\inst{2} \and M.~Ginolfi\inst{2} \and F.~Loiacono\inst{3,4} \and Y.~Khusanova\inst{1} \and P.~L.~Capak\inst{5,6,7} \and P.~Cassata\inst{8,9} \and A.~Faisst\inst{5} \and O.~Le Fèvre\inst{1} \and D.~Schaerer\inst{2,10} \and J.~D.~Silverman\inst{11,12} \and L.~Yan\inst{13} \and R.~Amorin\inst{14,15} \and S.~Bardelli\inst{4} \and M.~Boquien\inst{16} \and A.~Cimatti\inst{3,17} \and I.~Davidzon\inst{5,6} \and M.~Dessauges-Zavadsky\inst{2} \and S.~Fujimoto\inst{6,7} \and C.~Gruppioni\inst{4} \and N.~P.~Hathi\inst{18} \and E.~Ibar\inst{19} \and G.~C.~Jones\inst{20,21} \and A.~M. Koekemoer\inst{18} \and G.~Lagache\inst{1} \and B.~C.~Lemaux\inst{22} \and C.~Moreau\inst{1} \and P.~A.~Oesch\inst{2,6} \and F.~Pozzi\inst{3,4} \and D.~A.~Riechers\inst{23,24} \and M.~Talia\inst{3,4} \and S.~Toft\inst{6,7} \and L.~Vallini\inst{25} \and D.~Vergani\inst{4} \and G.~Zamorani\inst{4} \and E.~Zucca\inst{4}}

\institute{
Aix Marseille Univ, CNRS, LAM, Laboratoire d'Astrophysique de Marseille, Marseille, France, \email{matthieu.bethermin@lam.fr} \and
D\'epartement d'Astronomie, Universit\'e de Gen\`eve, 51 Ch. des Maillettes, 1290 Versoix, Switzerland \and 
Universit\`a di Bologna - Dipartimento di Fisica e Astronomia, Via Gobetti 93/2 - I-40129, Bologna, Italy \and
INAF - Osservatorio di Astrofisica e Scienza dello Spazio di Bologna, via Gobetti 93/3, I-40129, Bologna, Italy \and
IPAC, M/C 314-6, California Institute of Technology, 1200 East California Boulevard, Pasadena, CA 91125, USA \and
The Cosmic Dawn Center (DAWN), University of Copenhagen, Vibenshuset, Lyngbyvej 2, DK-2100 Copenhagen, Denmark \and
Niels Bohr Institute, University of Copenhagen, Lyngbyvej 2, DK-2100 Copenhagen, Denmark \and
Dipartimento di Fisica e Astronomia, Universit\`a di Padova, vicolo dell'Osservatorio, 3 I-35122 Padova, Italy \and 
INAF, Osservatorio Astronomico di Padova, vicolo dell'Osservatorio 5, I-35122 Padova, Italy \and 
Institut de Recherche en Astrophysique et Plan\'etologie $-$ IRAP, CNRS, Universit\'e de Toulouse, UPS-OMP, 14, avenue E. Belin, F31400 Toulouse, France \and
Kavli Institute for the Physics and Mathematics of the Universe, The University of Tokyo, Kashiwa, Japan 277-8583 (Kavli IPMU, WPI) \and
Department of Astronomy, School of Science, The University of Tokyo, 7-3-1 Hongo, Bunkyo, Tokyo 113-0033, Japan \and
The Caltech Optical Observatories, California Institute of Technology, Pasadena, CA 91125, USA \and
Instituto de Investigaci\'on Multidisciplinar en Ciencia y Tecnolog\'ia, Universidad de La Serena, Ra\'ul Bitr\'an 1305, La Serena, Chile \and
Departamento de Astronom\'ia, Universidad de La Serena, Av. Juan Cisternas 1200 Norte, La Serena, Chile \and
Centro de Astronom\'ia (CITEVA), Universidad de Antofagasta, Avenida Angamos 601, Antofagasta, Chile \and 
INAF - Osservatorio Astrofisico di Arcetri, Largo E. Fermi 5, I-50125, Firenze, Italy \and
Space Telescope Science Institute, 3700 San Martin Drive, Baltimore, MD 21218, USA \and
Instituto de F\'isica y Astronom\'ia, Universidad de Valpara\'iso, Avda. Gran Breta\~na 1111, Valpara\'iso, Chile \and
Cavendish Laboratory, University of Cambridge, 19 J. J. Thomson Ave., Cambridge CB3 0HE, UK \and
Kavli Institute for Cosmology, University of Cambridge, Madingley Road, Cambridge CB3 0HA, UK \and
Department of Physics, University of California, Davis, One Shields Ave., Davis, CA 95616, USA \and
Department of Astronomy, Cornell University, Space Sciences Building, Ithaca, NY 14853, USA \and
Max-Planck-Institut f\"ur Astronomie, K\"onigstuhl 17, D-69117 Heidelberg, Germany \and
Leiden Observatory, Leiden University, PO Box 9500, 2300 RA Leiden, The Netherlands
}

\date{Received ??? / Accepted ???}

\abstract{The Atacama Large Millimeter Array (ALMA) Large Program to INvestigate [CII] at Early times (ALPINE) targets the [CII] 158\,$\mu$m line and the far-infrared continuum in 118 spectroscopically confirmed star-forming galaxies between z=4.4 and z=5.9. It represents the first large [CII] statistical sample built in this redshift range. We present details regarding the data processing and the construction of the catalogs. We detected 23 of our targets in the continuum. To derive accurate infrared luminosities and obscured star formation rates (SFRs), we measured the conversion factor from the ALMA 158\,$\mu$m rest-frame dust continuum luminosity to the total infrared luminosity (L$_{\rm IR}$) after constraining the dust spectral energy distribution by stacking a photometric sample similar to ALPINE in ancillary single-dish far-infrared data. We found that our continuum detections have a median L$_{\rm IR}$ of 4.4$\times 10^{11}$\,L$_\odot$. We also detected 57 additional continuum sources in our ALMA pointings. They are at a lower redshift than the ALPINE targets, with a mean photometric redshift of 2.5$\pm$0.2. We measured the 850\,$\mu$m number counts between 0.35 and 3.5\,mJy, thus improving the current interferometric constraints in this flux density range. We found a slope break in the number counts around 3\,mJy with a shallower slope below this value. More than 40\,\% of the cosmic infrared background is emitted by sources brighter than 0.35\,mJy. Finally, we detected the [CII] line in 75 of our targets. Their median [CII] luminosity is 4.8$\times$10$^8$\,L$_\odot$ and their median full width at half maximum is 252\,km/s. After measuring the mean obscured SFR in various [CII] luminosity bins by stacking ALPINE continuum data, we find a good agreement between our data and the local and predicted SFR-L$_{\rm [CII]}$ relations. The ALPINE products are publicly available at https://cesam.lam.fr/a2c2s/.}



\titlerunning{ALPINE data processing, catalogs, and statistical source properties}

\authorrunning{B\'ethermin et al.}

\maketitle

\section{Introduction}

Understanding the early formation of the first massive galaxies is an important goal of modern astrophysics. At z$>$4, most of our constraints come from redshifted ultraviolet (UV) light, which probes the unobscured star formation rate (SFR). Except for a few very bright objects \citep[e.g.,][]{Walter2012,Riechers2013,Watson2015,Capak2015,Strandet2017,Zavala2018,Jin2019,Casey2019}, we have much less information about dust-obscured star formation, that is, the UV light absorbed by dust and re-emitted in the far infrared. To accurately measure the star formation history in the Universe, we need to know both the obscured and unobscured parts \citep[e.g.,][]{Madau2014,Maniyar2018}.

With its unprecedented sensitivity, the Atacama Large Millimeter Array (ALMA) is able to detect both the dust continuum and the brightest far-infrared and submillimeter lines in "normal" galaxies at z$>4$. However, this remains a difficult task for blind surveys. For instance, current deep field observations detect only a few continuum sources at z$>$4 after tens of hours of observations \citep[e.g.,][]{Dunlop2017,Aravena2016b,Franco2018,Hatsukade2018}. Targeted observations of known sources from optical and near-infrared spectroscopic surveys are usually more efficient. For instance, \citet{Capak2015} detected four objects at z$>$5 using a few hours of observations.

The [CII] fine structure line at 158um is mainly emitted by dense photodissociation regions, which are the outer layers of giant molecular clouds \citep{Hollenbach1999,Stacey2010,Gullberg2015}, although it can also trace the diffuse (cold and warm) neutral medium \citep{Wolfire2003}, and to a lesser degree the ionized medium \citep[e.g.,][]{Cormier2012}. It is one of the brightest galaxy lines across the electromagnetic spectrum. In addition, at z$>$4, it is conveniently redshifted to the $>$850\,$\mu$m atmospheric windows. This line has a variety of different scientific applications since it can be used to probe the interstellar medium \citep[e.g.,][]{Zanella2018}, the SFR \citep[e.g.,][]{De_Looze2014,Carniani2018b}, the gas dynamics \citep[e.g.,][]{De_Breuck2014,Jones2020}, or outflows \citep[e.g.,][]{Maiolino2012,Gallerani2018,Ginolfi2020}. It has now been detected in $\sim$35 galaxies at z$>$4, but most of them are magnified by lensing or starbursts and only one third of them are normal star-forming systems (see compilation in \citealt{Lagache2018}). 

Over the past several years, numerous theoretical studies have focused on the exact contribution of the various gas phases \citep[e.g.,][]{Olsen2017,Pallottini2019} and the effects of metallicity \citep{Vallini2015,Lagache2018}, gas dynamics \citep{Kohandel2019}, and star-formation feedback \citep{Katz2017,Vallini2017,Ferrara2019} on the [CII] emission, which nowadays is the most studied long-wavelength line at z$>$4.

The rest-frame $\sim$160\,$\mu$m dust continuum and the [CII] line can be observed simultaneously by ALMA and are the easiest and the most promising features to help gain an understanding of obscured star formation at z$>$4. The ALMA Large Program to INvestigate [CII] at Early times (ALPINE) aims to build the first large sample with a coherent selection process at z$>$4, increasing the size of the pioneering \citet{Capak2015} sample by an order of magnitude. \citet{Le_Fevre2019} describe the goals of the survey and \citet{Faisst2020} present the sample selection and the properties of galaxies in the sample, which were measured from ancillary data. In this paper, we present the task of processing the ALPINE data from the raw data to the catalogs and the immediate scientific results such as the basic dust and [CII] properties of the ALPINE targets together with the number counts and redshift distribution of the serendipitous continuum detections.

In Sect.\,\ref{sect:data_proc}, we describe the ALPINE data processing and the main products (maps and cubes). In Sect.\,\ref{sect:cont_cat}, we explain how we built the continuum source catalog and characterized the performance of our method (purity, completeness, and photometric accuracy). In Sect.\,\ref{sect:sed}, we derive a reliable conversion factor from the 158\,$\mu$m rest-frame dust continuum to the total infrared luminosity (L$_{\rm IR}$, 8--1000\,$\mu$m) and the infrared SFR (SFR$_{\rm IR}$) using the stacking of ancillary single-dish data at the position of photometric samples similar to ALPINE. In Sect.\,\ref{sect:props_cont}, we discuss the continuum properties of ALPINE detections and the statistical properties of nontarget sources found in the fields (redshift distribution, number counts). In Sect.\,\ref{sect:cii_cat}, we describe the procedure used to generate and validate the [CII] spectra and catalog. In Sect\,\ref{sect:props_cii}, we discuss the properties of the [CII] detections (luminosity, width, velocity offset) and we briefly discuss the correlation between SFR and [CII] luminosity. In this paper, we assume \citet{Chabrier2003} initial mass function (IMF) and a $\Lambda$CDM cosmology with $\Omega_\Lambda = 0.7$, $\Omega_m = 0.3$, and $H_0 = 70$\,km/s/Mpc.


\section{Data processing}

\label{sect:data_proc}

\subsection{Observations}

\label{sect:obs}

The ALPINE-ALMA large program (2017.1.00428.L, PI: Le F\`evre) targeted 122 individual $4.4<z_{\rm spec}<5.9$ and SFR$ \gtrsim$10\,M$_\odot$/yr galaxies with known spectroscopic redshifts from optical ground-based observations. The construction and the physical properties of the sample is described in \citet{Le_Fevre2019}  and \citet{Faisst2020}, respectively. The ALPINE sample contains sources from both the cosmic evolution survey (COSMOS) field and the \textit{Chandra} deep field south (CDFS).

In this redshift range, the [CII] line falls in the band 7 of ALMA (275 --373\,GHz). To avoid an atmospheric absorption feature, no source has been included between z=4.6 and 5.1. In order to minimize the calibration overheads, we created many groups of two sources with similar redshift, which are observed using the same spectral setting. In our sample, the typical optical line width is $\sigma \sim$100\,km/s (or FWHM$\sim$235\,km/s). At the targeted frequency, the coarse resolution ($\Delta \nu_{\rm channel}$ = 31.250\,MHz) offered by the Time Division Mode (TDM) is sufficient to resolve our lines ($\Delta {\rm v}_{\rm channel}$ = 25-35\,km/s) and results in a total size of our raw data below 3\,TB for the whole sample. The [CII] lines of the targeted sources are covered by two contiguous spectral windows (1.875\,GHz each), while we placed two remaining spectral windows in the other side band to optimize the bandwidth and thus the continuum sensitivity. To maximize the integrated flux sensitivity, we requested compact array configurations (C43-1 or C43-2) corresponding to a $>$0.7\,arcsec resolution to avoid diluting the flux of our sources into several synthesized beams.

We aimed for a 1-$\sigma$ sensitivity on the integrated [CII] luminosity L$_{\rm [CII]}$ of 0.4$\times$10$^8$\,L$_\odot$ assuming a line width of 235\,km/s. As shown in Sect.\,\ref{sect:FWHM_target_cii}, this sensitivity was reached on average by our observations. At higher redshift (lower frequency), we need to reach a lower noise in Jy/beam to obtain the same luminosity ($\sim$0.2\,mJy/beam in 235\,km/s band at z=5.8 versus $\sim$0.3\,mJy/beam in the same band at z=4.4). In contrast, at low frequency, the noise is lower because of the higher atmospheric transmission and the lower receiver temperature. The two effects compensate each other and the integration times are similar for our entire redshift range (15-25\,min on source). Each scheduling block containing the observations of the calibrators and two sources can be observed using a single 50\,min--1h15min execution. In total, we had 61 scheduling blocks (SBs) for a total of 69.3\,h including overheads.

ALPINE was selected in cycle 5 and most of the observations were completed during this period. Between 2018/05/08 and 2018/07/16, 102 of our sources were observed. Observations had to be stopped from mid-July to mid-August because of exceptional snowstorms. Two additional sources were observed after the snow storms (2018/08/20).  After that, the configuration was too extended and the 18 last sources were carried over in cycle 6. They were observed between 2019/01/09 and 2019/01/11.

We realized during the data analysis that four ALPINE sources were observed two times with different names: vuds\_cosmos\_5100822662 and DEIMOS\_COSMOS\_514583, vuds\_cosmos\_5101288969 and DEIMOS\_COSMOS\_679410, vuds\_cosmos\_510786441 and DEIMOS\_COSMOS\_455022, vuds\_efdcs\_530029038 and CANDELS\_GOODSS\_15. In the rest of the paper, we use the VIMOS ultra deep survey (VUDS) name of these objects. We thus combined the two ALPINE observations of each of these sources to obtain deeper cubes and maps. Our final sample contains 118 objects.

\subsection{Pipeline calibration and data quality}

The data were initially calibrated at the observatory using the standard ALMA pipeline of the Common Astronomy Software Applications package (CASA) software \citep{McMullin2007}. We checked the automatically-generated calibration reports and identified a few antennae with suspicious behaviors (e.g., phase drifts in the bandpass calibration, unstable phase or gain solutions, anomalously low gains or high system temperatures), which were not flagged by the pipeline. For example, we had to flag the DV19 antenna for all the cycle 6 observations, for which the bandpass phase solution drifted by $\sim$180\,deg/GHz in the \textit{XX} polarization. For half of the observations, no problems were found and we used directly the data calibrated by the observatory pipeline. Most of the other observations were usually good with only 1 or 2 antennae with possible problems. Four SBs have between 3 and 5 potentially problematic antennae. Considering the very low impact of a single antenna on the final sensitivity, we thus decided to be conservative and fully flag these suspicious antennae and subsequently excised them from our analysis.

While the reduction process was generally smooth, we encountered a couple minor issues. The pipeline sometimes flagged the channels of a spectral window overlapping with the noisy edge channels of another spectral window. It was solved by adding the \textit{fracspw}=0.03125 option to the \textit{hifa\_flagdata} task before re-running the pipeline script from the observatory. This option flags the edge channels corresponding to 3.125\,\% of the width of the spectral window, while the default is to flag two channels on each side in TDM mode, that is 4/128 = 0.0315. In theory, this command is equivalent to the default routine. In practice, it is not affected by the subtle bug flagging the channels of the other spectral windows when they overlap, which solves our problem. In a few cases, the pipeline used an inconsistent numbering of the spectral windows and we had to manually correct these problematic SBs.

\subsection{Flux calibrators variability and calibration uncertainties}

\label{sect:quasars}

\begin{figure}
\centering
\includegraphics[width=8.5cm]{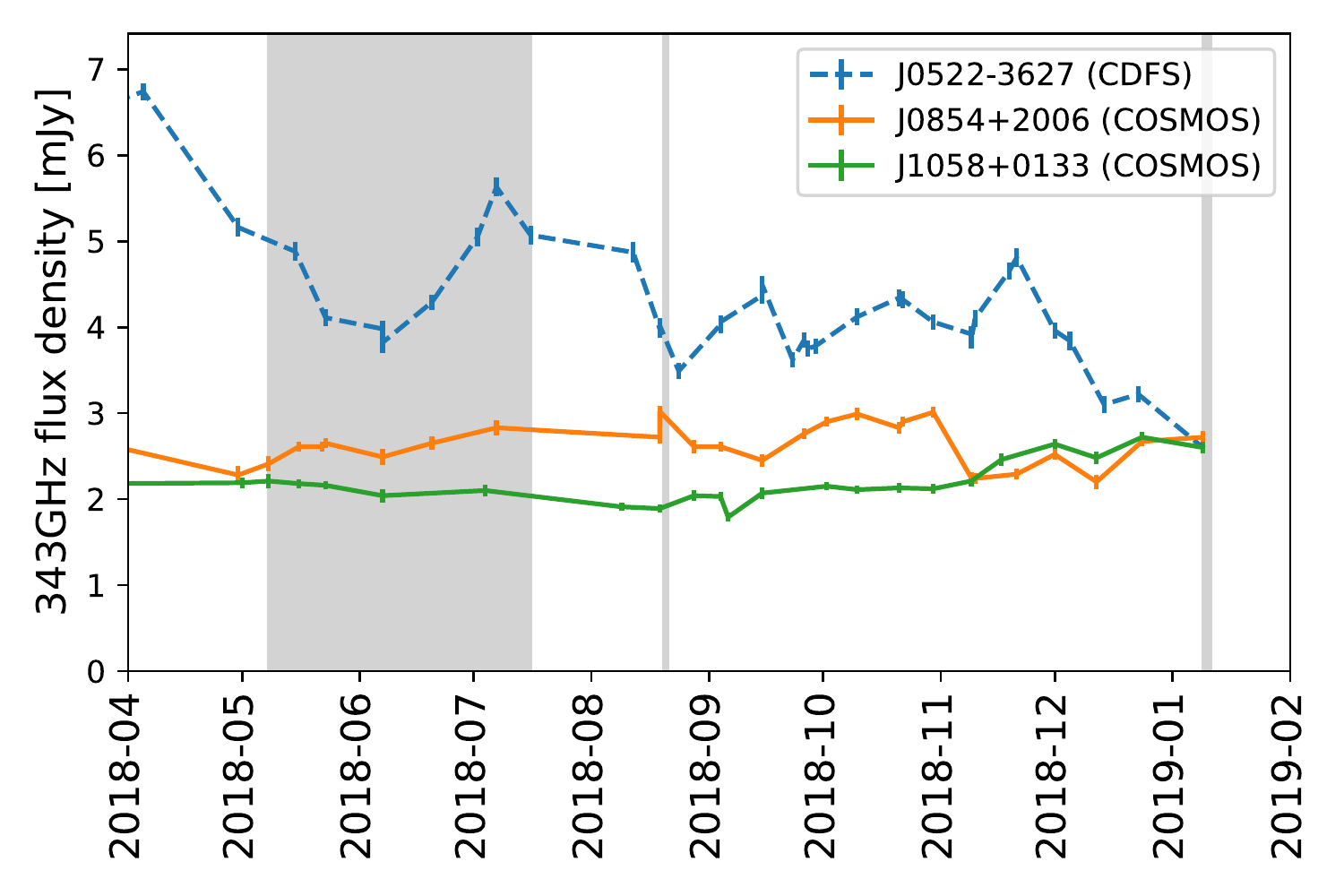}
\includegraphics[width=8.5cm]{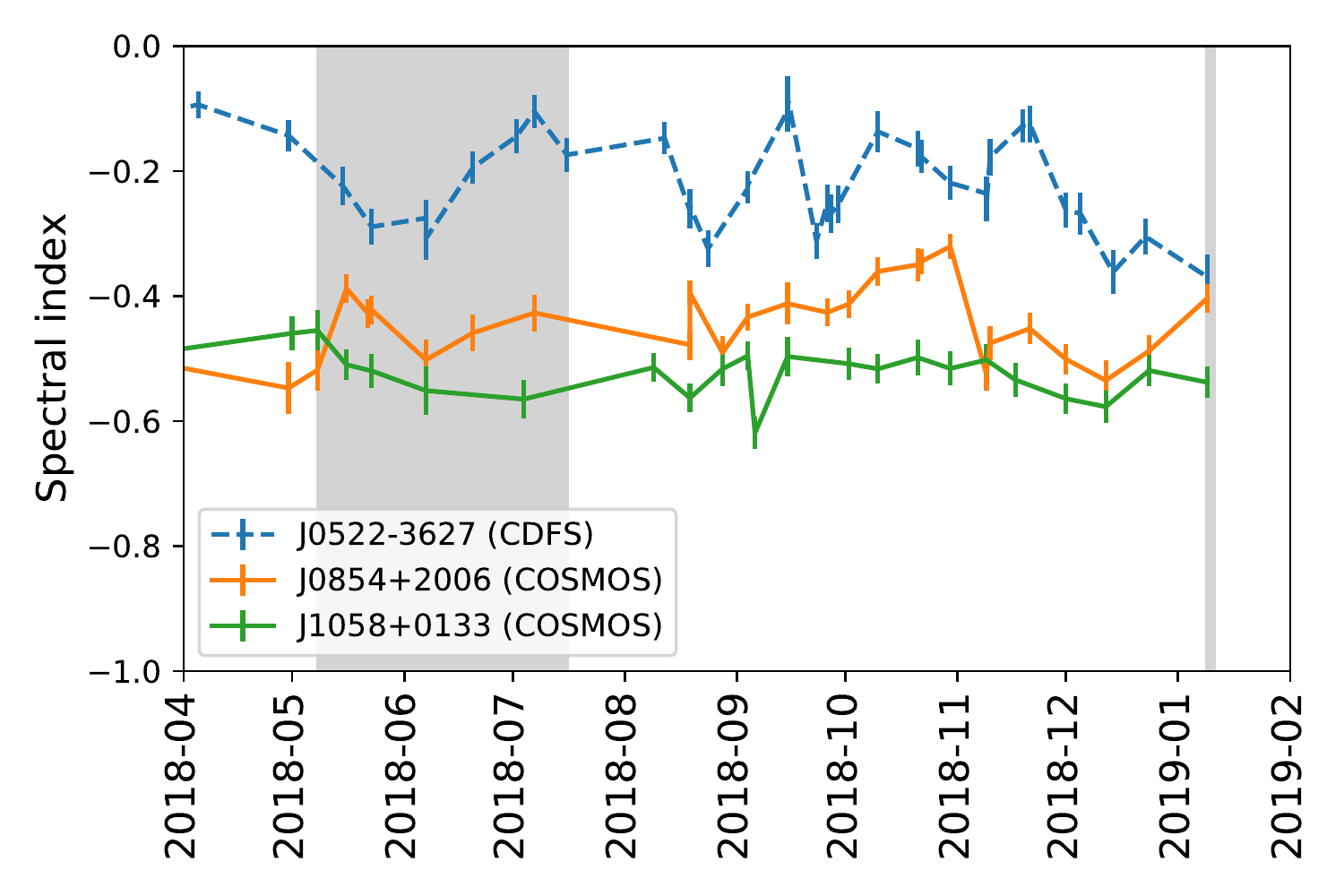}
\caption{\label{fig:quasars} Upper pannel: 345\,GHz flux density of the flux calibrators used by the ALPINE survey (the J1058+0133, J0854+2006, and J0522-3627 quasars) as a function of time. The gray areas indicate when the ALPINE targets were observed (see Sect.\,\ref{sect:obs}). The only quasar used for CDFS targets (J0522-3627) is plotted with a dashed line, while solid lines are used for the calibrators of COSMOS sources. Lower panel: spectral index versus time. This spectral index is estimated using the band-7 and band-3 flux from the calibrator monitoring performed by the observatory (see Sect.\,\ref{sect:quasars}).}
\end{figure}

The stability of the flux calibration over our entire survey is particularly important to interpret the sample statistically. We thus checked that the quasars used as secondary flux calibrators were reasonably stable across the ALPINE observations. These secondary calibrators are J1058+0133 and J0854+2006 for the targets in the COSMOS field and J0522-3627 for the ones in the CDFS. We downloaded the data from their flux monitoring by the observatory and calibrated using a well-known primary calibrator\footnote{\url{https://almascience.eso.org/sc/}}. In Fig.\,\ref{fig:quasars}, we present the evolution of their band-7 flux density and the spectral index determined using their measured band-7 and band-3 fluxes.

The three quasars are reasonably stable between two successive observations and in particular during the ALPINE observations (gray area in Fig.\,\ref{fig:quasars}). The standard deviation of the relative difference between two successive data points is only 0.059, 0.060 and 0.031 for J0522-3627, J0854+2006, and J1058+0133, respectively. In the figure, the variability of J0522-3627 could seem larger than J0854+2006. However, the actual relative variations between two successive observations are similar. The larger flux of J0522-3627 highlighting small relative variations in our linear-scale plot and the presence of long-term trends at the scale of several months can give this wrong impression. The maximum relative deviation between two successive visits is 0.20 and happened in J0854+2006 in November 2018, when ALPINE observations were not scheduled. Except this outlier, the maximal variation is 0.13. Usually, the last measurements performed by the quasar monitoring survey are used to determine the flux reference to calibrate a science observation. We can thus expect that the calibration uncertainty coming from the variability of the quasars is usually 6\% with 13\,\% outliers.

The frequency reference used for this monitoring is 345\,GHz. However, for the highest redshift object of our sample, the spectral setup is centered around 283\,GHz. The observatory uses the previously-measured spectral index measured using band-7 and band-3 data to derive the expected flux at the observed frequency. If this index varies too much between two monitorings, it could be a problem. The standard deviation of the spectral index between two successive monitorings is 0.075, 0.069, and 0.050 for J0522-3627, J0854+2006, and J1058+0133, respectively. This corresponds to an uncertainty of 1.5\,\%, 1.4\,\%, and 1.0\,\% on the extrapolation of the flux from 345\,GHz to 283\,GHz. The largest jump (0.22 in J0522-3627) corresponds to 4.5\,\%. The typical 1-$\sigma$ uncertainty of the calibration thus is 7.5\,\% for J0522-3627 and J0854+2006 and 4\,\% for J1058+0133 combining linearly the flux and spectral index uncertainties to be conservative and for the source requiring the most uncertain frequency interpolation. Our calibration uncertainty caused by quasar variability is thus slightly smaller than the typical 10\% of uncertainty of interferometric calibrations.

\subsection{Data cube imaging and production of [CII] moment-0 maps}

\label{sect:imaging_cubes}

The datacube were imaged using the \textit{tclean} CASA routine using 0.15\,arcsec pixels to well sample the synthesized beam (6 pixels per beam major axis in the field with the sharpest synthesized beam). The clean algorithm is run down to a flux threshold of 3\,$\sigma_{\rm noise}$, where $\sigma_{\rm noise}$ is the standard deviation measured in a previous nonprimary-beam-corrected cube after masking the sources. The determination of the final clean threshold is thus the result of an iterative process. The noise converges very quickly with negligible variations between the second and the third iteration. In practice, the exact choice of the clean threshold has a very low impact on the final flux measurements, since our pointings mostly contain one or a few sources, which are rarely bright. In addition, the natural weighting produces sidelobes and high signal-to-noise ratio (S/N) sources can produce nonnegligible artifacts in the dirty maps or unproperly cleaned maps. We checked that the amplitude of the largest sidelobes are below 10\,\% of the peak of the main beam. The sidelobe residuals after cleaning down to 3\,$\sigma$ should thus be below 0.3\,$\sigma$.

The standard ALPINE products were produced using a natural weighting of the visibilities. This choice maximizes the point-source sensitivity and produces a larger synthesized beam than other weighting schemes, which limits the flux spreading across several beams for slightly extended sources. These cubes are thus optimized to measure integrated properties of ALPINE targets.

We also produced continuum-free cubes. The continuum was subtracted in the uv-plane using the \textit{uvcontsub} CASA routine. This routine takes as input a user-provided range of channels containing line emission, and masks them before fitting a flat continuum model (order 0) to the visibilities. To identify the channels to mask, we used the line properties determined using the method presented in Sect.\,\ref{sect:cii_cat}. We use several iterations of the cube production and the line extraction to obtain the final version of these products. To avoid any line contamination, we chose to be conservative and excluded all the channels up to 3-$\sigma_{\rm v}$ from the central frequency of the best Gaussian fit of the line. When a [CII] spectrum exhibits a non-Gaussian excess in the wings, we masked manually an additional $\sim$0.1--0.2GHz to produce conservative continuum-free cubes.

Finally, we generated maps of the [CII] integrated intensity by summing all the channels containing the line emission, that is the moment-0 maps defined as $M(x,y) = \sum_{\rm k = 1}^{N_{\rm channel}} S_{\nu}(x,y,k) \, \Delta$v$_{\rm channel}(k)$, where $S_{\nu}(x,y,k)$ is flux density in the channel k at the position (x,y) and $\Delta$v$_{\rm channel}(k)$ is the velocity width of channel $k$. The integration windows were manually defined using the first extraction of the spectra as shown in Fig.\,\ref{fig:spec_1}, \ref{fig:spec_2}, and \ref{fig:spec_3}. Contrary to the continuum subtracted cubes, the integration window is not defined in a conservative way (see Sect.\,\ref{sect:cii_extr}), but designed to avoid adding noise from channels without signal in the moment-0 maps.

\subsection{Continuum imaging}

\label{sect:cont_maps}

We produced continuum maps using the similar method as for the cubes (same clean routine, pixel sizes, and weighting as in Sect.\,\ref{sect:imaging_cubes}), except that the continuum maps were produced using multi-frequency synthesis (MFS, \citealt{Conway1990}) rather than the channel-by-channel method used for the cubes. The MFS technique exploits the fact that various continuum channels probe various positions in the uv plane to better reconstruct 2-dimensional continuum maps. We excluded the same line-contaminated channels as for the uv-plane continuum subtraction used to produce the cubes. Only the lines of the ALPINE target sources were excluded. Some off-center continuum sources with lines were serendipitously detected in the field. A specific method has been used to measure their continuum flux (see Sect.\,\ref{sect:ser_with_line_phot}).

Some sources could be significantly more extended than the synthesized beam. To detect them, in addition to natural-weighted maps, we also produced lower-resolution uv-tapered maps, which are maps imaged assigning a lower weighting to the visibilities corresponding to small scales. We used a Gaussian 1.5-arcsec-diameter tapering. In Sect.\,\ref{sect:cont_ext}, we discuss the extraction of the sources using the normal and the tapered maps simultaneously. 

\subsection{Achieved beam sizes and sensitivities}

\label{sect:perf}

\begin{figure}
\centering
\includegraphics[width=8.5cm]{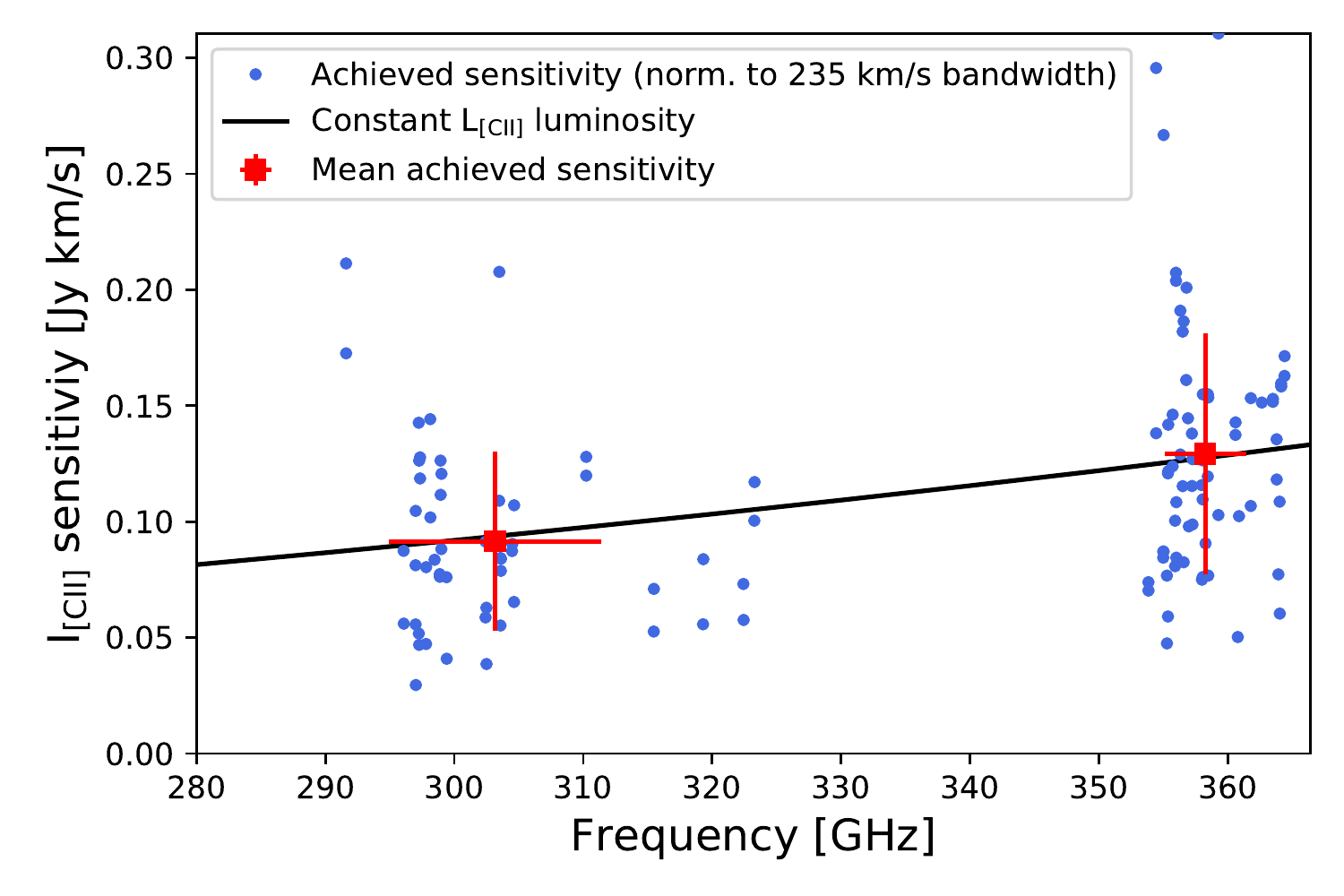}
\includegraphics[width=8.5cm]{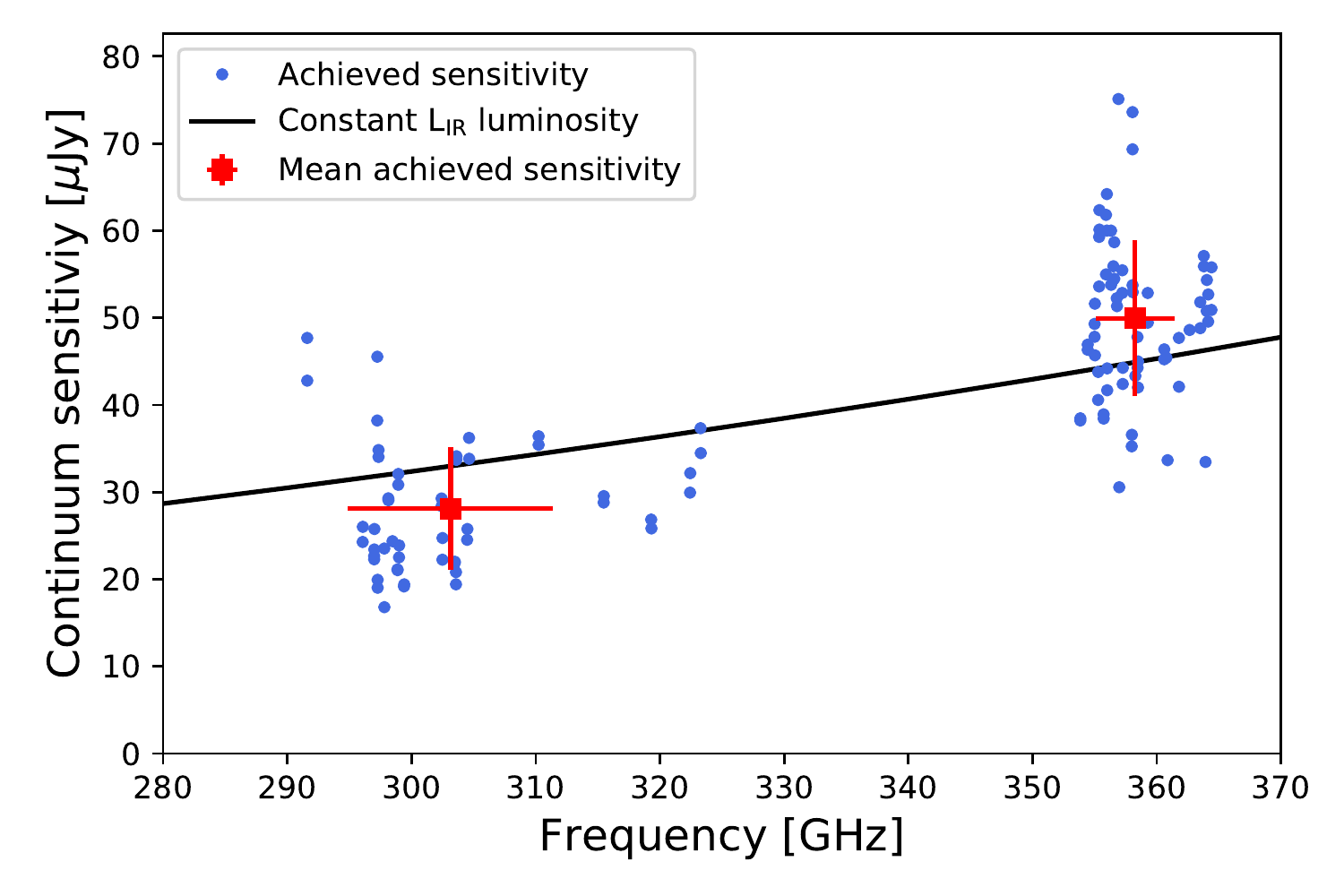}
\caption{\label{fig:sens} Achieved [CII] (upper panel) and continuum (lower panel) RMS  sensitivities. The blue dots indicate the values measured in individual fields and the red squares the mean values in the two redshift windows. The red error bars on the plots are the standard deviation in each redshift range. The actual uncertainties on the mean values are indeed $\sqrt{N}$ times smaller (central-limit theorem) and are smaller than the size of the squares. Since the [CII] sensitivities were measured using different bandwidths because of the different line widths, we normalized the measurements to a bandwidth of 235\,km/s by dividing our raw measurements by $\sqrt{\Delta {\rm v / (235\,km/s)}}$. The solid black lines indicate the trend of the [CII] flux I$_{\rm [CII]}$ and continuum flux S$_{\rm (1+z)158 \,\mu m}$ versus frequency (and thus redshift) at constant [CII] luminosity L$_{\rm [CII]}$ and fixed infrared luminosity L$_{\rm IR}$, respectively.}
\end{figure}

The achieved synthesized beam size varies with the frequency and the exact array configuration, when each source was observed. The average size of minor axis is 0.85\,arcsec (minimum of 0.72\,arcsec and maximum of 1.04\,arcsec), while the average major axis size is 1.13\,arcsec (minimum of 0.9\,arcsec and maximum of 1.6\,arcsec). Our data follow the requirements on the beam size ($>$0.7\,arcsec). The mean ratio between major and minor axis is 1.3 and the largest value is 1.8.

The [CII] sensitivity was measured on the moment-0 maps. The mean integrated line flux root mean square (RMS) sensitivity is 0.14\,Jy\,km/s. The mean sensitivity is better in the low-frequency range (283-315\,GHz, $5.1<z<5.9$) with 0.11$\pm$0.04\,Jy\,km/s than in the high-frequency range (345--356\,GHz, 4.3$<$z$<$4.6) with 0.17$\pm$0.04\,Jy\,km/s. A difference of sensitivity between fields observed at similar frequency can also be caused by different widths of the velocity window used to integrate the line fluxes. In Fig.\,\ref{fig:sens} (upper panel), we show the sensitivity versus frequency achieved in each field after renormalizing the effect caused by the different bandwidths used to produce the moment-0 maps of our targets. The mean sensitivities at the low and the high frequency (red squares) follow very well the trend expected from a constant [CII] luminosity. This is not surprising, since the survey was designed to have this property, but it is good to actually achieve it with the real data. However, beyond this very smooth overall trend, there is a large scatter around the mean behavior, since sources were observed under different weather conditions and variable number of good antennae.

The continuum sensitivity also varies with the frequency. For the sources in the 4.3$<$z$<$4.6 range (345--356\,GHz), the mean sensitivity is 50\,$\mu$Jy/beam. We obtained a better sensitivity for $5.1<z<5.9$ sources (283-315\,GHz) with an average value of 28\,$\mu$Jy/beam. The slope of the continuum sensitivity versus frequency is steeper than the continuum flux density versus redshift at fixed infrared luminosity L$_{\rm IR}$ (see the solid black line in Fig.\,\ref{fig:sens} upper panel, computed assuming the \citealt{Bethermin2017} spectral energy distribution template as discussed in Sect.\,\ref{sect:sed}). This means that our L$_{\rm [CII]}$-limited survey is paradoxically able to detect to detect galaxies with lower L$_{\rm IR}$ at higher redshift.

The performances obtained in each pointing are listed in Table\,\ref{tab:field_perf}.


\section{Continuum catalog}

\label{sect:cont_cat}

\subsection{Source extraction method, detection threshold and purity}

\label{sect:cont_ext}

\begin{figure}
\centering
\includegraphics[width=8.5cm]{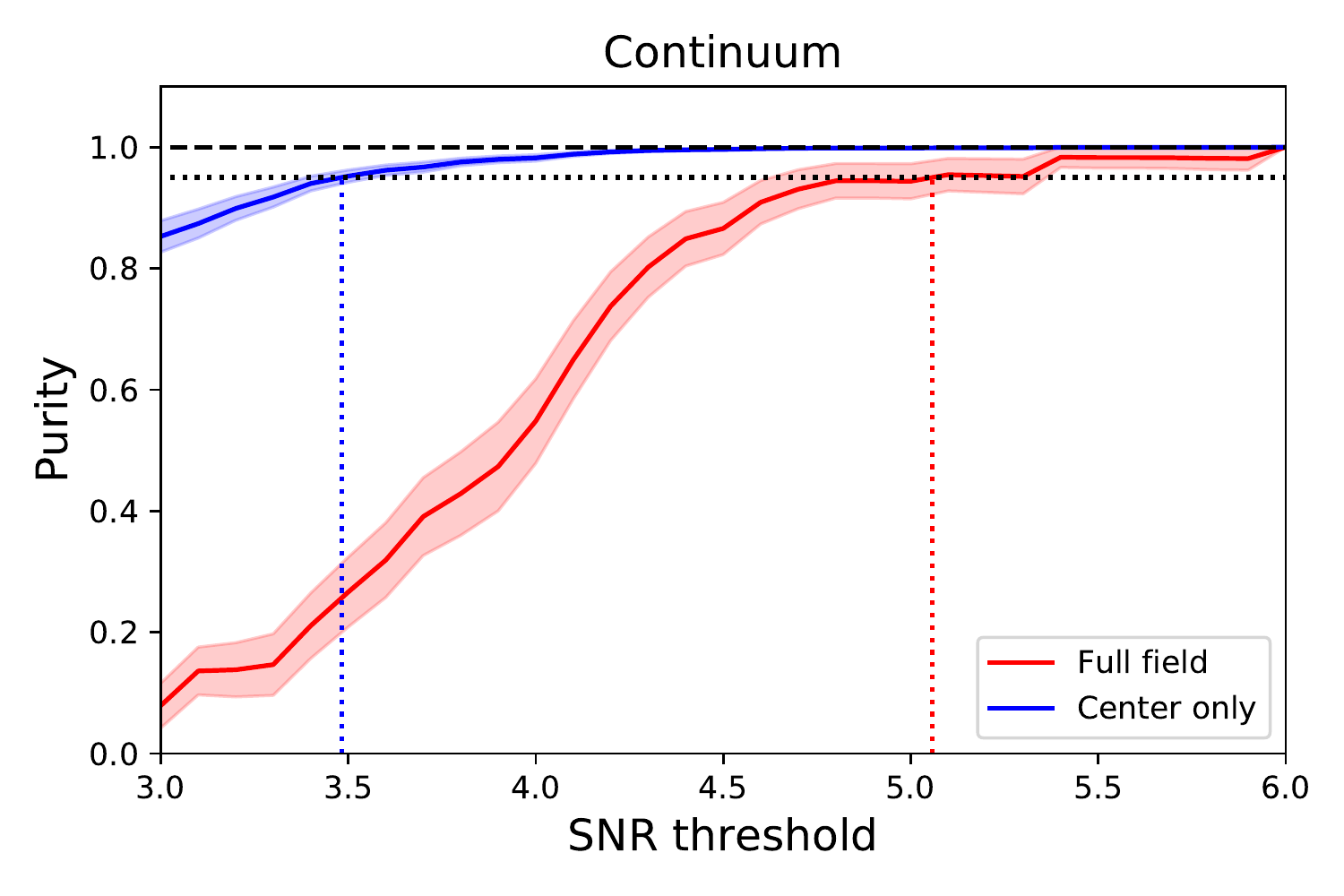}
\caption{\label{fig:cont_purity} Purity as a function of the S/N threshold. The results obtained around the center of the pointings (1\,arcsec radius) are in blue. The results in the full field are in red. The dotted lines show the S/N at which the 95\,\% is reached.}
\end{figure}

To extract the continuum sources, we created signal-to-noise-ratio (S/N) maps. We started from the nonprimary-beam-corrected map, which are maps not corrected for the low gain of the antennae far from the pointing center (normally the same as the phase center if the pointing is correct). These maps have the convenient property to have a similar noise level in the center and on the edge of the antennae field of view. We checked this by comparing the noise in the inner and outer regions of the maps (the border was set at 10\,arcsec from the center) and found only a 0.7\,\% higher noise in the center on average. This small excess may be caused by faint undetected sources in the central region, where the primary gain of the antennae is higher. The noise is computed using the standard deviation of the maps after excluding the pixels closer than 1\,arcsec to the phase center (possibly contaminated by our ALPINE target) and applying a 3-$\sigma$ clipping to avoid any noise overestimation due to serendipitous bright continuum sources. The final S/N maps are obtained by dividing the nonprimary-beam-corrected map by the estimated noise. The source are then extracted by searching for local maxima using the \textit{find\_peak} of \textit{astropy} \citep{astropy_paper}. To avoid missing extended sources, we apply the same procedure to the tapered continuum maps (Sect.\,\ref{sect:cont_maps}) and merge the two extracted catalogs. For the sources present in both catalogs, we use the position measured in the nontapered maps, where the synthesized beam is sharper. Practically, very few sources have a higher S/N in the tapered map due to their much higher noise.

The choice of the S/N threshold is crucial. If it is too low, the sample is contaminated by peaks of noise and the purity is very low. If it is too high, the faint sources are missed. We estimated the purity of the extracted sample as a function of the S/N by comparing the number of detections in the positive and the negative maps. The purity is computed using:
\begin{equation}
{\rm purity} = \frac{N_{\rm pos} - N_{\rm neg}}{N_{\rm pos}} = \frac{N_{\rm real}}{N_{\rm real} + N_{\rm spurious}},
\label{eq:purity}
\end{equation}
where $N_{\rm pos}$ is the number of detections in the positive map, which is also the sum of the $N_{\rm real}$ real and the $N_{\rm spurious}$ spurious sources. The average expected number of spurious sources in the positive and negative maps should be the same because the noise in our data is symmetrical. This is why we use the same $N_{\rm spurious}$ notation for both. $N_{\rm neg}$ is the number of detections in the negative map. Since we do not expect any real source with a negative flux in our data, this number is equal to the number of spurious sources ($N_{\rm neg}$ = $N_{\rm spurious}$). Of course, this is only true on average and Eq.\,\ref{eq:purity} is only valid when $N$ is large. The purity of the sample extracted from all the pointings as a function of the S/N threshold is presented in Fig.\,\ref{fig:cont_purity} (in red for the full field). The uncertainties are computed assuming Poisson statistics. The 95\,\% purity is reached for a S/N of 5.05 and we decided to cut our catalog at the standard 5\,$\sigma$.

Out of the 67 sources detected above 5\,$\sigma$, only 11 of them are close enough to the phase center to be potentially associated to an ALPINE target. However, when trying to detect a source close to the center of the field, we explore a much smaller number of synthesized beams (lower risk to detect high-S/N serendipitous sources) and a larger fraction of these beams are expected to contain a real source (higher ratio between real and spurious detections). Therefore, the S/N at which we reach 95\,\% completeness should be lower than in the entire field. We thus estimated the purity versus S/N considering only the central region of each pointing. The distribution of the distance of the detections to the phase center has a bump at small distance with a 1-$\sigma$ width of 0.4\,arcsec. Spatial offsets are discussed in \citet{Faisst2020}. We thus decided to use a 1\,arcsec radius to define the central region, which should contain 98.7\,\% of the ALMA continuum counterparts of our targets. In this small region, we found no S/N$>$5 source and only two S/N$>$3 sources in the negative map. To reduce the statistical uncertainties on $N_{\rm neg}$, we computed the number of sources in the total survey and rescaled by the ratio between the sum of the areas of the 118 central regions and the total imaged area of the survey. The final result is presented in Fig.\,\ref{fig:cont_purity} (blue curve). We reach a purity of 95\,\% for a S/N=3.5 cut. With this new threshold, we obtain 23 detections in the central regions, doubling the number of detected target sources.

We call target sample the sources extracted in the 1-arcsec central regions and nontarget sample the objects found outside of this area. The cutout images of these sources are shown in Fig.\,\ref{fig:cont_target_cutouts}, Fig.\,\ref{fig:cont_ser_cutouts}, and Fig.\,\ref{fig:cont_ser_cutouts2}. The position and the S/N of our target and nontarget detections are provided in Tables\,\ref{tab:cont_target_list} and \ref{tab:cont_ser_list}, respectively.

\subsection{Photometry}

\label{sect:cont_phot}

Many methods can be used to measure the flux of compact sources in interferometric data. These methods have various strengths and weaknesses. We thus decided to derive flux density values using four different map-based methods: peak flux, elliptical Gaussian fitting, aperture photometry, and integration of the signal in the 2-$\sigma$ contours. The first three methods are standard to analyze interferometric data. These four measurements are made automatically to allow us to perform easily Monte Carlo simulations to validate them. In Sect.\,\ref{sect:cont_phot_consistency}, we check the consistency between these methods.

All our measurements have been performed in the cleaned maps. Given that complex artifacts can appear during the cleaning process, as a test we performed the same measurements in the uncleaned (dirty) maps and found an excellent agreement in all the pointings, which do not contain bright sources producing side lobes.

The most basic method is to measure the peak flux of the source. The uncertainty is derived by dividing the noise measured in the nonprimary-beam-corrected map by the gain of the primary beam at the position of the source. While this method is optimal to measure point-source flux densities, it underestimates the flux of extended sources.

A simple way to measure the flux of compact marginally-resolved sources is fitting a two-dimension elliptical Gaussian. We used the \textit{astropy} fitting tools \citep{astropy:2013,astropy:2018} and chose a 3\,arcsec fitting box. The flux density of the source is just the integral of this Gaussian divided by the integral of the synthesized beam normalized to unity at its peak. The sources for which this method does not perform well are the extended clumpy or nonaxisymetrical sources, which are not well fit by an elliptical Gaussian. The uncertainties can be difficult to compute, since the noise in interferometric maps is correlated at the scale of the synthesized beam. We use the formalism of  \citet{Condon1997}, who proposed a simplified formalism to propagate the uncertainties.

Aperture photometry, that is the integration of flux in a circular aperture, relies on fewer assumptions than the previous method. We used the routine from the \textit{astropy photutils} package. The aperture radius needs to be chosen carefully. If it is too small, it will miss extended flux emission from the source. If it is too large, the relative contribution from the noise increases, which makes the measurements uncertain. By comparing the mean flux measured for our sample with different apertures, we showed that for most of them the flux converges for apertures around 3\,arcsec diameter. Beyond that, we do not gain flux anymore, but the measurements become noisier. We thus chose this aperture for the ALPINE catalog. We estimated the noise $\sigma_{\rm aper}$  using the following formula:
\begin{equation}
\sigma_{\rm aper} = \frac{\sigma_{\rm center}}{G_{\rm pb}} \, \sqrt{\frac{\pi \, D^2}{4 \, \Omega_{\rm beam}}},
\end{equation}
where $\sigma_{\rm center}$ is the RMS of the nonprimary beam corrected map, which is also the RMS expected at the center of a given pointing. $G_{\rm pb}$ is the gain of the primary beam at the position of the source, which is unity at the phase center and decreases when the distance from it increases. $\sigma_{\rm aper}$ is thus higher on the edge of the field than in the center. In theory, the gain slightly varies across the aperture, but we checked that using the value at the center of the aperture is a good approximation. $D$ is the diameter of the aperture and $\Omega_{\rm beam}$ is the solid angle of the synthesized beam.  The normalization of the noise by the square root of the ratio between the aperture area and $\Omega_{\rm beam}$ is equivalent to rescaling the noise by the square root of the number of independent primary beams in the aperture ($N_{\rm ind}$). We checked the validity of this approximation by measuring the aperture flux at random empty positions. $N_{\rm ind}$ varies from 4.2 to 9.2 in the various pointings with a mean value of 6.7. The flux uncertainties are thus on average 2.6 times higher for the aperture photometry than the peak measurement. This is the main weakness of this method.

Finally, we used another slightly less standard approach in millimeter interferometry, for which we define a S/N-based custom region, from which we integrate the source flux. This method has the advantage to produce smaller integration area for compact unblended sources than the large standard aperture described previously. It is similar to an isophotal magnitude measurement performed in optical astronomy, except that the integration area is defined in S/N instead of surface brightness. It is also better suited for sources with complex shapes. However, it does not deblend the close sources in multi-component systems, and tends to define very large areas encompassing the full blended systems (see Appendix\,\ref{sect:cii_multicomp}). Practically, we define our integration region as the contiguous area around the source where the S/N map is higher than 2. This value has been chosen after performing tests on a small subset of our sample. For point sources close to thnone S/N threshold, this region is smaller than the synthesized beam and the flux would be underestimated. We thus compute the correction to apply by measuring the synthesized beam map produced by CASA using a region with the exact same shape. Similarly to the aperture method, we compute the flux uncertainties by rescaling the noise by the square root of the number of independent synthesized beams in the region. For simplicity, this method will be called \textit{2-$\sigma$ clipped photometry} in this paper.

The flux densities measured for our target and nontarget detections can be found in Tables\,\ref{tab:cont_target_list} and \ref{tab:cont_ser_list}. Four of our continuum detections required a manual measurements of their flux because they are either multi-component or blended with a close bright neighbor. These peculiar systems are discussed in Sect.\,\ref{sect:DC881725}.

\subsection{Upper limits for nondetected target sources}

\label{sect:uplim}

A large fraction of the ALPINE targets are not detected in continuum (80\,\%), since our survey is able to detect only the most star-forming objects of our sample (see Sect.\,\ref{sect:disc_cont_target}). To produce 3-$\sigma$ upper limits, the easiest widely-used approach is to take 3 times the RMS of the noise. Since the target sources are at the phase center, it is just 3\,$\sigma_{\rm center}$ in our case (see the column called "aggressive" upper limits in Table\,\ref{tab:uplim_cont}). However, these upper limits are a bit too aggressive. If an intrinsic 2.999\,$\sigma_{\rm center}$ signal is present at the position of the source and if we assume a flat prior on the flux distribution of the sources, there is $\sim$50\% probability that the source is actually brighter than 3\,$\sigma_{\rm center}$. Therefore, we produced more robust upper limits by summing 3\,$\sigma_{\rm center}$ with the highest flux measured 1\,arcsec around the phase center ("normal" upper limits in Table\,\ref{tab:uplim_cont}). In the extreme case of a significantly extended source, the source could also be missed because its peak flux is a small fraction of the integrated source flux. We produced "secure" upper limits (Table\,\ref{tab:uplim_cont}) by applying the previous process to the tapered maps. We recommend to use these "secure" upper limits, except in the case of point sources for which the "normal" ones are appropriate.

\subsection{Line contamination of the continuum of nontarget sources}

\label{sect:ser_with_line_phot}

\begin{table}
\caption{\label{tab:line_cont} Continuum flux densities (2D-fit method) of nontarget sources contaminated by a line before and after re-imaging the maps without the contaminated channels (see Sect.\,\ref{sect:ser_with_line_phot}).}
\centering
\begin{tabular}{lcc}
\hline
\hline
Name of the nontarget source & S$_\nu$ & S$_\nu$\\
 & before & after \\
 & $\mu$Jy & $\mu$Jy \\
\hline
SC\_1\_DEIMOS\_COSMOS\_460378 & 838$\pm$128 & 680$\pm$117 \\ 
SC\_1\_DEIMOS\_COSMOS\_665626 & 486$\pm$85 & 392$\pm$87 \\ 
SC\_1\_DEIMOS\_COSMOS\_787780 & 938$\pm$120 & 398$\pm$106 \\ 
SC\_1\_DEIMOS\_COSMOS\_848185 & 7662$\pm$291 & 5983$\pm$227 \\ 
SC\_1\_vuds\_cosmos\_5101210235 & 1084$\pm$210 & 905$\pm$181 \\ 
SC\_1\_vuds\_cosmos\_5110377875 & 3773$\pm$169 & 3512$\pm$163 \\ 
SC\_2\_DEIMOS\_COSMOS\_773957 & 172$\pm$40 & 117$\pm$33 \\ 
SC\_2\_DEIMOS\_COSMOS\_818760 & 397$\pm$94 & 425$\pm$104 \\ 
SC\_2\_DEIMOS\_COSMOS\_842313 & 9898$\pm$99 & 8240$\pm$90 \\
\hline
\end{tabular}
\end{table}

Our continuum maps were produced excluding the channels contaminated by the [CII] line of the target sources only. The [CII] or another line can contaminate the flux density measurements of nontarget sources if it is outside of the excluded frequency range (Sect.\,\ref{sect:cont_phot}). To identify these problematic cases, we extracted their spectra and after visual inspection found 9 objects withnontarget a possible line contamination. The nature of these objects will be discussed in Loiacono et al. (in prep.). We generated new continuum maps, where we masked the line-contaminated channels of the nontarget source instead of the ALPINE target ones. We then remeasured the continuum flux using the same method as previously. Table\,\ref{tab:line_cont} summarizes the impact of this line decontamination. The relative impact of this correction can vary from a 58\,\% decrease of the flux density (SC\_1\_DEIMOS\_COSMOS\_787780) to a nonstatistically-significant increase of the flux (SC\_2\_DEIMOS\_COSMOS\_818760). It might be surprising that the line-free flux does not decrease significantly in some sources compared to the initial measurements (or even increase by a fraction of $\sigma$ in the case of SC\_2\_DEIMOS\_COSMOS\_818760), but the contaminating line can sometimes overlap with the [CII]-contaminated channels of the target source, which were masked initially.

\begin{figure*}
\centering
\begin{tabular}{cc}
\includegraphics[width=8.5cm]{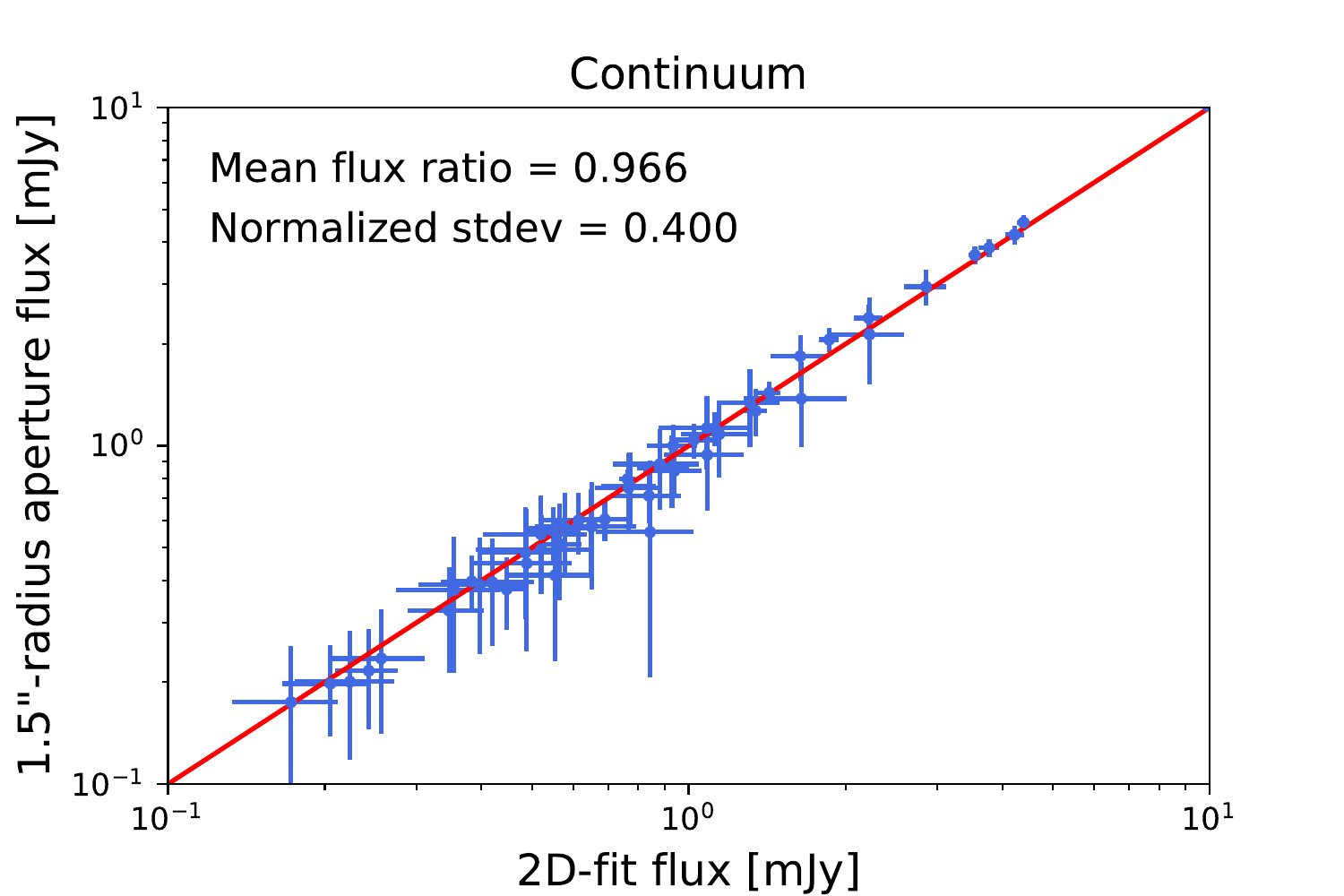} & \includegraphics[width=8.5cm]{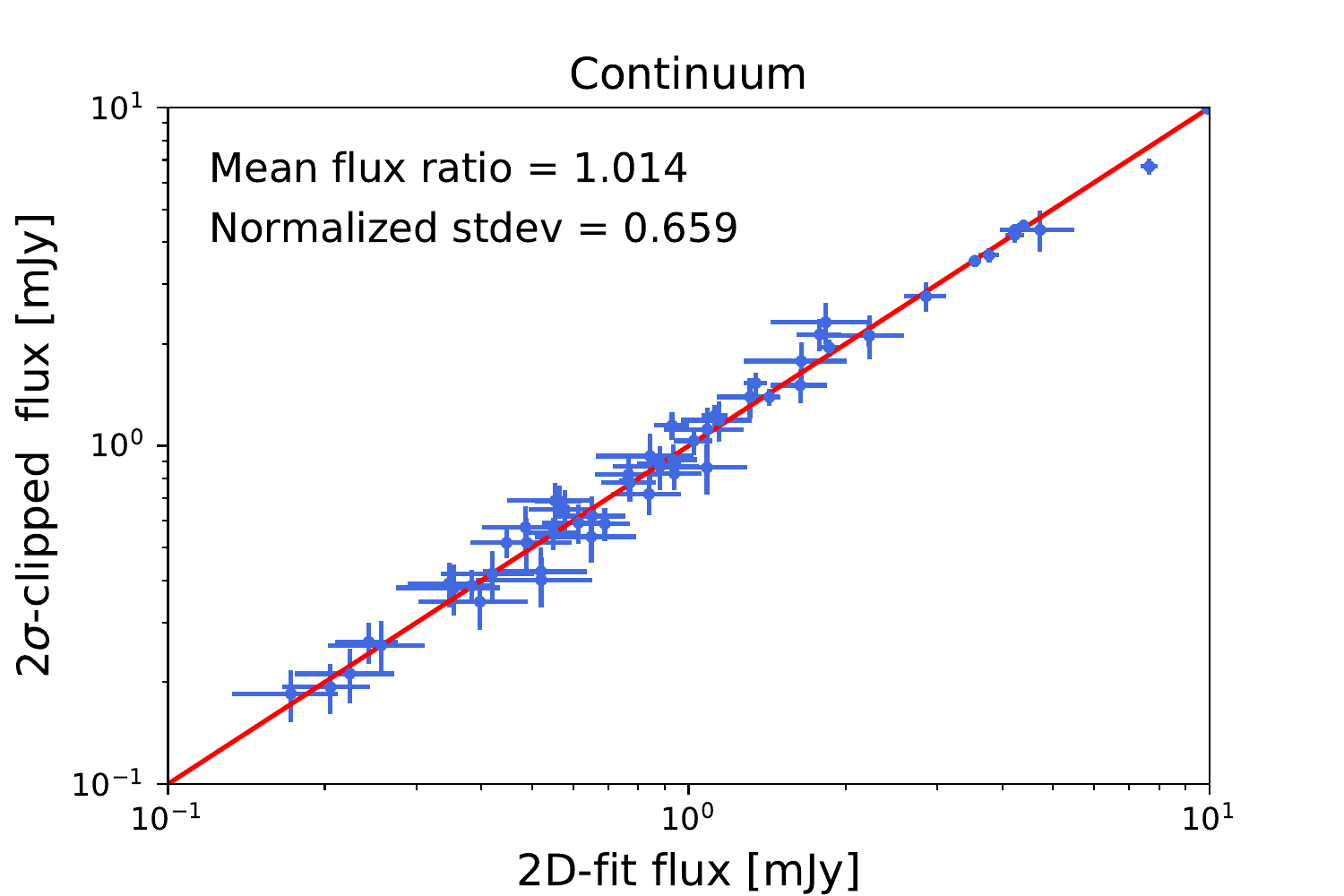}\\
\includegraphics[width=8.5cm]{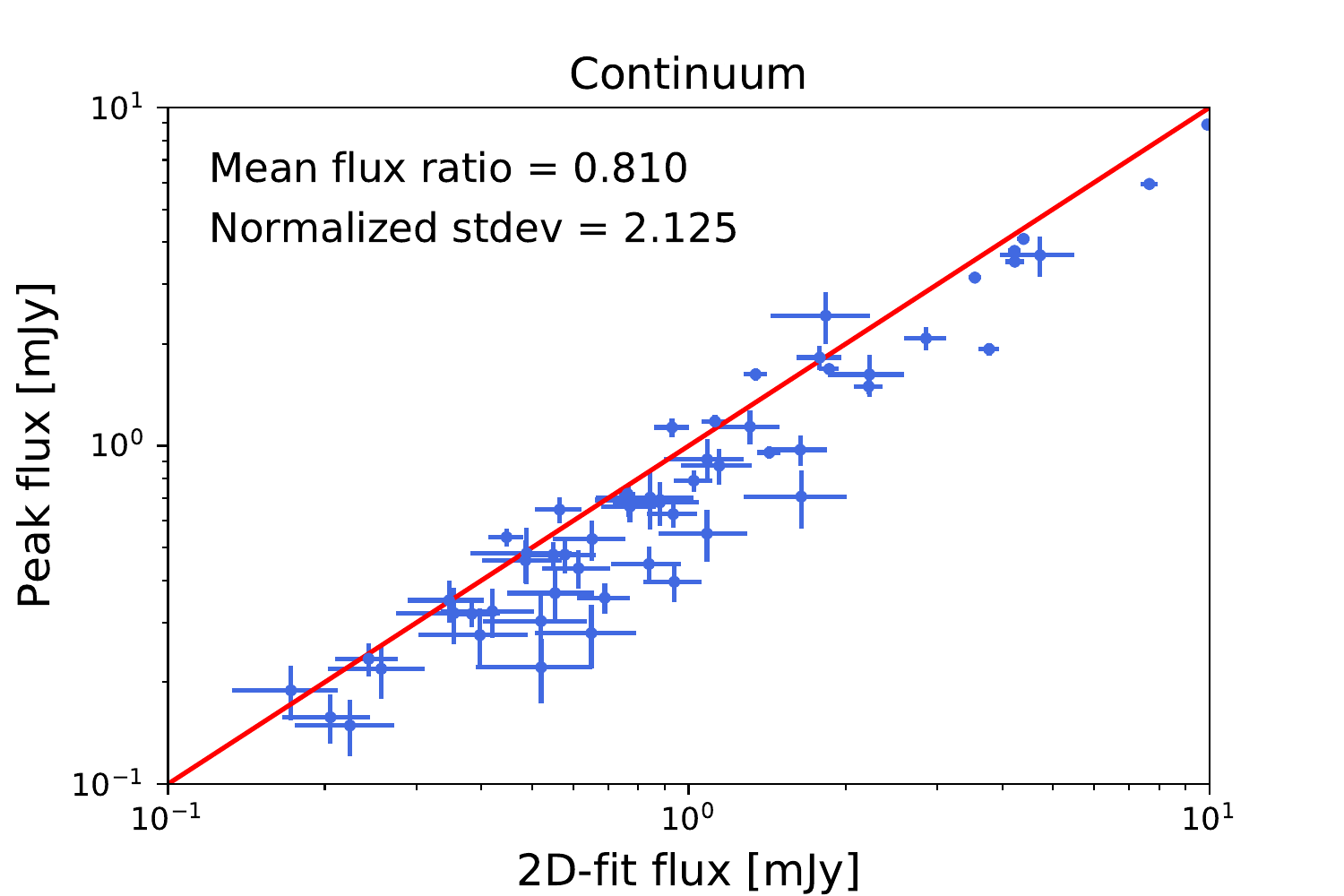} & \includegraphics[width=8.5cm]{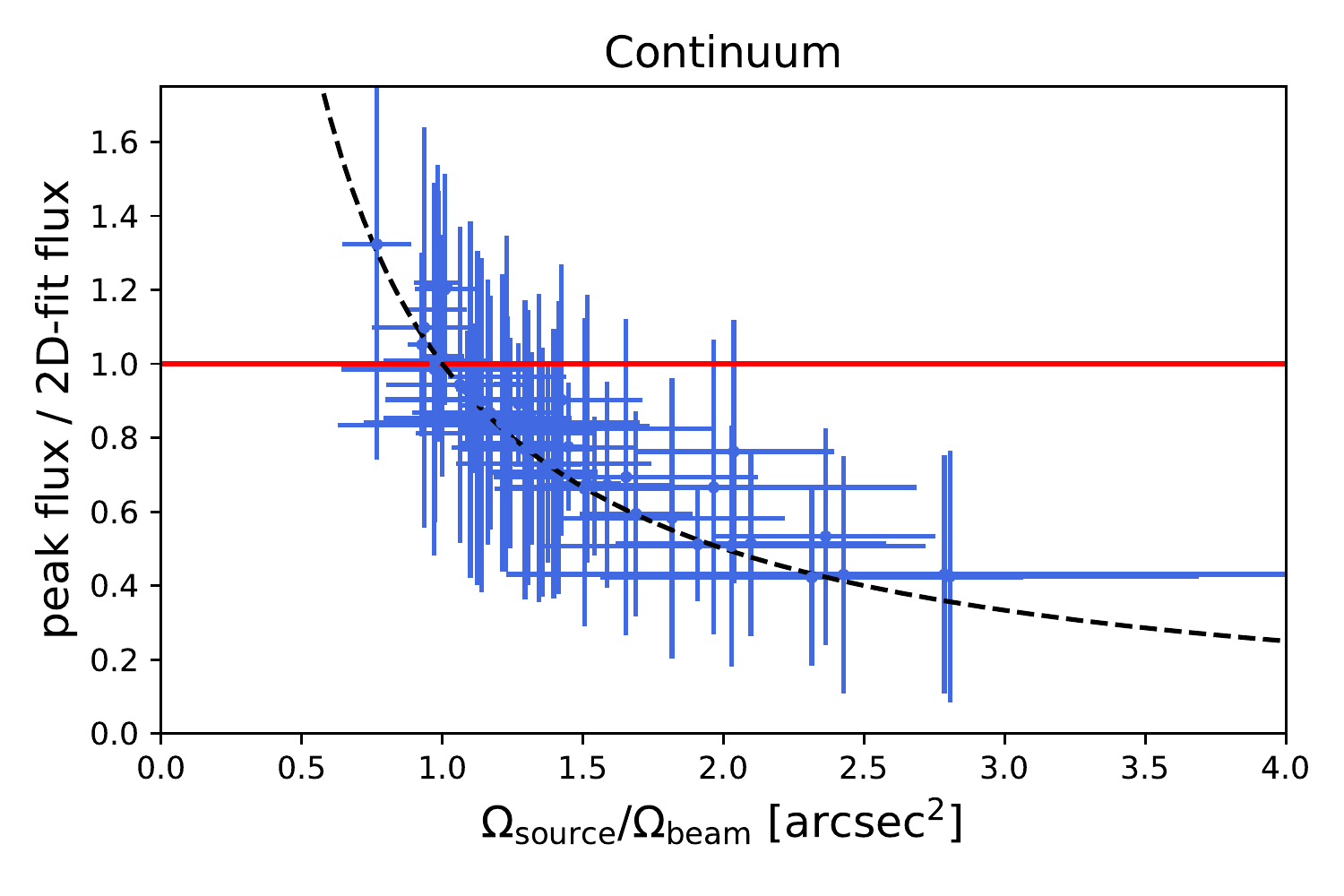}\\
\end{tabular}
\caption{\label{fig:phot_cont_compare} Comparison between our various photometric methods described in Sect.\,\ref{sect:cont_phot} for S/N$>$5 sources. The blue dots are our measurements and the red line is the one-to-one relation. The upper left, upper right, and lower left panels are the comparison between the 2D-fit flux density (x-axis) and the aperture, 2\,$\sigma$-clipped, and peak flux densities, respectively. The lower right panel shows the ratio between the peak flux and the 2D-fit flux as a function of the ratio between the source area (convolved by the synthesized beam) and the synthesized beam area. The dashed line indicates the expected trend (see Sect.\,\ref{sect:cont_phot_consistency}).}
\end{figure*}

\subsection{Consistency of the various photometric methods}

\label{sect:cont_phot_consistency}

Since the photometry of each source was determined using different methods, the consistency between these methods can be used as a robustness check (see Fig.\,\ref{fig:phot_cont_compare}). The 2D-fit, aperture, and 2\,$\sigma$-clipped measurements are overall in excellent agreement with each other (see the two upper panels of Fig.\,\ref{fig:phot_cont_compare}). Even if most of the measurements are compatible at 1\,$\sigma$ with each other, there is a small proportional offset of -3.4\% and +1.4\,\% between the aperture photometry and 2\,$\sigma$-clipped photometry, respectively, and the 2D-fit measurements. This remains negligible compared to the typical 10\,\% absolute calibration uncertainties of interferometric observations (see Sect.\,\ref{sect:quasars}). 

In order to check how consistent are our measurements, we computed the uncertainty-normalized difference between two measurements $S_\nu^{\rm method\,A}$ and $S_\nu^{\rm method\,B}$:
\begin{equation}
\frac{(S_\nu^{\rm method\,A}-S_\nu^{\rm method\,B})}{\sqrt{\sigma_{\rm method\,A}^2 + \sigma_{\rm method\,B}^2}},
\label{eq:phot_norm_diff}
\end{equation}
where $\sigma_{\rm method\,X}$ is the uncertainty derived for the method X. If the two measurements would be performed on independent realizations of the noise, the standard deviation of the normalized difference measured for a large sample should be close to unity. We found 0.40 and 0.66 for the comparison between aperture and clipped photometry, respectively, and 2D-fit measurements. It shows that the three methods are overall consistent at better than 1\,$\sigma$. It is not surprising to find a value below unity, since our methods are using the same realization of the noise. We did not expect to find zero either, since each method tends to weight the noise in the various pixels in a different way.

The peak photometry does not agree as well with the other methods and is on average 19\,\% lower than the 2D-fit flux (Fig.\,\ref{fig:phot_cont_compare}, lower left panel). This clearly indicates that our sources cannot be considered as point like and that the peak flux is not a good way to measure their integrated flux. In a Gaussian-profile case without noise, the ratio between the peak flux and the integrated flux directly depends on the source size and the synthesized beam size. If we note $\Omega_{\rm beam}$ the beam area defined as the integral of the synthesized beam and $\Omega_{\rm source}$ the integral of the profile of an extended source after normalizing its peak to unity, the peak flux $S_{\rm peak}$ is:
\begin{equation}
S_{\rm peak} = S_{\rm int} \, \frac{\Omega_{\rm beam}}{\Omega_{\rm source}},
\end{equation}
where $S_{\rm int}$ is the integrated flux. The $S_{\rm peak} / S_{\rm int}$ ratio should thus be inversely proportional to $\Omega_{\rm source} / \Omega_{\rm beam}$. In the lower right panel of Fig.\,\ref{fig:phot_cont_compare}, we show that this is exactly the trend followed by our measurements.

\begin{figure}
\centering
\includegraphics[width=8.5cm]{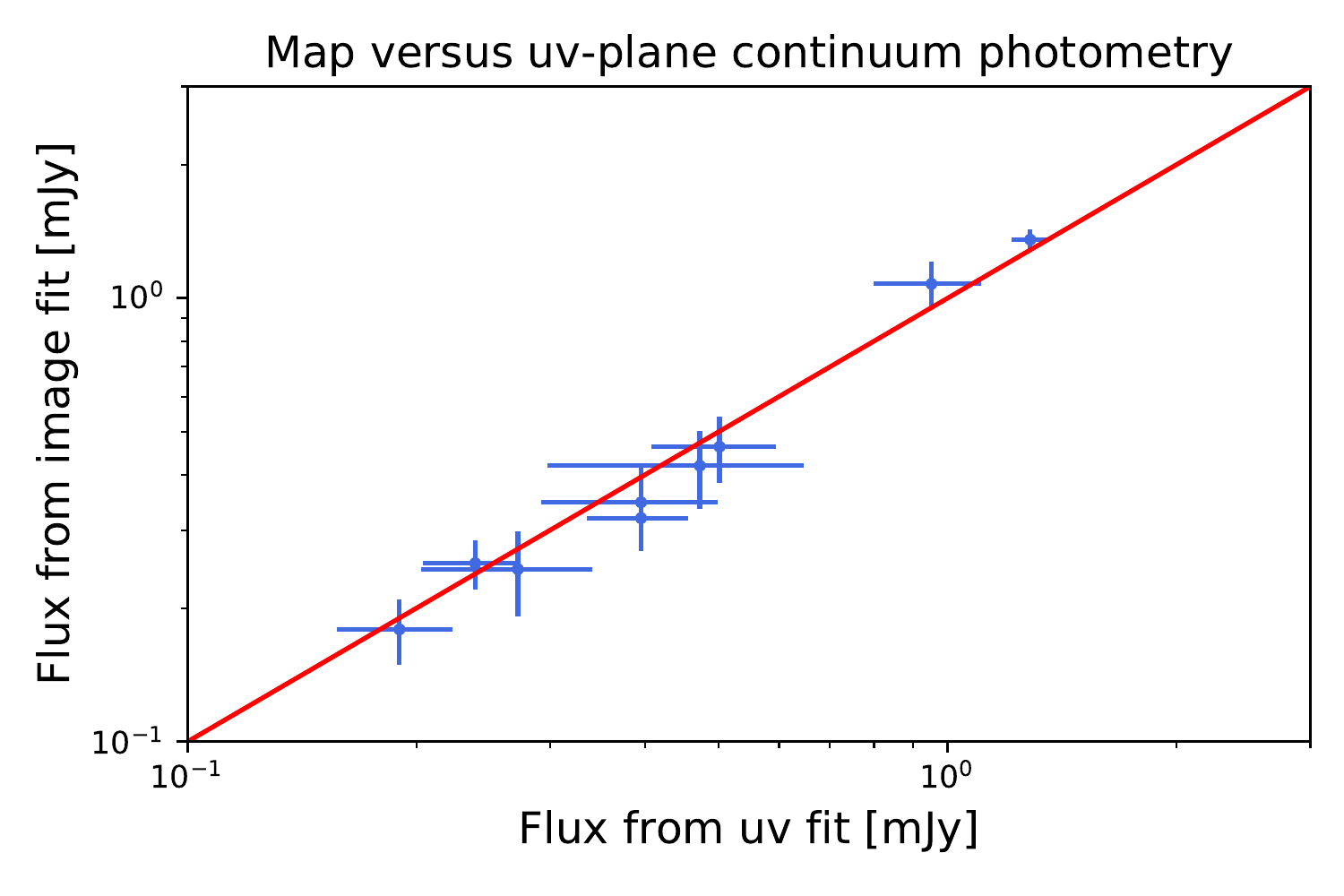}
\caption{\label{fig:uv_comp_cont} Comparison between the 2D-fit flux densities derived in map space (Sect.\,\ref{sect:cont_phot}) and the flux determined fitting an elliptical Gaussian model in the uv plane (Sect.\,\ref{sect:uv_comp_cont}). The blue dots are our measurements and the red line is the one-to-one relation.}
\end{figure}

\subsection{Comparison between map-based and uv-plane photometry}

\label{sect:uv_comp_cont}

In millimeter interferometry, we can also measure the flux of a source directly in the uv-plane. This technique is particularly powerful to deblend multiple sources and when the uv-coverage is limited. To perform the uv-fitting, we used the GILDAS\footnote{\url{http://www.iram.fr/IRAMFR/GILDAS}} software package MAPPING, which allows us to fit models directly to the uv visibilities. The use of GILDAS required beforehand to export our CASA measurement sets to uvfits tables and then to uvt tables, the GILDAS visibility table format\footnote{\url{https://www.iram.fr/IRAMFR/ARC/documents/filler/casa-gildas.pdf}}. We could successfully model nine continuum targets\footnote{In this analysis we focused on target sources since they are at the phase center and thus easier to model.} detected at $\geq 5\,\sigma$ and without any bright neighbor, using an elliptical Gaussian model for which the analytical Fourier transform could be fit to the merged visibilities of all channels of the 4 spectral windows and the two polarizations (excluding only channels contaminated by the [CII] emission line). We derived uv-based flux measurements for all these targets, marginally resolved in most cases. In Fig\,\ref{fig:uv_comp_cont}, we show the comparison between the map-based 2D-fit method and the uv-plane approach. All our sources are compatible at 1\,$\sigma$ with the one-to-one relation. This shows that measuring the flux in map space is sufficient in our case. However, uv-plane modeling is critical for size measurements and is presented in \citet{Fujimoto2020}.

\subsection{Monte-Carlo source injections}

\label{sect:mci}

To interpret the statistical properties of nontarget detections, we need to know the completeness in our various pointings as a function of the source flux density, the source size, and the distance to the phase center. We used Monte-Carlo source injections to estimate it, but also to test the reliability of our flux measurements.

We performed injections of sources using a grid of 4 different intrinsic sizes (FWHM = 0, 0.333, 0.666, and 1\,arcsec) and 18 different nonprimary beam corrected flux densities ranging from 0.02\,mJy to 1\,mJy spaced by 0.1\,dex. We injected 10 sources in any given pointing, which is sufficiently small to avoid overlap problems and sufficiently large to be efficient at getting a large number of injected sources in a reasonable computing time. We decided to repeat this task 10 times per set of properties (size and flux) in order to have 100\,objects per size and flux. Because of our limited computing resources, we limited our study to Gaussian circular sources and we injected sources directly in the image space. For each realization, we extracted the sources and measured their flux using the same exact method as for the real maps. We consider that a source is recovered if it is found less than 1\,arcsec from its injected position. We checked that the number of recovered sources are not significantly changed if we had used 0.5\,arcsec instead.

\subsection{Completeness}

\label{sect:completeness}

\begin{figure}
\centering
\includegraphics[width=8.5cm]{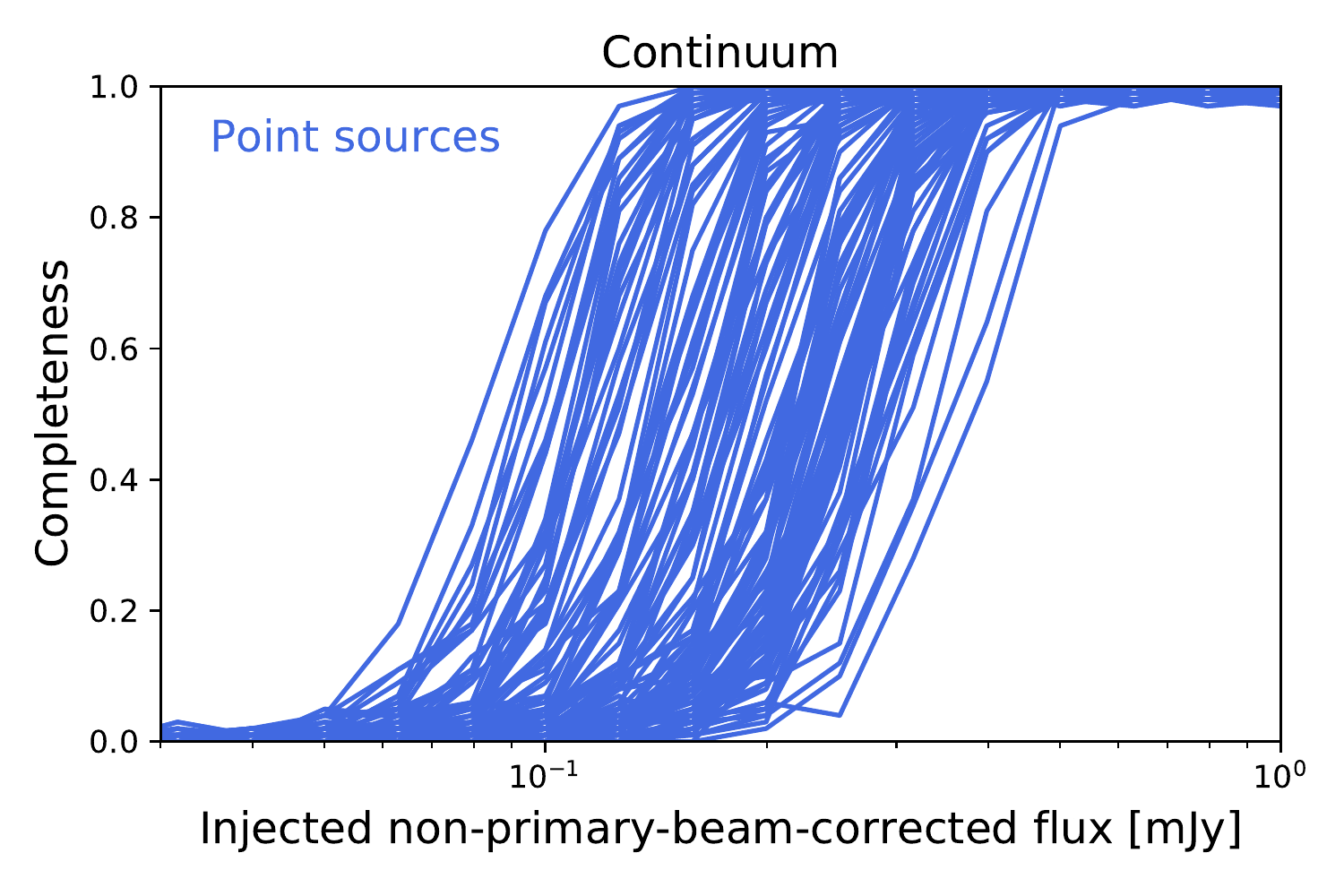}
\includegraphics[width=8.5cm]{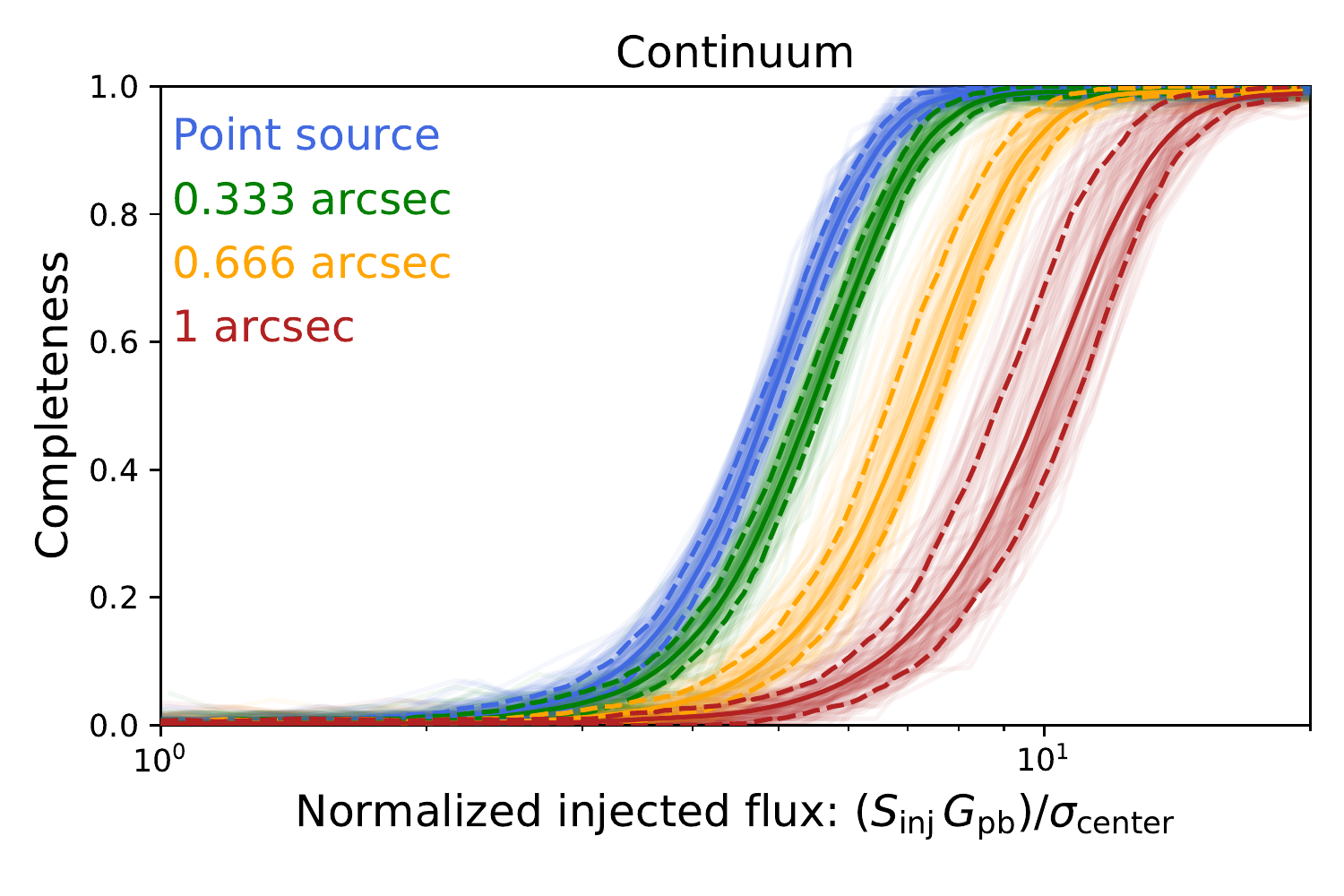}
\includegraphics[width=8.5cm]{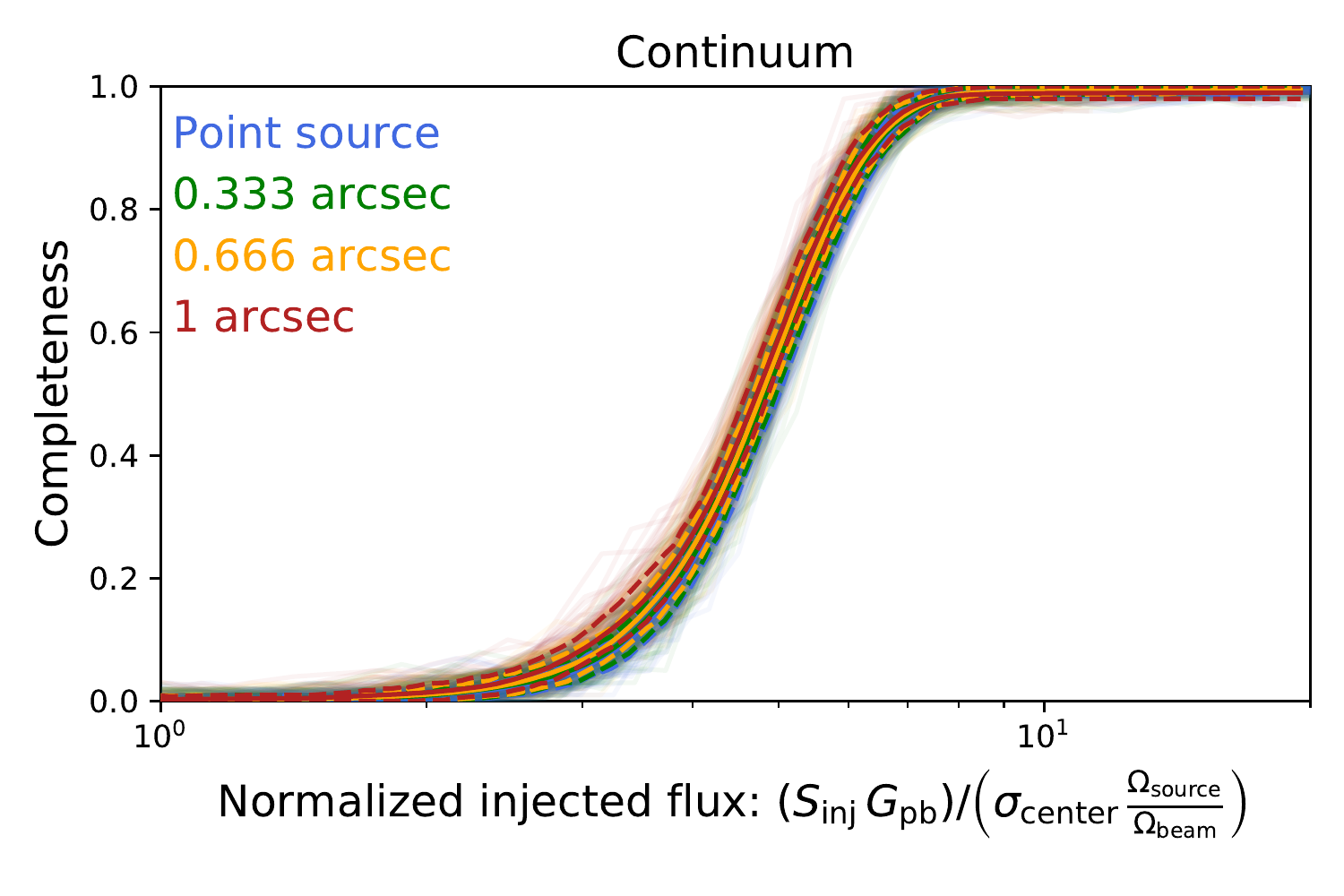}
\caption{\label{fig:completeness} Upper panel: completeness as a function of the continuum nonprimary-beam-corrected flux density ($G_{\rm pb} \, S_{\rm inj}$) achieved for point sources in various pointings. Middle panel: similar figure after having divided the nonprimary-beam-corrected flux by the noise at the center of each pointing ($\sigma_{\rm center}$). Various colors (blue, green, yellow, and red) corresponds  to various injected source sizes (FWHM = 0, 0.333, 0.666, and 1\,arcsec, respectively). The solid lines indicate the mean trend of the various pointings, while the dashed lines indicate the 1-$\sigma$ envelop. Lower panel: same plot after normalizing the injected flux by 1/$\sigma_{\rm center}$ and by the source area ($\Omega_{\rm source}$ / $\Omega_{\rm beam}$). These results are discussed in Sect.\,\ref{sect:completeness}.}
\end{figure}

\begin{figure*}
\centering
\begin{tabular}{cc}
\includegraphics[width=8.5cm]{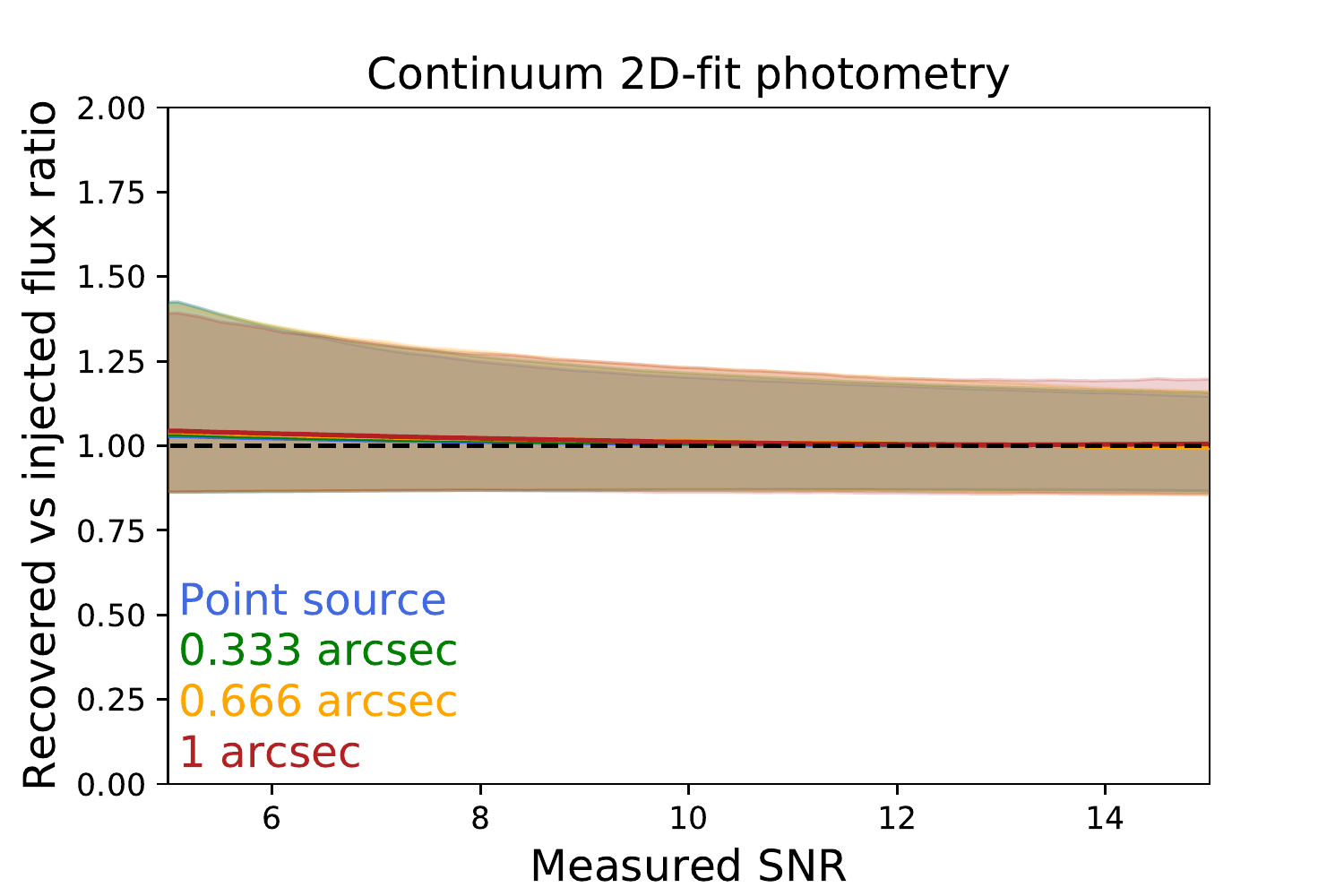} & \includegraphics[width=8.5cm]{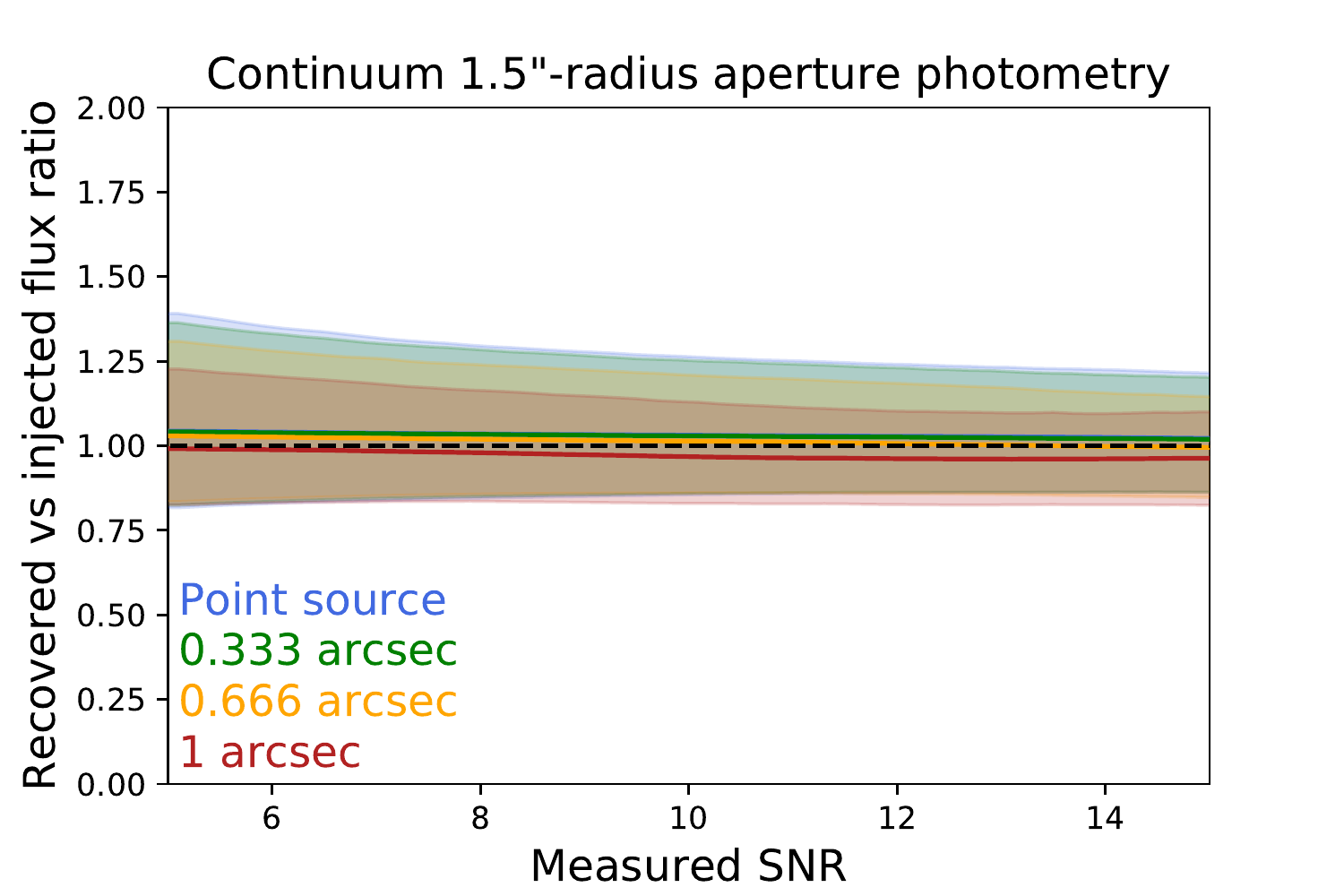}\\
\includegraphics[width=8.5cm]{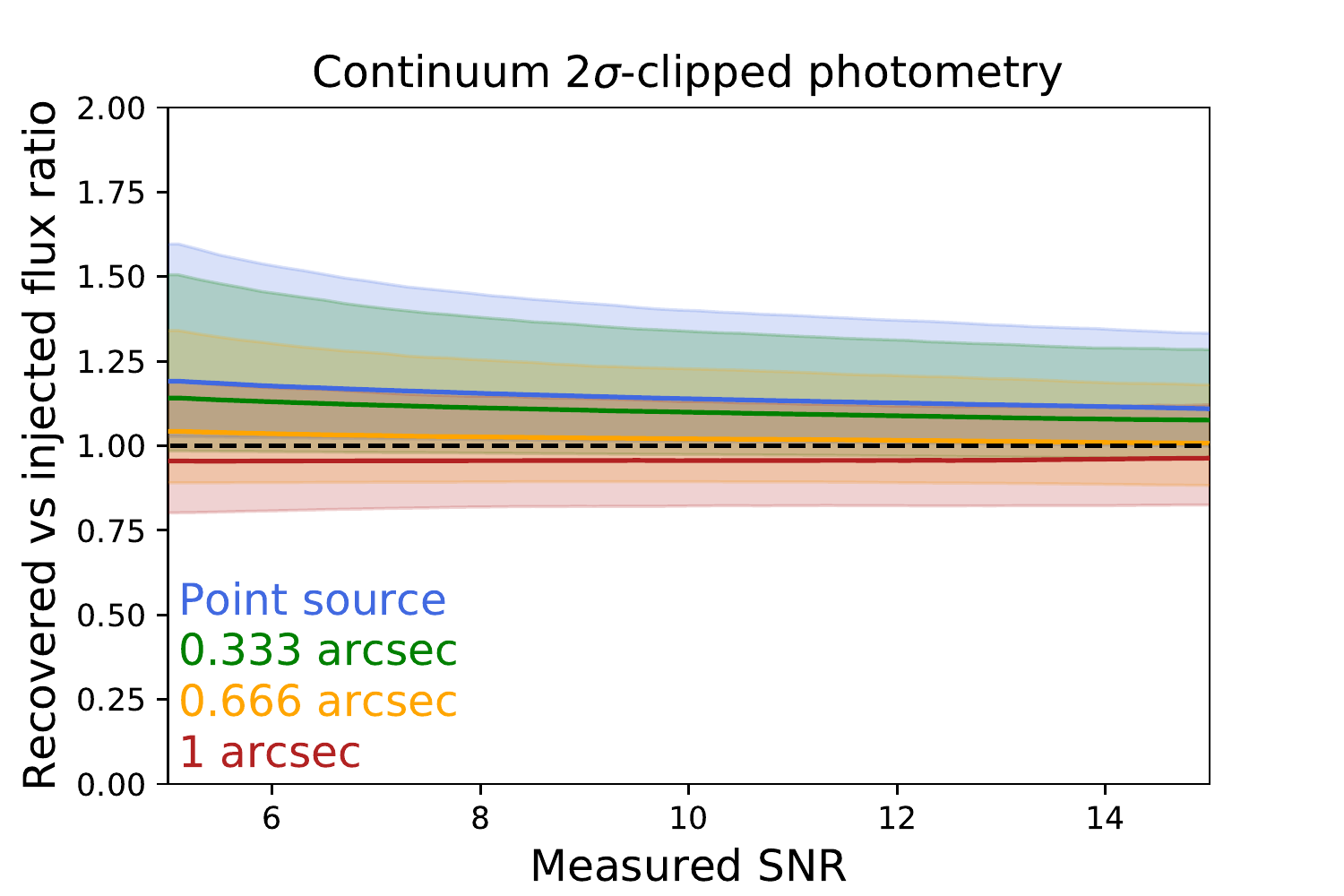} & \includegraphics[width=8.5cm]{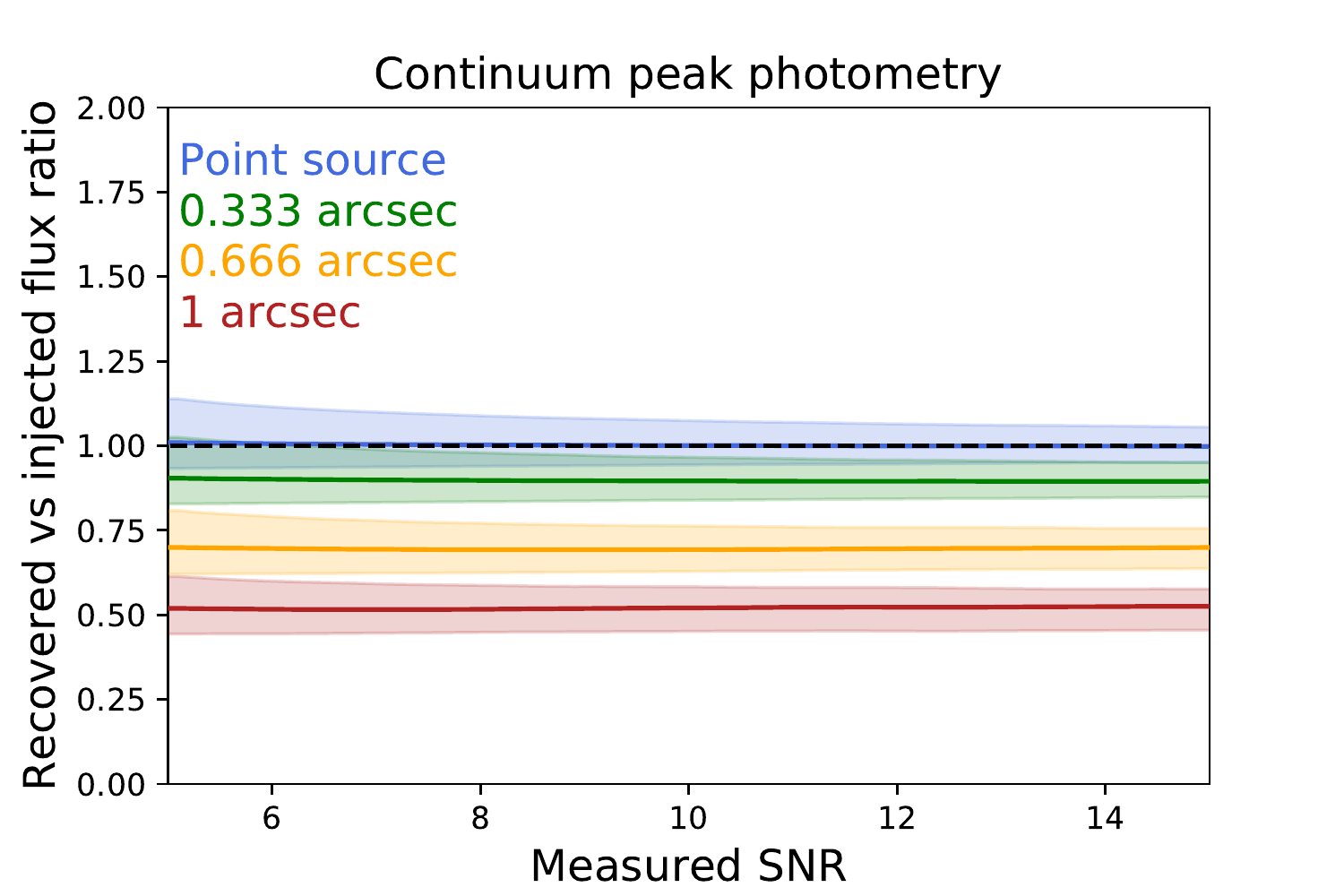}\\
\end{tabular}
\caption{\label{fig:flux_boosting} Ratio between the injected and recovered flux density as a function of measured S/N (see Sect.\,\ref{sect:flux_boosting}). The upper left, upper right, lower left, and lower right panels present the results obtained for the 2D-fit, aperture, 2\,$\sigma$-clipped, and peak photometry, respectively. The solid lines indicate the median and the shaded areas are the 1-$\sigma$ contours.} Various colors (blue, green, yellow, and red) are used to indicate the various sizes used (FWHM = 0, 0.333, 0.666, and 1\,arcsec, respectively). The dashed horizontal line indicate the one-to-one relation.
\end{figure*}

Using the Monte Carlo source injections described in Sect.\,\ref{sect:mci}, we can easily derive the completeness for a given injected flux and size by computing the fraction of recovered sources with this property. In practice, the primary beam gain ($G_{\rm pb}$) decreases quickly with the distance from the center and the noise is much larger on the edges of the maps. Consequently, the completeness depends strongly on the distance between the source and the center. However, the local noise can be easily computed by dividing the noise in the center $\sigma_{\rm center}$ (estimated in the nonprimary-beam corrected map) by the local primary-beam gain ($G_{\rm pb}$). If we inject sources with similar nonprimary-beam-corrected flux ($G_{\rm pb} \, S_{\rm inj}$), they will have similar S/N whatever their distance to the phase center. The actual flux density, which is corrected by the primary beam gain, of these injected sources will be larger on the edge than in the center. In Fig.\,\ref{fig:completeness} (upper panel), we present the completeness as a function of $G_{\rm pb} \, S_{\rm inj}$. For clarity, we only show the results for point sources. While the completeness tends to zero at low flux and unity at high flux, the flux at which the transition appears varies significantly from pointing to pointing. 

When we normalized the injected nonprimary-beam corrected flux by 1/$\sigma_{\rm center}$ (middle panel), all the pointings have a very similar completeness curve for point sources (in blue). However, the completeness is not the same for all source sizes. At fixed normalized flux $G_{\rm pb} \, S_{\rm inj} / \sigma_{\rm center}$, the completeness is lower for larger sources. A similar trend was found in the ALMA-GOODS deep field \citep{Franco2018}. We can also remark that a larger scatter from field to field is obtained for larger source size.

We used both the normal and the tapered maps to detect our sources. However, the S/N is usually higher in the normal map. The peak flux density in the normal map thus is a better proxy than the integrated flux to guess if a source will be detected or not by our algorithm searching for S/N peaks. We thus divided our previously-normalized flux densities by $\Omega_{\rm source} / \Omega_{\rm beam}$ (see Sect.\,\ref{sect:cont_phot_consistency}) to obtain a good proxy for the effect of the source size on the detectability. With this last correction, the completeness does not depend significantly on the source size and the scatter between pointings is highly reduced for the extended sources (see Fig.\,\ref{fig:completeness} lower panel). We derived the average curve for all sizes and pointings. The median distance to this average relation is only 1.2\% with a maximum of 4.7\,\%. We can thus reliably estimate the completeness based on this average relation from the source size, the primary-beam gain at its position, and its flux density.

\subsection{Photometric accuracy and flux boosting}

\label{sect:flux_boosting}

We also used our Monte Carlo simulations to test the accuracy of our photometry. In Fig.\,\ref{fig:flux_boosting}, we show the mean ratio between the recovered and injected flux density for our various photometric methods. For the 2D-fit photometry, the aperture photometry, and the peak flux in the case of point sources only, we observe the classical flux boosting effect at low S/N. Indeed, the sources with an injected flux density corresponding to an intrinsic S/N slightly lower than the detection threshold will be detected only if they are on a peak of noise. Their flux densities will thus be overestimated on average. In contrast, at high S/N, we expect that the output-versus-input flux density ratio will tend to unity, since sources located on both positive and negative fluctuations of the noise are detected. The 2$\sigma$-clipped method and the peak photometry of extended sources is more problematic and the results vary significantly with the size. In particular, even close to the S/N threshold, the flux densities are underestimated on average for a source size of 1\,arcsec. At high S/N, the 2$\sigma$-clipped method converges slowly to unity. As expected, there is no convergence for the peak photometry, since the flux of all extended sources is systematically underestimated even in absence of noise and thus at high S/N.

We used these results to compute the flux boosting correction to apply. We computed the flux boosting correction at the S/N of the source for the immediately lower and higher sizes and used a linear interpolation to derive the correction to adopt for our source size.

To summarize, the peak flux density systematically underestimates the actual flux density of extended sources. Concerning the 2$\sigma$-clipped method, the flux boosting converges very slowly at high S/N and the flux boosting is highly size-dependent. Both aperture and 2D-fit photometry provide good results. We decided to use the 2D-fit photometry, because of the very small impact of the size on the deboosting correction to apply. In the following sections of this paper, we use the 2D-fit measurements. The raw and deboosted flux densities obtained using the 2D-fit method are listed in Table\,\ref{tab:cont_ser_list}.

\subsection{Effective survey area associated with nontarget sources}

\label{sect:eff_area} 

To derive surface density of sources (also called number counts, see Sect.\,\ref{sect:counts}) or luminosity functions of nontarget sources (Gruppioni et al. in prep.), we need to know the effective surface area of our survey as a function of the source properties. Of course, it varies with the flux density, since only the brightest sources can be detected on the edges of the pointing. It also depends on source size, since compact sources have usually a better completeness at fixed flux density (Sect.\,\ref{sect:completeness} and Fig.\,\ref{fig:completeness}). 

In addition, each pointing is observed at a slightly different frequency. We thus have to take into account that a source detected at a given flux density in a pointing will have a slightly different flux density in another pointing because of the different observed frequency, and consequently a slightly different completeness. For this reason, we apply a frequency-dependent correction factor to convert all the flux densities to 850\,$\mu$m (353\,GHz) assuming the z=2.5 main-sequence spectral energy distribution (SED) template of the \citet{Bethermin2017} model (see the redshift distribution of nontarget sources in Sect.\,\ref{sect:Nz} and Fig.\,\ref{fig:Nz}). Since most of the nontarget sources are at z$<$4 and thus observed in the Rayleigh-Jeans part of their spectrum, the continuum slope around 850\,$\mu$m does not vary significantly with the redshift and it is thus a fair assumption to assume a single template.

The effective surface area $\Omega_{\rm eff}$ as a function of the source flux density $S_{850}$ and the source size $\theta_{\rm source}$ is derived from the completeness $C(S_{850}, \theta_{\rm source}, x, y)$ at a position (x,y) (see Sect.\,\ref{sect:completeness}) using:
\begin{equation}
\Omega_{\rm eff}(S_{850}, \theta_{\rm source}) = \sum_{\rm pointings} \iint C(S_{850}, \theta_{\rm source}, x, y) \, d \Omega.
\end{equation}
Since the nontarget sources are extracted outside the central 1\,arcsec-radius region, we exclude this area from the computation of the integral.

\begin{figure}
\centering
\includegraphics[width=8.5cm]{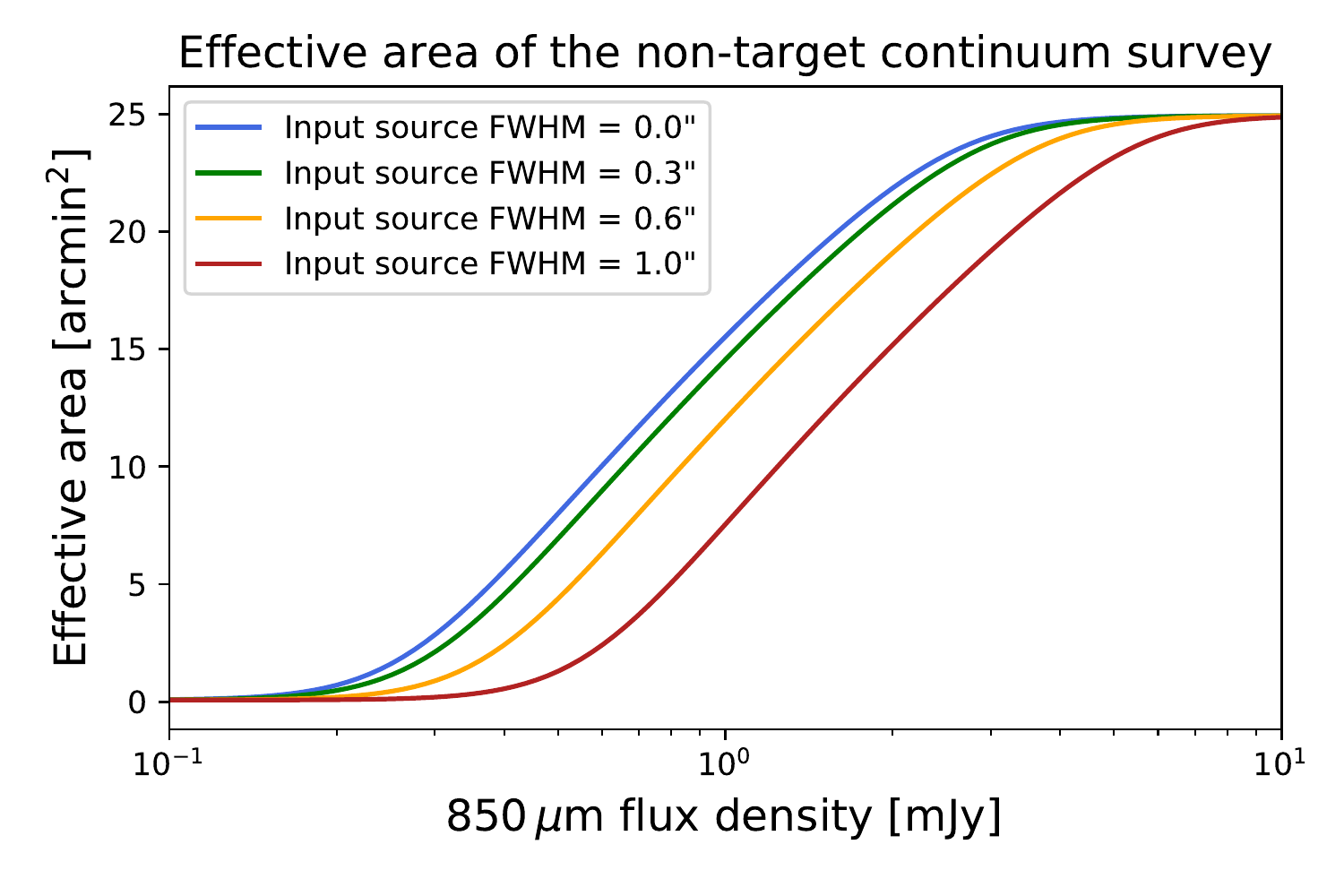}
\caption{\label{fig:eff_area} Effective surface area of ALPINE as a function of the 850\,$\mu$m flux after excluding the central 1-arcsec-radius area where target sources are extracted. The blue, green, gold, and red lines are the results obtained for a source size of 0, 0.3, 0.6, and 1\,arcsec, respectively. The method used to compute the surface area is described in Sect.\,\ref{sect:eff_area}.}
\end{figure}

The result is presented in Fig.\,\ref{fig:eff_area}. As expected, the surface area at intermediate flux densities varies significantly with the source size. At bright flux densities ($>$10\,mJy), the completeness tends to unity and the effective surface area is the total area of all our pointings\footnote{The pointings are imaged only in the region where the primary-beam gain is at least 0.2.} (24.92\,arcmin$^{2}$). Our survey is $\sim$3 times smaller than ALMA-GOODS \citep{Franco2018} for a similar sensitivity in mJy. However, typical galaxies are fainter by a factor of $\sim$2 at 1.1\,mm. Our band-7 serendipitous survey thus is a valuable complement to the band-6 deep fields \citep{Dunlop2017,Aravena2016b,Gonzalez-Lopez2017,Franco2018}.

\section{From rest-frame 158\,$\mu$m continuum fluxes to SFR}

\label{sect:sed}

\subsection{Dust spectral energy distribution variation from low-redshift to high-redshift Universe}

\label{sect:sed_intro}

The obscured star formation is directly related to the bolometric luminosity of the dust (${\rm SFR_{\rm IR}} = 1 \times 10^{-10}$\,M$_\odot$/yr/L$_\odot \times L_{\rm IR}$, \citealt{Kennicutt1998} after converting to \citealt{Chabrier2003} IMF). $L_{\rm IR}$ is usually defined as the total luminosity of a galaxy between 8 and 1000\,$\mu$m. ALPINE continuum photometry is only probing a narrow range of wavelength around 158\,$\mu$m rest-frame. Since we have only one photometric point available, we thus have to assume a spectral energy distribution (SED) to derive L$_{\rm IR}$. As discussed in \citet{Bouwens2016}, \citet{Fudamoto2017}, and \citet{Faisst2017}, for example, the assumption on the dust temperature of z$>$4 galaxies has a significant impact on the relation connecting the dust attenuation to the UV continuum slope $\beta$ or the stellar mass, which will be discussed in \citet{Fudamoto2020}.

While the SEDs of z$<$2 galaxies have been well studied thanks to \textit{Herschel} \citep[e.g.,][]{Elbaz2011,Dunne2011,Magdis2012b,Berta2013,Symeonidis2013,Magnelli2014}, we have fewer constraints on the SEDs at higher redshifts. These z$<$2 studies revealed that the temperature of normal, star-forming galaxies tends to increase with redshift, which agrees with the theoretical model predictions \citep[e.g.,][]{Cowley2017_SED,Imara2018,Behrens2018}. Because of the confusion noise, \textit{Herschel} can detect only the brightest galaxies \citep[e.g.,][]{Nguyen2010}. However, some interesting constraints up to z$\sim$4 were obtained using stacking analysis of galaxies selected using photometric redshifts \citep[e.g.,][]{Bethermin2015b,Schreiber2015}, Lyman-break selections \citep[e.g.,][]{Alvarez-Marquez2016}, and low-redshift analogs of z$>$5 galaxies \citep{Faisst2017}. According to these studies, temperature seems to continue to increase up to z$\sim$4. So far, we have very few constraints about what happens at z$>$4, which is critical to interpret the ALPINE survey.

In this section, we present a stacking analysis adapted from \citet{Bethermin2015a} to derive an average empirically-based conversion from the 158\,$\mu$m monochromatic continuum flux density to $L_{\rm IR}$ and SFR.

\begin{figure}
\centering
\includegraphics[width=8.5cm]{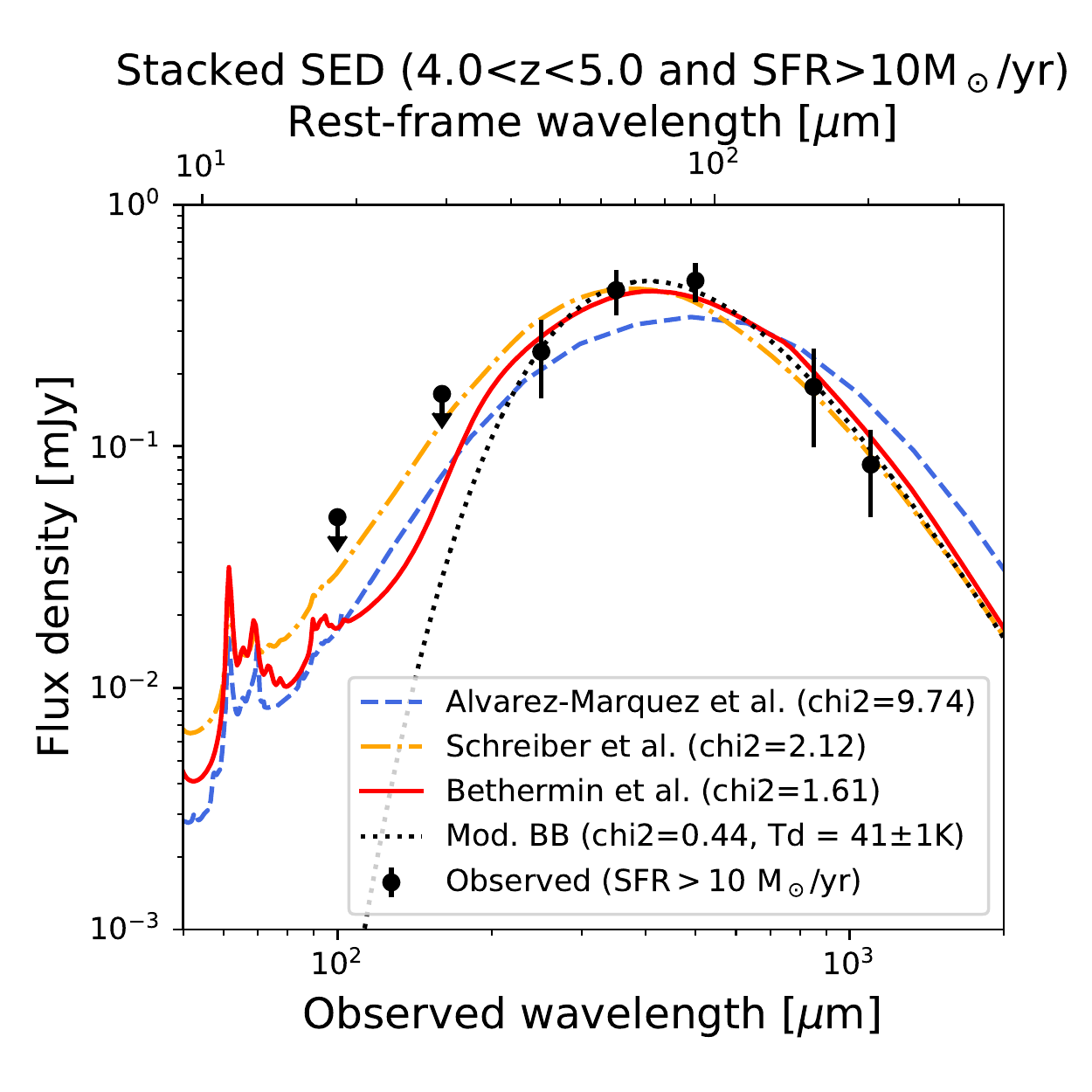}\\
\includegraphics[width=8.5cm]{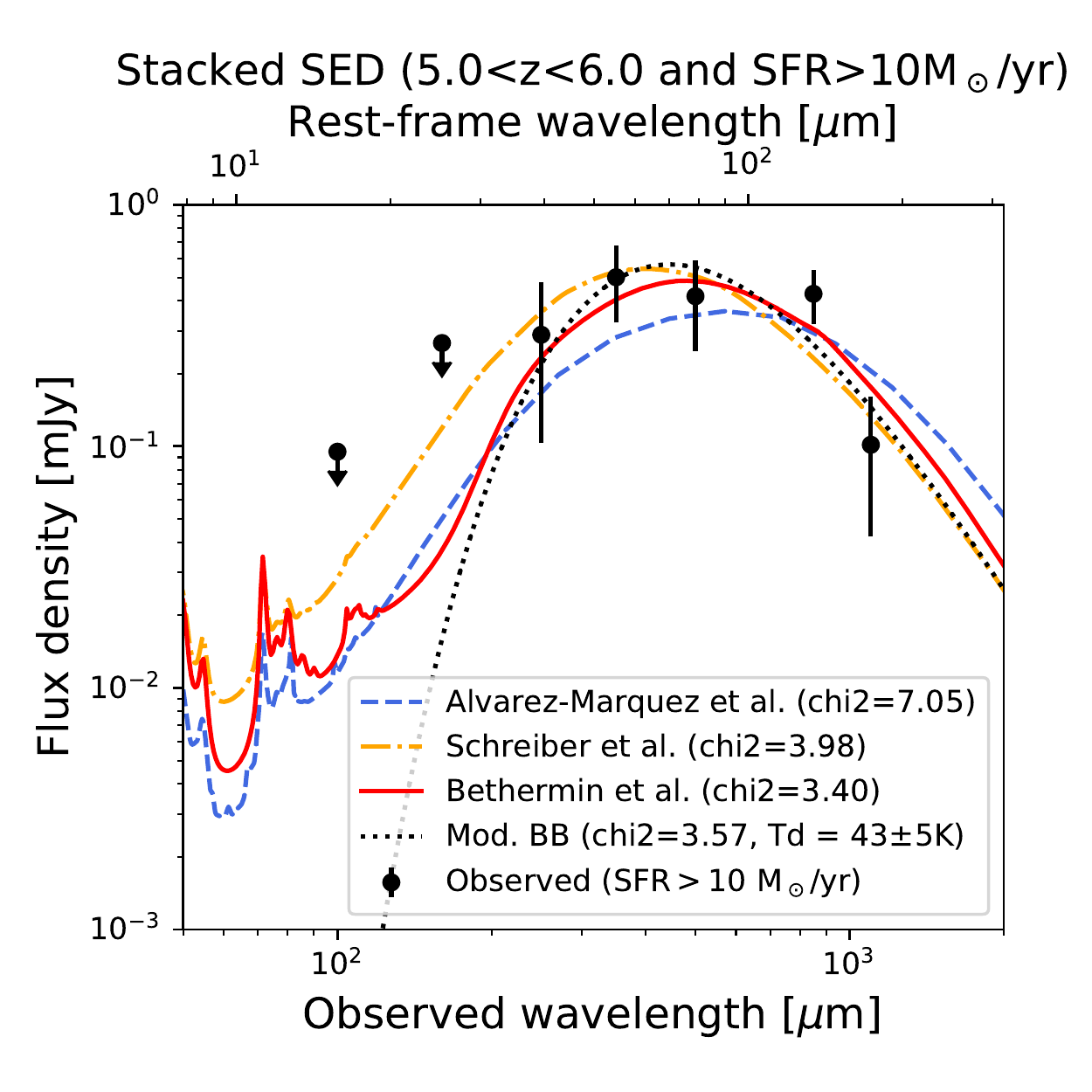}\\
\caption{\label{fig:seds} Comparison between the \citet[blue dashed line]{Alvarez-Marquez2016}, \citet[orange dot-dashed line]{Schreiber2018a}, and \citet[red solid line]{Bethermin2017} IR SED templates and the observed mean SEDs of SFR$>$10\,M$_\odot$/yr galaxies measured by stacking (black dots, see Sect.\,\ref{sect:stacking_sed}). The black dotted line is the best fit of the $\lambda_{\rm rest-frame} > 40\,\mu$m data points by a modified blackbody with $\beta$ fixed to 1.8 (the temperature in the legend is provided in the rest frame). The upper and lower panels correspond to 4$<$z$<$5 and 5$<$z$<$6, respectively.}
\end{figure}

\begin{table}
\caption{\label{tab:stacking_sed} Mean flux density of SFR$>$10\,M$_\odot$/yr measured by stacking in the COSMOS field (see Sect.\,\ref{sect:stacking_sed})}
\begin{tabular}{lcc}
\hline
\hline
Observed wavelength ($\mu$m) & \multicolumn{2}{c}{Mean flux density (mJy)} \\
 & 4$<$z$<$5 & 5$<$z$<$6 \\ 
\hline
100 & $<$0.05 & $<$0.09 \\
160 & $<$0.14 & $<$0.28 \\
250 & 0.25 $\pm$ 0.08 & 0.29 $\pm$ 0.17 \\
350 & 0.44 $\pm$ 0.10 & 0.50 $\pm$ 0.19 \\
500 & 0.48 $\pm$ 0.10 & 0.42 $\pm$ 0.15 \\
850 & 0.18 $\pm$ 0.07 & 0.43 $\pm$ 0.11 \\
1100 & 0.08 $\pm$ 0.04 & 0.10 $\pm$ 0.07 \\
\hline
\end{tabular}
\end{table}

\begin{table*}
\caption{\label{tab:conv_LIR} Ratio (without unit) between the monochromatic continuum luminosity $\nu L_\nu$ and the total infrared luminosity $L_{\rm IR}$ at different rest-frame wavelengths associated with important far-IR lines. These ratios were computed using the \citet[][B17]{Bethermin2017} z$>$4 and \citet[][,S18]{Schreiber2018a} main-sequence SED templates and modified blackbodies (MBBs) at various rest-frame temperatures ($\beta$ fixed to 1.8) for comparison.}
\centering
\begin{tabular}{lcccccc}
\hline
\hline
 & [OI]$_{63}$ & [OIII] & [NII]$_{122}$ & [OI]$_{145}$ & [CII] & [NII]$_{205}$\\
\hline
Rest-frame wavelength ($\mu$m) & 63 & 88 & 122 & 145 & 158 & 205 \\
\hline
$\nu L_\nu$ / $L_{\rm IR}$ from B17 & 0.69 & 0.50 & 0.27 & 0.18 & 0.133& 0.054 \\
$\nu L_\nu$ / $L_{\rm IR}$ from S18 & 0.64 & 0.43 & 0.20 & 0.12 & 0.093& 0.038\\
\hline
$\nu L_\nu$ / $L_{\rm IR}$ for 40\,K MBB & 0.93  & 0.69 & 0.34 & 0.20 & 0.155 & 0.062\\
$\nu L_\nu$ / $L_{\rm IR}$ for 45\,K MBB & 0.89 & 0.55 & 0.24 & 0.14 & 0.104 & 0.040\\
$\nu L_\nu$ / $L_{\rm IR}$ for 50\,K MBB & 0.81 & 0.44 &  0.17 &  0.10 & 0.071 &  0.026 \\
$\nu L_\nu$ / $L_{\rm IR}$ for 55\,K MBB & 0.71 & 0.34 & 0.13 & 0.07 & 0.050 & 0.018 \\
$\nu L_\nu$ / $L_{\rm IR}$ for 60\,K MBB & 0.60 & 0.27 & 0.09 &  0.05 & 0.036 & 0.013 \\
\hline
\end{tabular}
\end{table*}


\subsection{Mean stacked SEDs of ALPINE analogs in the COSMOS field}

\label{sect:stacking_sed}

\citet{Bethermin2015a} used a mean stacking analysis (without source weighting) of \textit{Herschel} and complementary ground-based measurements in the COSMOS field to derive the mean SEDs of z$<$4 galaxies. We used the same \textit{Herschel}\footnote{Herschel is an ESA space observatory with science instruments provided by European-led Principal Investigator consortia and with important participation from NASA.} \citep{Pilbratt2010} data from the PEP \citep{Lutz2011} and \textit{HerMES} \citep{Oliver2012} surveys and AzTEC/ASTE data of \citet{Aretxaga2011} at 1.1\,mm. At 850\,$\mu$m, we used the SCUBA2 data from \citet{Casey2013} instead of the shallower LABOCA ones used in the 2015 analysis.

The 2015 selection of the stacked targets was performed using a stellar mass cut of $ >3 \times 10^{10}$\,M$_\odot$ in the photometric \citet{Laigle2015} catalog. There are too few ALPINE sources to obtain a sufficiently high S/N in the stacked \textit{Herschel} data. We thus used a larger photometric sample with properties similar to ALPINE objects. We chose to select sources with an estimated SFR from an optical and near-infrared SED fitting higher than 10\,M$_\odot$/yr, which is approximately equivalent to the ALPINE SFR limit \citep{Le_Fevre2019,Faisst2020}.

 We also use higher redshift bins (4$<$z$<$5 and 5$<$z$<$6) to match the redshift range probed by ALPINE. Finally, we use the more recent COSMOS catalog of \citet{Davidzon2017} as input sample, since it has been optimized to provide more reliable photometric redshifts and physical parameters at z$>$4. Our stacked samples contain respectively 5749 and 1883 sources in the 4$<$z$<$5 and 5$<$z$<$6 ranges.

Our new stacking analysis was performed using the exact same procedure as in \citet{Bethermin2015a}. The uncertainties were derived using a bootstrap technique that takes into account both the photometric noise (instrumental and confusion) and the population variance. The contamination of the stacked flux by clustered neighbors is corrected using the method described in Appendix A of \citet{Bethermin2015a}. At z$>4$, these corrections are relatively small ($<$30\%) because of the lower global star formation rate density compared to z=2. Our results are presented in Fig.\,\ref{fig:seds} and Table\,\ref{tab:stacking_sed}. The 5$<$z$<$6 SED is $\sim$2 times noisier mainly because of the smaller number of stacked objects.

\subsection{SED template and conversion factors}

\label{sect:lir_conv}

The final step to compute the conversion factor from monochromatic luminosity to L$_{\rm IR}$ is to find an SED model or a parametric description fitting the data. Using an agnostic model as a spline is difficult, since we have few constraints on the mid-infrared ($\lambda_{\rm rest}<30 \,\mu$m). In Fig.\,\ref{fig:seds}, the SEDs are represented in flux density units (S$_\nu$ = dS/d$\nu$), which can give the wrong impression that the contribution at short wavelength is negligible, while it contains $\sim$\,15\% of the energy\footnote{Computed using the \citet{Bethermin2017} template}. For this reason, although fitting well the \textit{Herschel} data points, a modified blackbody ($\nu^\beta \, B_\nu(\nu,T)$, where $B_\nu$ is a blackbody law) tends to underestimate the L$_{\rm IR}$ because of the very low emission in the mid-infrared. For information, we show in Fig.\,\ref{fig:seds} the best fit of our SEDs by a modified blackbody with a fixed $\beta$ of 1.8, but a free amplitude and temperature. We excluded rest-frame wavelengths below 40\,$\mu$m from the fit, since the greybody model does not take into account the warm dust and the polycyclic aromatic hydrocarbon (PAH) features dominating in this wavelength range. The fit is excellent at 4$<$z$<$5 ($\chi^2$ = 0.44 for 3 degrees of freedoms) and acceptable at 5$<$z$<$6 ($\chi^2$ = 3.55 for 2 degrees of freedoms).

We thus chose to use empirical template libraries. We compare our observed SEDs with three different templates. \citet{Alvarez-Marquez2016} template\footnote{An update of this work has been published by \citet{Alvarez-Marquez2019}, but became public too late to be used in our analysis.} is based on the stacking of 2.5$<$z$<$3.5 Lyman-break galaxies. The SED templates for main sequence galaxies of the \citet{Bethermin2017} model evolves with redshift up to z$\sim$4. Above this redshift, no evolution is assumed ($\langle U \rangle$ = 50). These templates are an update of the \citet{Magdis2012b} templates calibrated using the \textit{Herschel} stacking up to z$\sim$4 \citep{Bethermin2015a}. Finally, \citet{Schreiber2018a} also built a template evolving with redshift and calibrated it using another independent \textit{Herschel} stacking analysis. Contrary to the previous templates, they assume an evolution of the rest-frame dust temperature above z=4 (4.6\,K per unit of redshift). Because of the nature of the ALPINE sample \citep{Faisst2020}, we only consider the templates corresponding to galaxies on the main sequence.

In Fig.\,\ref{fig:seds}, we show the comparison between our measured SED and the templates described above. We renormalized the templates to fit the data. This is the only free parameter in our analysis. While the \citet{Alvarez-Marquez2016} template is too cold for both redshift bins, both \citet{Schreiber2018a} and \citet{Bethermin2017} templates well fit the data ($\chi^2$<4 with 4 degrees of freedom for both templates in both redshift bins). Since the $\chi^2$ of \citet{Bethermin2017} is marginally better, we decided to use this template. In Table\,\ref{tab:conv_LIR}, we provide the ratio between the monochromatic luminosity ($\nu$L$_\nu$ units) and L$_{\rm IR}$ computed using this template at wavelengths associated with bright fine-structure lines, which can be targeted by ALMA. In practice, for the ALPINE catalog (Appendix\,\ref{app:cont_cat}), we use the exact effective wavelength of the ALMA continuum.

\begin{figure}
\centering
\includegraphics[width=8.5cm]{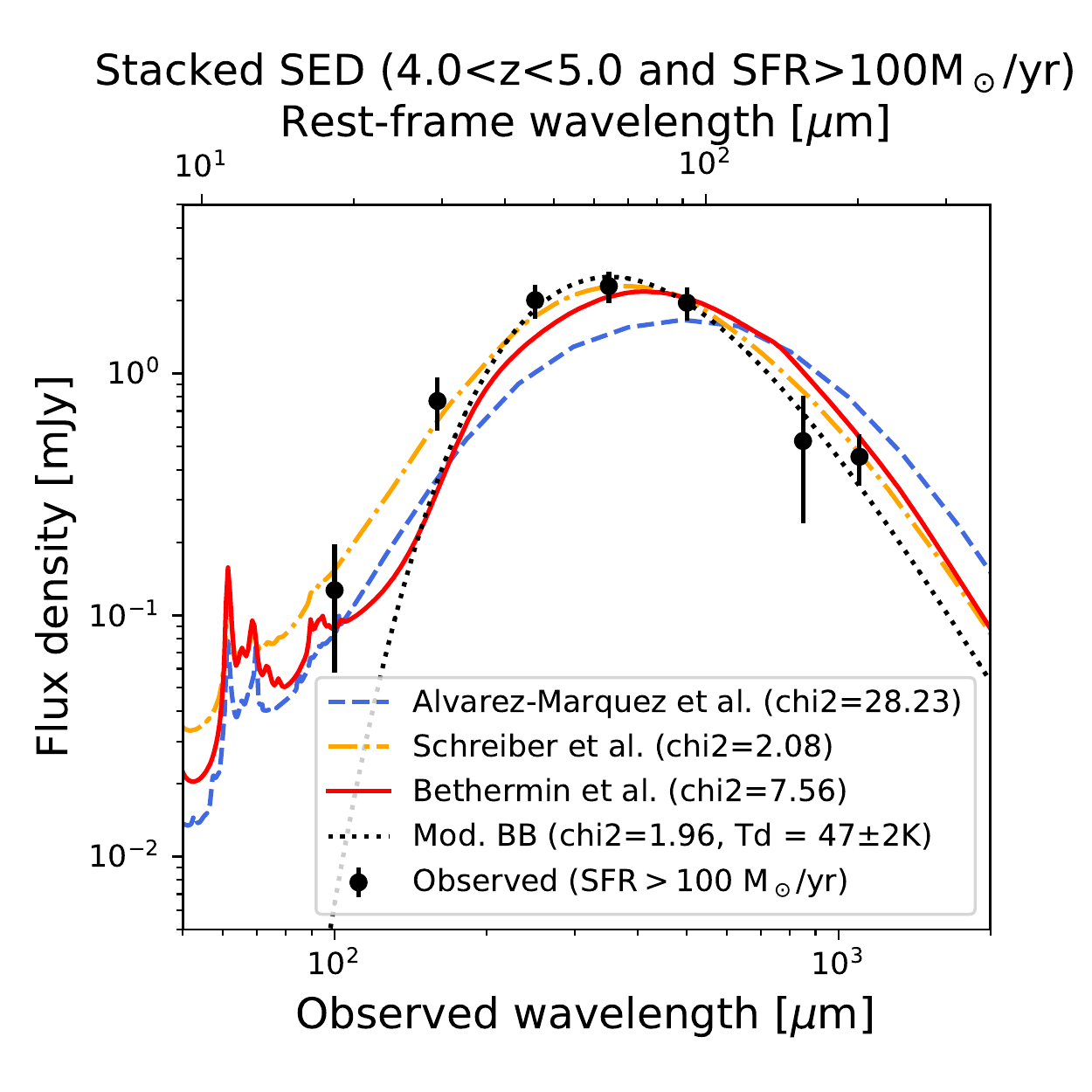}\\
\caption{\label{fig:seds_bright} Same figure as Fig.\,\ref{fig:seds} (upper panel), but using a SFR$>$100\,M$_\odot$/yr cut. }
\end{figure}

\subsection{Caveats}

\label{sect:seds_caveats}

The conversion factors derived previously are based on the best effort, but they are clearly not the final answer about this complex topic. First of all, the selection of the stacked sample is not perfect and based on photometric redshifts and SFRs derived from rest-frame UV to near-IR SED fitting. It is also difficult to estimate how similar this SFR selection is compared to the actual ALPINE sample. Even if it is not likely, we could imagine that a population with very peculiar dust SEDs is missing in one of the two samples. 

The stacked SED was obtained by averaging all the galaxies from our stacked sample. The derived conversion factors could thus be largely inaccurate for outliers with extreme dusty SEDs. Finally, even if the same weight is attributed to each source, stacking provides luminosity-weighted mean SEDs, since brighter sources will have a larger relative contribution to the final signal. We could imagine that a population, which represents a significant fraction of the sample in number but contributes little to the luminosity, has an extreme SED. The stacking analysis would miss such objects and their individual L$_{\rm IR}$ estimates could be incorrect.

In Fig.\,\ref{fig:seds_bright}, we present the stacking for a larger SFR cut of 100\,M$_\odot$/yr. According to optical and near-infrared SED fitting \citep{Faisst2020}, only 11 out of our 118 sources are following this criterion. This analysis is only possible in the 4$<$z$<$5 bin, since there is no detection at higher redshift.  For these objects, the dust temperature is warmer (47\,K versus 41\,K) and the \citet{Schreiber2018a} template fits better the data. The consequences of a slightly warmer dust at higher SFR will be discussed in Sect.\,\ref{sect:cont_vs_cii}.
       

\section{Continuum source properties}

\label{sect:props_cont}

In this section, we discuss the properties of our continuum detections. In Sect.\,\ref{sect:disc_cont_target}, we discuss briefly the basic properties of the detected target sources. In the following sections, we focus on the properties of the nontarget detections: redshift distribution (Sect.\,\ref{sect:Nz}), number counts (Sect.\,\ref{sect:counts}), and contribution to the cosmic infrared background (CIB, Sect.\,\ref{sect:cib}).

\begin{figure*}
\centering
\begin{tabular}{cc}
\includegraphics[width=8.5cm]{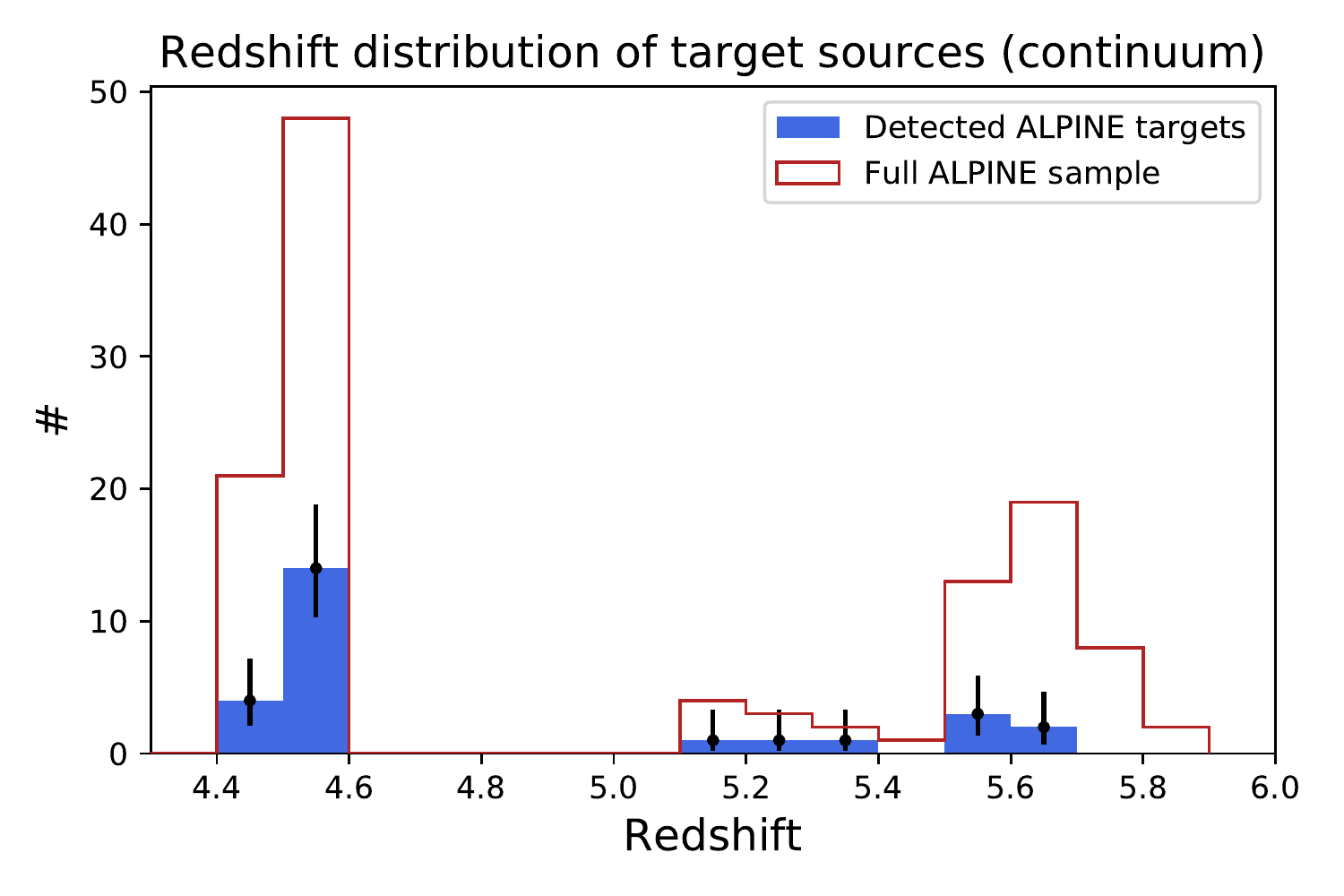} & \includegraphics[width=8.5cm]{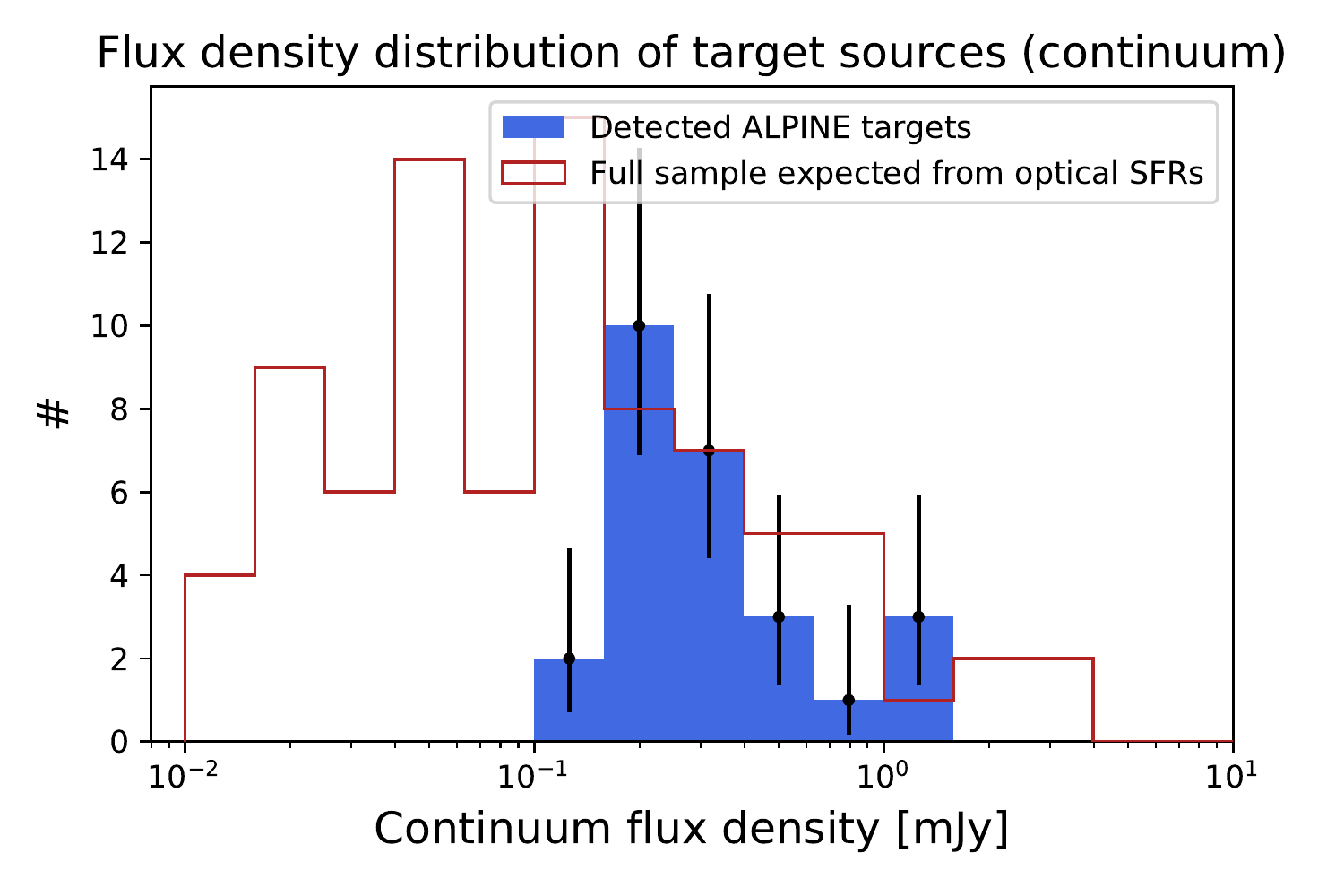}\\
\includegraphics[width=8.5cm]{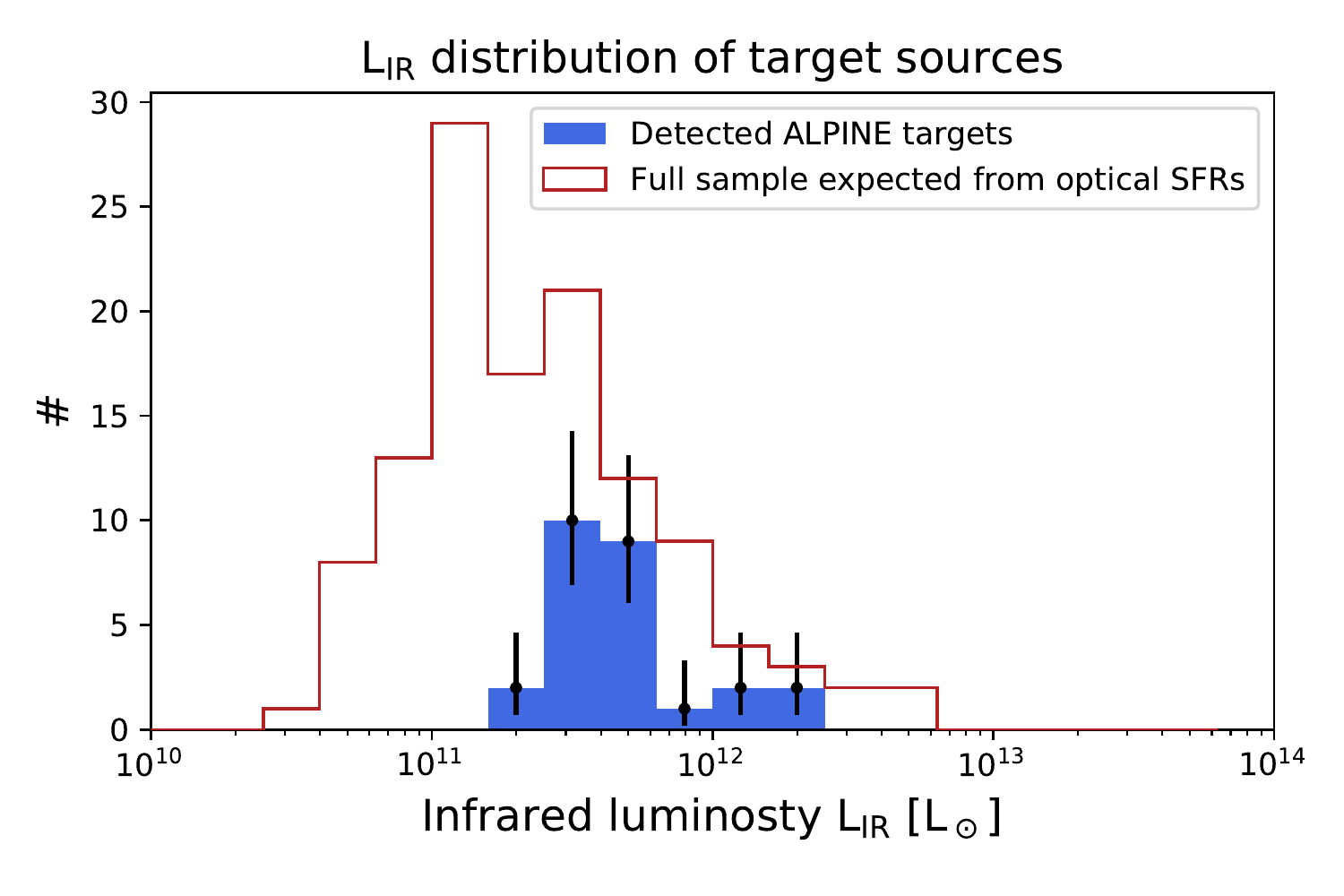} & \includegraphics[width=8.5cm]{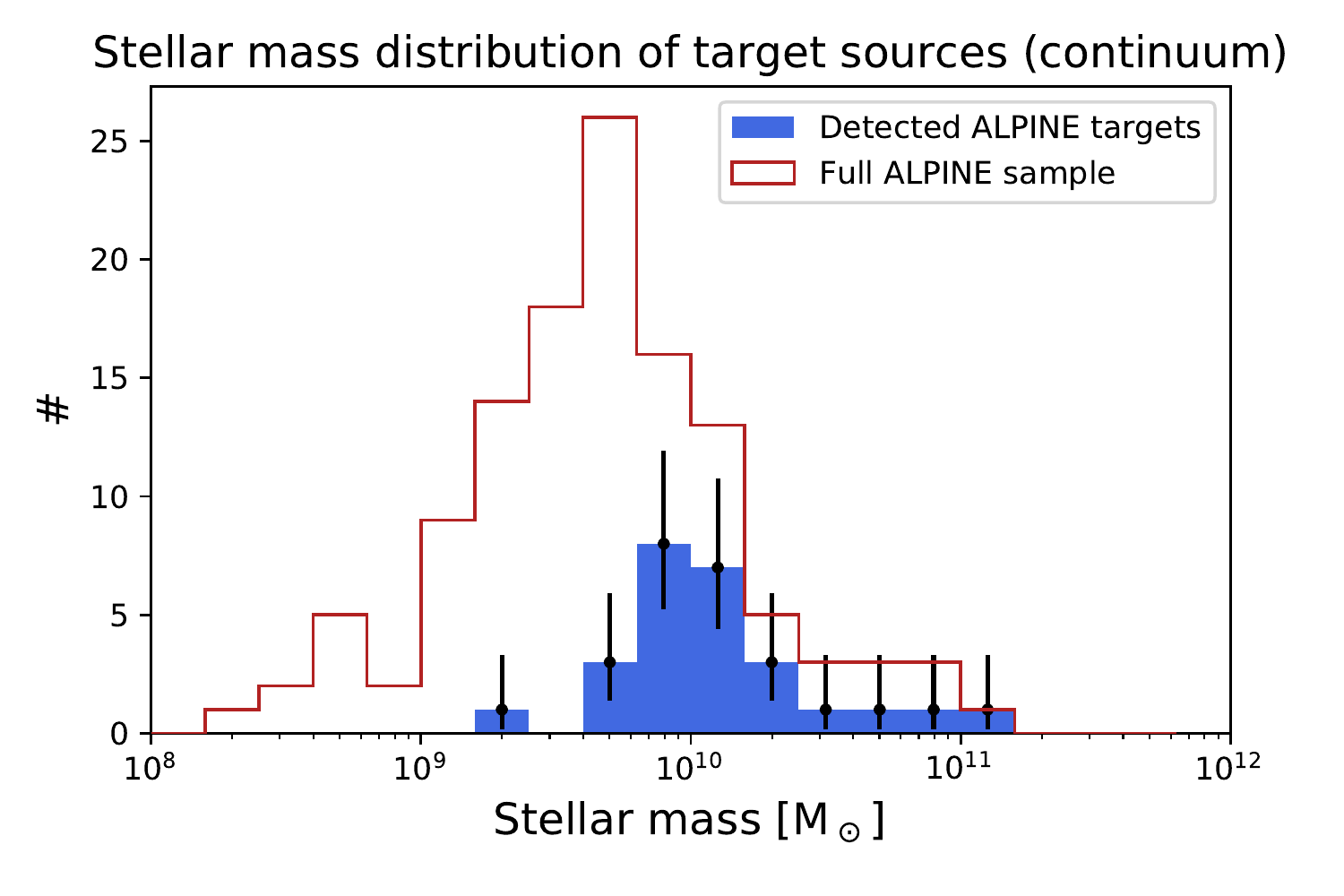}\\
\end{tabular}
\caption{\label{fig:cont_target_props} Upper left panel: redshift distribution of the ALPINE target sources. The blue and red histograms are the distribution of detected sources only and the full sample, respectively. Upper right panel: distribution of the continuum flux densities. The red histogram indicates the distribution expected from the optical and near-IR SED-derived SFR assuming the long-wavelength SED presented in Sect.\,\ref{sect:sed}. Lower left panel: same figure as previously but for L$_{\rm IR}$. Lower right panel: stellar mass distribution of detected (blue) and all (red) sources.}
\end{figure*}

\subsection{Properties of the target sources}

\label{sect:disc_cont_target}

The redshift distribution of the detected target sources is presented in Fig.\,\ref{fig:cont_target_props} (upper left panel). While the detections are distributed across most of the redshift range of the total sample, the detection rate is slightly better in the lower redshift window (26$\pm$6\,\%) than in the high redshift window (15$\pm$5\,\%). Since the sensitivity at fixed luminosity is better in the z$>$5 redshift window (Sect.\,\ref{sect:perf}), we could have expected the opposite trend. However, this is only a 1.4\,$\sigma$ difference and the dust content could be lower at higher redshift. The dust attenuation of our detections will be discussed in \citet{Fudamoto2020}.

We can also compare the flux density distribution of our detections and the expected distribution from the ancillary data \citep[][version including \textit{Spitzer} photometry in the SED fitting]{Faisst2020}. To produce the expected ALPINE flux densities from ancillary data, we estimated the expected $L_{\rm IR}$ from the SFR based on optical and near-infrared SED fitting assuming a $1 \times 10^{-10}$\,L$_\odot$/(M$_\odot$/yr) conversion factor (see Sect.\,\ref{sect:sed_intro}). By doing so, we assume implicitly that the infrared traces the entire star formation. Finally, we use the long-wavelength SED template presented in Sect.\,\ref{sect:stacking_sed} to predict the flux density. The results are shown in Fig.\,\ref{fig:cont_target_props} (upper right panel). The most extreme predicted flux densities ($>$2\,mJy) are not found in the real sample. These very high SFR are almost certainly due to overestimated dust-attenuation corrections. In contrast, all the detected objects are above the mode of the predicted distribution. This shows that we are sensitive only to the highest SFRs. However, this is not a sharp cutoff. This demonstrates that the measured distribution could not have been predicted from the ancillary data and that submillimeter data are important to derive reliable SFRs. A similar trend is found for the infrared luminosity L$_{\rm IR}$ (see lower left panel).

Finally, we compared the stellar mass distribution of the full sample and of the detections only (lower right panel). The mass distributions of the full sample and of the detections are significantly different according to the Kolmogorov–Smirnov test (p-value = $3.8\times10^{-5}$) and only two detections are below the median stellar mass of our full sample. This is an expected consequence of the correlation between the stellar mass and the star formation rate often called main sequence \citep[e.g.][Khusanova et al. in prep.]{Schreiber2015,Tasca2015}. 

\subsection{Redshift distribution of the nontarget continuum detections}

\begin{figure}
\centering
\includegraphics[width=8.5cm]{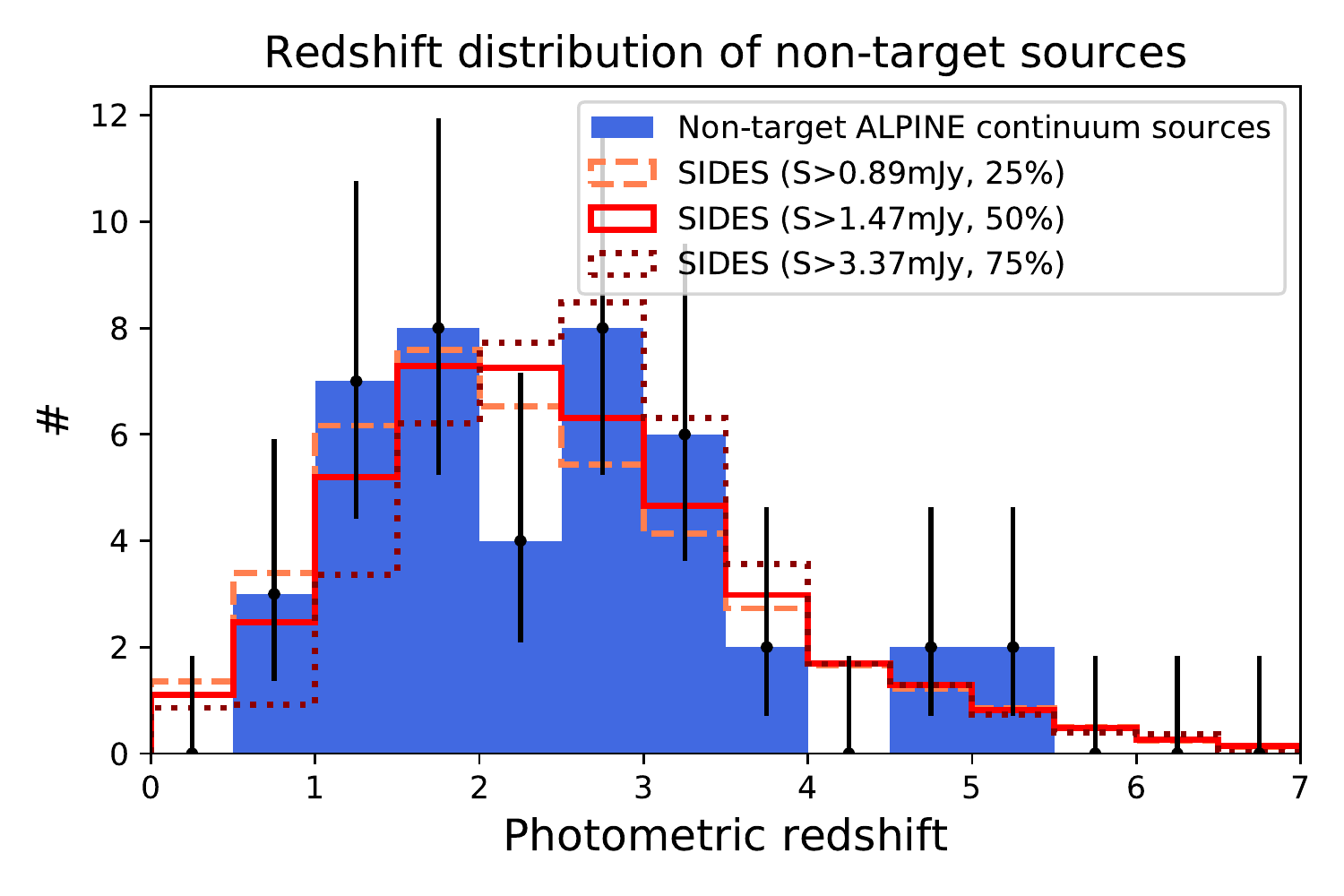}
\caption{\label{fig:Nz} Photometric redshift distribution of the nontarget ALPINE sources (blue filled histogram). The dashed, solid, and dotted histograms are the predictions from the \citet{Bethermin2017} SIDES simulation for a flux cut corresponding to the first-quartile, the median, and the third-quartile of the observed sample, respectively.}
\end{figure}

\label{sect:Nz}

Contrary to the target sample, determining the redshift of nontarget sources is not trivial. We have to identify the optical/near-infrared counterparts and use photometric redshifts when spectroscopic redshift are not available. Fortunately, this sample lies in survey areas with rich ancillary data (fully described in \citealt{Faisst2020}) drawn primarily from COSMOS \citep{Scoville2007}, GOODS \citep{Giavalisco2004} and CANDELS \citep{Grogin2011,Koekemoer2011}. The counterparts of 42 of our 57 nontarget continuum detections were identified in the \citet[][COSMOS]{Laigle2015} or the \citet[][3DHST]{Momcheva2016} catalogs. The detailed identification of each source and the sources without counterpart in the previously cited catalogs will be discussed in details in Gruppioni et al. (in prep.). 

The redshift distribution of our nontarget sources is presented in Fig.\,\ref{fig:Nz}. The mean redshift of our sample is z=2.5$\pm$0.2 (median = 2.3$\pm$0.3) with a tail up to z=6. Our uncertainties are computed using a bootstrap technique and thus include sample variance. This is 1-$\sigma$ lower than the median redshift (z = 2.65$\pm$0.13) found by \citet{Simpson2017} following up $>$1\,mJy sources selected in a single-dish survey using ALMA at the same wavelength. As shown in \citet[][see also \citealt{Hodge2020}]{Bethermin2015b}, fainter submillimeter sources are paradoxically expected to have a lower mean redshift. This small difference is thus not surprising. To test if this trend is also found inside our own sample, we split it into two equally populated subsamples containing the faint ($<$1.47\,mJy) and the bright sources ($>$1.47\,mJy). The faint sources have a mean redshift of z=2.6$\pm$0.3, while we found z=2.3$\pm$0.2 for the bright ones. It agrees with the trend predicted by \citet{Bethermin2015b}, but our sample is too small to provide a statistically significant result. We also expect that longer wavelengths probe higher redshifts. As expected, our median redshift is smaller than what is found at 1.1\,mm by \citet[][z$_{\rm med}$=2.9]{Franco2018} and \citet[][z$_{\rm med}$=2.48$\pm$0.05]{Brisbin2017} or 1.4\,mm by  \citet[][z$_{\rm med}$=3.9]{Strandet2016}. This last sample is lensed and it might push the median redshift to higher values. In contrast, our median redshift is higher than the very faint 1.1\,mm sample of \citet[][down to 0.05\,mJy, z$_{\rm med}$=1.9$\pm$0.4]{Aravena2016b}. As shown in Fig.\,3 of \citet{Bethermin2015b}, it is expected that $<$0.1\,mJy 1.1\,mm sources are at lower redshift than $\sim$1\,mJy 850\,$\mu$m sources.

Finally, we compare our measured distribution with the predictions of the simulated infrared dusty infrared sky (SIDES) simulation\footnote{\url{http://cesam.lam.fr/sides/}} \citep{Bethermin2017}. Since the depth of our various pointings is not homogenous, our sample is not flux limited. We thus computed the redshift distribution for different flux cuts corresponding to the first quartile, median, and third quartile of the observed sample. The three predicted distributions are compatible with our measurements at 1-$\sigma$. Because of galaxy clustering, we could have expected an excess of sources at the same redshift as the ALPINE targets, but we observe only a 1-$\sigma$ excess between z=5 and z=6. We can thus assume that the sample of nontarget sources with optical counterparts is statistically similar to a sample, which would have been obtained using random pointings.

\begin{table*}
\caption{\label{tab:int_counts} Integral number counts at 850\,$\mu$m derived from ALPINE for the full nontarget sample (all) and the nontarget sources with an optical counterpart at z$<$4 (secure, z$<$4).}
\centering
\begin{tabular}{lcccc}
\hline
\hline
S$_{\rm cut}$ & N$_{\rm all}$ & N($>$S$_{\rm cut}$) (all) & N$_{\rm secure, \, z<4}$ & N($>$S$_{\rm cut}$) (secure, z$<$4) \\
mJy & & deg$^{-2}$ &  & deg$^{-2}$ \\
\hline
\vspace{0.15cm} 
\vspace{0.15cm} 
0.35 & 54 & 31000$_{-7000}^{+43000}$  & 37 & 14000$_{-3000}^{+15000}$\\
\vspace{0.15cm} 
0.56 & 47 & 14000$_{-3000}^{+13000}$  & 33 & 8800$_{-1800}^{+6000}$\\
\vspace{0.15cm} 
0.89 & 38 & 7600$_{-1300}^{+2500}$  & 27 & 5400$_{-1100}^{+1900}$\\
\vspace{0.15cm} 
1.41 & 30 & 5500$_{-1000}^{+1800}$  & 21 & 3700$_{-800}^{+1300}$\\
\vspace{0.15cm} 
2.24 & 19 & 2900$_{-700}^{+1000}$  & 14 & 2200$_{-600}^{+800}$\\
\vspace{0.15cm} 
3.55 & 10 & 1500$_{-500}^{+600}$  & 8 & 1200$_{-400}^{+600}$\\
\hline
\end{tabular}
\end{table*}

\begin{table*}
\caption{\label{tab:diff_counts} Euclidian-normalized differential number counts at 850\,$\mu$m derived from ALPINE for the full nontarget sample (all) and the nontarget sources with an optical counterpart at z$<$4 (secure, z$<$4).}
\centering
\begin{tabular}{lcccc}
\hline
\hline
Flux density & N$_{\rm all}$ & dN/dS\,S$^{2.5}$ (all) & N$_{\rm secure, \, z<4}$ & dN/dS\,S$^{2.5}$ (secure, z$<$4) \\
mJy & & Jy$^{1.5}$/sr & & Jy$^{1.5}$/sr \\
\hline
\vspace{0.15cm} 
0.46 (0.35--0.56) & 7 & 1200$_{-600}^{+2600}$ & 4 & 340$_{-200}^{+830}$\\
\vspace{0.15cm} 
0.73 (0.56--0.89) & 9 & 980$_{-410}^{+1750}$ & 6 & 490$_{-220}^{+770}$\\
\vspace{0.15cm} 
1.15 (0.89--1.41) & 8 & 600$_{-220}^{+420}$ & 6 & 480$_{-200}^{+390}$\\
\vspace{0.15cm} 
1.83 (1.41--2.24) & 11 & 1400$_{-400}^{+800}$ & 7 & 870$_{-330}^{+590}$\\
\vspace{0.15cm} 
2.89 (2.24--3.55) & 9 & 1700$_{-600}^{+900}$ & 6 & 1100$_{-500}^{+800}$\\
\vspace{0.15cm} 
4.59 (3.55--5.62) & 5 & 1600$_{-700}^{+1100}$ & 4 & 1300$_{-630}^{+1100}$\\
\hline
\end{tabular}
\end{table*}

\begin{figure*}
\centering
\begin{tabular}{cc}
\includegraphics[width=8.5cm]{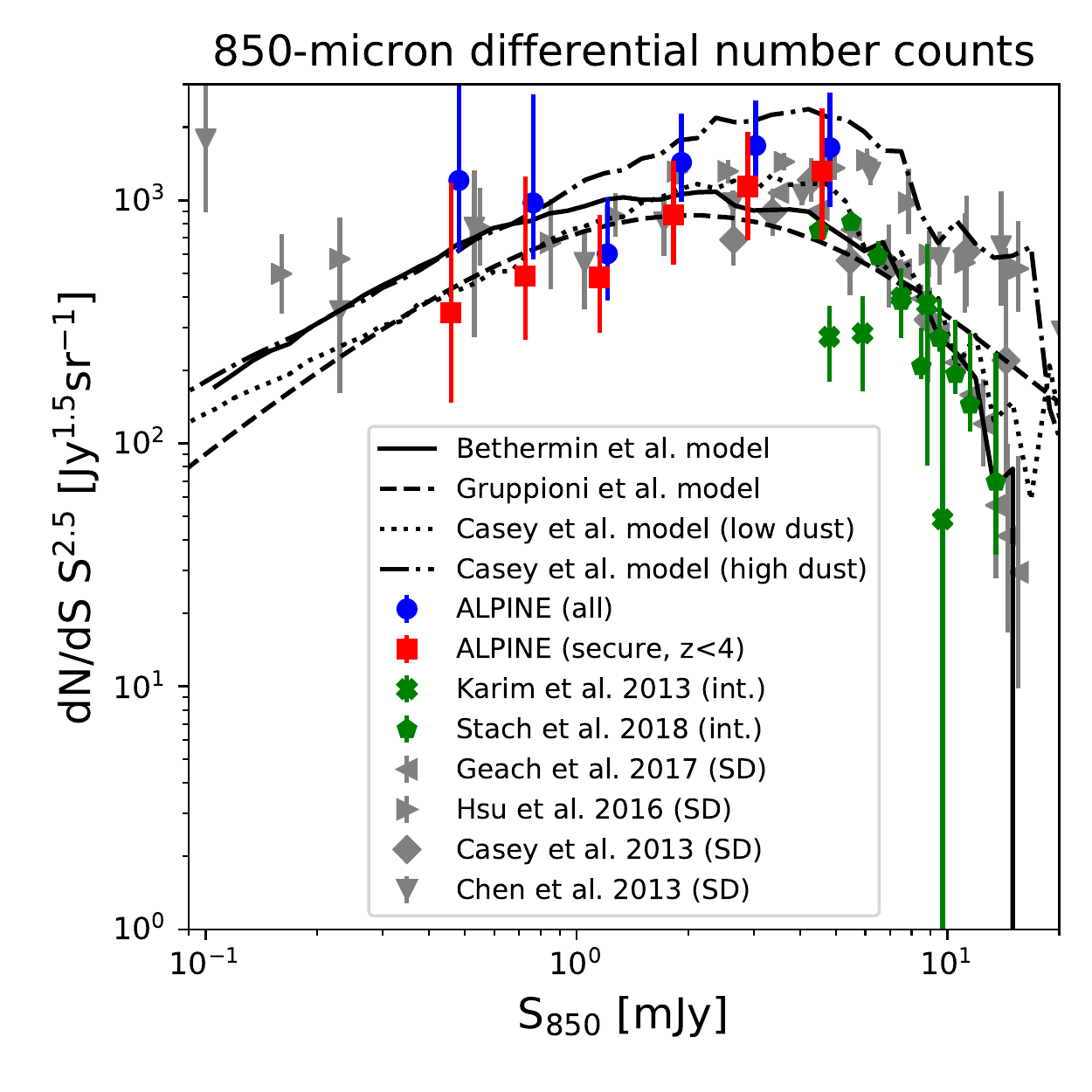} & \includegraphics[width=8.5cm]{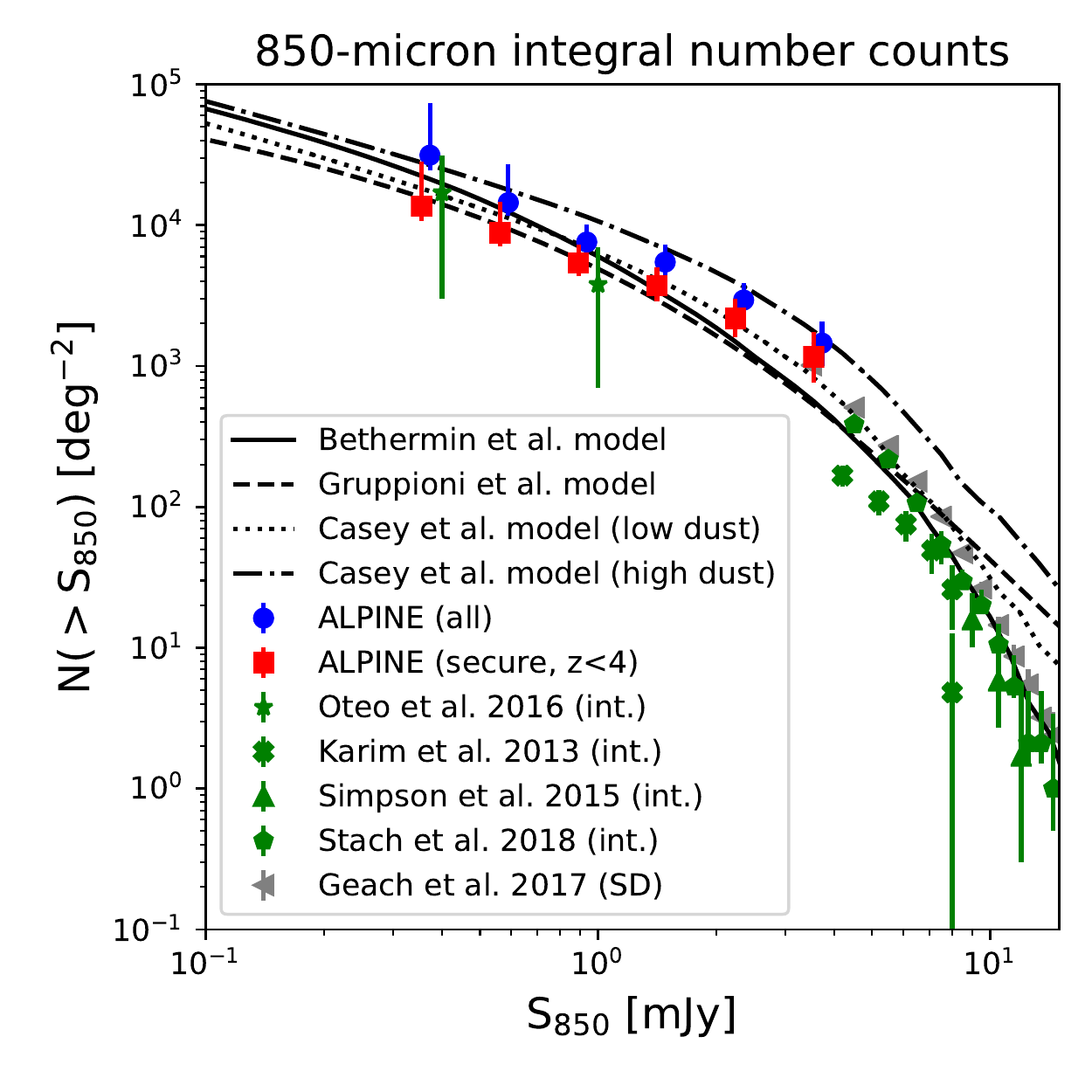} \\
\end{tabular}
\caption{\label{fig:counts} Left panel: Differential number counts at 850\,$\mu$m. The blue circles are the counts derived from the full ALPINE nontarget sample. The red squares are computed from the ALPINE nontarget sources with a confirmed optical or near-IR counterpart below z=4 (secure z$<$4 sample). It is thus a secure lower limit, since sources clustered with ALPINE sources are excluded. The green crosses and pentagons are the interferometric measurements of \citet{Karim2013} and \citet{Stach2018}, respectively. The gray left-facing triangles, right-facing triangles, diamonds, and down-facing triangles are the measurements performed using single-dish data with a lower angular resolution by \citet{Geach2017}, \citet{Hsu2016}, \citet{Casey2013}, and \citet{Chen2013}, respectively. The solid, dashed, dotted, and dash-dot lines are the models of \citet[][SIDES simulation]{Bethermin2017}, \citet{Gruppioni2011}, \citet[][low dust]{Casey2018}, and \citet[][high dust]{Casey2018}. Right panel: Integral number counts at 850\,$\mu$m. The symbols are the same as in the left panel. The interferometric measurements of \citet{Oteo2016} and \citet{Simpson2015}, respectively, are represented by green stars and triangles.} 
\end{figure*}
 
\subsection{Number counts}

\label{sect:counts}

The area probed by the ALPINE survey is sufficiently large to produce new meaningful constraints on the faint galaxy number counts at 850\,$\mu$m. While the ALMA band 6 (1.1--1.4\,mm) was extensively used for deep surveys, the band 7 ($\sim$850\,$\mu$m) has been much less explored, especially below 3\,mJy. \citet{Oteo2016} provided constraints based on the ALMA calibrator survey, but with large uncertainties. In contrast, this wavelength has been widely explored with single-dish instruments \citep[e.g.,][]{Coppin2006,Casey2013,Chen2013,Hsu2016,Geach2017}. However, because of their limited spatial resolution, several galaxies can be blended in the same beam, biasing the bright number counts toward higher values \citep{Hayward2013,Karim2013,Bussmann2015,Cowley2017,Scudder2016,Bethermin2017}. Above 3\,mJy, interferometric observing campaigns had followed up single-dish sources to correct for this effect \citep{Karim2013,Simpson2015,Stach2018}.

We derived the integral number counts dN/d$\Omega$, which is the surface density of sources above a certain flux cut, by summing the inverse of the effective area $\Omega_{\rm eff}$ for each nontarget source\footnote{As explained in Sect.\,\ref{sect:eff_area}, our definition of the effective area already takes into account the completeness. The central 1\,arcsec radius region around the target sources is excluded from our analysis and thus in particular from the computation of the effective area.}:
\begin{equation}
\frac{dN}{d\Omega}(>S) = \sum_{i=1}^{N_{\rm source}} \frac{1}{\Omega_{\rm eff}(S_i, \theta_i)},
\end{equation}
where $S_i$ and $\theta_i$ are the deboosted flux density (see Sect.\,\ref{sect:flux_boosting}) and size of the i-th source. All the flux densities have been converted to 850\,$\mu$m at which the effective area (see Sect\,\ref{sect:eff_area}) was computed. As shown in Sect.\,\ref{sect:flux_boosting} and Fig.\,\ref{fig:flux_boosting}, the deboosting factor, which is necessary to apply here, can have a 30\,\% uncertainty. We thus computed the difference between the number counts derived using the 1-$\sigma$ lower and upper envelopes of the flux boosting curve to estimate the associated uncertainties. These uncertainties are combined with the Poissonian error bars. Finally, SC\_1\_DEIMOS\_COSMOS\_787780, SC\_1\_vuds\_cosmos\_5101210235, and SC\_2\_DEIMOS\_COSMOS\_773957 are detected by our algorithm only because their flux density is boosted by a line. Else, their S/N without line contimination falls below our threshold of 5. These sources are thus excluded from our continuum number count computation.

A similar method was used to derive the differential number counts. We summed the inverse of the effective area of all the sources in a given bin and divided by the bin size. To reduce the dynamical range on the figures, we normalized the differential counts by S$^{2.5}$. With this normalization, the number counts in an Euclidian nonevolving Universe are flat. This is usually the case for very bright fluxes ($>$100\,mJy) in the submillimeter domain \citep{Planck_eucl}, where the detected sources are mainly local. The deviations from this trend at fainter flux densities provide important constraints for galaxy evolution models. It has also the convenient property to reduce the dynamical scale of the plot and help the visual comparison between the models and the data. 

We estimated the integral number counts for various thresholds spaced by 0.2\,dex. We chose to use 0.35\,mJy for the lowest threshold, which corresponds to the deboosted 850\,$\mu$m-converted flux density of the second faintest object. Concerning the differential number counts, we used the intervals delimited by this list of thresholds. We do not use fainter bins, since the faintest object (0.30\,mJy after conversion to 850\,$\mu$m, SC\_2\_DEIMOS\_COSMOS\_773957) is associated to a completeness of 13\,\% and the correction to apply is thus very large. The mean completeness in the faintest bin (0.35--0.56\,mJy) is 50\,\%. In all the other bins, the mean completeness is above 80\,\%.

Our measurements are summarized in Table\,\ref{tab:int_counts} and \ref{tab:diff_counts} and shown in Fig.\,\ref{fig:counts}. Even if the redshift distribution of the nontarget sources provides no firm evidence for it (see Sect.\,\ref{sect:Nz}), we cannot formally exclude a small overdensity of sources at the same redshift as ALPINE targets compared to a random position in the sky. We thus estimated the number counts using both the full sample and a secure z$<$4 sample, where only the sources with identified optical or near-IR counterparts below the ALPINE redshift range are kept. These two samples provide respectively an upper and a lower limit on the number counts, which would be derived at a random position in the sky. The values derived using these two samples agree at a 1\,$\sigma$ level.

Our new measurements exhibit a shallower slope below $\sim$3\,mJy than what was measured by previous surveys at higher flux density. This is the first time that we probe so well the regime below this slope break, since \citet{Oteo2016} data points suffer from an order of magnitude of uncertainties and the \citet{Hsu2016} analysis is affected by their low angular resolution and rely on complex methods to invert the lensing to recover the intrinsic counts. Our results are compatible at 1\,$\sigma$ with \citet[][only shown for integral number counts, i.e. the right panel of Fig.\,\ref{fig:counts}]{Oteo2016} and \citet{Hsu2016}.

We can also compare our new measurements with various models of galaxy evolution. The \citet{Bethermin2017} model is an update of the \citet{Bethermin2012c} models, which decomposes star-forming galaxies into main-sequence and starbursts galaxies. Each type has a different SED evolving with redshift. It starts from the observed stellar mass function and the measured evolution of the main sequence of star-forming galaxies to predict infrared and (sub-)millimeter observables. This model includes clustering and can reproduce the single-dish low-resolution data and interferometric data simultaneously. However, there were few constraints below 3\,mJy at 850\,$\mu$m, when it was published. The \citet{Gruppioni2011} model updated in \citet{Gruppioni2019} is based on the observed luminosity functions \citep{Gruppioni2013}. This model contains five different populations, whose luminosity functions evolve with redshift (spirals, starbursts, low-luminosity AGN, type-1 AGN, and type-2 AGN). The \citet{Casey2018} model assumes an evolution of both the infrared luminosity function and the SEDs. Two versions were proposed. The first one has a low number of dusty objects at high redshift and the other assumes a large volume density of bright obscured galaxies.

At the spatial resolution of the ALPINE data ($~\sim$1\,arcsec), we can directly compare the source counts with the galaxy counts in the model, since it is unlikely to have two physically-separated galaxies in the same beam. Overall, all models agree with our data, since they are between the 1\,$\sigma$ lower limit of the secure z$<$4 and the 1\,$\sigma$ upper limit of the full sample. However, the high-dust model of \citet{Casey2018} is higher than the \citet{Stach2018} measurements at high flux density. Below the $\sim$3\,mJy slope break, the predictions of the Gruppioni et al. and the low-dust Casey et al. models are a factor of 2 lower than the high-dust Casey et al. and the B\'ethermin et al. models. Unfortunately, our uncertainties are still too large to identify the correct scenario.

\begin{figure}
\centering
\includegraphics[width=8.5cm]{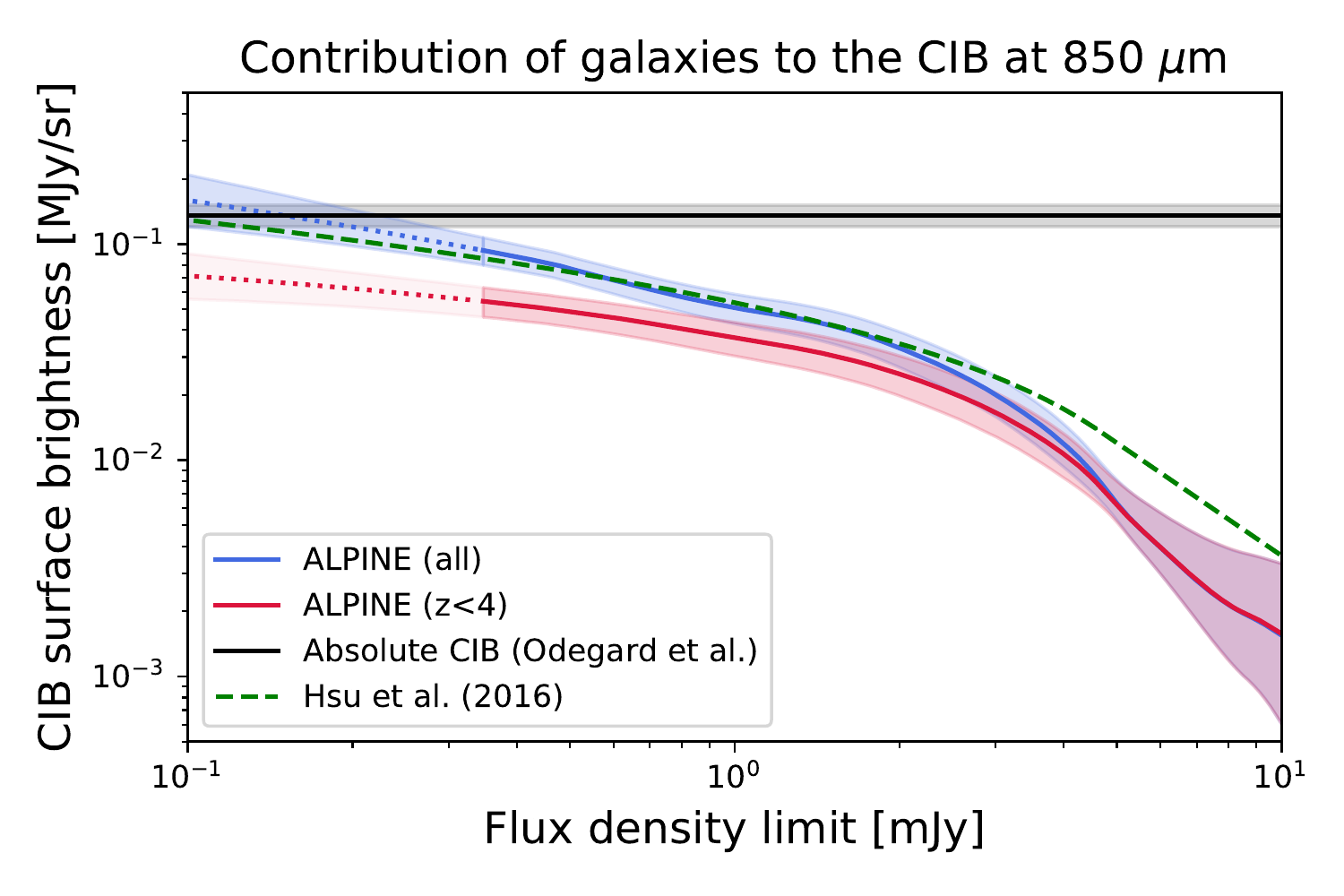}
\caption{\label{fig:cib}Contribution of galaxies to the cosmic infrared background at 850\,$\mu$m as a function of the flux density limit. The blue and red solid lines are our results based on the full and the secure z$<$4 samples, respectively. The shaded area is the 1-$\sigma$ confidence region. The dashed green line is the measurement of \citet{Hsu2016} using SCUBA2 data of cluster fields. The black line is the absolute measurement of the total CIB combining COBE/FIRAS and \textit{Planck} data \citep{Odegard2019}.}
\end{figure}

\subsection{Cosmic infrared background}

\label{sect:cib}

The cosmic infrared background (CIB) is the relic of all dust emission by galaxies across cosmic times \citep[e.g.,][]{Dole2006}. Its absolute brightness was measured in the nineties by the COBE/FIRAS instrument \citep{Puget1996,Hauser1998,Fixsen1998,Gispert2000,Lagache2000}. By combining FIRAS and \textit{Planck}, \citet{Odegard2019} estimated the absolute CIB level with a better precision than historical analyses. Their values are used as the reference CIB level in this paper. The CIB SED provides a budget of far-infrared photons that galaxies emit during their evolution. While \textit{Herschel} identified the galaxy populations (luminosity, redshift) emitting the CIB below 500\,$\mu$m, there were fewer constraints at longer wavelength before ALMA. However, new number counts extracted from band-6 surveys ($\sim$1.3\,mm) can explain 50-100\,\% of the CIB absolute level \citep{Carniani2015,Aravena2016b,Fujimoto2016}. At 850\,$\mu$m, using single-dish SCUBA2 data of cluster fields, \citet{Hsu2016} found that the full CIB can be explained by galaxies brighter than 0.1\,mJy.

The CIB brightness produced by all the sources above a certain flux density threshold is:
\begin{equation}
B_\nu = \int_{S_{\rm lim}}^{\infty} S_\nu \frac{dN}{dS_\nu} \, dS_\nu,
\end{equation}
where $B_\nu$ is the surface brightness density of the CIB produced by sources above the flux density limit $S_{\rm lim}$. To compute this integral, we assume a power-law to connect our data points. We extrapolated the contribution of sources fainter than 0.35\,mJy by fitting a power-law to the five faintest data points. The slope is poorly constrained and is responsible for large uncertainties on this extrapolation: $dN/dS_{\nu} \propto S_{\nu}^{-2.2\pm0.3}$ for the full sample and $dN/dS_{\nu} \propto S_{\nu}^{-1.8\pm0.4}$ for the secure sample. Above our brightest data point, we use the Euclidian plateau level measured by \citet[][dN/dS S$^{2.5}$ = 15\,Jy$^{1.5}/sr$]{Planck_eucl}. The uncertainties are determined by recomputing 100\,000 times the integral of the number counts after randomly offsetting the data points according to their error bars. 

In Fig.\,\ref{fig:cib}, we present the contribution to the CIB as a function of the flux density limit. The contribution of sources brighter than the limit of our number counts (0.35\,mJy) is 0.093$\pm$0.013\,MJy/sr for the full sample and 0.054$\pm$0.009\,MJy/sr for the secure sample. This is 69\,\% and 40\,\% of the full CIB measured by \citet[][0.135\,MJy/sr assuming a CIB spectrum for the color correction]{Odegard2019}. Below 2\,mJy, our measurements agree with \citet{Hsu2016}, since their data are between the values determined from our secure z$<$4 measurements (lower limit) and from our full sample (upper limit).\\


\section{[CII] catalog}

In this paper, we focus only on the [CII] line detections of the ALPINE targets. The serendipitous detections and [CII]-emitting close companions of the targets will be discussed in Loiacono et al. (in prep.).

\label{sect:cii_cat}

\subsection{Extraction of the candidates}

\label{sect:cii_extr}

Extracting lines from a cube can be significantly more difficult than the continuum when the redshift of the source is not well known, since we also have to explore the spectral dimension. This is the case for ALPINE, since we found significant offsets between the [CII] lines and our reference redshifts derived from optical spectroscopy (see Sect.\,\ref{sect:z_offset}). To perform this task, we used the semi-automatic procedure described below.

We first ran a customized line finder algorithm (see comparison with the \textit{findclumps} algorithm in Sect.\,\ref{sect:cii_purity}) to search for [CII] emission in a 3D area defined as the cylinder with a 1" aperture around the phase center extending over the full bandwidth, to allow for spatial and spectral offsets. In practice, we produced running averaged cubes using various numbers of channels. The minimum allowed number of collapsed channels is two (50 km/s) to avoid single-channel spikes of noise, and the maximum is 40 channels (1000\,km/s). In other words, we produce running moment-0 cubes across the full bandwidth. We then computed RMS of these nonprimary-beam-corrected maps using all pixels at $>1''$ from the phase-center and searched for S/N$>$3 peaks in the maps. Finally, we saved the positions, frequencies and S/Ns of each of these [CII]-emitter candidates.

For each [CII]-emitter candidate, we extracted a spectrum from the continuum subtracted cubes at the position of the brightest pixel of the moment-0 map and had a further quality assessment based on visual inspection. For the good candidates, we fit a Gaussian profile of the spectrum and computed central frequency (f$_{\rm cen}$) and FWHM. We then produced a new moment-0 map collapsing the channels included in the [f$_{\rm cen}$ - FWHM, f$_{\rm cen}$ + FWHM] spectral range. The choice of such frequency interval is an optimal compromise between including most of the line profile and excluding any noise at the edge of the line. This step is needed because the moment-0 map computed during the first step is the one that maximizes the S/N, and therefore, does not necessarily cover the full line profile. We then re-extracted a spectrum inside the 2-$\sigma$ region measured in this moment-0 map. This new spectrum is more informative than the first one in the case of spatially-resolved sources, since it is extracted from the entire region of emission and not just at the peak. These steps were repeated several times until the shape of the 2-$\sigma$ contour, in which the spectrum is extracted, is stable. Few iterations were necessary, since the flux measured in both the spectra and in the moment-0 maps converged within the uncertainties after only one iteration. The final spectra are presented in Figs.\,\ref{fig:spec_1}, \ref{fig:spec_2}, and \ref{fig:spec_3}.

\subsection{Detection threshold and reliability}

\label{sect:cii_purity}

\begin{figure}
\centering
\includegraphics[width=8.5cm]{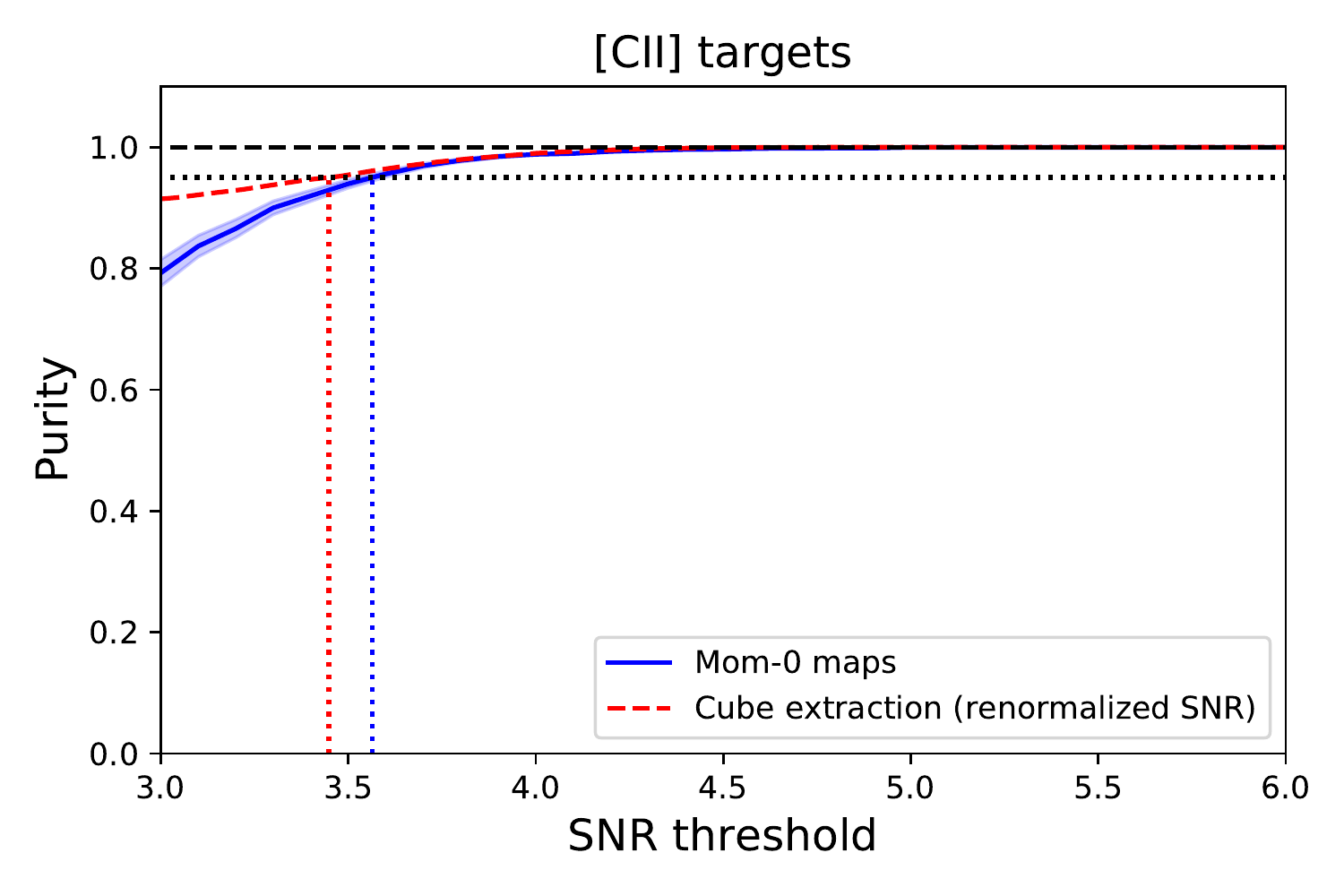}
\caption{\label{fig:purity_cii} Purity of [CII] detections as a function of the S/N threshold in the moment-0 map. The purity is provided only for the 1\,arcsec central region and was estimated using the methods described in Sect.\,\ref{sect:cii_purity}. The blue solid line is derived using the method based on the moment-0 maps and the red dashed line using an extraction in the negative cubes (see Sect.\,\ref{sect:cii_purity}). The dotted vertical lines are the S/N at which the 95\,\% purity is reached.}
\end{figure}

To assess the reliability of the line emitter candidates, we use the same method based on the negative maps as for the continuum (Sect.\,\ref{sect:cont_ext}), but using the final moment-0 maps produced after optimizing the integration frequency window (see Sect.\,\ref{sect:cii_extr}). Contrary to continuum maps, we have to take into account that we searched the line over a large velocity range, which could contain several times the line width. We thus normalized the number of detections in the negative map by the ratio between the velocity range in which we expect a line ($\sim$1500\,km/s, see Sect.\,\ref{sect:z_offset}) and the actual width of the velocity window used to produce the moment-0 map. We estimated that the 95\% purity in the central 1" region is reached for a S/N$>$3.56 threshold. The purity as a function of the S/N is shown in Fig.\,\ref{fig:purity_cii} (blue solid line).

Since this normalization is an approximation, we cross-checked our number with an independent approach. We extracted the sources in the negative and the positive cubes using the  \textit{findclumps} \citep{Decarli2016,Walter2016} routine. The S/N determined by \textit{findclump} is on average 15\% higher than in our moment-0 map. This is expected since \textit{findclump} chooses an integration window to maximize the S/N. It is usually narrower and only includes the high-S/N central channels of the line, while our moment-0 maps include also the noisier tails. A similar difference of S/N estimates was also found by the ASPECS survey team \citep{Gonzalez-Lopez2019}. After correcting for this 15\% systematic difference to agree with our moment-0 map S/Ns, we find a 95\% purity limit for S/N=3.45 in the 1\,arcsec central region. The two methods are thus in excellent agreement. We also compared the FWHM of the detections in the positive cube and the spurious sources found in the negative cubes close from the detection threshold (3.5$<$S/N$<$4.5). We found a mean value of 290$\pm$40\,km/s and 216$\pm$1\,km/s, respectively\footnote{The error bars on the mean FWHM are the uncertainty on the mean of the sample and not the scatter (see caption of Fig.\,\ref{fig:sens}). The spurious sources are extracted from the full cubes and not the 1\,arcsec-radius central regions. They are thus much more numerous. The uncertainty on their mean FWHM is thus smaller than the detections in the central region}. This 2-$\sigma$ difference is a reassuring hint that the low S/N sources are not spurious.

We chose to cut our catalog at S/N$>$3.56 to be conservative. The purity estimated using this method is presented in Fig.\,\ref{fig:purity_cii}. Since the number density of line emitters in the rest of the field is much lower (Sect.\,\ref{sect:cont_ext}), a much higher S/N threshold is necessary in the rest of the field. This will be discussed in Loiacono et al. (in prep.). Finally, we checked that the two extraction methods provide compatible source lists and found that this is true except for a couple of objects close to the S/N detection threshold.

It could seem counterintuitive that the S/N threshold to reach 95\% purity for target sources is the same for the continuum and lines. Indeed, this is a coincidence. If the surface number density of continuum and line sources were the same at fixed S/N, the purity would be much lower for the lines, since line spurious sources can come from several velocity channels and are thus more numerous at fixed S/N. However, the number of S/N$>$3.56 [CII] sources is three times higher than the number of continuum emitters. The higher number of real sources thus compensates the higher number of spurious sources in Eq.\,\ref{eq:purity}.

We obtained 75 [CII] detections out of our 118 targets. The moment-0 cutouts of our detected [CII] targets is presented in Fig.\,\ref{fig:target_mom0_1}, \ref{fig:target_mom0_2}, and \ref{fig:target_mom0_3}. An extensive discussion of their morphology is presented in \citet{Le_Fevre2019}.

\subsection{[CII] flux measurements and consistency}

\label{sect:phot_cii}

\begin{figure*}
\centering
\begin{tabular}{cc}
\includegraphics[width=8.5cm]{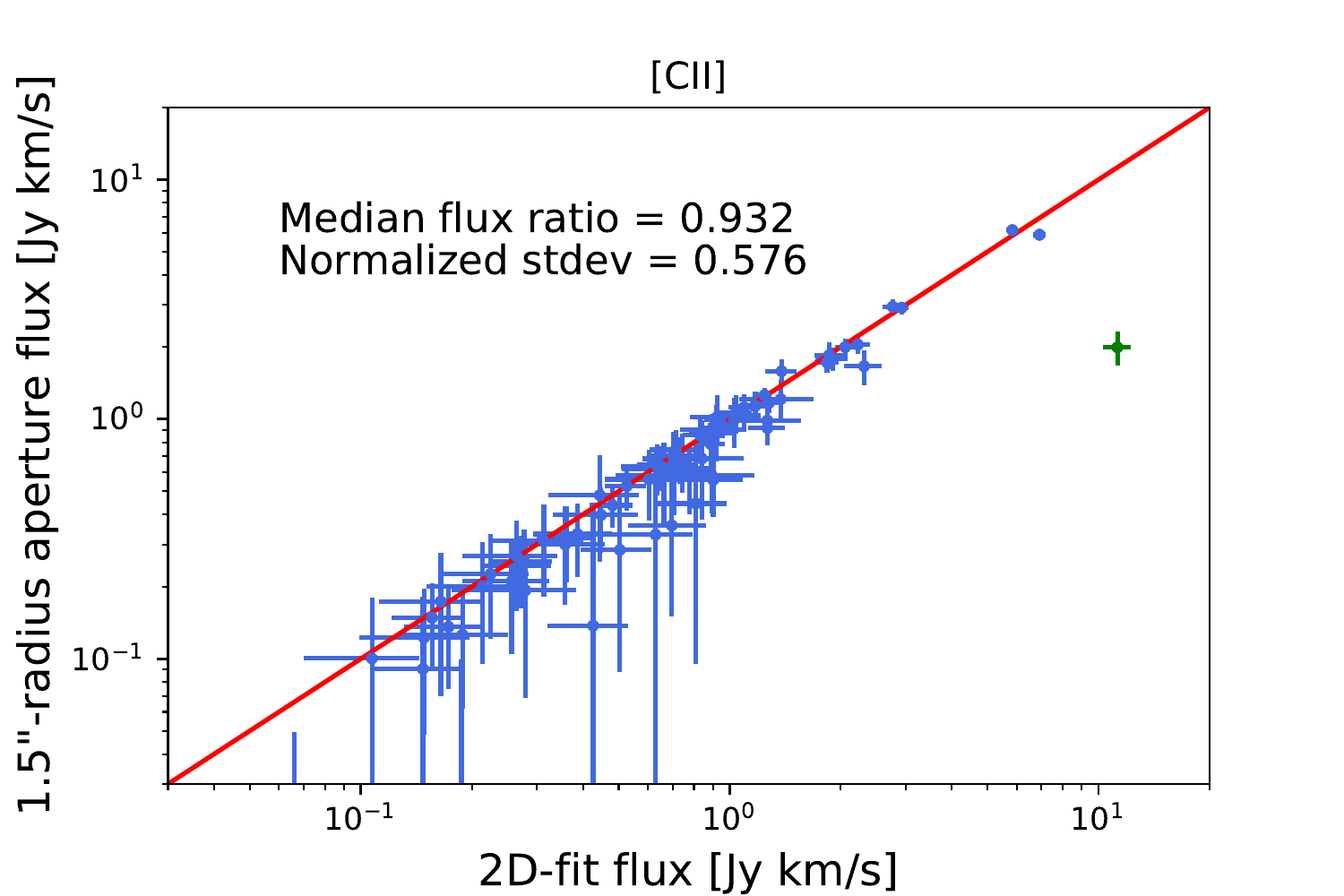} & \includegraphics[width=8.5cm]{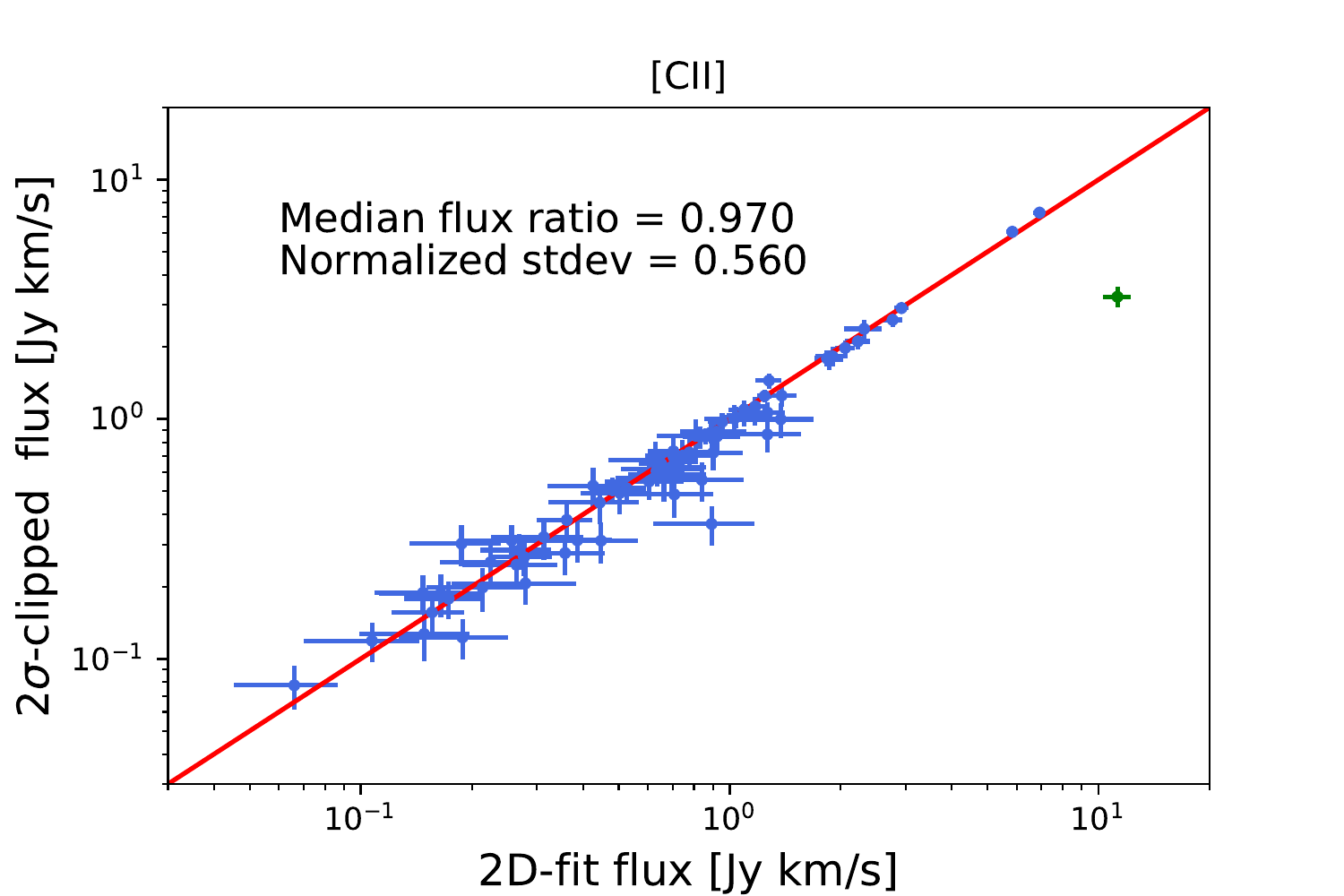} \\
\includegraphics[width=8.5cm]{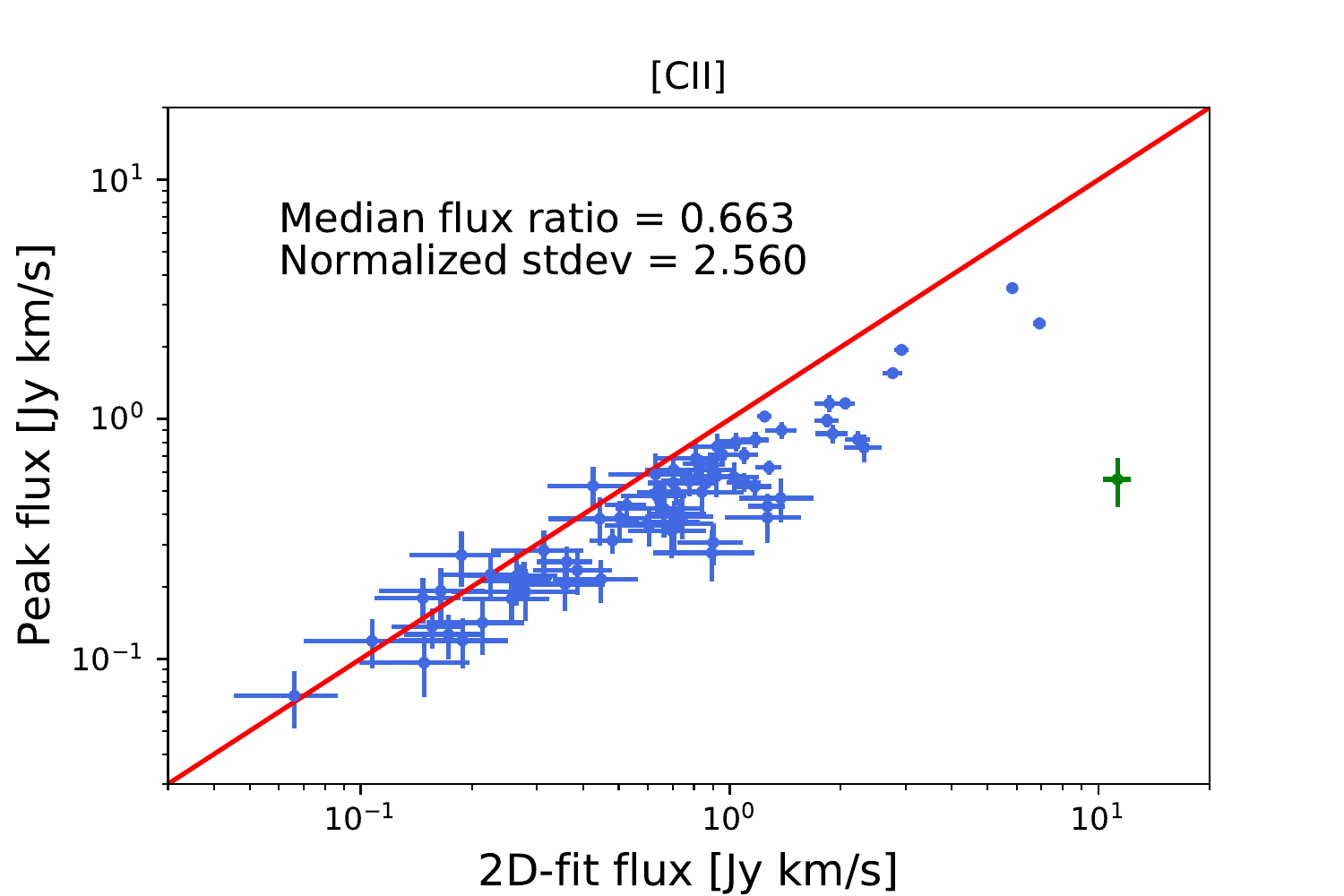} & \includegraphics[width=8.5cm]{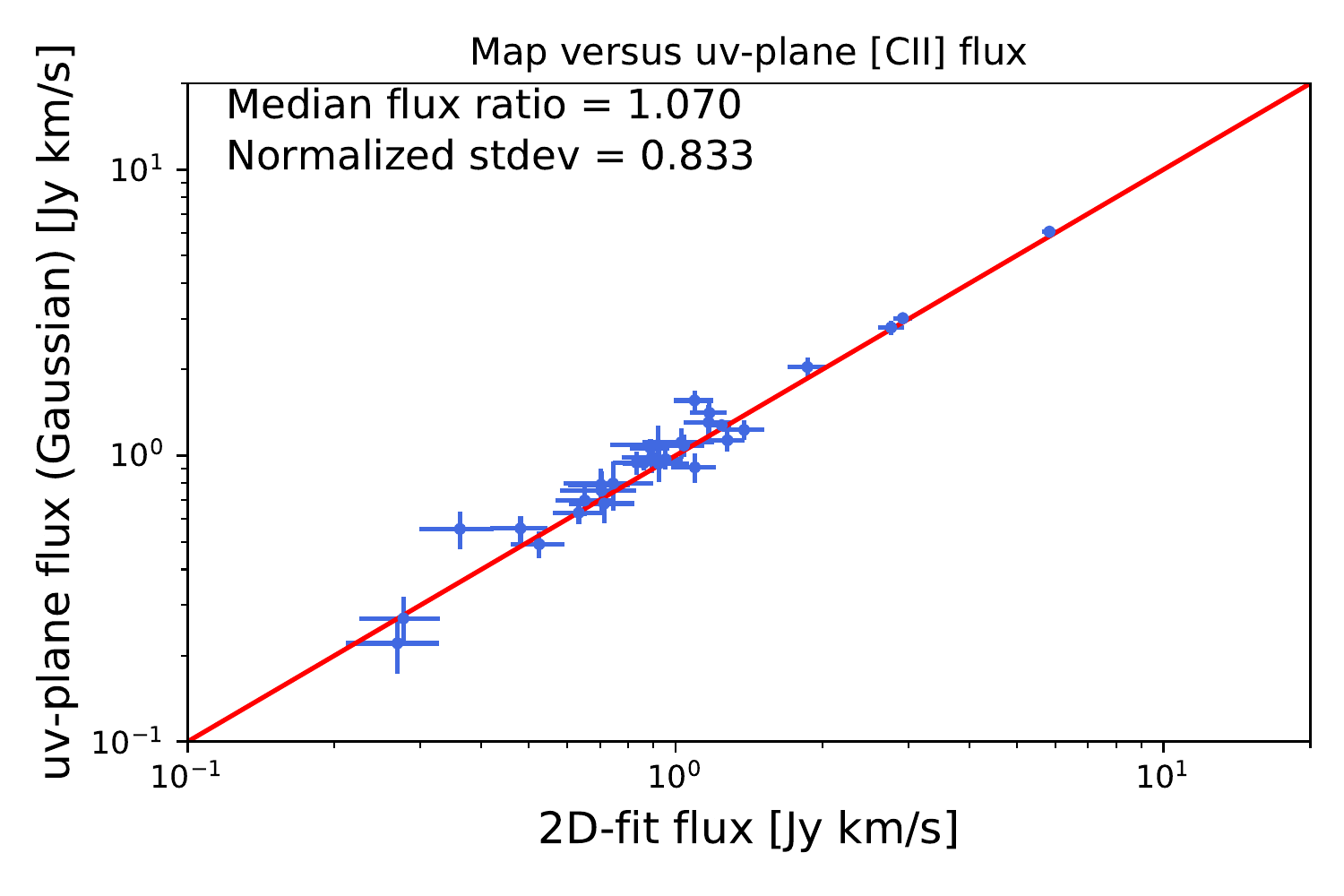} \\
\end{tabular}
\caption{\label{fig:comp_cii_phot} Comparison between our various [CII] line flux measurement methods described in Sect.\,\ref{sect:phot_cii}. The blue dots are our measurements and the red line is the one-to-one relation. The upper left, upper right, and lower left panels are the comparison between the 2D-fit flux (x-axis) and the aperture, 2-$\sigma$-clipped, and peak flux densities, respectively (y-axis). The vuds\_cosmos\_5101209780 object (in green) has a complex geometry and is discussed in Appendix\,\ref{sect:cii_multicomp}. The lower right panel shows the comparison between uv-plane fluxes and 2D-fit fluxes for single-component systems.}
\end{figure*}

The line flux of the detected [CII] sources was measured using the same methods as for the continuum (see Sect.\,\ref{sect:cont_phot}), but using the moment-0 maps instead of continuum maps. This allows us to reliably measure the line flux of spatially-resolved sources. As discussed in Sect.\,\ref{sect:cii_extr}, the conservative velocity windows used to build the moment-0 maps  minimize the flux loss from the high-velocity tails.

In Fig.\,\ref{fig:comp_cii_phot}, we compare the results obtained by our various photometric methods. Except the peak flux method, which systematically underestimates the flux of extended sources mainly located at the bright end of the sample (see Sect.\,\ref{sect:cont_phot}), we find a good overall agreement  between the 2D-fit, aperture, and 2-$\sigma$-clipped fluxes. However, we identified a small systematic offset of 7\,\% and 3\,\% compared to the 2D-fit flux for the aperture and 2-$\sigma$-clipped flux, respectively. The uncertainty-normalized difference defined in Eq.\,\ref{eq:phot_norm_diff} is 0.57 and 0.56, respectively. The systematic uncertainty between the methods is thus smaller than the typical 1-$\sigma$ flux uncertainties. Finally, we identified a strong outlier for which our various methods disagree with each others. It is a source with a close bright neighbor, which requires manual deblending (see Appendix\,\ref{sect:cii_multicomp} and \citet{Ginolfi2020b}.

We also compared our flux measurements performed in the map space with uv-derived values. These measurements were performed using the $uvmodelfit$ CASA routine. The full measurement process and the comparison of the results obtained assuming various profile models will be presented in Fujimoto et al. (in prep.). In this paper, we compare our measurements with the results obtained in the uv plane using an elliptical Gaussian model. In Fig.\,\ref{fig:comp_cii_phot} (lower right panel), we compare the fluxes from the map space and the uv plane for the sources properly fit in the uv space by a single component. The two methods agree within an offset of 7\,\%.

The results are presented in Table\,\ref{tab:cii_target}. The fluxes in the table are estimated using the 2D-fit method. For the non detections, we derived upper limits using the same methods as for the continuum (Sect.\,\ref{tab:uplim_cont}) but derived from the moment-0 maps. Since the line is not detected, we have to decide which channels to use to produce a moment-0 map. We chose to use a 300\,km/s window centered on the optical redshift, which is slightly above the median FWHM of the detected sample (see Sect.\,\ref{sect:FWHM_target_cii}). Of course, these upper limits do not apply if the line is particularly broad or if there is a catastrophic error on the reference optical redshift. The results are provided in Table\,\ref{tab:cii_uplims}.


\section{[CII] target properties}

\label{sect:props_cii}

In this section, we describe the basic [CII] properties of the target sources. In Sect\,\ref{sect:Nz_target_cii}, we present the redshift distribution of the detections and discuss the detection rate. The velocity offsets between [CII] and optical redshifts are presented in Sect.\,\ref{sect:z_offset}. We discuss the line width and the luminosities in Sect.\,\ref{sect:FWHM_target_cii} and \ref{sect:lum_cii}, respectively. Finally, we use the ALPINE sample to constrain the relation between the SFR and the [CII] luminosity (Sect.\,\ref{sect:cont_vs_cii}).

\begin{figure*}
\centering
\begin{tabular}{cc}
\includegraphics[width=8.5cm]{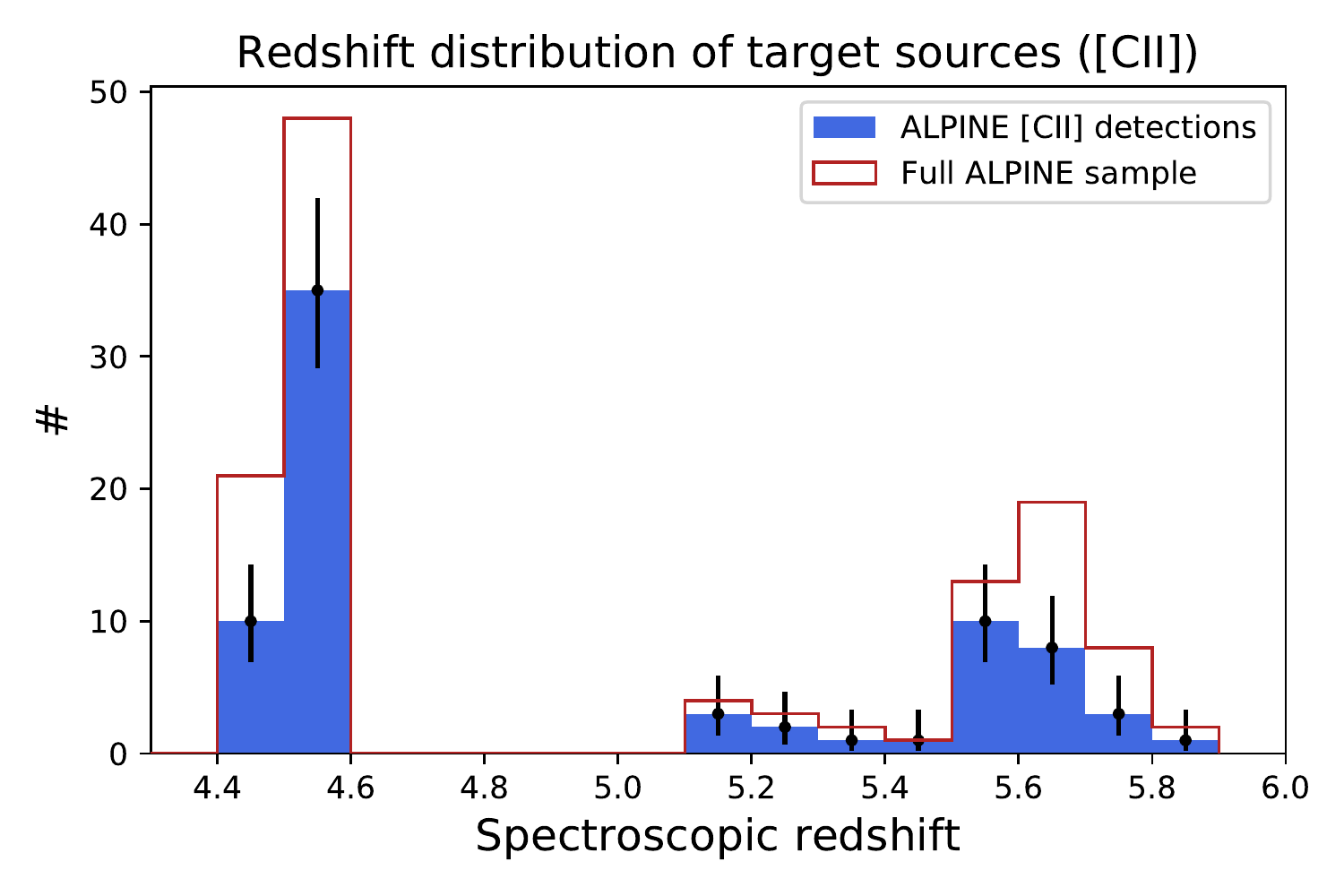} & \includegraphics[width=8.5cm]{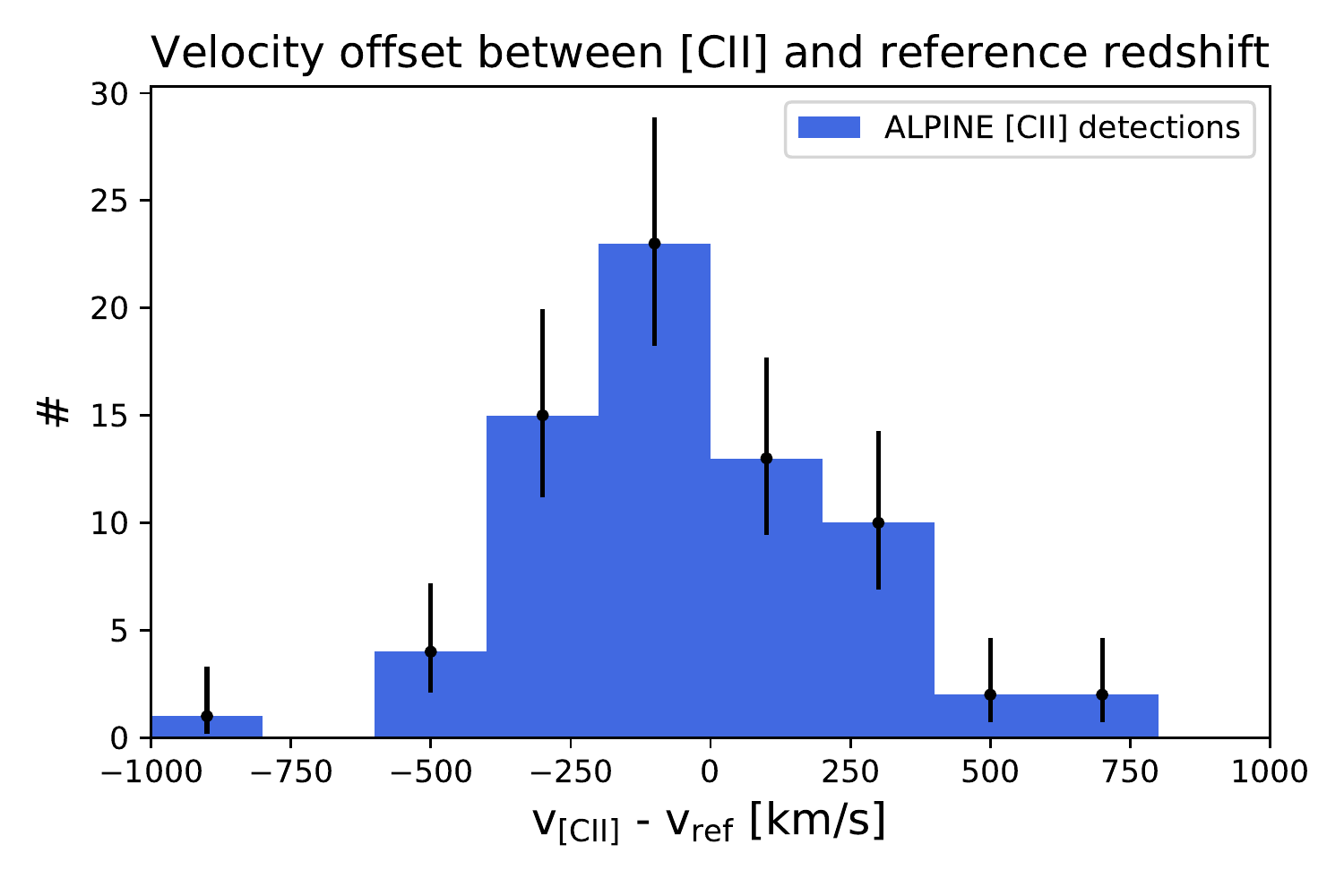} \\
\includegraphics[width=8.5cm]{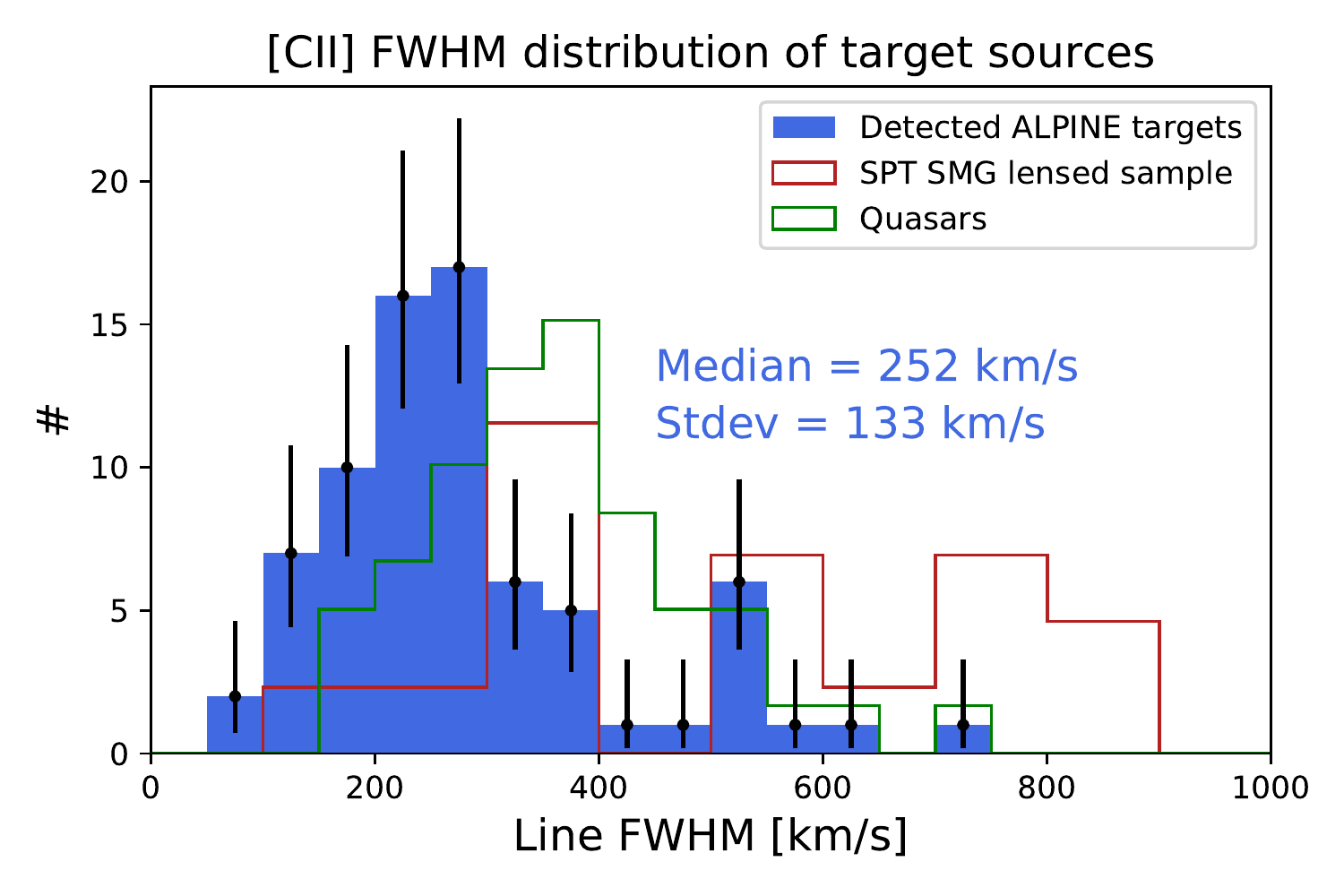} & \includegraphics[width=8.5cm]{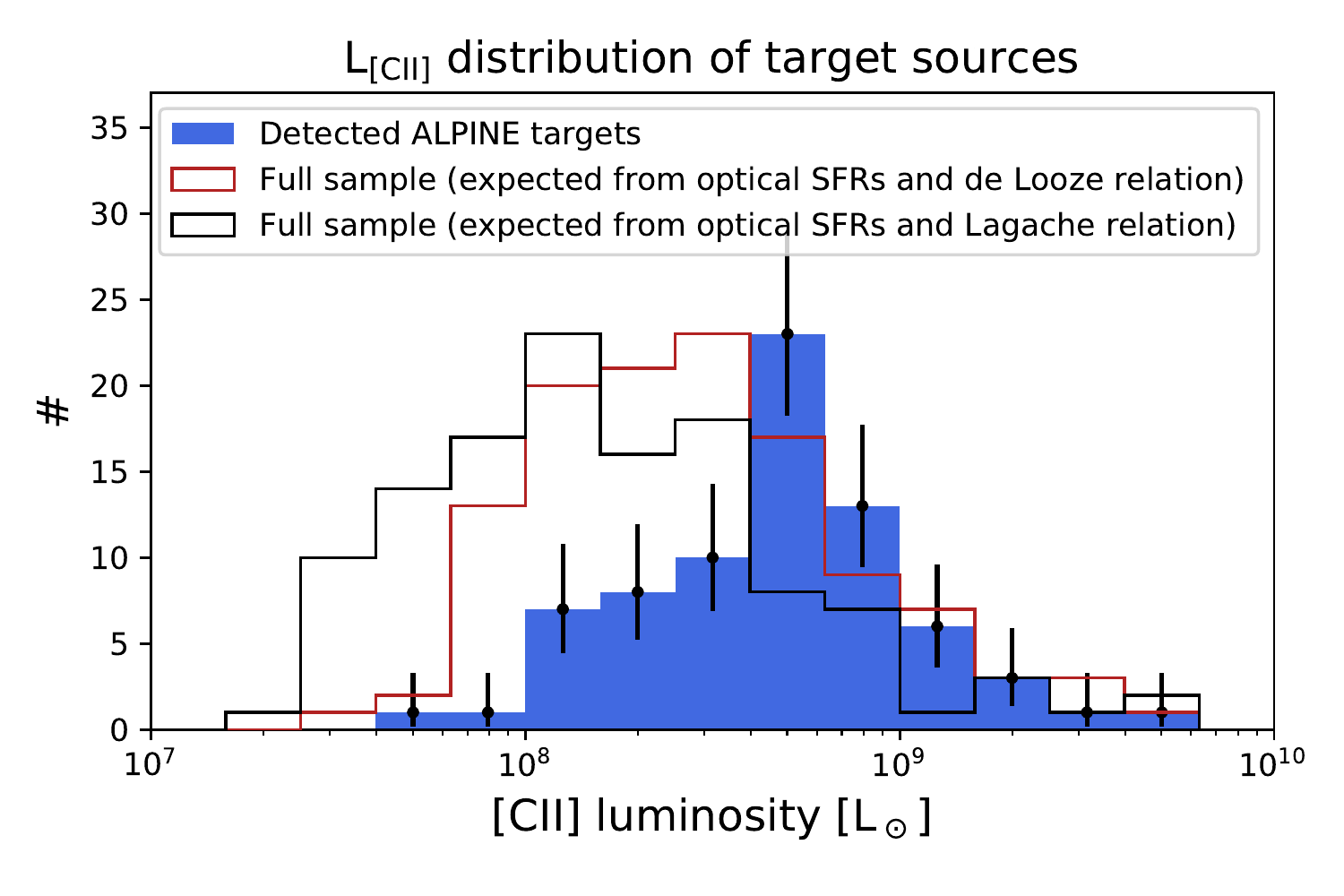} \\
\end{tabular}
\caption{\label{fig:cii_props} Upper left panel: Redshift distribution of the full sample (red open histogram) and the [CII] detections (blue filled histogram). Upper right panel: distribution of the velocity offset between the [CII] and the reference optical redshift. Lower left panel: Distribution of the [CII] line FWHM of the detected ALPINE sources (blue filled histogram) and comparison with the lensed galaxy sample of \citet[][red]{Gullberg2015} and the quasar sample of \citet[][green]{Decarli2018}. Lower right panel: Comparison between the [CII] luminosity distribution of the detections (blue filled histogram) and the expected values based on the SFRs determined by optical and near-infrared SED fitting and the \citet[][red open histogram]{De_Looze2014} and \citet[][black open histogram]{Lagache2018} relations.}
\end{figure*}

\subsection{Redshift distribution and detection rate}
\label{sect:Nz_target_cii}

In Fig.\,\ref{fig:cii_props} (upper left panel), we show the distribution of the reference optical redshifts of the full ALPINE sample and the redshift distribution of the subsample detected in [CII]. Overall, the distribution of the detections follows the full sample distribution. The detection rate below and above z=5 is 66\,\% and 58\,\%, respectively. This corresponds to a 1-$\sigma$ statistical fluctuation and cannot be considered as significant. This is not surprising, since our survey was built to be at constant [CII]-luminosity sensitivity and no variation with redshift of the average sensitivity in luminosity was found in the real data (Sect.\,\ref{sect:perf}).

\subsection{Offsets between optical and [CII] redshifts}

\label{sect:z_offset}

For the detected [CII] sources, we also compared our optical reference redshift and our new [CII] redshift (see Fig.\,\ref{fig:cii_props}, upper right panel). We found non negligible offsets up to 1000\,km/s. A posteriori, this justifies our choice to search for the line not only at the optical velocity, but in the entire side band (see Sect.\,\ref{sect:cii_extr}). These offsets cannot be explained by the uncertainties on the [CII] redshift, since they are usually smaller than 100\,km/s. The optical redshifts could suffer from several effects leading to a small inaccuracy. First, for sources without bright emission lines, redshifts are determined using the continuum and/or weak absorption features and could have a lack of precision \citep{Le_Fevre2015,Hasinger2018}. Other sources can have a very bright Lyman $\alpha$ line, which has a lot of weight in the determination of the redshift. While [CII] traces the gas in the galaxy and is very close to the systemic velocity, Lyman $\alpha$ radiative transfer is complex and can produce significant offsets \citep[e.g.,][]{Steidel2010,Faisst2016,Matthee2017,Carniani2018a,Verhamme2018,Matthee2019,Pahl2019, Behrens2019}. A complete analysis of the origins of these offsets is presented in \citet{Faisst2020}. The physics of the velocity offsets between [CII] and Lyman $\alpha$ is discussed in \citet{Cassata2020}.

\subsection{[CII] line width}

\label{sect:FWHM_target_cii}

The [CII] line width is also an interesting physical constraint, since it is linked to the dynamical mass and the size. The constraints on the dynamical masses will be discussed in more details in Dessauges-Zavadsky et al. (in prep.). Before ALPINE, studies of the [CII] line width have been mainly performed using lensed galaxy samples \citep[e.g.,][]{Gullberg2015}. However, these samples are biased toward more star-forming systems, which could also be more massive, and possibly more compact systems. ALPINE is probing less extreme galaxies and we thus expect a narrower line width on average. 

In Fig.\,\ref{fig:cii_props} (lower left panel), we compare our [CII] FWHM distribution with the SPT SMG lensed sample of \citet{Gullberg2015}. As expected, our sources have much narrower lines with a median FWHM of 252$\pm$13\,km/s versus 541$\pm$110\,km/s for SPT SMGs. At fixed integrated flux, we could be slightly biased against broader lines. For instance, \citet{Kohandel2019} showed using numerical simulations that edge-on systems tend to be more difficult to detect because of their broader lines, but it is unlikely to be the cause of the large difference between the ALPINE sources and quasars or SMGs. However, the continuum stacking of non detections tends to indicate that they are mainly lower SFR systems (see Sect.\,\ref{sect:cont_vs_cii}) and thus have probably a low mass and low FWHM. The average FHWM of our sample is also smaller than the 355$\pm$18\,km/s measured by \citet{Decarli2018} in z$>$5.94 in quasar hosts, which are also expected to be particularly massive systems. 

We detected two sources with particularly narrow lines: CANDELS\_GOODSS\_42 with a FWHM of 63\,km/s and vuds\_cosmos\_510596653 with 62\,km/s. CANDELS\_GOODSS\_42 is a low-mass object (log(M$_\star$/M$_\odot$) = 9.3$\pm$0.3, \citealt{Faisst2020}). The line is close to the edge of the spectral window, but the signal seems to go back to the baseline level before the very last channel. However, we cannot firmly exclude an edge effect. vuds\_cosmos\_510596653 is too faint at short wavelength to reliably estimate a stellar mass from ancillary data, which is compatible with a low-mass and thus low-dispersion system.

\subsection{[CII] luminosity distribution}

\label{sect:lum_cii}

The [CII] luminosity L$_{\rm [CII]}$ of our targets was computed using the formula provided in \citet{Carilli2013}:
\begin{equation}
L_{\rm [CII]} = 1.04 \times 10^{-3} \times I_{\rm [CII]} D_L^2 \, \nu_{\rm obs} \, \frac{L_\odot}{\textrm{Jy\,km/s} \, \textrm{Mpc}^2 \, \textrm{GHz}}, 
\end{equation}
where $I_{\rm [CII]}$ is the [CII] flux, $D_L$ is the luminosity distance, and $\nu_{\rm obs}$ is the observed frequency. The CMB effect on the [CII] line measurements is expected to be negligible in the ALPINE redshift range \citep{Vallini2015,Lagache2018}. We found a median luminosity of $(4.8 \pm 0.4) \times 10^{8}$\,L$_\odot$, which is very close to $3.6\times10^{8}$\,L$_\odot$ found for the pilot sample of 10 sources of \citet{Capak2015}. We can also compare the results with the expected luminosity distributions based on empirical SFR-[CII] relations. We used the relation of \citet[][fit of HII/starburst galaxies]{De_Looze2014} measured in the local Universe and the relation of \citet{Lagache2018} calibrated combining the pre-ALPINE [CII] detections and semi-analytical modeling. The SFRs were taken from the UV to near-IR SED fitting of the ancillary data \citep{Faisst2020}. We did not take into account any scatter in the relations nor the formal uncertainties on SFR from SED fitting for this simple comparison. The results are presented in Fig.\,\ref{fig:cii_props} (lower right panel). The bulk of our detections are more luminous than the median expected value of the sample. As expected, we thus tend to miss mainly the faintest systems. We observe a significant excess of luminous [CII] objects compared to the distribution expected from the combination of the \citet{Lagache2018} relation and the SFRs based on optical and near-infrared SED fitting. This excess decreases to 1.2\,$\sigma$ if we use the \citet{De_Looze2014} relation instead.


\subsection{Obscured SFR as a function of [CII] luminosity}

\label{sect:cont_vs_cii}

\begin{figure}
\centering
\includegraphics[width=9cm]{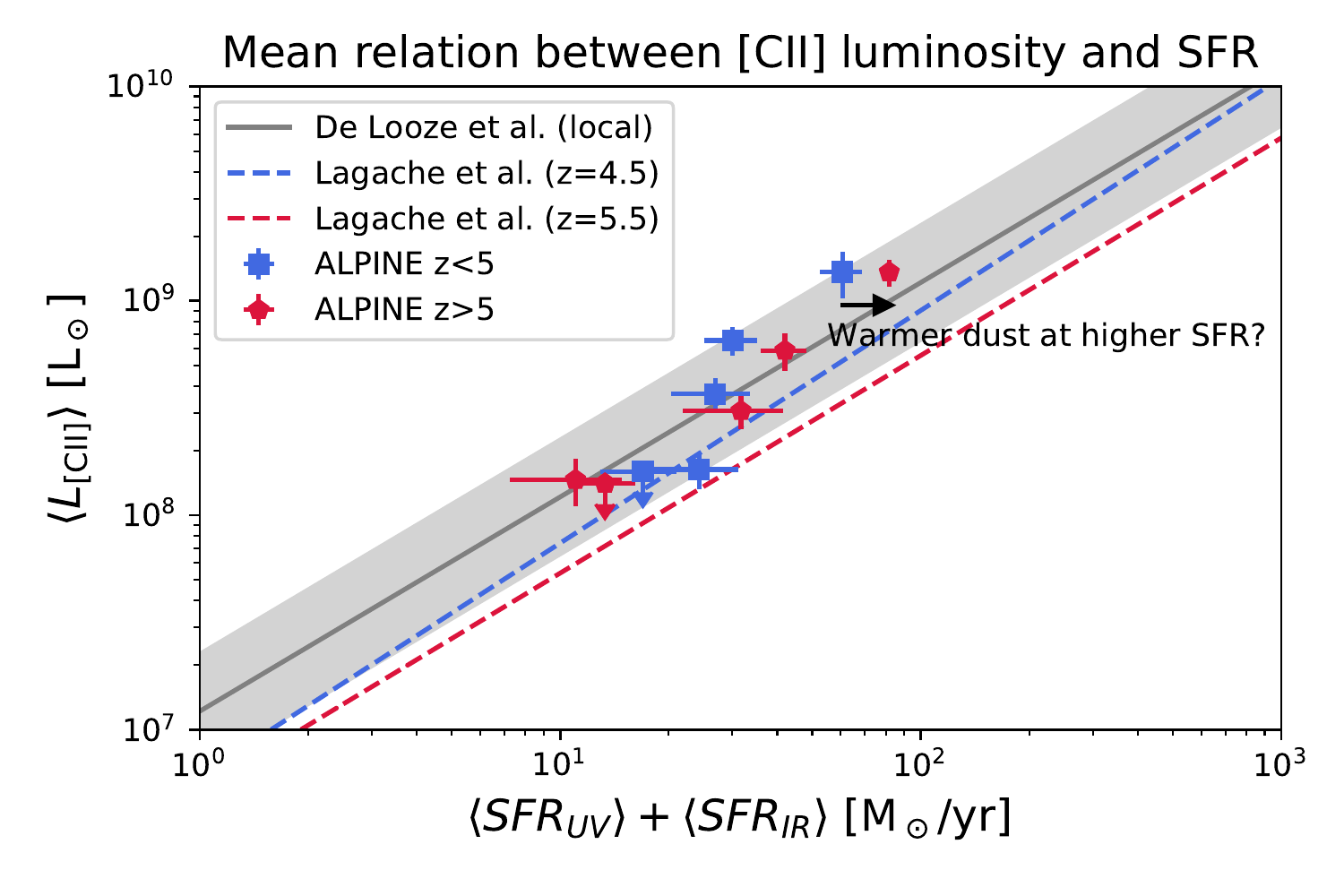}
\caption{\label{fig:sfr_cii_stack} Mean [CII] luminosity as a function of SFR. The obscured SFR is measured by stacking the ALPINE continuum maps for various subsamples selected in redshift and [CII] luminosity. It is combined with the mean uncorrected UV SFR from ancillary data \citep{Faisst2020}. The method is described in Sect.\,\ref{sect:cont_vs_cii}. The blue squares are our measurements at z$<$5 and the red diamonds are for z$>$5. The y-axis upper limits correspond to the mean total SFR of nondetected [CII] samples (the y position corresponds to mean of the secure [CII] upper limits). The blue and red dashed lines indicate the relation of \citet{Lagache2018} at z=4.5 and z=5.5, respectively. The gray shaded area is the 1-$\sigma$ region of the relation of \citet{De_Looze2014}. The black arrow represents the shift in SFR if we use the warmer \citet{Schreiber2018a} SED templates instead of the \citet{Bethermin2017} ones to determine the infrared luminosity.}
\end{figure}

\begin{table*}
\caption{\label{tab:sfr_cii_stack} Mean total SFR measured combining ALPINE continuum stacking and ancillary UV data (see Sect\,\ref{sect:cont_vs_cii}) in various redshift and [CII] luminosity bins. }
\centering
\begin{tabular}{ccccccc}
\hline
\hline
 & $\langle L_{\rm [CII]} \rangle$ & $\langle z \rangle$ & $\langle S_{\nu} \rangle$ & $\langle {\rm SFR_{\rm IR}} \rangle$ & $\langle {\rm SFR_{\rm UV}} \rangle$ & $\langle {\rm SFR_{\rm tot}} \rangle$ \\
 & 10$^8$ L$_\odot$ &  & mJy & M$_\odot$/yr & M$_\odot$/yr  & M$_\odot$/yr  \\
\hline
\multicolumn{7}{c}{$z<5$} \\
\hline
Non detections & $<$1.6 & 4.49 & 0.048$\pm$0.028 & 7$\pm$4 & 10$\pm$1 & 17$\pm$4\\
8.00<log($L_{\rm [CII]}$)<8.33 & 1.6 & 4.53 & 0.051$\pm$0.039 & 7$\pm$6 & 16$\pm$4 & 24$\pm$7\\
8.33<log($L_{\rm [CII]}$)<8.67 & 3.7 & 4.51 & 0.087$\pm$0.041 & 13$\pm$6 & 15$\pm$1 & 27$\pm$6\\
8.67<log($L_{\rm [CII]}$)<9.00 & 6.5 & 4.51 & 0.079$\pm$0.033 & 11$\pm$5 & 19$\pm$2 & 30$\pm$5\\
9.00<log($L_{\rm [CII]}$)<9.33 & 13.6 & 4.54 & 0.289$\pm$0.049 & 42$\pm$7 & 18$\pm$3 & 61$\pm$8\\
\hline
\multicolumn{7}{c}{$z>5$}\\
\hline
Non detections & $<$1.4 & 5.61 & 0.014$\pm$0.012 & 3$\pm$2 & 10$\pm$1 & 13$\pm$3\\
8.00<log($L_{\rm [CII]}$)<8.33 & 1.5 & 5.63 & $<$0.048 & $<$9 & 11$\pm$2 & 11$\pm$4\\
8.33<log($L_{\rm [CII]}$)<8.67 & 3.1 & 5.54 & 0.031$\pm$0.030 & 6$\pm$6 & 25$\pm$8 & 31$\pm$10\\
8.67<log($L_{\rm [CII]}$)<9.00 & 5.9 & 5.45 & 0.109$\pm$0.028 & 21$\pm$5 & 21$\pm$3 & 42$\pm$6\\
9.00<log($L_{\rm [CII]}$)<9.33 & 13.6 & 5.49 & 0.269$\pm$0.008 & 53$\pm$2 & 29$\pm$4 & 82$\pm$4\\
\hline

\end{tabular}

\end{table*}

To understand the excess of luminous [CII] objects in our sample, we derived the average SFR in various [CII] luminosity and redshift bins. These mean SFRs are determined by combining the UV rest-frame data \citep{Faisst2020} and the mean obscured SFRs determined using a stacking of ALPINE continuum data. This analysis is based only on stacked ALMA data and does not explore the scatter. A more comprehensive study combining ALPINE data and ancillary data is presented in Schaerer et al. (2020). 

We first define four [CII] luminosity bins between 10$^8$ and 10$^{9.33}$\,L$_\odot$ with the same logarithmic width. We chose to stack by bins of [CII] luminosity and not by bins of continuum flux density, since we have many more [CII] detections. In addition, the continuum is not affected by the velocity offsets presented in Sect.\, \ref{sect:z_offset} and can be stacked knowing a priori the exact redshift and line width. The stacking of the undetected [CII] lines would have been more difficult, since we do not know in which frequency range it is located and using a very broad frequency window would lead to poor S/Ns. A few outliers are below or above the chosen [CII] luminosity bins, but the stacking of less than three objects would not provide meaningful results. We also split the sample into two redshift subsamples, below and above z=5. We then stacked the continuum of all the sources of a given luminosity bin in image space after masking the continuum nontarget detections. We also stacked the ALMA continuum of the [CII] non detections to estimate their mean SFR. Since the beam size and the source size can vary significantly from one source to the other, the resulting stacked profile cannot be well fit by a single elliptical Gaussian. We thus decided to use the aperture method to derive the flux density (see Sect.\,\ref{sect:cont_phot}). To be consistent, we used the [CII] luminosities derived using the same technique. Finally, we derived the uncertainties using a bootstrap technique, which takes into account both the noise and the intrinsic population variance of the sources \citep{Bethermin2012b}. The full description of the stacking technique and its validation will be described in Khusanova et al. (in prep.). 

To derive the mean SFR of each subsample, we convert the stacked continuum fluxes into L$_{\rm IR}$ using the method described in Sect.\,\ref{sect:sed} assuming the mean redshift of the subsample. The far-UV luminosity, L$_{FUV}$, is provided by the ancillary ALPINE catalog presented in \citet{Faisst2020}. The full SFR is derived combining the UV and infrared SFRs \citep{Madau2014}:
\begin{equation}
{\rm SFR}_{\rm tot} = {\rm SFR}_{\rm IR} + {\rm SFR}_{\rm UV} = \kappa_{\rm IR} L_{\rm IR} + \kappa_{\rm FUV} L_{\rm FUV},
\end{equation}
where SFR$_{\rm tot}$ is the total SFR, that is the sum of the dust-obscured component SFR$_{\rm IR}$ and the unobscured one SFR$_{\rm UV}$. $\kappa_{\rm IR}$ and $\kappa_{\rm FUV}$ are conversion factors from luminosity to SFR and their values are 1.02$\times$10$^{-10}$\,M$_\odot$yr$^{-1}$L$_\odot^{-1}$ and 1.47$\times$10$^{-10}$\,M$_\odot$yr$^{-1}$L$_\odot^{-1}$, respectively.\footnote{The $\kappa$ coefficients from \citet{Madau2014} have been converted from Salpeter to Chabrier IMF by dividing them by a factor of 1.7.}

Our results are presented in Fig.\,\ref{fig:sfr_cii_stack} and in Table\,\ref{tab:sfr_cii_stack}. In addition to detections, we performed the same stacking procedure on the non detections and computed the mean of the secure upper limits on their luminosity. All our results are within the scatter of the \citet{De_Looze2014} and \citet{Lagache2018} relations\footnote{The scatter on the \citet{Lagache2018} is not shown on the figure for clarity, but it is larger than \citet{De_Looze2014} ($\sim$0.55\,dex).}. At low luminosity (log(L$_{\rm CII}$)$<$8.66), our mean data points nicely agree with the center of their relation. However, at high [CII] fluxes (log(L$_{\rm CII}$)$>$8.66), our mean stacked z$<$5 data point is on the 1-$\sigma$ upper envelop of these relations. There is thus a mild average [CII] excess in the most luminous sources of our sample. Even if it is rare, [CII]-excesses have been observed in the local Universe \citep{Smirnova-Pinchukova2019}, but we usually observe a deficit in bright sources \citep{Diaz-Santos2013}. This apparent [CII] excess could also be an SED effect. Indeed, if brighter sources have warmer dust than the average stacked value, their obscured SFR would be underestimated. This explanation is compatible with the warmer dust SED measured by stacking for the SFR$>$100\,M$_\odot$/yr sources at the same redshift in the COSMOS field (see Sect.\,\ref{sect:seds_caveats}). If we use the \citet{Schreiber2018a} SED template, which fits better the high-SFR stacked SED, instead of the \citet{Bethermin2017} one, we find that SFR$_{\rm IR}$ is a factor of 1.4 higher. In Fig.\,\ref{fig:sfr_cii_stack}, we illustrate the impact of a warmer dust using a black arrow, showing that it can explain this small [CII] excess. Finally, we tested if this excess could be caused by merging systems by excluding objects classified as merger (type=2) in \citet{Le_Fevre2019} and redoing the full procedure, but the excess remained. These results and a detailed analysis of the [CII]--SFR relation at high redshift are discussed in depth in \citet{Schaerer2020}. 


\section{Conclusion}

In this paper, we presented the data processing and  the catalog construction of the ALPINE ALMA large program. The performance of our survey is fully compatible with our initial goal. Our main technical results are: 
\begin{itemize}
\item After flagging a few badly calibrated antennae, we produced continuum and [CII] moment-0 maps and reached a typical sensitivity of 30\,$\mu$Jy RMS in continuum and 0.14\,Jy\,km/s for [CII] in a 235\,km/s bandwidth. The average beam size is 1.13"$\times$0.85". 
\item We investigated the stability of quasars used as flux calibrators during the observation campaign and found that the fluctuations between two successive calibrator survey observations are lower than the standard 10\,\% calibration uncertainty.
\item We detected 23 of our 118 targets in continuum above 3.5\,$\sigma$ threshold corresponding to a 95\,\% purity. In the rest of the field, we had to extract sources above 5\,$\sigma$ because of the lower number density of emitters and detected 57 nontarget additional continuum sources.
\item We measured the flux densities of our detections using five different methods. They well agree with each other, except the one based on peak flux density because of the large fraction of marginally resolved objects in our sample. 
\item We performed Monte Carlo simulations to estimate the completeness and the flux boosting for the nontarget continuum sources and proposed a simple way to derive them as a function of the source flux density, size, and the local noise.
\item After adjusting our extraction algorithm to reach 95\,\% purity, we detected 75 of our 118 targets in [CII]. Similarly to the continuum, we checked the robustness of the photometry by comparing five different methods.
\end{itemize}

In addition to these technical results, we obtained promising first scientific results. Our main findings are:
\begin{itemize}
\item To determine the conversion factor from the 158\,$\mu$m continuum luminosity to the total infrared luminosity (L$_{158\,\mu m}$ = 0.093 L$_{\rm IR}$), we measured average dust SEDs by stacking single-dish data at the position of COSMOS photometric sources similar to ALPINE ones.
\item The target sources detected in continuum have a median flux density of 0.26\,mJy, a median L$_{\rm IR}$ of $4.4\times10^{11}$\,L$_\odot$, and a median stellar mass of $1.1\times10^{10}$\,M$_\odot$. As expected, the detections are among the most massive and star-forming systems of our sample.
\item The nontarget continuum detections have a mean redshift of 2.5$\pm$0.2 (median = 2.3) and are mainly lower-redshift sources without a physical connection with the ALPINE targets. 
\item We derived number counts probing the 0.35 to 5.6\,mJy range. We identified a slope break in the counts around 3\,mJy and estimated that the contribution of $>$0.35\,mJy sources to the CIB is 40--69\,\%.
\item The detected [CII] targets have a median [CII] luminosity of $4.8\times10^{8}$\,L$_\odot$ and a median FWHM of 252\,km/s. We also observed significant offsets between the optical and [CII] redshifts (up to 1000\,km/s).
\item We measured the mean obscured SFR in various [CII] luminosity bins by stacking ALPINE continuum data and combined it with the unobscured SFR from ancillary rest-frame UV data. Our data agrees with the local and predicted SFR-L$_{\rm [CII]}$ relations of \citet{De_Looze2014} and \citet{Lagache2018}.
\end{itemize}
A series of companion papers discuss the other scientific results of ALPINE \citep[e.g.,][]{Le_Fevre2019,Jones2020,Ginolfi2020b,Ginolfi2020,Faisst2019,Cassata2020,Schaerer2020,Fudamoto2020,Fujimoto2020,Dessauges2020,Yan2020,Loiacono2020,Gruppioni2020,Romano2020}. The ALPINE products are publicly available at \url{https://cesam.lam.fr/a2c2s/}.\\

\begin{acknowledgements}

We thanks the ALMA observatory staff for their support and in particular Edwige Chapillon, our ALMA contact scientist, for her very useful advice. MB acknowledges Mladan Novak for his excellent advice about the photometry in interferometric surveys. This paper makes use of the following ALMA data: ADS/JAO.ALMA\#2017.1.00428.L. ALMA is a partnership of ESO (representing its member states), NSF (USA) and NINS (Japan), together with NRC (Canada), MOST and ASIAA (Taiwan), and KASI (Republic of Korea), in cooperation with the Republic of Chile. The Joint ALMA Observatory is operated by ESO, AUI/NRAO and NAOJ. This research made use of Astropy,\footnote{http://www.astropy.org} a community-developed core Python package for Astronomy \citep{astropy:2013,astropy:2018}. This program receives funding from the CNRS national program Cosmology and Galaxies. A.C., C.G., F.L., F.P. and M.T. acknowledge the support from grant PRINMIUR 2017 20173ML3WW\_001. DR acknowledges support from the National Science Foundation under grant numbers AST-1614213 and AST-1910107 and from the Alexander von Humboldt Foundation through a Humboldt Research Fellowship for Experienced Researchers. G.C.J. acknowledges ERC Advanced Grant 695671 "QUENCH'' and support by the Science and Technology Facilities Council (STFC). E.I.\ acknowledges partial support from FONDECYT through grant N$^\circ$\,1171710. GL acknowledges support from the European Research Council (ERC) under the European Union’s Horizon 2020 research and innovation programme (project CONCERTO, grant agreement No 788212) and from the Excellence Initiative of Aix-Marseille University-A*Midex, a French “Investissements d’Avenir” programme. This paper is dedicated to the memory of Olivier Le F\`evre, PI of the ALPINE survey.

\end{acknowledgements}

\bibliographystyle{aa}

\bibliography{biblio}

\begin{thebibliography}{140}
\expandafter\ifx\csname natexlab\endcsname\relax\def\natexlab#1{#1}\fi

\bibitem[{{{\'A}lvarez-M{\'a}rquez} {et~al.}(2019){{\'A}lvarez-M{\'a}rquez},
  {Burgarella}, {Buat}, {Ilbert}, \&
  {P{\'e}rez-Gonz{\'a}lez}}]{Alvarez-Marquez2019}
{{\'A}lvarez-M{\'a}rquez}, J., {Burgarella}, D., {Buat}, V., {Ilbert}, O., \&
  {P{\'e}rez-Gonz{\'a}lez}, P.~G. 2019, \aap, 630, A153

\bibitem[{{{\'A}lvarez-M{\'a}rquez} {et~al.}(2016){{\'A}lvarez-M{\'a}rquez},
  {Burgarella}, {Heinis}, {Buat}, {Lo Faro}, {B{\'e}thermin},
  {L{\'o}pez-Fort{\'\i}n}, {Cooray}, {Farrah}, {Hurley}, {Ibar}, {Ilbert},
  {Koekemoer}, {Lemaux}, {P{\'e}rez-Fournon}, {Rodighiero}, {Salvato}, {Scott},
  {Taniguchi}, {Vieira}, \& {Wang}}]{Alvarez-Marquez2016}
{{\'A}lvarez-M{\'a}rquez}, J., {Burgarella}, D., {Heinis}, S., {et~al.} 2016,
  \aap, 587, A122

\bibitem[{{Aravena} {et~al.}(2016){Aravena}, {Decarli}, {Walter}, {Da Cunha},
  {Bauer}, {Carilli}, {Daddi}, {Elbaz}, {Ivison}, {Riechers}, {Smail},
  {Swinbank}, {Weiss}, {Anguita}, {Assef}, {Bell}, {Bertoldi}, {Bacon},
  {Bouwens}, {Cortes}, {Cox}, {G{\'o}nzalez-L{\'o}pez}, {Hodge}, {Ibar},
  {Inami}, {Infante}, {Karim}, {Le Le F{\`e}vre}, {Magnelli}, {Ota}, {Popping},
  {Sheth}, {van der Werf}, \& {Wagg}}]{Aravena2016b}
{Aravena}, M., {Decarli}, R., {Walter}, F., {et~al.} 2016, \apj, 833, 68

\bibitem[{{Aretxaga} {et~al.}(2011){Aretxaga}, {Wilson}, {Aguilar}, {Alberts},
  {Scott}, {Scoville}, {Yun}, {Austermann}, {Downes}, {Ezawa}, {Hatsukade},
  {Hughes}, {Kawabe}, {Kohno}, {Oshima}, {Perera}, {Tamura}, \&
  {Zeballos}}]{Aretxaga2011}
{Aretxaga}, I., {Wilson}, G.~W., {Aguilar}, E., {et~al.} 2011, \mnras, 415,
  3831

\bibitem[{{Astropy Collaboration} {et~al.}(2013{\natexlab{a}}){Astropy
  Collaboration}, {Robitaille}, {Tollerud}, {Greenfield}, {Droettboom}, {Bray},
  {Aldcroft}, {Davis}, {Ginsburg}, {Price-Whelan}, {Kerzendorf}, {Conley},
  {Crighton}, {Barbary}, {Muna}, {Ferguson}, {Grollier}, {Parikh}, {Nair},
  {Unther}, {Deil}, {Woillez}, {Conseil}, {Kramer}, {Turner}, {Singer}, {Fox},
  {Weaver}, {Zabalza}, {Edwards}, {Azalee Bostroem}, {Burke}, {Casey},
  {Crawford}, {Dencheva}, {Ely}, {Jenness}, {Labrie}, {Lim}, {Pierfederici},
  {Pontzen}, {Ptak}, {Refsdal}, {Servillat}, \& {Streicher}}]{astropy_paper}
{Astropy Collaboration}, {Robitaille}, T.~P., {Tollerud}, E.~J., {et~al.}
  2013{\natexlab{a}}, \aap, 558, A33

\bibitem[{{Astropy Collaboration} {et~al.}(2013{\natexlab{b}}){Astropy
  Collaboration}, {Robitaille}, {Tollerud}, {Greenfield}, {Droettboom}, {Bray},
  {Aldcroft}, {Davis}, {Ginsburg}, {Price-Whelan}, {Kerzendorf}, {Conley},
  {Crighton}, {Barbary}, {Muna}, {Ferguson}, {Grollier}, {Parikh}, {Nair},
  {Unther}, {Deil}, {Woillez}, {Conseil}, {Kramer}, {Turner}, {Singer}, {Fox},
  {Weaver}, {Zabalza}, {Edwards}, {Azalee Bostroem}, {Burke}, {Casey},
  {Crawford}, {Dencheva}, {Ely}, {Jenness}, {Labrie}, {Lim}, {Pierfederici},
  {Pontzen}, {Ptak}, {Refsdal}, {Servillat}, \& {Streicher}}]{astropy:2013}
{Astropy Collaboration}, {Robitaille}, T.~P., {Tollerud}, E.~J., {et~al.}
  2013{\natexlab{b}}, \aap, 558, A33

\bibitem[{{Behrens} {et~al.}(2018){Behrens}, {Pallottini}, {Ferrara},
  {Gallerani}, \& {Vallini}}]{Behrens2018}
{Behrens}, C., {Pallottini}, A., {Ferrara}, A., {Gallerani}, S., \& {Vallini},
  L. 2018, \mnras, 477, 552

\bibitem[{{Behrens} {et~al.}(2019){Behrens}, {Pallottini}, {Ferrara},
  {Gallerani}, \& {Vallini}}]{Behrens2019}
{Behrens}, C., {Pallottini}, A., {Ferrara}, A., {Gallerani}, S., \& {Vallini},
  L. 2019, \mnras, 486, 2197

\bibitem[{{Berta} {et~al.}(2013){Berta}, {Lutz}, {Santini}, {Wuyts}, {Rosario},
  {Brisbin}, {Cooray}, {Franceschini}, {Gruppioni}, {Hatziminaoglou}, {Hwang},
  {Le Floc'h}, {Magnelli}, {Nordon}, {Oliver}, {Page}, {Popesso}, {Pozzetti},
  {Pozzi}, {Riguccini}, {Rodighiero}, {Roseboom}, {Scott}, {Symeonidis},
  {Valtchanov}, {Viero}, \& {Wang}}]{Berta2013}
{Berta}, S., {Lutz}, D., {Santini}, P., {et~al.} 2013, \aap, 551, A100

\bibitem[{{B{\'e}thermin} {et~al.}(2015{\natexlab{a}}){B{\'e}thermin}, {Daddi},
  {Magdis}, {Lagos}, {Sargent}, {Albrecht}, {Aussel}, {Bertoldi}, {Buat},
  {Galametz}, {Heinis}, {Ilbert}, {Karim}, {Koekemoer}, {Lacey}, {Le Floc'h},
  {Navarrete}, {Pannella}, {Schreiber}, {Smol{\v c}i{\'c}}, {Symeonidis}, \&
  {Viero}}]{Bethermin2015a}
{B{\'e}thermin}, M., {Daddi}, E., {Magdis}, G., {et~al.} 2015{\natexlab{a}},
  \aap, 573, A113

\bibitem[{{B{\'e}thermin} {et~al.}(2012{\natexlab{a}}){B{\'e}thermin}, {Daddi},
  {Magdis}, {Sargent}, {Hezaveh}, {Elbaz}, {Le Borgne}, {Mullaney}, {Pannella},
  {Buat}, {Charmandaris}, {Lagache}, \& {Scott}}]{Bethermin2012c}
{B{\'e}thermin}, M., {Daddi}, E., {Magdis}, G., {et~al.} 2012{\natexlab{a}},
  \apjl, 757, L23

\bibitem[{{B{\'e}thermin} {et~al.}(2015{\natexlab{b}}){B{\'e}thermin}, {De
  Breuck}, {Sargent}, \& {Daddi}}]{Bethermin2015b}
{B{\'e}thermin}, M., {De Breuck}, C., {Sargent}, M., \& {Daddi}, E.
  2015{\natexlab{b}}, \aap, 576, L9

\bibitem[{{B{\'e}thermin} {et~al.}(2012{\natexlab{b}}){B{\'e}thermin}, {Le
  Floc'h}, {Ilbert}, {Conley}, {Lagache}, {Amblard}, {Arumugam}, {Aussel},
  {Berta}, {Bock}, {Boselli}, {Buat}, {Casey}, {Castro-Rodr{\'{\i}}guez},
  {Cava}, {Clements}, {Cooray}, {Dowell}, {Eales}, {Farrah}, {Franceschini},
  {Glenn}, {Griffin}, {Hatziminaoglou}, {Heinis}, {Ibar}, {Ivison},
  {Kartaltepe}, {Levenson}, {Magdis}, {Marchetti}, {Marsden}, {Nguyen},
  {O'Halloran}, {Oliver}, {Omont}, {Page}, {Panuzzo}, {Papageorgiou},
  {Pearson}, {P{\'e}rez-Fournon}, {Pohlen}, {Rigopoulou}, {Roseboom},
  {Rowan-Robinson}, {Salvato}, {Schulz}, {Scott}, {Seymour}, {Shupe}, {Smith},
  {Symeonidis}, {Trichas}, {Tugwell}, {Vaccari}, {Valtchanov}, {Vieira},
  {Viero}, {Wang}, {Xu}, \& {Zemcov}}]{Bethermin2012b}
{B{\'e}thermin}, M., {Le Floc'h}, E., {Ilbert}, O., {et~al.}
  2012{\natexlab{b}}, \aap, 542, A58

\bibitem[{{Bethermin} {et~al.}(2017){Bethermin}, {Wu}, {Lagache}, {Dor{\'e}},
  {Wang}, \& {Cousin}}]{Bethermin2017}
{Bethermin}, M., {Wu}, H.-Y., {Lagache}, G., {et~al.} 2017, \aap, to be sub.

\bibitem[{{Bouwens} {et~al.}(2016){Bouwens}, {Aravena}, {Decarli}, {Walter},
  {da Cunha}, {Labb{\'e}}, {Bauer}, {Bertoldi}, {Carilli}, {Chapman}, {Daddi},
  {Hodge}, {Ivison}, {Karim}, {Le Fevre}, {Magnelli}, {Ota}, {Riechers},
  {Smail}, {van der Werf}, {Weiss}, {Cox}, {Elbaz}, {Gonzalez-Lopez},
  {Infante}, {Oesch}, {Wagg}, \& {Wilkins}}]{Bouwens2016}
{Bouwens}, R.~J., {Aravena}, M., {Decarli}, R., {et~al.} 2016, \apj, 833, 72

\bibitem[{{Brisbin} {et~al.}(2017){Brisbin}, {Miettinen}, {Aravena},
  {Smol{\v{c}}i{\'c}}, {Delvecchio}, {Jiang}, {Magnelli}, {Albrecht},
  {Arancibia}, {Aussel}, {Baran}, {Bertoldi}, {B{\'e}thermin}, {Capak},
  {Casey}, {Civano}, {Hayward}, {Ilbert}, {Karim}, {Le Fevre}, {Marchesi},
  {McCracken}, {Navarrete}, {Novak}, {Riechers}, {Padilla}, {Salvato}, {Scott},
  {Schinnerer}, {Sheth}, \& {Tasca}}]{Brisbin2017}
{Brisbin}, D., {Miettinen}, O., {Aravena}, M., {et~al.} 2017, \aap, 608, A15

\bibitem[{{Bussmann} {et~al.}(2015){Bussmann}, {Riechers}, {Fialkov},
  {Scudder}, {Hayward}, {Cowley}, {Bock}, {Calanog}, {Chapman}, {Cooray}, {De
  Bernardis}, {Farrah}, {Fu}, {Gavazzi}, {Hopwood}, {Ivison}, {Jarvis},
  {Lacey}, {Loeb}, {Oliver}, {P{\'e}rez-Fournon}, {Rigopoulou}, {Roseboom},
  {Scott}, {Smith}, {Vieira}, {Wang}, \& {Wardlow}}]{Bussmann2015}
{Bussmann}, R.~S., {Riechers}, D., {Fialkov}, A., {et~al.} 2015, \apj, 812, 43

\bibitem[{{Capak} {et~al.}(2015){Capak}, {Carilli}, {Jones}, {Casey},
  {Riechers}, {Sheth}, {Carollo}, {Ilbert}, {Karim}, {Lefevre}, {Lilly},
  {Scoville}, {Smolcic}, \& {Yan}}]{Capak2015}
{Capak}, P.~L., {Carilli}, C., {Jones}, G., {et~al.} 2015, \nat, 522, 455

\bibitem[{{Carilli} \& {Walter}(2013)}]{Carilli2013}
{Carilli}, C.~L. \& {Walter}, F. 2013, \araa, 51, 105

\bibitem[{{Carniani} {et~al.}(2018{\natexlab{a}}){Carniani}, {Maiolino},
  {Amorin}, {Pentericci}, {Pallottini}, {Ferrara}, {Willott}, {Smit},
  {Matthee}, {Sobral}, {Santini}, {Castellano}, {De Barros}, {Fontana},
  {Grazian}, \& {Guaita}}]{Carniani2018b}
{Carniani}, S., {Maiolino}, R., {Amorin}, R., {et~al.} 2018{\natexlab{a}},
  \mnras, 478, 1170

\bibitem[{{Carniani} {et~al.}(2015){Carniani}, {Maiolino}, {De Zotti},
  {Negrello}, {Marconi}, {Bothwell}, {Capak}, {Carilli}, {Castellano},
  {Cristiani}, {Ferrara}, {Fontana}, {Gallerani}, {Jones}, {Ohta}, {Ota},
  {Pentericci}, {Santini}, {Sheth}, {Vallini}, {Vanzella}, {Wagg}, \&
  {Williams}}]{Carniani2015}
{Carniani}, S., {Maiolino}, R., {De Zotti}, G., {et~al.} 2015, \aap, 584, A78

\bibitem[{{Carniani} {et~al.}(2018{\natexlab{b}}){Carniani}, {Maiolino},
  {Smit}, \& {Amor{\'\i}n}}]{Carniani2018a}
{Carniani}, S., {Maiolino}, R., {Smit}, R., \& {Amor{\'\i}n}, R.
  2018{\natexlab{b}}, \apjl, 854, L7

\bibitem[{{Casey} {et~al.}(2013){Casey}, {Chen}, {Cowie}, {Barger}, {Capak},
  {Ilbert}, {Koss}, {Lee}, {Le Floc'h}, {Sanders}, \& {Williams}}]{Casey2013}
{Casey}, C.~M., {Chen}, C.-C., {Cowie}, L.~L., {et~al.} 2013, \mnras, 436, 1919

\bibitem[{{Casey} {et~al.}(2019){Casey}, {Zavala}, {Aravena}, {Bethermin},
  {Caputi}, {Champagne}, {Clements}, {Da Cunha}, {Drew}, {Finkelstein},
  {Hayward}, {Kartaltepe}, {Knudsen}, {Koekemoer}, {Magdis}, {Man}, {Manning},
  {Scoville}, {Sheth}, {Spilker}, {Staguhn}, {Talia}, {Taniguchi}, {Toft},
  {Treister}, \& {Yun}}]{Casey2019}
{Casey}, C.~M., {Zavala}, J.~A., {Aravena}, M., {et~al.} 2019, arXiv e-prints,
  arXiv:1910.13331

\bibitem[{{Casey} {et~al.}(2018){Casey}, {Zavala}, {Spilker}, {da Cunha},
  {Hodge}, {Hung}, {Staguhn}, {Finkelstein}, \& {Drew}}]{Casey2018}
{Casey}, C.~M., {Zavala}, J.~A., {Spilker}, J., {et~al.} 2018, \apj, 862, 77

\bibitem[{{Cassata} {et~al.}(2020){Cassata}, {Morselli}, {Faisst}, {Ginolfi},
  {Bethermin}, {Capak}, {Le Fevre}, {Schaerer}, {Silverman}, {Yan}, {Lemaux},
  {Romano}, {Talia}, {Bardelli}, {Boquien}, {Cimatti}, {Dessauges-Zavadsky},
  {Fudamoto}, {Fujimoto}, {Giavalisco}, {Hathi}, {Ibar}, {Jones}, {Koekemoer},
  {Mendez-Hernandez}, {Mancini}, {Oesch}, {Pozzi}, {Riechers}, {Rodighiero},
  {Vergani}, {Zamorani}, \& {Zucca}}]{Cassata2020}
{Cassata}, P., {Morselli}, L., {Faisst}, A., {et~al.} 2020, arXiv e-prints,
  arXiv:2002.00967

\bibitem[{{Chabrier}(2003)}]{Chabrier2003}
{Chabrier}, G. 2003, \pasp, 115, 763

\bibitem[{{Chen} {et~al.}(2013){Chen}, {Cowie}, {Barger}, {Casey}, {Lee},
  {Sanders}, {Wang}, \& {Williams}}]{Chen2013}
{Chen}, C.-C., {Cowie}, L.~L., {Barger}, A.~J., {et~al.} 2013, \apj, 762, 81

\bibitem[{{Condon}(1997)}]{Condon1997}
{Condon}, J.~J. 1997, \pasp, 109, 166

\bibitem[{{Conway} {et~al.}(1990){Conway}, {Cornwell}, \&
  {Wilkinson}}]{Conway1990}
{Conway}, J.~E., {Cornwell}, T.~J., \& {Wilkinson}, P.~N. 1990, \mnras, 246,
  490

\bibitem[{{Coppin} {et~al.}(2006){Coppin}, {Chapin}, {Mortier}, {Scott},
  {Borys}, {Dunlop}, {Halpern}, {Hughes}, {Pope}, {Scott}, {Serjeant}, {Wagg},
  {Alexander}, {Almaini}, {Aretxaga}, {Babbedge}, {Best}, {Blain}, {Chapman},
  {Clements}, {Crawford}, {Dunne}, {Eales}, {Edge}, {Farrah}, {Gazta{\~n}aga},
  {Gear}, {Granato}, {Greve}, {Fox}, {Ivison}, {Jarvis}, {Jenness}, {Lacey},
  {Lepage}, {Mann}, {Marsden}, {Martinez-Sansigre}, {Oliver}, {Page},
  {Peacock}, {Pearson}, {Percival}, {Priddey}, {Rawlings}, {Rowan-Robinson},
  {Savage}, {Seigar}, {Sekiguchi}, {Silva}, {Simpson}, {Smail}, {Stevens},
  {Takagi}, {Vaccari}, {van Kampen}, \& {Willott}}]{Coppin2006}
{Coppin}, K., {Chapin}, E.~L., {Mortier}, A.~M.~J., {et~al.} 2006, \mnras, 372,
  1621

\bibitem[{{Cormier} {et~al.}(2012){Cormier}, {Lebouteiller}, {Madden}, {Abel},
  {Hony}, {Galliano}, {Baes}, {Barlow}, {Cooray}, {De Looze}, {Galametz},
  {Karczewski}, {Parkin}, {R{\'e}my}, {Sauvage}, {Spinoglio}, {Wilson}, \&
  {Wu}}]{Cormier2012}
{Cormier}, D., {Lebouteiller}, V., {Madden}, S.~C., {et~al.} 2012, \aap, 548,
  A20

\bibitem[{{Cowley} {et~al.}(2017{\natexlab{a}}){Cowley}, {B{\'e}thermin},
  {Lagos}, {Lacey}, {Baugh}, \& {Cole}}]{Cowley2017_SED}
{Cowley}, W.~I., {B{\'e}thermin}, M., {Lagos}, C. d.~P., {et~al.}
  2017{\natexlab{a}}, \mnras, 467, 1231

\bibitem[{{Cowley} {et~al.}(2017{\natexlab{b}}){Cowley}, {Lacey}, {Baugh},
  {Cole}, \& {Wilkinson}}]{Cowley2017}
{Cowley}, W.~I., {Lacey}, C.~G., {Baugh}, C.~M., {Cole}, S., \& {Wilkinson}, A.
  2017{\natexlab{b}}, \mnras, 469, 3396

\bibitem[{{Davidzon} {et~al.}(2017){Davidzon}, {Ilbert}, {Laigle}, {Coupon},
  {McCracken}, {Delvecchio}, {Masters}, {Capak}, {Hsieh}, {Le F{\`e}vre},
  {Tresse}, {Bethermin}, {Chang}, {Faisst}, {Le Floc'h}, {Steinhardt}, {Toft},
  {Aussel}, {Dubois}, {Hasinger}, {Salvato}, {Sanders}, {Scoville}, \&
  {Silverman}}]{Davidzon2017}
{Davidzon}, I., {Ilbert}, O., {Laigle}, C., {et~al.} 2017, \aap, 605, A70

\bibitem[{{De Breuck} {et~al.}(2014){De Breuck}, {Williams}, {Swinbank},
  {Caselli}, {Coppin}, {Davis}, {Maiolino}, {Nagao}, {Smail}, {Walter},
  {Wei{\ss}}, \& {Zwaan}}]{De_Breuck2014}
{De Breuck}, C., {Williams}, R.~J., {Swinbank}, M., {et~al.} 2014, The
  Messenger, 156, 38

\bibitem[{{De Looze} {et~al.}(2014){De Looze}, {Cormier}, {Lebouteiller},
  {Madden}, {Baes}, {Bendo}, {Boquien}, {Boselli}, {Clements}, {Cortese},
  {Cooray}, {Galametz}, {Galliano}, {Graci{\'a}-Carpio}, {Isaak}, {Karczewski},
  {Parkin}, {Pellegrini}, {R{\'e}my-Ruyer}, {Spinoglio}, {Smith}, \&
  {Sturm}}]{De_Looze2014}
{De Looze}, I., {Cormier}, D., {Lebouteiller}, V., {et~al.} 2014, \aap, 568,
  A62

\bibitem[{{Decarli} {et~al.}(2016){Decarli}, {Walter}, {Aravena}, {Carilli},
  {Bouwens}, {da Cunha}, {Daddi}, {Ivison}, {Popping}, {Riechers}, {Smail},
  {Swinbank}, {Weiss}, {Anguita}, {Assef}, {Bauer}, {Bell}, {Bertoldi},
  {Chapman}, {Colina}, {Cortes}, {Cox}, {Dickinson}, {Elbaz},
  {G{\'o}nzalez-L{\'o}pez}, {Ibar}, {Infante}, {Hodge}, {Karim}, {Le Fevre},
  {Magnelli}, {Neri}, {Oesch}, {Ota}, {Rix}, {Sargent}, {Sheth}, {van der Wel},
  {van der Werf}, \& {Wagg}}]{Decarli2016}
{Decarli}, R., {Walter}, F., {Aravena}, M., {et~al.} 2016, \apj, 833, 69

\bibitem[{{Decarli} {et~al.}(2018){Decarli}, {Walter}, {Venemans},
  {Ba{\~n}ados}, {Bertoldi}, {Carilli}, {Fan}, {Farina}, {Mazzucchelli},
  {Riechers}, {Rix}, {Strauss}, {Wang}, \& {Yang}}]{Decarli2018}
{Decarli}, R., {Walter}, F., {Venemans}, B.~P., {et~al.} 2018, \apj, 854, 97

\bibitem[{{Dessauges-Zavadsky} {et~al.}(2020){Dessauges-Zavadsky}, {Ginolfi},
  {Pozzi}, {B{\'e}thermin}, {Le F{\`e}vre}, {Fujimoto}, {Silverman}, {Jones},
  {Schaerer}, {Faisst}, {Khusanova}, {Fudamoto}, {Cassata}, {Loiacono},
  {Capak}, {Yan}, {Amorin}, {Bardelli}, {Boquien}, {Cimatti}, {Gruppioni},
  {Hathi}, {Ibar}, {Koekemoer}, {Lemaux}, {Narayanan}, {Oesch}, {Rodighiero},
  {Romano}, {Talia}, {Toft}, {Vallini}, {Vergani}, {Zamorani}, \&
  {Zucca}}]{Dessauges2020}
{Dessauges-Zavadsky}, M., {Ginolfi}, M., {Pozzi}, F., {et~al.} 2020, arXiv
  e-prints, arXiv:2004.10771

\bibitem[{{D{\'\i}az-Santos} {et~al.}(2013){D{\'\i}az-Santos}, {Armus},
  {Charmandaris}, {Stierwalt}, {Murphy}, {Haan}, {Inami}, {Malhotra},
  {Meijerink}, {Stacey}, {Petric}, {Evans}, {Veilleux}, {van der Werf}, {Lord},
  {Lu}, {Howell}, {Appleton}, {Mazzarella}, {Surace}, {Xu}, {Schulz},
  {Sanders}, {Bridge}, {Chan}, {Frayer}, {Iwasawa}, {Melbourne}, \&
  {Sturm}}]{Diaz-Santos2013}
{D{\'\i}az-Santos}, T., {Armus}, L., {Charmandaris}, V., {et~al.} 2013, \apj,
  774, 68

\bibitem[{{Dole} {et~al.}(2006){Dole}, {Lagache}, {Puget}, {Caputi},
  {Fern{\'a}ndez-Conde}, {Le Floc'h}, {Papovich}, {P{\'e}rez-Gonz{\'a}lez},
  {Rieke}, \& {Blaylock}}]{Dole2006}
{Dole}, H., {Lagache}, G., {Puget}, J., {et~al.} 2006, \aap, 451, 417

\bibitem[{{Dunlop} {et~al.}(2017){Dunlop}, {McLure}, {Biggs}, {Geach},
  {Micha{\l}owski}, {Ivison}, {Rujopakarn}, {van Kampen}, {Kirkpatrick},
  {Pope}, {Scott}, {Swinbank}, {Targett}, {Aretxaga}, {Austermann}, {Best},
  {Bruce}, {Chapin}, {Charlot}, {Cirasuolo}, {Coppin}, {Ellis}, {Finkelstein},
  {Hayward}, {Hughes}, {Ibar}, {Jagannathan}, {Khochfar}, {Koprowski},
  {Narayanan}, {Nyland}, {Papovich}, {Peacock}, {Rieke}, {Robertson},
  {Vernstrom}, {Werf}, {Wilson}, \& {Yun}}]{Dunlop2017}
{Dunlop}, J.~S., {McLure}, R.~J., {Biggs}, A.~D., {et~al.} 2017, Monthly
  Notices of the Royal Astronomical Society, 466, 861

\bibitem[{{Dunne} {et~al.}(2011){Dunne}, {Gomez}, {da Cunha}, {Charlot}, {Dye},
  {Eales}, {Maddox}, {Rowlands}, {Smith}, {Auld}, {Baes}, {Bonfield}, {Bourne},
  {Buttiglione}, {Cava}, {Clements}, {Coppin}, {Cooray}, {Dariush}, {de Zotti},
  {Driver}, {Fritz}, {Geach}, {Hopwood}, {Ibar}, {Ivison}, {Jarvis}, {Kelvin},
  {Pascale}, {Pohlen}, {Popescu}, {Rigby}, {Robotham}, {Rodighiero}, {Sansom},
  {Serjeant}, {Temi}, {Thompson}, {Tuffs}, {van der Werf}, \&
  {Vlahakis}}]{Dunne2011}
{Dunne}, L., {Gomez}, H.~L., {da Cunha}, E., {et~al.} 2011, \mnras, 417, 1510

\bibitem[{{Elbaz} {et~al.}(2011){Elbaz}, {Dickinson}, {Hwang},
  {D{\'{\i}}az-Santos}, {Magdis}, {Magnelli}, {Le Borgne}, {Galliano},
  {Pannella}, {Chanial}, {Armus}, {Charmandaris}, {Daddi}, {Aussel}, {Popesso},
  {Kartaltepe}, {Altieri}, {Valtchanov}, {Coia}, {Dannerbauer}, {Dasyra},
  {Leiton}, {Mazzarella}, {Alexander}, {Buat}, {Burgarella}, {Chary}, {Gilli},
  {Ivison}, {Juneau}, {Le Floc'h}, {Lutz}, {Morrison}, {Mullaney}, {Murphy},
  {Pope}, {Scott}, {Brodwin}, {Calzetti}, {Cesarsky}, {Charlot}, {Dole},
  {Eisenhardt}, {Ferguson}, {F{\"o}rster Schreiber}, {Frayer}, {Giavalisco},
  {Huynh}, {Koekemoer}, {Papovich}, {Reddy}, {Surace}, {Teplitz}, {Yun}, \&
  {Wilson}}]{Elbaz2011}
{Elbaz}, D., {Dickinson}, M., {Hwang}, H.~S., {et~al.} 2011, \aap, 533, A119

\bibitem[{{Faisst} {et~al.}(2016){Faisst}, {Capak}, {Davidzon}, {Salvato},
  {Laigle}, {Ilbert}, {Onodera}, {Hasinger}, {Kakazu}, {Masters}, {McCracken},
  {Mobasher}, {Sanders}, {Silverman}, {Yan}, \& {Scoville}}]{Faisst2016}
{Faisst}, A.~L., {Capak}, P.~L., {Davidzon}, I., {et~al.} 2016, \apj, 822, 29

\bibitem[{{Faisst} {et~al.}(2017){Faisst}, {Capak}, {Yan}, {Pavesi},
  {Riechers}, {Bari{\v{s}}i{\'c}}, {Cooke}, {Kartaltepe}, \&
  {Masters}}]{Faisst2017}
{Faisst}, A.~L., {Capak}, P.~L., {Yan}, L., {et~al.} 2017, \apj, 847, 21

\bibitem[{{Faisst} {et~al.}(2020){Faisst}, {Schaerer}, {Lemaux}, {Oesch},
  {Fudamoto}, {Cassata}, {B{\'e}thermin}, {Capak}, {Le F{\`e}vre}, {Silverman},
  {Yan}, {Ginolfi}, {Koekemoer}, {Morselli}, {Amor{\'\i}n}, {Bardelli},
  {Boquien}, {Brammer}, {Cimatti}, {Dessauges-Zavadsky}, {Fujimoto},
  {Gruppioni}, {Hathi}, {Hemmati}, {Ibar}, {Jones}, {Khusanova}, {Loiacono},
  {Pozzi}, {Talia}, {Tasca}, {Riechers}, {Rodighiero}, {Romano}, {Scoville},
  {Toft}, {Vallini}, {Vergani}, {Zamorani}, \& {Zucca}}]{Faisst2020}
{Faisst}, A.~L., {Schaerer}, D., {Lemaux}, B.~C., {et~al.} 2020, \apjs, 247, 61

\bibitem[{{Faisst et al.}(2019)}]{Faisst2019}
{Faisst et al.} 2019, to be sub.

\bibitem[{{Ferrara} {et~al.}(2019){Ferrara}, {Vallini}, {Pallottini},
  {Gallerani}, {Carniani}, {Kohandel}, {Decataldo}, \& {Behrens}}]{Ferrara2019}
{Ferrara}, A., {Vallini}, L., {Pallottini}, A., {et~al.} 2019, \mnras, 489, 1

\bibitem[{{Fixsen} {et~al.}(1998){Fixsen}, {Dwek}, {Mather}, {Bennett}, \&
  {Shafer}}]{Fixsen1998}
{Fixsen}, D.~J., {Dwek}, E., {Mather}, J.~C., {Bennett}, C.~L., \& {Shafer},
  R.~A. 1998, \apj, 508, 123

\bibitem[{{Franco} {et~al.}(2018){Franco}, {Elbaz}, {B{\'e}thermin},
  {Magnelli}, {Schreiber}, {Ciesla}, {Dickinson}, {Nagar}, {Silverman},
  {Daddi}, {Alexander}, {Wang}, {Pannella}, {Le Floc'h}, {Pope}, {Giavalisco},
  {Maury}, {Bournaud}, {Chary}, {Demarco}, {Ferguson}, {Finkelstein}, {Inami},
  {Iono}, {Juneau}, {Lagache}, {Leiton}, {Lin}, {Magdis}, {Messias},
  {Motohara}, {Mullaney}, {Okumura}, {Papovich}, {Pforr}, {Rujopakarn},
  {Sargent}, {Shu}, \& {Zhou}}]{Franco2018}
{Franco}, M., {Elbaz}, D., {B{\'e}thermin}, M., {et~al.} 2018, Astronomy and
  Astrophysics, 620, A152

\bibitem[{{Fudamoto} {et~al.}(2020){Fudamoto}, {Oesch}, {Faisst}, {Bethermin},
  {Ginolfi}, {Khusanova}, {Loiacono}, {Le Fevre}, {Capak}, {Schaerer},
  {Silverman}, {Cassata}, {Yan}, {Amorin}, {Bardelli}, {Boquien}, {Cimatti},
  {Dessauges-Zavadsky}, {Fujimoto}, {Gruppioni}, {Hathi}, {Ibar}, {Jones},
  {Koekemoer}, {Lagache}, {Lemaux}, {Maiolino}, {Narayanan}, {Pozzi},
  {Riechers}, {Rodighiero}, {Talia}, {Toft}, {Vallini}, {Vergani}, {Zamorani},
  \& {Zucca}}]{Fudamoto2020}
{Fudamoto}, Y., {Oesch}, P.~A., {Faisst}, A., {et~al.} 2020, arXiv e-prints,
  arXiv:2004.10760

\bibitem[{{Fudamoto} {et~al.}(2017){Fudamoto}, {Oesch}, {Schinnerer}, {Groves},
  {Karim}, {Magnelli}, {Sargent}, {Cassata}, {Lang}, {Liu}, {Le F{\`e}vre},
  {Leslie}, {Smol{\v{c}}i{\'c}}, \& {Tasca}}]{Fudamoto2017}
{Fudamoto}, Y., {Oesch}, P.~A., {Schinnerer}, E., {et~al.} 2017, \mnras, 472,
  483

\bibitem[{{Fujimoto} {et~al.}(2016){Fujimoto}, {Ouchi}, {Ono}, {Shibuya},
  {Ishigaki}, {Nagai}, \& {Momose}}]{Fujimoto2016}
{Fujimoto}, S., {Ouchi}, M., {Ono}, Y., {et~al.} 2016, \apjs, 222, 1

\bibitem[{{Fujimoto} {et~al.}(2020){Fujimoto}, {Silverman}, {Bethermin},
  {Ginolfi}, {Jones}, {Le F{\`e}vre}, {Dessauges-Zavadsky}, {Rujopakarn},
  {Faisst}, {Fudamoto}, {Cassata}, {Morselli}, {Maiolino}, {Schaerer}, {Capak},
  {Yan}, {Vallini}, {Toft}, {Loiacono}, {Zamorani}, {Talia}, {Narayanan},
  {Hathi}, {Lemaux}, {Boquien}, {Amorin}, {Ibar}, {Koekemoer},
  {M{\'e}ndez-Hern{\'a}ndez}, {Bardelli}, {Vergani}, {Zucca}, {Romano}, \&
  {Cimatti}}]{Fujimoto2020}
{Fujimoto}, S., {Silverman}, J.~D., {Bethermin}, M., {et~al.} 2020, arXiv
  e-prints, arXiv:2003.00013

\bibitem[{{Gallerani} {et~al.}(2018){Gallerani}, {Pallottini}, {Feruglio},
  {Ferrara}, {Maiolino}, {Vallini}, {Riechers}, \& {Pavesi}}]{Gallerani2018}
{Gallerani}, S., {Pallottini}, A., {Feruglio}, C., {et~al.} 2018, \mnras, 473,
  1909

\bibitem[{{Geach} {et~al.}(2017){Geach}, {Dunlop}, {Halpern}, {Smail}, {van der
  Werf}, {Alexander}, {Almaini}, {Aretxaga}, {Arumugam}, {Asboth}, {Banerji},
  {Beanlands}, {Best}, {Blain}, {Birkinshaw}, {Chapin}, {Chapman}, {Chen},
  {Chrysostomou}, {Clarke}, {Clements}, {Conselice}, {Coppin}, {Cowley},
  {Danielson}, {Eales}, {Edge}, {Farrah}, {Gibb}, {Harrison}, {Hine}, {Hughes},
  {Ivison}, {Jarvis}, {Jenness}, {Jones}, {Karim}, {Koprowski}, {Knudsen},
  {Lacey}, {Mackenzie}, {Marsden}, {McAlpine}, {McMahon}, {Meijerink},
  {Micha{\l}owski}, {Oliver}, {Page}, {Peacock}, {Rigopoulou}, {Robson},
  {Roseboom}, {Rotermund}, {Scott}, {Serjeant}, {Simpson}, {Simpson}, {Smith},
  {Spaans}, {Stanley}, {Stevens}, {Swinbank}, {Targett}, {Thomson}, {Valiante},
  {Wake}, {Webb}, {Willott}, {Zavala}, \& {Zemcov}}]{Geach2017}
{Geach}, J.~E., {Dunlop}, J.~S., {Halpern}, M., {et~al.} 2017, \mnras, 465,
  1789

\bibitem[{{Giavalisco} {et~al.}(2004){Giavalisco}, {Ferguson}, {Koekemoer},
  {Dickinson}, {Alexander}, {Bauer}, {Bergeron}, {Biagetti}, {Brandt},
  {Casertano}, {Cesarsky}, {Chatzichristou}, {Conselice}, {Cristiani}, {Da
  Costa}, {Dahlen}, {de Mello}, {Eisenhardt}, {Erben}, {Fall}, {Fassnacht},
  {Fosbury}, {Fruchter}, {Gardner}, {Grogin}, {Hook}, {Hornschemeier}, {Idzi},
  {Jogee}, {Kretchmer}, {Laidler}, {Lee}, {Livio}, {Lucas}, {Madau},
  {Mobasher}, {Moustakas}, {Nonino}, {Padovani}, {Papovich}, {Park},
  {Ravindranath}, {Renzini}, {Richardson}, {Riess}, {Rosati}, {Schirmer},
  {Schreier}, {Somerville}, {Spinrad}, {Stern}, {Stiavelli}, {Strolger},
  {Urry}, {Vandame}, {Williams}, \& {Wolf}}]{Giavalisco2004}
{Giavalisco}, M., {Ferguson}, H.~C., {Koekemoer}, A.~M., {et~al.} 2004, \apjl,
  600, L93

\bibitem[{{Ginolfi} {et~al.}(2020{\natexlab{a}}){Ginolfi}, {Jones},
  {Bethermin}, {Faisst}, {Lemaux}, {Schaerer}, {Fudamoto}, {Oesch},
  {Dessauges-Zavadsky}, {Fujimoto}, {Carniani}, {Le Fevre}, {Cassata},
  {Silverman}, {Capak}, {Yan}, {Bardelli}, {Cucciati}, {Gal}, {Gruppioni},
  {Hathi}, {Lubin}, {Maiolino}, {Morselli}, {Pelliccia}, {Talia}, {Vergani}, \&
  {Zamorani}}]{Ginolfi2020b}
{Ginolfi}, M., {Jones}, G.~C., {Bethermin}, M., {et~al.} 2020{\natexlab{a}},
  arXiv e-prints, arXiv:2004.13737

\bibitem[{{Ginolfi} {et~al.}(2020{\natexlab{b}}){Ginolfi}, {Jones},
  {B{\'e}thermin}, {Fudamoto}, {Loiacono}, {Fujimoto}, {Le F{\'e}vre},
  {Faisst}, {Schaerer}, {Cassata}, {Silverman}, {Yan}, {Capak}, {Bardelli},
  {Boquien}, {Carraro}, {Dessauges-Zavadsky}, {Giavalisco}, {Gruppioni},
  {Ibar}, {Khusanova}, {Lemaux}, {Maiolino}, {Narayanan}, {Oesch}, {Pozzi},
  {Rodighiero}, {Talia}, {Toft}, {Vallini}, {Vergani}, \&
  {Zamorani}}]{Ginolfi2020}
{Ginolfi}, M., {Jones}, G.~C., {B{\'e}thermin}, M., {et~al.}
  2020{\natexlab{b}}, \aap, 633, A90

\bibitem[{{Gispert} {et~al.}(2000){Gispert}, {Lagache}, \&
  {Puget}}]{Gispert2000}
{Gispert}, R., {Lagache}, G., \& {Puget}, J.~L. 2000, \aap, 360, 1

\bibitem[{{Gonz{\'a}lez-L{\'o}pez} {et~al.}(2017){Gonz{\'a}lez-L{\'o}pez},
  {Bauer}, {Romero-Ca{\~n}izales}, {Kneissl}, {Villard}, {Carvajal}, {Kim},
  {Laporte}, {Anguita}, {Aravena}, {Bouwens}, {Bradley}, {Carrasco}, {Demarco},
  {Ford}, {Ibar}, {Infante}, {Messias}, {Mu{\~n}oz Arancibia}, {Nagar},
  {Padilla}, {Treister}, {Troncoso}, \& {Zitrin}}]{Gonzalez-Lopez2017}
{Gonz{\'a}lez-L{\'o}pez}, J., {Bauer}, F.~E., {Romero-Ca{\~n}izales}, C.,
  {et~al.} 2017, \aap, 597, A41

\bibitem[{{Gonz{\'a}lez-L{\'o}pez} {et~al.}(2019){Gonz{\'a}lez-L{\'o}pez},
  {Decarli}, {Pavesi}, {Walter}, {Aravena}, {Carilli}, {Boogaard}, {Popping},
  {Weiss}, {Assef}, {Bauer}, {Bertoldi}, {Bouwens}, {Contini}, {Cortes}, {Cox},
  {da Cunha}, {Daddi}, {D{\'\i}az-Santos}, {Inami}, {Hodge}, {Ivison}, {Le
  F{\`e}vre}, {Magnelli}, {Oesch}, {Riechers}, {Rix}, {Smail}, {Swinbank},
  {Somerville}, {Uzgil}, \& {van der Werf}}]{Gonzalez-Lopez2019}
{Gonz{\'a}lez-L{\'o}pez}, J., {Decarli}, R., {Pavesi}, R., {et~al.} 2019, \apj,
  882, 139

\bibitem[{{Grogin} {et~al.}(2011){Grogin}, {Kocevski}, {Faber}, {Ferguson},
  {Koekemoer}, {Riess}, {Acquaviva}, {Alexander}, {Almaini}, {Ashby}, {Barden},
  {Bell}, {Bournaud}, {Brown}, {Caputi}, {Casertano}, {Cassata}, {Castellano},
  {Challis}, {Chary}, {Cheung}, {Cirasuolo}, {Conselice}, {Roshan Cooray},
  {Croton}, {Daddi}, {Dahlen}, {Dav{\'e}}, {de Mello}, {Dekel}, {Dickinson},
  {Dolch}, {Donley}, {Dunlop}, {Dutton}, {Elbaz}, {Fazio}, {Filippenko},
  {Finkelstein}, {Fontana}, {Gardner}, {Garnavich}, {Gawiser}, {Giavalisco},
  {Grazian}, {Guo}, {Hathi}, {H{\"a}ussler}, {Hopkins}, {Huang}, {Huang},
  {Jha}, {Kartaltepe}, {Kirshner}, {Koo}, {Lai}, {Lee}, {Li}, {Lotz}, {Lucas},
  {Madau}, {McCarthy}, {McGrath}, {McIntosh}, {McLure}, {Mobasher},
  {Moustakas}, {Mozena}, {Nandra}, {Newman}, {Niemi}, {Noeske}, {Papovich},
  {Pentericci}, {Pope}, {Primack}, {Rajan}, {Ravindranath}, {Reddy}, {Renzini},
  {Rix}, {Robaina}, {Rodney}, {Rosario}, {Rosati}, {Salimbeni}, {Scarlata},
  {Siana}, {Simard}, {Smidt}, {Somerville}, {Spinrad}, {Straughn}, {Strolger},
  {Telford}, {Teplitz}, {Trump}, {van der Wel}, {Villforth}, {Wechsler},
  {Weiner}, {Wiklind}, {Wild}, {Wilson}, {Wuyts}, {Yan}, \& {Yun}}]{Grogin2011}
{Grogin}, N.~A., {Kocevski}, D.~D., {Faber}, S.~M., {et~al.} 2011, \apjs, 197,
  35

\bibitem[{{Gruppioni} {et~al.}(2020){Gruppioni}, {Bethermin}, {Loiacono}, {Le
  Fevre}, {Capak}, {Cassata}, {Faisst}, {Schaerer}, {Silverman}, {Yan},
  {Bardelli}, {Boquien}, {Carraro}, {Cimatti}, {Dessauges-Zavadsky}, {Ginolfi},
  {Fujimoto}, {Hathi}, {Jones}, {Khusanova}, {Koekemoer}, {Lagache}, {Lemaux},
  {Oesch}, {Pozzi}, {Riechers}, {Rodighiero}, {Romano}, {Talia}, {Vallini},
  {Vergani}, {Zamorani}, \& {Zucca}}]{Gruppioni2020}
{Gruppioni}, C., {Bethermin}, M., {Loiacono}, F., {et~al.} 2020, arXiv
  e-prints, arXiv:2006.04974

\bibitem[{{Gruppioni} \& {Pozzi}(2019)}]{Gruppioni2019}
{Gruppioni}, C. \& {Pozzi}, F. 2019, \mnras, 483, 1993

\bibitem[{{Gruppioni} {et~al.}(2013){Gruppioni}, {Pozzi}, {Rodighiero},
  {Delvecchio}, {Berta}, {Pozzetti}, {Zamorani}, {Andreani}, {Cimatti},
  {Ilbert}, {Le Floc'h}, {Lutz}, {Magnelli}, {Marchetti}, {Monaco}, {Nordon},
  {Oliver}, {Popesso}, {Riguccini}, {Roseboom}, {Rosario}, {Sargent},
  {Vaccari}, {Altieri}, {Aussel}, {Bongiovanni}, {Cepa}, {Daddi},
  {Dom{\'{\i}}nguez-S{\'a}nchez}, {Elbaz}, {F{\"o}rster Schreiber}, {Genzel},
  {Iribarrem}, {Magliocchetti}, {Maiolino}, {Poglitsch}, {P{\'e}rez
  Garc{\'{\i}}a}, {Sanchez-Portal}, {Sturm}, {Tacconi}, {Valtchanov},
  {Amblard}, {Arumugam}, {Bethermin}, {Bock}, {Boselli}, {Buat}, {Burgarella},
  {Castro-Rodr{\'{\i}}guez}, {Cava}, {Chanial}, {Clements}, {Conley}, {Cooray},
  {Dowell}, {Dwek}, {Eales}, {Franceschini}, {Glenn}, {Griffin},
  {Hatziminaoglou}, {Ibar}, {Isaak}, {Ivison}, {Lagache}, {Levenson}, {Lu},
  {Madden}, {Maffei}, {Mainetti}, {Nguyen}, {O'Halloran}, {Page}, {Panuzzo},
  {Papageorgiou}, {Pearson}, {P{\'e}rez-Fournon}, {Pohlen}, {Rigopoulou},
  {Rowan-Robinson}, {Schulz}, {Scott}, {Seymour}, {Shupe}, {Smith}, {Stevens},
  {Symeonidis}, {Trichas}, {Tugwell}, {Vigroux}, {Wang}, {Wright}, {Xu},
  {Zemcov}, {Bardelli}, {Carollo}, {Contini}, {Le F{\'e}vre}, {Lilly},
  {Mainieri}, {Renzini}, {Scodeggio}, \& {Zucca}}]{Gruppioni2013}
{Gruppioni}, C., {Pozzi}, F., {Rodighiero}, G., {et~al.} 2013, \mnras, 432, 23

\bibitem[{{Gruppioni} {et~al.}(2011){Gruppioni}, {Pozzi}, {Zamorani}, \&
  {Vignali}}]{Gruppioni2011}
{Gruppioni}, C., {Pozzi}, F., {Zamorani}, G., \& {Vignali}, C. 2011, \mnras,
  416, 70

\bibitem[{{Gullberg} {et~al.}(2015){Gullberg}, {De Breuck}, {Vieira},
  {Wei{\ss}}, {Aguirre}, {Aravena}, {B{\'e}thermin}, {Bradford}, {Bothwell},
  {Carlstrom}, {Chapman}, {Fassnacht}, {Gonzalez}, {Greve}, {Hezaveh},
  {Holzapfel}, {Husband}, {Ma}, {Malkan}, {Marrone}, {Menten}, {Murphy},
  {Reichardt}, {Spilker}, {Stark}, {Strandet}, \& {Welikala}}]{Gullberg2015}
{Gullberg}, B., {De Breuck}, C., {Vieira}, J.~D., {et~al.} 2015, \mnras, 449,
  2883

\bibitem[{{Hasinger} {et~al.}(2018){Hasinger}, {Capak}, {Salvato}, {Barger},
  {Cowie}, {Faisst}, {Hemmati}, {Kakazu}, {Kartaltepe}, {Masters}, {Mobasher},
  {Nayyeri}, {Sanders}, {Scoville}, {Suh}, {Steinhardt}, \&
  {Yang}}]{Hasinger2018}
{Hasinger}, G., {Capak}, P., {Salvato}, M., {et~al.} 2018, \apj, 858, 77

\bibitem[{{Hatsukade} {et~al.}(2018){Hatsukade}, {Kohno}, {Yamaguchi},
  {Umehata}, {Ao}, {Aretxaga}, {Caputi}, {Dunlop}, {Egami}, {Espada},
  {Fujimoto}, {Hayatsu}, {Hughes}, {Ikarashi}, {Iono}, {Ivison}, {Kawabe},
  {Kodama}, {Lee}, {Matsuda}, {Nakanishi}, {Ohta}, {Ouchi}, {Rujopakarn},
  {Suzuki}, {Tamura}, {Ueda}, {Wang}, {Wang}, {Wilson}, {Yoshimura}, \&
  {Yun}}]{Hatsukade2018}
{Hatsukade}, B., {Kohno}, K., {Yamaguchi}, Y., {et~al.} 2018, \pasj, 70, 105

\bibitem[{{Hauser} {et~al.}(1998){Hauser}, {Arendt}, {Kelsall}, {Dwek},
  {Odegard}, {Weiland}, {Freudenreich}, {Reach}, {Silverberg}, {Moseley},
  {Pei}, {Lubin}, {Mather}, {Shafer}, {Smoot}, {Weiss}, {Wilkinson}, \&
  {Wright}}]{Hauser1998}
{Hauser}, M.~G., {Arendt}, R.~G., {Kelsall}, T., {et~al.} 1998, \apj, 508, 25

\bibitem[{{Hayward} {et~al.}(2013){Hayward}, {Behroozi}, {Somerville},
  {Primack}, {Moreno}, \& {Wechsler}}]{Hayward2013}
{Hayward}, C.~C., {Behroozi}, P.~S., {Somerville}, R.~S., {et~al.} 2013,
  \mnras, 434, 2572

\bibitem[{{Hodge} \& {da Cunha}(2020)}]{Hodge2020}
{Hodge}, J.~A. \& {da Cunha}, E. 2020, arXiv e-prints, arXiv:2004.00934

\bibitem[{{Hollenbach} \& {Tielens}(1999)}]{Hollenbach1999}
{Hollenbach}, D.~J. \& {Tielens}, A.~G.~G.~M. 1999, Reviews of Modern Physics,
  71, 173

\bibitem[{{Hsu} {et~al.}(2016){Hsu}, {Cowie}, {Chen}, {Barger}, \&
  {Wang}}]{Hsu2016}
{Hsu}, L.-Y., {Cowie}, L.~L., {Chen}, C.-C., {Barger}, A.~J., \& {Wang}, W.-H.
  2016, \apj, 829, 25

\bibitem[{{Imara} {et~al.}(2018){Imara}, {Loeb}, {Johnson}, {Conroy}, \&
  {Behroozi}}]{Imara2018}
{Imara}, N., {Loeb}, A., {Johnson}, B.~D., {Conroy}, C., \& {Behroozi}, P.
  2018, \apj, 854, 36

\bibitem[{{Jin} {et~al.}(2019){Jin}, {Daddi}, {Magdis}, {Liu}, {Schinnerer},
  {Papadopoulos}, {Gu}, {Gao}, \& {Calabro}}]{Jin2019}
{Jin}, S., {Daddi}, E., {Magdis}, G.~E., {et~al.} 2019, arXiv e-prints,
  arXiv:1906.00040

\bibitem[{{Jones} {et~al.}(2020){Jones}, {B{\'e}thermin}, {Fudamoto},
  {Ginolfi}, {Capak}, {Cassata}, {Faisst}, {Le F{\`e}vre}, {Schaerer},
  {Silverman}, {Yan}, {Bardelli}, {Boquien}, {Cimatti}, {Dessauges-Zavadsky},
  {Giavalisco}, {Gruppioni}, {Ibar}, {Khusanova}, {Koekemoer}, {Lemaux},
  {Loiacono}, {Maiolino}, {Oesch}, {Pozzi}, {Riechers}, {Rodighiero}, {Talia},
  {Vallini}, {Vergani}, {Zamorani}, \& {Zucca}}]{Jones2020}
{Jones}, G.~C., {B{\'e}thermin}, M., {Fudamoto}, Y., {et~al.} 2020, \mnras,
  491, L18

\bibitem[{{Karim} {et~al.}(2013){Karim}, {Swinbank}, {Hodge}, {Smail},
  {Walter}, {Biggs}, {Simpson}, {Danielson}, {Alexander}, {Bertoldi}, {de
  Breuck}, {Chapman}, {Coppin}, {Dannerbauer}, {Edge}, {Greve}, {Ivison},
  {Knudsen}, {Menten}, {Schinnerer}, {Wardlow}, {Wei{\ss}}, \& {van der
  Werf}}]{Karim2013}
{Karim}, A., {Swinbank}, A.~M., {Hodge}, J.~A., {et~al.} 2013, \mnras, 432, 2

\bibitem[{{Katz} {et~al.}(2017){Katz}, {Kimm}, {Sijacki}, \&
  {Haehnelt}}]{Katz2017}
{Katz}, H., {Kimm}, T., {Sijacki}, D., \& {Haehnelt}, M.~G. 2017, \mnras, 468,
  4831

\bibitem[{{Kennicutt}(1998)}]{Kennicutt1998}
{Kennicutt}, Jr., R.~C. 1998, \apj, 498, 541

\bibitem[{{Koekemoer} {et~al.}(2011){Koekemoer}, {Faber}, {Ferguson}, {Grogin},
  {Kocevski}, {Koo}, {Lai}, {Lotz}, {Lucas}, {McGrath}, {Ogaz}, {Rajan},
  {Riess}, {Rodney}, {Strolger}, {Casertano}, {Castellano}, {Dahlen},
  {Dickinson}, {Dolch}, {Fontana}, {Giavalisco}, {Grazian}, {Guo}, {Hathi},
  {Huang}, {van der Wel}, {Yan}, {Acquaviva}, {Alexander}, {Almaini}, {Ashby},
  {Barden}, {Bell}, {Bournaud}, {Brown}, {Caputi}, {Cassata}, {Challis},
  {Chary}, {Cheung}, {Cirasuolo}, {Conselice}, {Roshan Cooray}, {Croton},
  {Daddi}, {Dav{\'e}}, {de Mello}, {de Ravel}, {Dekel}, {Donley}, {Dunlop},
  {Dutton}, {Elbaz}, {Fazio}, {Filippenko}, {Finkelstein}, {Frazer}, {Gardner},
  {Garnavich}, {Gawiser}, {Gruetzbauch}, {Hartley}, {H{\"a}ussler},
  {Herrington}, {Hopkins}, {Huang}, {Jha}, {Johnson}, {Kartaltepe},
  {Khostovan}, {Kirshner}, {Lani}, {Lee}, {Li}, {Madau}, {McCarthy},
  {McIntosh}, {McLure}, {McPartland}, {Mobasher}, {Moreira}, {Mortlock},
  {Moustakas}, {Mozena}, {Nandra}, {Newman}, {Nielsen}, {Niemi}, {Noeske},
  {Papovich}, {Pentericci}, {Pope}, {Primack}, {Ravindranath}, {Reddy},
  {Renzini}, {Rix}, {Robaina}, {Rosario}, {Rosati}, {Salimbeni}, {Scarlata},
  {Siana}, {Simard}, {Smidt}, {Snyder}, {Somerville}, {Spinrad}, {Straughn},
  {Telford}, {Teplitz}, {Trump}, {Vargas}, {Villforth}, {Wagner}, {Wand ro},
  {Wechsler}, {Weiner}, {Wiklind}, {Wild}, {Wilson}, {Wuyts}, \&
  {Yun}}]{Koekemoer2011}
{Koekemoer}, A.~M., {Faber}, S.~M., {Ferguson}, H.~C., {et~al.} 2011, \apjs,
  197, 36

\bibitem[{{Kohandel} {et~al.}(2019){Kohandel}, {Pallottini}, {Ferrara},
  {Zanella}, {Behrens}, {Carniani}, {Gallerani}, \& {Vallini}}]{Kohandel2019}
{Kohandel}, M., {Pallottini}, A., {Ferrara}, A., {et~al.} 2019, \mnras, 487,
  3007

\bibitem[{{Lagache} {et~al.}(2018){Lagache}, {Cousin}, \&
  {Chatzikos}}]{Lagache2018}
{Lagache}, G., {Cousin}, M., \& {Chatzikos}, M. 2018, Astronomy and
  Astrophysics, 609, A130

\bibitem[{{Lagache} {et~al.}(2000){Lagache}, {Haffner}, {Reynolds}, \&
  {Tufte}}]{Lagache2000}
{Lagache}, G., {Haffner}, L.~M., {Reynolds}, R.~J., \& {Tufte}, S.~L. 2000,
  \aap, 354, 247

\bibitem[{{Laigle} {et~al.}(2016){Laigle}, {McCracken}, {Ilbert}, {Hsieh},
  {Davidzon}, {Capak}, {Hasinger}, {Silverman}, {Pichon}, {Coupon}, {Aussel},
  {Le Borgne}, {Caputi}, {Cassata}, {Chang}, {Civano}, {Dunlop}, {Fynbo},
  {Kartaltepe}, {Koekemoer}, {Le F{\`e}vre}, {Le Floc'h}, {Leauthaud}, {Lilly},
  {Lin}, {Marchesi}, {Milvang-Jensen}, {Salvato}, {Sanders}, {Scoville},
  {Smolcic}, {Stockmann}, {Taniguchi}, {Tasca}, {Toft}, {Vaccari}, \&
  {Zabl}}]{Laigle2015}
{Laigle}, C., {McCracken}, H.~J., {Ilbert}, O., {et~al.} 2016, \apjs, 224, 24

\bibitem[{{Le F{\`e}vre} {et~al.}(2019){Le F{\`e}vre}, {B{\'e}thermin},
  {Faisst}, {Capak}, {Cassata}, {Silverman}, {Schaerer}, \&
  {Yan}}]{Le_Fevre2019}
{Le F{\`e}vre}, O., {B{\'e}thermin}, M., {Faisst}, A., {et~al.} 2019, arXiv
  e-prints, arXiv:1910.09517

\bibitem[{{Le F{\`e}vre} {et~al.}(2015){Le F{\`e}vre}, {Tasca}, {Cassata},
  {Garilli}, {Le Brun}, {Maccagni}, {Pentericci}, {Thomas}, {Vanzella},
  {Zamorani}, {Zucca}, {Amorin}, {Bardelli}, {Capak}, {Cassar{\`a}},
  {Castellano}, {Cimatti}, {Cuby}, {Cucciati}, {de la Torre}, {Durkalec},
  {Fontana}, {Giavalisco}, {Grazian}, {Hathi}, {Ilbert}, {Lemaux}, {Moreau},
  {Paltani}, {Ribeiro}, {Salvato}, {Schaerer}, {Scodeggio}, {Sommariva},
  {Talia}, {Taniguchi}, {Tresse}, {Vergani}, {Wang}, {Charlot}, {Contini},
  {Fotopoulou}, {L{\'o}pez-Sanjuan}, {Mellier}, \& {Scoville}}]{Le_Fevre2015}
{Le F{\`e}vre}, O., {Tasca}, L.~A.~M., {Cassata}, P., {et~al.} 2015, \aap, 576,
  A79

\bibitem[{{Loiacono} {et~al.}(2020){Loiacono}, {Decarli}, {Gruppioni}, {Talia},
  {Cimatti}, {Zamorani}, {Pozzi}, {Yan}, {Lemaux}, {Riechers}, {Le F{\`e}vre},
  {B{\'e}thermin}, {Capak}, {Cassata}, {Faisst}, {Schaerer}, {Silverman},
  {Bardelli}, {Boquien}, {Burkutean}, {Dessauges-Zavadsky}, {Fudamoto},
  {Fujimoto}, {Ginolfi}, {Hathi}, {Jones}, {Khusanova}, {Koekemoer}, {Lagache},
  {Massardi}, {Oesch}, {Romano}, {Vallini}, {Vergani}, \&
  {Zucca}}]{Loiacono2020}
{Loiacono}, F., {Decarli}, R., {Gruppioni}, C., {et~al.} 2020, arXiv e-prints,
  arXiv:2006.04837

\bibitem[{{Lutz} {et~al.}(2011){Lutz}, {Poglitsch}, {Altieri}, {Andreani},
  {Aussel}, {Berta}, {Bongiovanni}, {Brisbin}, {Cava}, {Cepa}, {Cimatti},
  {Daddi}, {Dominguez-Sanchez}, {Elbaz}, {F{\"o}rster Schreiber}, {Genzel},
  {Grazian}, {Gruppioni}, {Harwit}, {Le Floc'h}, {Magdis}, {Magnelli},
  {Maiolino}, {Nordon}, {P{\'e}rez Garc{\'{\i}}a}, {Popesso}, {Pozzi},
  {Riguccini}, {Rodighiero}, {Saintonge}, {Sanchez Portal}, {Santini}, {Shao},
  {Sturm}, {Tacconi}, {Valtchanov}, {Wetzstein}, \& {Wieprecht}}]{Lutz2011}
{Lutz}, D., {Poglitsch}, A., {Altieri}, B., {et~al.} 2011, \aap, 532, A90

\bibitem[{{Madau} \& {Dickinson}(2014)}]{Madau2014}
{Madau}, P. \& {Dickinson}, M. 2014, \araa, 52, 415

\bibitem[{{Magdis} {et~al.}(2012){Magdis}, {Daddi}, {B{\'e}thermin}, {Sargent},
  {Elbaz}, {Pannella}, {Dickinson}, {Dannerbauer}, {da Cunha}, {Walter},
  {Rigopoulou}, {Charmandaris}, {Hwang}, \& {Kartaltepe}}]{Magdis2012b}
{Magdis}, G.~E., {Daddi}, E., {B{\'e}thermin}, M., {et~al.} 2012, \apj, 760, 6

\bibitem[{{Magnelli} {et~al.}(2014){Magnelli}, {Lutz}, {Saintonge}, {Berta},
  {Santini}, {Symeonidis}, {Altieri}, {Andreani}, {Aussel}, {B{\'e}thermin},
  {Bock}, {Bongiovanni}, {Cepa}, {Cimatti}, {Conley}, {Daddi}, {Elbaz},
  {F{\"o}rster Schreiber}, {Genzel}, {Ivison}, {Le Floc'h}, {Magdis},
  {Maiolino}, {Nordon}, {Oliver}, {Page}, {P{\'e}rez Garc{\'{\i}}a},
  {Poglitsch}, {Popesso}, {Pozzi}, {Riguccini}, {Rodighiero}, {Rosario},
  {Roseboom}, {Sanchez-Portal}, {Scott}, {Sturm}, {Tacconi}, {Valtchanov},
  {Wang}, \& {Wuyts}}]{Magnelli2014}
{Magnelli}, B., {Lutz}, D., {Saintonge}, A., {et~al.} 2014, \aap, 561, A86

\bibitem[{{Maiolino} {et~al.}(2012){Maiolino}, {Gallerani}, {Neri}, {Cicone},
  {Ferrara}, {Genzel}, {Lutz}, {Sturm}, {Tacconi}, {Walter}, {Feruglio},
  {Fiore}, \& {Piconcelli}}]{Maiolino2012}
{Maiolino}, R., {Gallerani}, S., {Neri}, R., {et~al.} 2012, \mnras, 425, L66

\bibitem[{{Maniyar} {et~al.}(2018){Maniyar}, {B{\'e}thermin}, \&
  {Lagache}}]{Maniyar2018}
{Maniyar}, A.~S., {B{\'e}thermin}, M., \& {Lagache}, G. 2018, Astronomy and
  Astrophysics, 614, A39

\bibitem[{{Matthee} {et~al.}(2019){Matthee}, {Sobral}, {Boogaard},
  {R{\"o}ttgering}, {Vallini}, {Ferrara}, {Paulino-Afonso}, {Boone},
  {Schaerer}, \& {Mobasher}}]{Matthee2019}
{Matthee}, J., {Sobral}, D., {Boogaard}, L.~A., {et~al.} 2019, \apj, 881, 124

\bibitem[{{Matthee} {et~al.}(2017){Matthee}, {Sobral}, {Darvish}, {Santos},
  {Mobasher}, {Paulino-Afonso}, {R{\"o}ttgering}, \& {Alegre}}]{Matthee2017}
{Matthee}, J., {Sobral}, D., {Darvish}, B., {et~al.} 2017, \mnras, 472, 772

\bibitem[{{McMullin} {et~al.}(2007){McMullin}, {Waters}, {Schiebel}, {Young},
  \& {Golap}}]{McMullin2007}
{McMullin}, J.~P., {Waters}, B., {Schiebel}, D., {Young}, W., \& {Golap}, K.
  2007, ASPC, 376, 127

\bibitem[{{Momcheva} {et~al.}(2016){Momcheva}, {Brammer}, {van Dokkum},
  {Skelton}, {Whitaker}, {Nelson}, {Fumagalli}, {Maseda}, {Leja}, {Franx},
  {Rix}, {Bezanson}, {Da Cunha}, {Dickey}, {F{\"o}rster Schreiber},
  {Illingworth}, {Kriek}, {Labb{\'e}}, {Ulf Lange}, {Lundgren}, {Magee},
  {Marchesini}, {Oesch}, {Pacifici}, {Patel}, {Price}, {Tal}, {Wake}, {van der
  Wel}, \& {Wuyts}}]{Momcheva2016}
{Momcheva}, I.~G., {Brammer}, G.~B., {van Dokkum}, P.~G., {et~al.} 2016, \apjs,
  225, 27

\bibitem[{{Nguyen} {et~al.}(2010){Nguyen}, {Schulz}, {Levenson}, {Amblard},
  {Arumugam}, {Aussel}, {Babbedge}, {Blain}, {Bock}, {Boselli}, {Buat},
  {Castro-Rodriguez}, {Cava}, {Chanial}, {Chapin}, {Clements}, {Conley},
  {Conversi}, {Cooray}, {Dowell}, {Dwek}, {Eales}, {Elbaz}, {Fox},
  {Franceschini}, {Gear}, {Glenn}, {Griffin}, {Halpern}, {Hatziminaoglou},
  {Ibar}, {Isaak}, {Ivison}, {Lagache}, {Lu}, {Madden}, {Maffei}, {Mainetti},
  {Marchetti}, {Marsden}, {Marshall}, {O'Halloran}, {Oliver}, {Omont}, {Page},
  {Panuzzo}, {Papageorgiou}, {Pearson}, {Perez Fournon}, {Pohlen}, {Rangwala},
  {Rigopoulou}, {Rizzo}, {Roseboom}, {Rowan-Robinson}, {Scott}, {Seymour},
  {Shupe}, {Smith}, {Stevens}, {Symeonidis}, {Trichas}, {Tugwell}, {Vaccari},
  {Valtchanov}, {Vigroux}, {Wang}, {Ward}, {Wiebe}, {Wright}, {Xu}, \&
  {Zemcov}}]{Nguyen2010}
{Nguyen}, H.~T., {Schulz}, B., {Levenson}, L., {et~al.} 2010, \aap, 518, L5

\bibitem[{{Odegard} {et~al.}(2019){Odegard}, {Weiland}, {Fixsen}, {Chuss},
  {Dwek}, {Kogut}, \& {Switzer}}]{Odegard2019}
{Odegard}, N., {Weiland}, J.~L., {Fixsen}, D.~J., {et~al.} 2019, \apj, 877, 40

\bibitem[{{Oliver} {et~al.}(2012){Oliver}, {Bock}, {Altieri}, {Amblard},
  {Arumugam}, {Aussel}, {Babbedge}, {Beelen}, {B{\'e}thermin}, {Blain},
  {Boselli}, {Bridge}, {Brisbin}, {Buat}, {Burgarella},
  {Castro-Rodr{\'{\i}}guez}, {Cava}, {Chanial}, {Cirasuolo}, {Clements},
  {Conley}, {Conversi}, {Cooray}, {Dowell}, {Dubois}, {Dwek}, {Dye}, {Eales},
  {Elbaz}, {Farrah}, {Feltre}, {Ferrero}, {Fiolet}, {Fox}, {Franceschini},
  {Gear}, {Giovannoli}, {Glenn}, {Gong}, {Gonz{\'a}lez Solares}, {Griffin},
  {Halpern}, {Harwit}, {Hatziminaoglou}, {Heinis}, {Hurley}, {Hwang}, {Hyde},
  {Ibar}, {Ilbert}, {Isaak}, {Ivison}, {Lagache}, {Le Floc'h}, {Levenson},
  {Faro}, {Lu}, {Madden}, {Maffei}, {Magdis}, {Mainetti}, {Marchetti},
  {Marsden}, {Marshall}, {Mortier}, {Nguyen}, {O'Halloran}, {Omont}, {Page},
  {Panuzzo}, {Papageorgiou}, {Patel}, {Pearson}, {P{\'e}rez-Fournon}, {Pohlen},
  {Rawlings}, {Raymond}, {Rigopoulou}, {Riguccini}, {Rizzo}, {Rodighiero},
  {Roseboom}, {Rowan-Robinson}, {S{\'a}nchez Portal}, {Schulz}, {Scott},
  {Seymour}, {Shupe}, {Smith}, {Stevens}, {Symeonidis}, {Trichas}, {Tugwell},
  {Vaccari}, {Valtchanov}, {Vieira}, {Viero}, {Vigroux}, {Wang}, {Ward},
  {Wardlow}, {Wright}, {Xu}, \& {Zemcov}}]{Oliver2012}
{Oliver}, S.~J., {Bock}, J., {Altieri}, B., {et~al.} 2012, \mnras, 424, 1614

\bibitem[{{Olsen} {et~al.}(2017){Olsen}, {Greve}, {Narayanan}, {Thompson},
  {Dav{\'e}}, {Niebla Rios}, \& {Stawinski}}]{Olsen2017}
{Olsen}, K., {Greve}, T.~R., {Narayanan}, D., {et~al.} 2017, \apj, 846, 105

\bibitem[{{Oteo} {et~al.}(2016){Oteo}, {Zwaan}, {Ivison}, {Smail}, \&
  {Biggs}}]{Oteo2016}
{Oteo}, I., {Zwaan}, M.~A., {Ivison}, R.~J., {Smail}, I., \& {Biggs}, A.~D.
  2016, \apj, 822, 36

\bibitem[{{Pahl} {et~al.}(2019){Pahl}, {Shapley}, {Faisst}, {Capak}, {Du},
  {Reddy}, {Laursen}, \& {Topping}}]{Pahl2019}
{Pahl}, A.~J., {Shapley}, A., {Faisst}, A.~L., {et~al.} 2019, arXiv e-prints,
  arXiv:1910.04179

\bibitem[{{Pallottini} {et~al.}(2019){Pallottini}, {Ferrara}, {Decataldo},
  {Gallerani}, {Vallini}, {Carniani}, {Behrens}, {Kohandel}, \&
  {Salvadori}}]{Pallottini2019}
{Pallottini}, A., {Ferrara}, A., {Decataldo}, D., {et~al.} 2019, \mnras, 487,
  1689

\bibitem[{{Pilbratt} {et~al.}(2010){Pilbratt}, {Riedinger}, {Passvogel},
  {Crone}, {Doyle}, {Gageur}, {Heras}, {Jewell}, {Metcalfe}, {Ott}, \&
  {Schmidt}}]{Pilbratt2010}
{Pilbratt}, G.~L., {Riedinger}, J.~R., {Passvogel}, T., {et~al.} 2010, \aap,
  518, L1+

\bibitem[{{Planck Collaboration} {et~al.}(2013){Planck Collaboration}, {Ade},
  {Aghanim}, {Arg{\"u}eso}, {Arnaud}, {Ashdown}, {Atrio-Barandela}, {Aumont},
  {Baccigalupi}, {Balbi}, {Banday}, {Barreiro}, {Battaner}, {Benabed},
  {Beno{\^\i}t}, {Bernard}, {Bersanelli}, {Bethermin}, {Bhatia}, {Bonaldi},
  {Bond}, {Borrill}, {Bouchet}, {Burigana}, {Cabella}, {Cardoso}, {Catalano},
  {Cay{\'o}n}, {Chamballu}, {Chary}, {Chen}, {Chiang}, {Christensen},
  {Clements}, {Colafrancesco}, {Colombi}, {Colombo}, {Coulais}, {Crill},
  {Cuttaia}, {Danese}, {Davis}, {de Bernardis}, {de Gasperis}, {de Zotti},
  {Delabrouille}, {Dickinson}, {Diego}, {Dole}, {Donzelli}, {Dor{\'e}},
  {D{\"o}rl}, {Douspis}, {Dupac}, {Efstathiou}, {En{\ss}lin}, {Eriksen},
  {Finelli}, {Forni}, {Fosalba}, {Frailis}, {Franceschi}, {Galeotta}, {Ganga},
  {Giard}, {Giardino}, {Giraud-H{\'e}raud}, {Gonz{\'a}lez-Nuevo}, {G{\'o}rski},
  {Gregorio}, {Gruppuso}, {Hansen}, {Harrison}, {Henrot-Versill{\'e}},
  {Hern{\'a}ndez-Monteagudo}, {Herranz}, {Hildebrandt}, {Hivon}, {Hobson},
  {Holmes}, {Jaffe}, {Jaffe}, {Jagemann}, {Jones}, {Juvela}, {Keih{\"a}nen},
  {Kisner}, {Kneissl}, {Knoche}, {Knox}, {Kunz}, {Kurinsky}, {Kurki-Suonio},
  {Lagache}, {L{\"a}hteenm{\"a}ki}, {Lamarre}, {Lasenby}, {Lawrence},
  {Leonardi}, {Lilje}, {L{\'o}pez-Caniego}, {Mac{\'{\i}}as-P{\'e}rez}, {Maino},
  {Mandolesi}, {Maris}, {Marshall}, {Mart{\'{\i}}nez-Gonz{\'a}lez}, {Masi},
  {Massardi}, {Matarrese}, {Mazzotta}, {Melchiorri}, {Mendes}, {Mennella},
  {Mitra}, {Miville-Desch{\`e}nes}, {Moneti}, {Montier}, {Morgante},
  {Mortlock}, {Munshi}, {Murphy}, {Naselsky}, {Nati}, {Natoli},
  {N{\o}rgaard-Nielsen}, {Noviello}, {Novikov}, {Novikov}, {Osborne}, {Pajot},
  {Paladini}, {Paoletti}, {Partridge}, {Pasian}, {Patanchon}, {Perdereau},
  {Perotto}, {Perrotta}, {Piacentini}, {Piat}, {Pierpaoli}, {Plaszczynski},
  {Pointecouteau}, {Polenta}, {Ponthieu}, {Popa}, {Poutanen}, {Pratt},
  {Prunet}, {Puget}, {Rachen}, {Reach}, {Rebolo}, {Reinecke}, {Renault},
  {Ricciardi}, {Riller}, {Ristorcelli}, {Rocha}, {Rosset}, {Rowan-Robinson},
  {Rubi{\~n}o-Mart{\'{\i}}n}, {Rusholme}, {Sajina}, {Sandri}, {Savini},
  {Scott}, {Smoot}, {Starck}, {Sudiwala}, {Suur-Uski}, {Sygnet}, {Tauber},
  {Terenzi}, {Toffolatti}, {Tomasi}, {Tristram}, {Tucci}, {T{\"u}rler},
  {Valenziano}, {Van Tent}, {Vielva}, {Villa}, {Vittorio}, {Wade}, {Wandelt},
  {White}, {Yvon}, {Zacchei}, \& {Zonca}}]{Planck_eucl}
{Planck Collaboration}, {Ade}, P.~A.~R., {Aghanim}, N., {et~al.} 2013, \aap,
  550, A133

\bibitem[{{Price-Whelan} {et~al.}(2018){Price-Whelan}, {Sip{\H{o}}cz},
  {G{\"u}nther}, {Lim}, {Crawford}, {Conseil}, {Shupe}, {Craig}, {Dencheva},
  {Ginsburg}, {VanderPlas}, {Bradley}, {P{\'e}rez-Su{\'a}rez}, {de Val-Borro},
  {Paper Contributors}, {Aldcroft}, {Cruz}, {Robitaille}, {Tollerud},
  {Coordination Committee}, {Ardelean}, {Babej}, {Bach}, {Bachetti}, {Bakanov},
  {Bamford}, {Barentsen}, {Barmby}, {Baumbach}, {Berry}, {Biscani}, {Boquien},
  {Bostroem}, {Bouma}, {Brammer}, {Bray}, {Breytenbach}, {Buddelmeijer},
  {Burke}, {Calderone}, {Cano Rodr{\'\i}guez}, {Cara}, {Cardoso}, {Cheedella},
  {Copin}, {Corrales}, {Crichton}, {D{\textquoteright}Avella}, {Deil},
  {Depagne}, {Dietrich}, {Donath}, {Droettboom}, {Earl}, {Erben}, {Fabbro},
  {Ferreira}, {Finethy}, {Fox}, {Garrison}, {Gibbons}, {Goldstein}, {Gommers},
  {Greco}, {Greenfield}, {Groener}, {Grollier}, {Hagen}, {Hirst}, {Homeier},
  {Horton}, {Hosseinzadeh}, {Hu}, {Hunkeler}, {Ivezi{\'c}}, {Jain}, {Jenness},
  {Kanarek}, {Kendrew}, {Kern}, {Kerzendorf}, {Khvalko}, {King}, {Kirkby},
  {Kulkarni}, {Kumar}, {Lee}, {Lenz}, {Littlefair}, {Ma}, {Macleod},
  {Mastropietro}, {McCully}, {Montagnac}, {Morris}, {Mueller}, {Mumford},
  {Muna}, {Murphy}, {Nelson}, {Nguyen}, {Ninan}, {N{\"o}the}, {Ogaz}, {Oh},
  {Parejko}, {Parley}, {Pascual}, {Patil}, {Patil}, {Plunkett}, {Prochaska},
  {Rastogi}, {Reddy Janga}, {Sabater}, {Sakurikar}, {Seifert}, {Sherbert},
  {Sherwood-Taylor}, {Shih}, {Sick}, {Silbiger}, {Singanamalla}, {Singer},
  {Sladen}, {Sooley}, {Sornarajah}, {Streicher}, {Teuben}, {Thomas},
  {Tremblay}, {Turner}, {Terr{\'o}n}, {van Kerkwijk}, {de la Vega}, {Watkins},
  {Weaver}, {Whitmore}, {Woillez}, {Zabalza}, \& {Contributors}}]{astropy:2018}
{Price-Whelan}, A.~M., {Sip{\H{o}}cz}, B.~M., {G{\"u}nther}, H.~M., {et~al.}
  2018, \aj, 156, 123

\bibitem[{{Puget} {et~al.}(1996){Puget}, {Abergel}, {Bernard}, {Boulanger},
  {Burton}, {Desert}, \& {Hartmann}}]{Puget1996}
{Puget}, J., {Abergel}, A., {Bernard}, J., {et~al.} 1996, \aap, 308, L5+

\bibitem[{{Riechers} {et~al.}(2013){Riechers}, {Bradford}, {Clements},
  {Dowell}, {P{\'e}rez-Fournon}, {Ivison}, {Bridge}, {Conley}, {Fu}, {Vieira},
  {Wardlow}, {Calanog}, {Cooray}, {Hurley}, {Neri}, {Kamenetzky}, {Aguirre},
  {Altieri}, {Arumugam}, {Benford}, {B{\'e}thermin}, {Bock}, {Burgarella},
  {Cabrera-Lavers}, {Chapman}, {Cox}, {Dunlop}, {Earle}, {Farrah}, {Ferrero},
  {Franceschini}, {Gavazzi}, {Glenn}, {Solares}, {Gurwell}, {Halpern},
  {Hatziminaoglou}, {Hyde}, {Ibar}, {Kov{\'a}cs}, {Krips}, {Lupu}, {Maloney},
  {Martinez-Navajas}, {Matsuhara}, {Murphy}, {Naylor}, {Nguyen}, {Oliver},
  {Omont}, {Page}, {Petitpas}, {Rangwala}, {Roseboom}, {Scott}, {Smith},
  {Staguhn}, {Streblyanska}, {Thomson}, {Valtchanov}, {Viero}, {Wang},
  {Zemcov}, \& {Zmuidzinas}}]{Riechers2013}
{Riechers}, D.~A., {Bradford}, C.~M., {Clements}, D.~L., {et~al.} 2013, \nat,
  496, 329

\bibitem[{{Romano} {et~al.}(2020){Romano}, {Cassata}, {Morselli}, {Lemaux},
  {B{\'e}thermin}, {Capak}, {Faisst}, {Le F{\`e}vre}, {Schaerer}, {Silverman},
  {Yan}, {Bardelli}, {Boquien}, {Cimatti}, {Dessauges-Zavadsky}, {Enia},
  {Fudamoto}, {Fujimoto}, {Ginolfi}, {Gruppioni}, {Hathi}, {Ibar}, {Jones},
  {Koekemoer}, {Loiacono}, {Mancini}, {Riechers}, {Rodighiero},
  {Rodr{\'\i}guez-Mu{\~n}oz}, {Talia}, {Vallini}, {Vergani}, {Zamorani}, \&
  {Zucca}}]{Romano2020}
{Romano}, M., {Cassata}, P., {Morselli}, L., {et~al.} 2020, \mnras
  [\eprint[arXiv]{2002.00961}]

\bibitem[{{Schaerer} {et~al.}(2020){Schaerer}, {Ginolfi}, {Bethermin},
  {Fudamoto}, {Oesch}, {Le Fevre}, {Faisst}, {Capak}, {Cassata}, {Silverman},
  {Yan}, {Jones}, {Amorin}, {Bardelli}, {Boquien}, {Cimatti},
  {Dessauges-Zavadsky}, {Giavalisco}, {Hathi}, {Fujimoto}, {Ibar}, {Koekemoer},
  {Lagache}, {Lemaux}, {Loiacono}, {Maiolino}, {Narayanan}, {Morselli},
  {Mendez-Hernandez}, {Pozzi}, {Riechers}, {Talia}, {Toft}, {Vallini},
  {Vergani}, {Zamorani}, \& {Zucca}}]{Schaerer2020}
{Schaerer}, D., {Ginolfi}, M., {Bethermin}, M., {et~al.} 2020, arXiv e-prints,
  arXiv:2002.00979

\bibitem[{{Schinnerer} {et~al.}(2008){Schinnerer}, {Carilli}, {Capak},
  {Martinez-Sansigre}, {Scoville}, {Smol{\v{c}}i{\'c}}, {Taniguchi}, {Yun},
  {Bertoldi}, {Le Fevre}, \& {de Ravel}}]{Schinnerer2008}
{Schinnerer}, E., {Carilli}, C.~L., {Capak}, P., {et~al.} 2008, \apjl, 689, L5

\bibitem[{{Schreiber} {et~al.}(2018){Schreiber}, {Elbaz}, {Pannella}, {Ciesla},
  {Wang}, \& {Franco}}]{Schreiber2018a}
{Schreiber}, C., {Elbaz}, D., {Pannella}, M., {et~al.} 2018, Astronomy and
  Astrophysics, 609, A30

\bibitem[{{Schreiber} {et~al.}(2015){Schreiber}, {Pannella}, {Elbaz},
  {B{\'e}thermin}, {Inami}, {Dickinson}, {Magnelli}, {Wang}, {Aussel}, {Daddi},
  {Juneau}, {Shu}, {Sargent}, {Buat}, {Faber}, {Ferguson}, {Giavalisco},
  {Koekemoer}, {Magdis}, {Morrison}, {Papovich}, {Santini}, \&
  {Scott}}]{Schreiber2015}
{Schreiber}, C., {Pannella}, M., {Elbaz}, D., {et~al.} 2015, \aap, 575, A74

\bibitem[{{Scoville} {et~al.}(2007){Scoville}, {Aussel}, {Benson}, {Blain},
  {Calzetti}, {Capak}, {Ellis}, {El-Zant}, {Finoguenov}, {Giavalisco}, {Guzzo},
  {Hasinger}, {Koda}, {Le F{\`e}vre}, {Massey}, {McCracken}, {Mobasher},
  {Renzini}, {Rhodes}, {Salvato}, {Sanders}, {Sasaki}, {Schinnerer}, {Sheth},
  {Shopbell}, {Taniguchi}, {Taylor}, \& {Thompson}}]{Scoville2007}
{Scoville}, N., {Aussel}, H., {Benson}, A., {et~al.} 2007, \apjs, 172, 150

\bibitem[{{Scudder} {et~al.}(2016){Scudder}, {Oliver}, {Hurley}, {Griffin},
  {Sargent}, {Scott}, {Wang}, \& {Wardlow}}]{Scudder2016}
{Scudder}, J.~M., {Oliver}, S., {Hurley}, P.~D., {et~al.} 2016, \mnras, 460,
  1119

\bibitem[{{Simpson} {et~al.}(2015){Simpson}, {Smail}, {Swinbank}, {Chapman},
  {Geach}, {Ivison}, {Thomson}, {Aretxaga}, {Blain}, {Cowley}, {Chen},
  {Coppin}, {Dunlop}, {Edge}, {Farrah}, {Ibar}, {Karim}, {Knudsen},
  {Meijerink}, {Micha{\l}owski}, {Scott}, {Spaans}, \& {van der
  Werf}}]{Simpson2015}
{Simpson}, J.~M., {Smail}, I., {Swinbank}, A.~M., {et~al.} 2015, \apj, 807, 128

\bibitem[{{Simpson} {et~al.}(2017){Simpson}, {Smail}, {Swinbank}, {Ivison},
  {Dunlop}, {Geach}, {Almaini}, {Arumugam}, {Bremer}, {Chen}, {Conselice},
  {Coppin}, {Farrah}, {Ibar}, {Hartley}, {Ma}, {Micha{\l}owski}, {Scott},
  {Spaans}, {Thomson}, \& {van der Werf}}]{Simpson2017}
{Simpson}, J.~M., {Smail}, I., {Swinbank}, A.~M., {et~al.} 2017, \apj, 839, 58

\bibitem[{{Smirnova-Pinchukova} {et~al.}(2019){Smirnova-Pinchukova},
  {Husemann}, {Busch}, {Appleton}, {Bethermin}, {Combes}, {Croom}, {Davis},
  {Fischer}, {Gaspari}, {Groves}, {Klein}, {O'Dea}, {P{\'e}rez-Torres},
  {Scharw{\"a}chter}, {Singha}, {Tremblay}, \&
  {Urrutia}}]{Smirnova-Pinchukova2019}
{Smirnova-Pinchukova}, I., {Husemann}, B., {Busch}, G., {et~al.} 2019, \aap,
  626, L3

\bibitem[{{Stacey} {et~al.}(2010){Stacey}, {Hailey-Dunsheath}, {Ferkinhoff},
  {Nikola}, {Parshley}, {Benford}, {Staguhn}, \& {Fiolet}}]{Stacey2010}
{Stacey}, G.~J., {Hailey-Dunsheath}, S., {Ferkinhoff}, C., {et~al.} 2010, \apj,
  724, 957

\bibitem[{{Stach} {et~al.}(2018){Stach}, {Smail}, {Swinbank}, {Simpson},
  {Geach}, {An}, {Almaini}, {Arumugam}, {Blain}, {Chapman}, {Chen},
  {Conselice}, {Cooke}, {Coppin}, {Dunlop}, {Farrah}, {Gullberg}, {Hartley},
  {Ivison}, {Maltby}, {Micha{\l}owski}, {Scott}, {Simpson}, {Thomson},
  {Wardlow}, \& {van der Werf}}]{Stach2018}
{Stach}, S.~M., {Smail}, I., {Swinbank}, A.~M., {et~al.} 2018, \apj, 860, 161

\bibitem[{{Steidel} {et~al.}(2010){Steidel}, {Erb}, {Shapley}, {Pettini},
  {Reddy}, {Bogosavljevi{\'c}}, {Rudie}, \& {Rakic}}]{Steidel2010}
{Steidel}, C.~C., {Erb}, D.~K., {Shapley}, A.~E., {et~al.} 2010, \apj, 717, 289

\bibitem[{{Strandet} {et~al.}(2017){Strandet}, {Weiss}, {De Breuck}, {Marrone},
  {Vieira}, {Aravena}, {Ashby}, {B{\'e}thermin}, {Bothwell}, {Bradford},
  {Carlstrom}, {Chapman}, {Cunningham}, {Chen}, {Fassnacht}, {Gonzalez},
  {Greve}, {Gullberg}, {Hayward}, {Hezaveh}, {Litke}, {Ma}, {Malkan}, {Menten},
  {Miller}, {Murphy}, {Narayanan}, {Phadke}, {Rotermund}, {Spilker}, \&
  {Sreevani}}]{Strandet2017}
{Strandet}, M.~L., {Weiss}, A., {De Breuck}, C., {et~al.} 2017, The
  Astrophysical Journal, 842, L15

\bibitem[{{Strandet} {et~al.}(2016){Strandet}, {Weiss}, {Vieira}, {de Breuck},
  {Aguirre}, {Aravena}, {Ashby}, {B{\'e}thermin}, {Bradford}, {Carlstrom},
  {Chapman}, {Crawford}, {Everett}, {Fassnacht}, {Furstenau}, {Gonzalez},
  {Greve}, {Gullberg}, {Hezaveh}, {Kamenetzky}, {Litke}, {Ma}, {Malkan},
  {Marrone}, {Menten}, {Murphy}, {Nadolski}, {Rotermund}, {Spilker}, {Stark},
  \& {Welikala}}]{Strandet2016}
{Strandet}, M.~L., {Weiss}, A., {Vieira}, J.~D., {et~al.} 2016, \apj, 822, 80

\bibitem[{{Symeonidis} {et~al.}(2013){Symeonidis}, {Vaccari}, {Berta}, {Page},
  {Lutz}, {Arumugam}, {Aussel}, {Bock}, {Boselli}, {Buat}, {Capak}, {Clements},
  {Conley}, {Conversi}, {Cooray}, {Dowell}, {Farrah}, {Franceschini},
  {Giovannoli}, {Glenn}, {Griffin}, {Hatziminaoglou}, {Hwang}, {Ibar},
  {Ilbert}, {Ivison}, {Le Floc'h}, {Lilly}, {Kartaltepe}, {Magnelli}, {Magdis},
  {Marchetti}, {Nguyen}, {Nordon}, {O'Halloran}, {Oliver}, {Omont},
  {Papageorgiou}, {Patel}, {Pearson}, {P{\'e}rez-Fournon}, {Pohlen}, {Popesso},
  {Pozzi}, {Rigopoulou}, {Riguccini}, {Rosario}, {Roseboom}, {Rowan-Robinson},
  {Salvato}, {Schulz}, {Scott}, {Seymour}, {Shupe}, {Smith}, {Valtchanov},
  {Wang}, {Xu}, {Zemcov}, \& {Wuyts}}]{Symeonidis2013}
{Symeonidis}, M., {Vaccari}, M., {Berta}, S., {et~al.} 2013, \mnras, 431, 2317

\bibitem[{{Tasca} {et~al.}(2015){Tasca}, {Le F{\`e}vre}, {Hathi}, {Schaerer},
  {Ilbert}, {Zamorani}, {Lemaux}, {Cassata}, {Garilli}, {Le Brun}, {Maccagni},
  {Pentericci}, {Thomas}, {Vanzella}, {Zucca}, {Amorin}, {Bardelli},
  {Cassar{\`a}}, {Castellano}, {Cimatti}, {Cucciati}, {Durkalec}, {Fontana},
  {Giavalisco}, {Grazian}, {Paltani}, {Ribeiro}, {Scodeggio}, {Sommariva},
  {Talia}, {Tresse}, {Vergani}, {Capak}, {Charlot}, {Contini}, {de la Torre},
  {Dunlop}, {Fotopoulou}, {Koekemoer}, {L{\'o}pez-Sanjuan}, {Mellier}, {Pforr},
  {Salvato}, {Scoville}, {Taniguchi}, \& {Wang}}]{Tasca2015}
{Tasca}, L.~A.~M., {Le F{\`e}vre}, O., {Hathi}, N.~P., {et~al.} 2015, \aap,
  581, A54

\bibitem[{{Vallini} {et~al.}(2017){Vallini}, {Ferrara}, {Pallottini}, \&
  {Gallerani}}]{Vallini2017}
{Vallini}, L., {Ferrara}, A., {Pallottini}, A., \& {Gallerani}, S. 2017,
  \mnras, 467, 1300

\bibitem[{{Vallini} {et~al.}(2015){Vallini}, {Gallerani}, {Ferrara},
  {Pallottini}, \& {Yue}}]{Vallini2015}
{Vallini}, L., {Gallerani}, S., {Ferrara}, A., {Pallottini}, A., \& {Yue}, B.
  2015, The Astrophysical Journal, 813, 36

\bibitem[{{Verhamme} {et~al.}(2018){Verhamme}, {Garel}, {Ventou}, {Contini},
  {Bouch{\'e}}, {Herenz}, {Richard}, {Bacon}, {Schmidt}, {Maseda}, {Marino},
  {Brinchmann}, {Cantalupo}, {Caruana}, {Cl{\'e}ment}, {Diener}, {Drake},
  {Hashimoto}, {Inami}, {Kerutt}, {Kollatschny}, {Leclercq}, {Patr{\'\i}cio},
  {Schaye}, {Wisotzki}, \& {Zabl}}]{Verhamme2018}
{Verhamme}, A., {Garel}, T., {Ventou}, E., {et~al.} 2018, \mnras, 478, L60

\bibitem[{{Walter} {et~al.}(2016){Walter}, {Decarli}, {Aravena}, {Carilli},
  {Bouwens}, {da Cunha}, {Daddi}, {Ivison}, {Riechers}, {Smail}, {Swinbank},
  {Weiss}, {Anguita}, {Assef}, {Bacon}, {Bauer}, {Bell}, {Bertoldi}, {Chapman},
  {Colina}, {Cortes}, {Cox}, {Dickinson}, {Elbaz}, {G{\'o}nzalez-L{\'o}pez},
  {Ibar}, {Inami}, {Infante}, {Hodge}, {Karim}, {Le Fevre}, {Magnelli}, {Neri},
  {Oesch}, {Ota}, {Popping}, {Rix}, {Sargent}, {Sheth}, {van der Wel}, {van der
  Werf}, \& {Wagg}}]{Walter2016}
{Walter}, F., {Decarli}, R., {Aravena}, M., {et~al.} 2016, \apj, 833, 67

\bibitem[{{Walter} {et~al.}(2012){Walter}, {Decarli}, {Carilli}, {Bertoldi},
  {Cox}, {da Cunha}, {Daddi}, {Dickinson}, {Downes}, {Elbaz}, {Ellis}, {Hodge},
  {Neri}, {Riechers}, {Weiss}, {Bell}, {Dannerbauer}, {Krips}, {Krumholz},
  {Lentati}, {Maiolino}, {Menten}, {Rix}, {Robertson}, {Spinrad}, {Stark}, \&
  {Stern}}]{Walter2012}
{Walter}, F., {Decarli}, R., {Carilli}, C., {et~al.} 2012, \nat, 486, 233

\bibitem[{{Watson} {et~al.}(2015){Watson}, {Christensen}, {Knudsen}, {Richard},
  {Gallazzi}, \& {Micha{\l}owski}}]{Watson2015}
{Watson}, D., {Christensen}, L., {Knudsen}, K.~K., {et~al.} 2015, \nat, 519,
  327

\bibitem[{{Wolfire} {et~al.}(2003){Wolfire}, {McKee}, {Hollenbach}, \&
  {Tielens}}]{Wolfire2003}
{Wolfire}, M.~G., {McKee}, C.~F., {Hollenbach}, D., \& {Tielens}, A.~G.~G.~M.
  2003, \apj, 587, 278

\bibitem[{{Yan} {et~al.}(2020){Yan}, {Sajina}, {Loiacono}, {Lagache},
  {B{\`e}thermin}, {Faisst}, {Ginolfi}, {Le F{\`e}vre}, {Gruppioni}, {Capak},
  {Cassata}, {Schaerer}, {Silverman}, {Bardelli}, {Dessauges-Zavadsky},
  {Cimatti}, {Hathi}, {Lemaux}, {Ibar}, {Jones}, {Koekemoer}, {Oesch}, {Talia},
  {Pozzi}, {Riechers}, {Tasca}, {Toft}, {Vallini}, {Vergani}, {Zamorani}, \&
  {Zucca}}]{Yan2020}
{Yan}, L., {Sajina}, A., {Loiacono}, F., {et~al.} 2020, arXiv e-prints,
  arXiv:2006.04835

\bibitem[{{Zanella} {et~al.}(2018){Zanella}, {Daddi}, {Magdis}, {Diaz Santos},
  {Cormier}, {Liu}, {Cibinel}, {Gobat}, {Dickinson}, {Sargent}, {Popping},
  {Madden}, {Bethermin}, {Hughes}, {Valentino}, {Rujopakarn}, {Pannella},
  {Bournaud}, {Walter}, {Wang}, {Elbaz}, \& {Coogan}}]{Zanella2018}
{Zanella}, A., {Daddi}, E., {Magdis}, G., {et~al.} 2018, \mnras, 481, 1976

\bibitem[{{Zavala} {et~al.}(2018){Zavala}, {Monta{\~n}a}, {Hughes}, {Yun},
  {Ivison}, {Valiante}, {Wilner}, {Spilker}, {Aretxaga}, {Eales},
  {Avila-Reese}, {Ch{\'a}vez}, {Cooray}, {Dannerbauer}, {Dunlop}, {Dunne},
  {G{\'o}mez-Ruiz}, {Micha{\l}owski}, {Narayanan}, {Nayyeri}, {Oteo}, {Rosa
  Gonz{\'a}lez}, {S{\'a}nchez-Arg{\"u}elles}, {Schloerb}, {Serjeant}, {Smith},
  {Terlevich}, {Vega}, {Villalba}, {van der Werf}, {Wilson}, \&
  {Zeballos}}]{Zavala2018}
{Zavala}, J.~A., {Monta{\~n}a}, A., {Hughes}, D.~H., {et~al.} 2018, Nature
  Astronomy, 2, 56

\end{thebibliography}

\clearpage

\begin{appendix}

\onecolumn

\section{Pointing list, beam size, and depth}

\label{sect:field_perf}

\begin{longtable}{lccccccc}
\caption{\label{tab:field_perf} Summary of the characteristics of our observations in the field around each of our target source. The coordinates correspond to the phase center of the ALMA pointing, which is also the optical position of the target. The sensitivities obtained in the moment-0 maps are normalized to a velocity window width of 235\,km/s (see Sect.\,\ref{sect:perf}). These results are discussed in Sect.\,\ref{sect:perf}.}
\\\hline
\hline
Target source & RA & Dec & Frequency & beam size & $\sigma_{\rm cont}$  & $\sigma_{mom0} \sqrt{(235 {\rm km/s}) / {\rm \Delta v}}$ \\
& h:m:s & deg:min:s & GHz & & $\mu$Jy/beam & Jy\,km/s/beam\\
\hline
\endfirsthead
\caption{continued.}\\
\hline\hline
Target source & RA & Dec & Frequency & beam size & $\sigma_{\rm cont}$  & $\sigma_{mom0}  \sqrt{(235 {\rm km/s}) / {\rm \Delta v}}$ \\
& h:min:s & deg:min:s & GHz & & $\mu$Jy/beam & Jy\,km/s/beam\\
\hline
\endhead
\hline
\endfoot
CANDELS\_GOODSS\_12 & 3:32:54.03 &  -27:50:00.82 & 356.2 & 1.04"$\times$0.79" & 54 & 0.232 \\ 
CANDELS\_GOODSS\_14 & 3:32:18.92 &  -27:53:02.75 & 294.6 & 1.22"$\times$0.92" & 22 & 0.054 \\ 
CANDELS\_GOODSS\_19 & 3:32:22.97 &  -27:46:29.02 & 350.6 & 1.01"$\times$0.77" & 45 & 0.099 \\ 
CANDELS\_GOODSS\_21 & 3:32:11.93 &  -27:41:57.08 & 295.6 & 1.15"$\times$0.98" & 22 & 0.029 \\ 
CANDELS\_GOODSS\_32 & 3:32:17.00 &  -27:41:13.72 & 356.2 & 1.04"$\times$0.79" & 51 & 0.104 \\ 
CANDELS\_GOODSS\_37 & 3:32:41.61 &  -27:49:05.89 & 350.6 & 1.01"$\times$0.78" & 42 & 0.049 \\ 
CANDELS\_GOODSS\_38 & 3:32:15.90 &  -27:41:23.95 & 294.5 & 1.41"$\times$0.90" & 29 & 0.098 \\ 
CANDELS\_GOODSS\_42 & 3:32:39.82 &  -27:52:58.08 & 295.6 & 1.15"$\times$0.98" & 22 & 0.014 \\ 
CANDELS\_GOODSS\_47 & 3:32:45.23 &  -27:49:09.84 & 294.5 & 1.41"$\times$0.90" & 28 & 0.062 \\ 
CANDELS\_GOODSS\_57 & 3:32:39.03 &  -27:52:23.09 & 296.6 & 1.17"$\times$1.01" & 26 & 0.041 \\ 
CANDELS\_GOODSS\_75 & 3:32:32.61 &  -27:47:54.02 & 294.6 & 1.22"$\times$0.92" & 25 & 0.086 \\ 
CANDELS\_GOODSS\_8 & 3:32:37.63 &  -27:50:22.41 & 296.6 & 1.17"$\times$1.01" & 25 & 0.039 \\ 
DEIMOS\_COSMOS\_206253 & 10:01:07.03 &  +1:35:36.91 & 353.9 & 1.11"$\times$0.77" & 42 & 0.069 \\ 
DEIMOS\_COSMOS\_224751 & 9:59:52.13 &  +1:37:23.10 & 289.4 & 1.17"$\times$0.98" & 19 & 0.016 \\ 
DEIMOS\_COSMOS\_274035 & 9:59:32.47 &  +1:42:05.97 & 351.5 & 1.06"$\times$0.89" & 53 & 0.132 \\ 
DEIMOS\_COSMOS\_298678 & 10:01:26.90 &  +1:44:30.16 & 291.0 & 1.26"$\times$0.97" & 32 & 0.050 \\ 
DEIMOS\_COSMOS\_308643 & 10:01:26.69 &  +1:45:26.21 & 348.7 & 1.16"$\times$0.86" & 59 & 0.172 \\ 
DEIMOS\_COSMOS\_328419 & 9:59:55.01 &  +1:47:20.68 & 289.3 & 1.57"$\times$0.88" & 38 & 0.064 \\ 
DEIMOS\_COSMOS\_336830 & 10:01:36.10 &  +1:48:06.43 & 288.1 & 1.07"$\times$0.79" & 24 & 0.039 \\ 
DEIMOS\_COSMOS\_351640 & 10:01:29.09 &  +1:49:29.37 & 290.3 & 1.19"$\times$1.03" & 29 & 0.038 \\ 
DEIMOS\_COSMOS\_357722 & 9:59:52.03 &  +1:50:05.52 & 289.0 & 0.96"$\times$0.82" & 23 & 0.028 \\ 
DEIMOS\_COSMOS\_372292 & 9:59:39.17 &  +1:51:28.11 & 314.5 & 1.11"$\times$0.89" & 30 & 0.063 \\ 
DEIMOS\_COSMOS\_378903 & 10:01:11.42 &  +1:52:06.06 & 302.3 & 1.40"$\times$0.92" & 35 & 0.058 \\ 
DEIMOS\_COSMOS\_396844 & 10:00:59.64 &  +1:53:47.45 & 348.0 & 1.28"$\times$0.77" & 55 & 0.154 \\ 
DEIMOS\_COSMOS\_400160 & 10:01:04.10 &  +1:54:05.00 & 350.2 & 1.42"$\times$0.80" & 69 & 0.229 \\ 
DEIMOS\_COSMOS\_403030 & 10:00:06.58 &  +1:54:21.19 & 348.1 & 1.26"$\times$0.78" & 64 & 0.092 \\ 
DEIMOS\_COSMOS\_406956 & 10:02:08.81 &  +1:54:45.12 & 289.5 & 1.40"$\times$0.92" & 34 & 0.053 \\ 
DEIMOS\_COSMOS\_412589 & 10:01:36.82 &  +1:55:16.85 & 354.8 & 1.05"$\times$0.78" & 49 & 0.068 \\ 
DEIMOS\_COSMOS\_416105 & 10:02:45.67 &  +1:55:35.89 & 291.5 & 1.22"$\times$0.97" & 19 & 0.029 \\ 
DEIMOS\_COSMOS\_417567 & 10:02:04.10 &  +1:55:44.09 & 289.9 & 0.95"$\times$0.72" & 24 & 0.061 \\ 
DEIMOS\_COSMOS\_420065 & 10:01:23.83 &  +1:56:00.24 & 289.1 & 0.98"$\times$0.79" & 23 & 0.036 \\ 
DEIMOS\_COSMOS\_421062 & 10:00:08.54 &  +1:56:04.52 & 295.7 & 1.01"$\times$0.85" & 19 & 0.038 \\ 
DEIMOS\_COSMOS\_422677 & 10:01:59.47 &  +1:56:12.92 & 355.9 & 0.98"$\times$0.80" & 56 & 0.126 \\ 
DEIMOS\_COSMOS\_430951 & 10:01:18.43 &  +1:57:03.64 & 289.3 & 1.56"$\times$0.88" & 46 & 0.209 \\ 
DEIMOS\_COSMOS\_431067 & 10:01:28.46 &  +1:57:03.84 & 356.5 & 0.93"$\times$0.84" & 51 & 0.073 \\ 
DEIMOS\_COSMOS\_432340 & 10:02:09.55 &  +1:57:05.77 & 356.5 & 0.93"$\times$0.84" & 56 & 0.077 \\ 
DEIMOS\_COSMOS\_434239 & 10:01:17.11 &  +1:57:19.12 & 352.9 & 1.01"$\times$0.85" & 45 & 0.214 \\ 
DEIMOS\_COSMOS\_442844 & 9:59:59.81 &  +1:58:13.29 & 352.8 & 0.90"$\times$0.82" & 46 & 0.064 \\ 
DEIMOS\_COSMOS\_454608 & 10:02:43.37 &  +1:59:20.84 & 347.5 & 1.06"$\times$0.81" & 60 & 0.130 \\ 
DEIMOS\_COSMOS\_460378 & 10:00:11.38 &  +1:59:54.39 & 302.3 & 1.41"$\times$0.93" & 36 & 0.054 \\ 
DEIMOS\_COSMOS\_470116 & 9:59:44.06 &  +2:00:50.72 & 289.1 & 0.98"$\times$0.79" & 22 & 0.059 \\ 
DEIMOS\_COSMOS\_471063 & 10:00:15.67 &  +2:00:56.13 & 289.5 & 1.41"$\times$0.92" & 35 & 0.057 \\ 
DEIMOS\_COSMOS\_472215 & 10:00:19.97 &  +2:01:03.42 & 291.0 & 1.26"$\times$0.98" & 31 & 0.057 \\ 
DEIMOS\_COSMOS\_488399 & 10:03:01.15 &  +2:02:35.98 & 291.5 & 1.22"$\times$0.97" & 19 & 0.063 \\ 
DEIMOS\_COSMOS\_493583 & 10:00:23.35 &  +2:03:04.37 & 349.4 & 1.03"$\times$0.78" & 53 & 0.083 \\ 
DEIMOS\_COSMOS\_494057 & 9:58:28.51 &  +2:03:06.80 & 295.7 & 1.01"$\times$0.85" & 21 & 0.042 \\ 
DEIMOS\_COSMOS\_494763 & 10:00:05.11 &  +2:03:12.11 & 311.4 & 1.03"$\times$0.89" & 27 & 0.072 \\ 
DEIMOS\_COSMOS\_503575 & 9:58:53.26 &  +2:04:01.36 & 290.9 & 1.19"$\times$0.75" & 21 & 0.034 \\ 
DEIMOS\_COSMOS\_510660 & 9:59:53.16 &  +2:04:37.68 & 347.1 & 1.28"$\times$0.77" & 49 & 0.152 \\ 
DEIMOS\_COSMOS\_519281 & 9:59:00.91 &  +2:05:27.60 & 295.8 & 1.28"$\times$0.98" & 34 & 0.111 \\ 
DEIMOS\_COSMOS\_536534 & 9:59:53.26 &  +2:07:05.42 & 289.0 & 0.95"$\times$0.82" & 26 & 0.126 \\ 
DEIMOS\_COSMOS\_539609 & 9:59:07.27 &  +2:07:21.31 & 314.5 & 1.12"$\times$0.89" & 32 & 0.088 \\ 
DEIMOS\_COSMOS\_549131 & 10:00:42.94 &  +2:08:12.36 & 296.7 & 1.18"$\times$0.94" & 34 & 0.064 \\ 
DEIMOS\_COSMOS\_550156 & 10:01:00.48 &  +2:08:17.86 & 355.6 & 0.98"$\times$0.80" & 49 & 0.069 \\ 
DEIMOS\_COSMOS\_552206 & 9:58:26.78 &  +2:08:27.32 & 296.7 & 1.17"$\times$0.95" & 36 & 0.109 \\ 
DEIMOS\_COSMOS\_567070 & 10:01:05.95 &  +2:09:48.75 & 348.4 & 1.11"$\times$0.79" & 54 & 0.086 \\ 
DEIMOS\_COSMOS\_576372 & 9:59:54.77 &  +2:10:39.26 & 290.3 & 1.19"$\times$1.04" & 29 & 0.046 \\ 
DEIMOS\_COSMOS\_586681 & 10:00:08.78 &  +2:11:36.49 & 283.7 & 1.61"$\times$0.91" & 43 & 0.078 \\ 
DEIMOS\_COSMOS\_592644 & 10:00:26.62 &  +2:12:05.96 & 350.6 & 0.99"$\times$0.79" & 44 & 0.070 \\ 
DEIMOS\_COSMOS\_627939 & 10:01:04.87 &  +2:15:14.03 & 350.1 & 1.03"$\times$0.85" & 37 & 0.097 \\ 
DEIMOS\_COSMOS\_628063 & 10:00:52.25 &  +2:15:15.42 & 348.1 & 1.28"$\times$0.78" & 60 & 0.079 \\ 
DEIMOS\_COSMOS\_628137 & 9:59:54.53 &  +2:15:16.38 & 291.1 & 1.27"$\times$0.89" & 23 & 0.040 \\ 
DEIMOS\_COSMOS\_629750 & 10:00:34.32 &  +2:15:24.47 & 315.4 & 1.25"$\times$0.88" & 34 & 0.053 \\ 
DEIMOS\_COSMOS\_630594 & 10:00:32.62 &  +2:15:28.40 & 353.9 & 1.13"$\times$0.77" & 48 & 0.103 \\ 
DEIMOS\_COSMOS\_665509 & 9:58:56.45 &  +2:18:39.24 & 348.6 & 1.06"$\times$0.84" & 56 & 0.148 \\ 
DEIMOS\_COSMOS\_665626 & 10:01:14.23 &  +2:18:42.34 & 347.2 & 1.06"$\times$0.83" & 46 & 0.035 \\ 
DEIMOS\_COSMOS\_680104 & 10:01:10.15 &  +2:19:56.28 & 350.2 & 0.99"$\times$0.83" & 53 & 0.081 \\ 
DEIMOS\_COSMOS\_683613 & 10:00:09.43 &  +2:20:13.86 & 295.8 & 1.29"$\times$0.98" & 34 & 0.072 \\ 
DEIMOS\_COSMOS\_709575 & 9:59:47.06 &  +2:22:32.79 & 355.9 & 0.99"$\times$0.80" & 57 & 0.092 \\ 
DEIMOS\_COSMOS\_722679 & 9:59:44.90 &  +2:23:46.36 & 288.1 & 1.07"$\times$0.79" & 26 & 0.072 \\ 
DEIMOS\_COSMOS\_733857 & 10:01:19.92 &  +2:24:47.48 & 347.5 & 1.13"$\times$0.80" & 54 & 0.104 \\ 
DEIMOS\_COSMOS\_742174 & 10:00:39.12 &  +2:25:32.43 & 291.1 & 1.27"$\times$0.89" & 24 & 0.026 \\ 
DEIMOS\_COSMOS\_743730 & 10:01:12.50 &  +2:25:42.78 & 349.4 & 1.17"$\times$0.77" & 42 & 0.057 \\ 
DEIMOS\_COSMOS\_761315 & 9:59:48.53 &  +2:27:20.35 & 347.4 & 0.96"$\times$0.85" & 41 & 0.099 \\ 
DEIMOS\_COSMOS\_773957 & 10:01:10.06 &  +2:28:29.07 & 289.4 & 1.17"$\times$0.99" & 20 & 0.055 \\ 
DEIMOS\_COSMOS\_787780 & 9:59:56.66 &  +2:29:48.08 & 351.5 & 1.06"$\times$0.90" & 49 & 0.040 \\ 
DEIMOS\_COSMOS\_790930 & 10:01:27.77 &  +2:30:06.06 & 290.9 & 1.18"$\times$0.75" & 21 & 0.035 \\ 
DEIMOS\_COSMOS\_803480 & 9:59:57.24 &  +2:31:12.88 & 347.8 & 0.93"$\times$0.83" & 38 & 0.039 \\ 
DEIMOS\_COSMOS\_814483 & 10:01:27.17 &  +2:32:10.17 & 347.1 & 1.26"$\times$0.77" & 48 & 0.133 \\ 
DEIMOS\_COSMOS\_818760 & 10:01:54.86 &  +2:32:31.54 & 347.2 & 1.05"$\times$0.82" & 52 & 0.157 \\ 
DEIMOS\_COSMOS\_834764 & 9:59:35.74 &  +2:34:00.51 & 350.2 & 1.45"$\times$0.81" & 74 & 0.163 \\ 
DEIMOS\_COSMOS\_838532 & 10:01:43.70 &  +2:34:21.10 & 348.9 & 1.22"$\times$0.78" & 51 & 0.053 \\ 
DEIMOS\_COSMOS\_842313 & 10:00:54.53 &  +2:34:35.16 & 349.1 & 1.25"$\times$0.78" & 75 & 0.075 \\ 
DEIMOS\_COSMOS\_843045 & 10:00:12.36 &  +2:34:43.68 & 283.7 & 1.62"$\times$0.91" & 48 & 0.056 \\ 
DEIMOS\_COSMOS\_845652 & 10:00:51.60 &  +2:34:57.55 & 307.6 & 0.92"$\times$0.85" & 30 & 0.061 \\ 
DEIMOS\_COSMOS\_848185 & 10:00:21.50 &  +2:35:11.08 & 307.6 & 0.92"$\times$0.85" & 29 & 0.110 \\ 
DEIMOS\_COSMOS\_859732 & 10:00:00.50 &  +2:36:19.14 & 348.4 & 1.12"$\times$0.79" & 60 & 0.111 \\ 
DEIMOS\_COSMOS\_869970 & 10:00:24.98 &  +2:37:18.17 & 311.4 & 1.04"$\times$0.89" & 26 & 0.038 \\ 
DEIMOS\_COSMOS\_873321 & 10:00:04.06 &  +2:37:35.90 & 315.4 & 1.26"$\times$0.88" & 37 & 0.077 \\ 
DEIMOS\_COSMOS\_873756 & 10:00:02.71 &  +2:37:40.20 & 347.4 & 0.96"$\times$0.85" & 44 & 0.245 \\ 
DEIMOS\_COSMOS\_880016 & 9:59:55.18 &  +2:38:08.21 & 348.1 & 1.22"$\times$0.83" & 44 & 0.109 \\ 
DEIMOS\_COSMOS\_881725 & 10:00:13.56 &  +2:38:16.92 & 346.6 & 1.19"$\times$0.75" & 47 & 0.083 \\ 
DEIMOS\_COSMOS\_910650 & 10:00:22.51 &  +2:41:03.38 & 290.6 & 0.97"$\times$0.76" & 24 & 0.038 \\ 
DEIMOS\_COSMOS\_920848 & 9:59:33.82 &  +2:41:56.12 & 348.9 & 1.24"$\times$0.78" & 52 & 0.072 \\ 
DEIMOS\_COSMOS\_926434 & 10:00:26.18 &  +2:42:23.24 & 355.6 & 0.99"$\times$0.80" & 52 & 0.068 \\ 
DEIMOS\_COSMOS\_933876 & 9:59:36.29 &  +2:43:09.54 & 356.3 & 0.97"$\times$0.81" & 53 & 0.071 \\ 
vuds\_cosmos\_5100537582 & 10:01:33.52 &  +1:50:20.39 & 349.4 & 1.16"$\times$0.77" & 44 & 0.085 \\ 
vuds\_cosmos\_5100541407 & 10:01:00.91 &  +1:48:33.85 & 347.5 & 1.06"$\times$0.81" & 62 & 0.109 \\ 
vuds\_cosmos\_5100559223 & 10:00:53.14 &  +1:51:53.41 & 348.7 & 1.17"$\times$0.86" & 54 & 0.071 \\ 
vuds\_cosmos\_5100822662 & 9:58:57.91 &  +2:04:51.45 & 349.8 & 1.09"$\times$0.83" & 31 & 0.054 \\ 
vuds\_cosmos\_5100969402 & 10:01:20.10 &  +2:17:01.06 & 346.0 & 0.98"$\times$0.81" & 38 & 0.090 \\ 
vuds\_cosmos\_5100994794 & 10:00:41.15 &  +2:17:14.13 & 346.0 & 0.97"$\times$0.82" & 38 & 0.079 \\ 
vuds\_cosmos\_5101013812 & 10:00:50.37 &  +2:09:47.25 & 356.3 & 0.96"$\times$0.81" & 50 & 0.072 \\ 
vuds\_cosmos\_5101209780 & 10:01:33.45 &  +2:22:10.23 & 348.0 & 1.29"$\times$0.76" & 62 & 0.248 \\ 
vuds\_cosmos\_5101210235 & 10:01:31.60 &  +2:21:57.95 & 347.8 & 0.93"$\times$0.82" & 39 & 0.056 \\ 
vuds\_cosmos\_5101218326 & 10:01:12.50 &  +2:18:52.72 & 348.1 & 1.20"$\times$0.83" & 42 & 0.104 \\ 
vuds\_cosmos\_5101244930 & 10:00:47.66 &  +2:18:02.32 & 347.5 & 1.12"$\times$0.80" & 59 & 0.291 \\ 
vuds\_cosmos\_5101288969 & 9:59:30.65 &  +2:19:53.50 & 290.3 & 0.93"$\times$0.77" & 17 & 0.034 \\ 
vuds\_cosmos\_510148750 & 9:59:22.27 &  +1:59:34.55 & 350.1 & 1.03"$\times$0.85" & 35 & 0.052 \\ 
vuds\_cosmos\_510327576 & 10:00:26.55 &  +1:47:06.05 & 348.6 & 1.06"$\times$0.84" & 54 & 0.082 \\ 
vuds\_cosmos\_510581738 & 9:59:38.11 &  +1:52:39.22 & 350.6 & 0.99"$\times$0.79" & 48 & 0.081 \\ 
vuds\_cosmos\_510596653 & 9:59:18.29 &  +1:56:17.19 & 346.6 & 1.18"$\times$0.75" & 46 & 0.028 \\ 
vuds\_cosmos\_510605533 & 9:59:24.61 &  +1:52:43.05 & 350.2 & 0.99"$\times$0.82" & 54 & 0.194 \\ 
vuds\_cosmos\_510786441 & 10:00:34.30 &  +1:59:21.14 & 353.9 & 1.01"$\times$0.85" & 34 & 0.061 \\ 
vuds\_cosmos\_5110377875 & 10:01:32.33 &  +2:24:30.41 & 349.4 & 1.03"$\times$0.79" & 55 & 0.123 \\ 
vuds\_cosmos\_5131465996 & 10:00:52.80 &  +2:27:55.94 & 352.8 & 0.90"$\times$0.83" & 45 & 0.062 \\ 
vuds\_cosmos\_5180966608 & 10:01:37.46 &  +2:08:23.69 & 350.4 & 1.26"$\times$0.75" & 43 & 0.096 \\ 
vuds\_efdcs\_530029038 & 3:32:19.03 &  -27:52:38.14 & 356.0 & 1.11"$\times$0.77" & 33 & 0.082 \\ 
\end{longtable}

\section{Continuum cutouts and catalogs}

\label{app:cont_cat}

\begin{figure*}[!h]
\centering
\includegraphics[width=18cm]{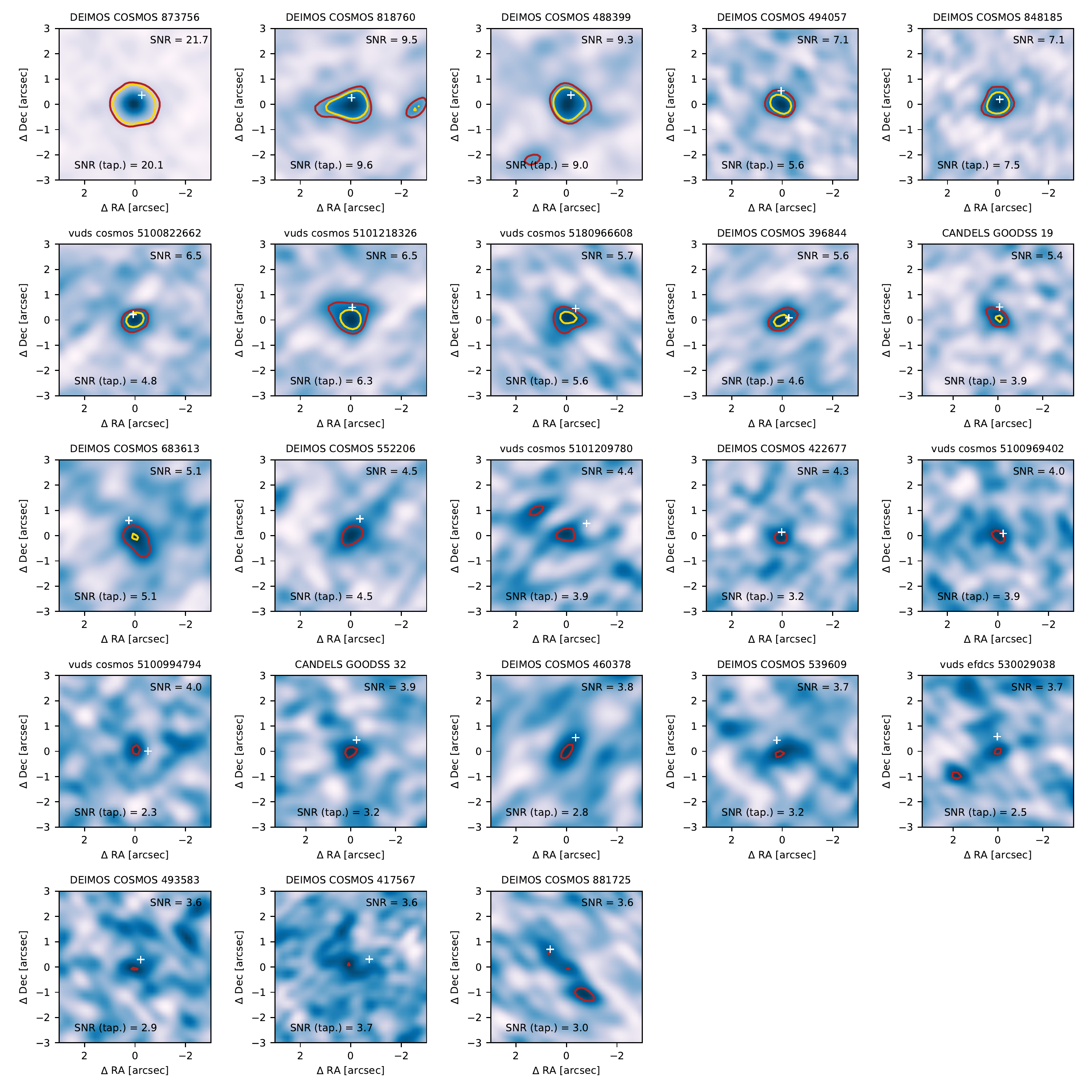}
\caption{\label{fig:cont_target_cutouts} Cutout images of the continuum-detected target sources. The red and the gold contours indicate the 3.5\,$\sigma$ and 5\,$\sigma$ levels, respectively. The white crosshair indicate the phase center.}
\end{figure*}

\begin{table*}[!h]
\caption{\label{tab:cont_target_list} Characteristics of the continuum-detected target sources. The frequency is the mean of the channels used to derive the continuum. The flux S$_\nu$ was measured using the 2D-fit approached described in Sect.\,\ref{sect:cont_ext}. The monochromatic luminosity $\nu$L$_\nu$ is derived at $(1+z) \, \nu_{\rm obs}$ and does not need any assumption on the galaxy SEDs. The total infrared luminosity L$_{\rm IR}$ (8-1000\,$\mu$m) is derived assuming \citet{Bethermin2017} SED (see discussion in Sect.\,\ref{sect:sed}). The measurement uncertainties do not include the calibration uncertainties ($\sim$4.5\,\% on average and up to 10\,\%, see Sect.\,\ref{sect:quasars}).}
\centering
\begin{tabular}{lcccrrrrr}
\hline
\hline
Name & RA & Dec & Freq. & S/N & S/N & S$_\nu$ & $\nu$L$_\nu$ & L$_{\rm IR}$ \\
 & h:min:s & deg:min:s & GHz & &  (tap.) & $\mu$Jy & \multicolumn{2}{c}{log$_{10}$(L/L$_\odot$)} \\
\hline
DEIMOS\_COSMOS\_873756 & 10:00:02.72 &  +02:37:39.99 & 347.4 & 21.7 & 20.1 & 1355$\pm$76 & 11.41 & 12.26 \\ 
DEIMOS\_COSMOS\_818760 & 10:01:54.86 &  +02:32:31.43 & 347.2 & 9.5 & 9.6 & 1077$\pm$130 & 11.31 & 12.16 \\ 
DEIMOS\_COSMOS\_488399 & 10:03:01.15 &  +02:02:35.77 & 291.5 & 9.3 & 9.0 & 253$\pm$32 & 10.83 & 11.67 \\ 
DEIMOS\_COSMOS\_494057 & 09:58:28.50 &  +02:03:06.42 & 295.7 & 7.1 & 5.6 & 179$\pm$30 & 10.66 & 11.51 \\ 
DEIMOS\_COSMOS\_848185 & 10:00:21.50 &  +02:35:11.03 & 307.6 & 7.1 & 7.5 & 319$\pm$50 & 10.88 & 11.73 \\ 
vuds\_cosmos\_5100822662 & 09:58:57.90 &  +02:04:51.38 & 349.8 & 6.5 & 4.8 & 210$\pm$38 & 10.60 & 11.45 \\ 
vuds\_cosmos\_5101218326 & 10:01:12.50 &  +02:18:52.37 & 348.1 & 6.5 & 6.3 & 462$\pm$79 & 10.95 & 11.79 \\ 
vuds\_cosmos\_5180966608 & 10:01:37.48 &  +02:08:23.39 & 350.4 & 5.7 & 5.6 & 419$\pm$84 & 10.90 & 11.74 \\ 
DEIMOS\_COSMOS\_396844 & 10:00:59.65 &  +01:53:47.52 & 348.0 & 5.6 & 4.6 & 346$\pm$69 & 10.82 & 11.67 \\ 
CANDELS\_GOODSS\_19 & 03:32:22.97 &  -27:46:29.32 & 350.6 & 5.4 & 3.9 & 278$\pm$61 & 10.71 & 11.57 \\ 
DEIMOS\_COSMOS\_683613 & 10:00:09.41 &  +02:20:13.41 & 295.8 & 5.1 & 5.1 & 245$\pm$54 & 10.80 & 11.65 \\ 
DEIMOS\_COSMOS\_552206 & 09:58:26.80 &  +02:08:26.80 & 296.7 & 4.5 & 4.5 & 285$\pm$73 & 10.86 & 11.71 \\ 
vuds\_cosmos\_5101209780 & \multicolumn{8}{c}{Multi-component object, see Appendix\,\ref{sect:DC881725}}\\ 
DEIMOS\_COSMOS\_422677 & 10:01:59.46 &  +01:56:12.93 & 355.9 & 4.3 & 3.2 & 375$\pm$123 & 10.84 & 11.68 \\ 
vuds\_cosmos\_5100969402 & 10:01:20.11 &  +02:17:01.13 & 346.0 & 4.0 & 3.9 & 327$\pm$99 & 10.80 & 11.65 \\ 
vuds\_cosmos\_5100994794 & 10:00:41.18 &  +02:17:14.27 & 346.0 & 4.0 & 2.3 & 117$\pm$36 & 10.35 & 11.20 \\ 
CANDELS\_GOODSS\_32 & 03:32:17.01 &  -27:41:13.97 & 356.2 & 3.9 & 3.2 & 230$\pm$65 & 10.62 & 11.47 \\ 
DEIMOS\_COSMOS\_460378 & 10:00:11.39 &  +01:59:54.00 & 302.3 & 3.8 & 2.8 & 116$\pm$35 & 10.46 & 11.31 \\ 
DEIMOS\_COSMOS\_539609 & 09:59:07.25 &  +02:07:21.02 & 314.5 & 3.7 & 3.2 & 187$\pm$54 & 10.64 & 11.48 \\ 
vuds\_efdcs\_530029038 & \multicolumn{8}{c}{Multi-component object, see Appendix\,\ref{sect:DC881725}}\\ 
DEIMOS\_COSMOS\_493583 & 10:00:23.36 &  +02:03:04.23 & 349.4 & 3.6 & 2.9 & 235$\pm$81 & 10.64 & 11.50 \\ 
DEIMOS\_COSMOS\_417567 & 10:02:04.14 &  +01:55:43.93 & 289.9 & 3.6 & 3.7 & 201$\pm$60 & 10.73 & 11.58 \\ 
DEIMOS\_COSMOS\_881725 & \multicolumn{8}{c}{Multi-component object, see Appendix\,\ref{sect:DC881725}}\\ 
\hline
\end{tabular}
\end{table*}

\begin{longtable}{lccccccccc}
\caption{\label{tab:uplim_cont} Upper limits on the flux densities and luminosities for target continuum non detections (see explanation in Sect.\,\ref{sect:uplim}). Three methods are used to compute them (aggressive, normal, and secure) and are described in Sect.\,\ref{sect:uplim}). In the absence of any external information, we recommend to use the secure upper limits. If the source is known to be point-like, the normal upper limits can be used. The luminosities are derived using the same method as in Table\,\ref{tab:cont_target_list}.}
\\\hline
\hline
Name & \multicolumn{3}{c}{3-$\sigma$ upper limit on flux density} & \multicolumn{3}{c}{3-$\sigma$ upper limit on $\nu$L$_\nu$} & \multicolumn{3}{c}{3-$\sigma$ upper limit on L$_{\rm IR}$} \\
 & aggr.& norm. & sec.& aggr.& norm. & sec. & aggr.& norm. & sec.\\
 & \multicolumn{3}{c}{mJy} &   \multicolumn{3}{c}{log$_{10}$(L/L$_\odot$)} & \multicolumn{3}{c}{log$_{10}$(L/L$_\odot$)} \\
\\\hline
\endfirsthead
\caption{continued.}\\
\hline\hline
 Name & \multicolumn{3}{c}{3-$\sigma$ upper limit on flux density} & \multicolumn{3}{c}{3-$\sigma$ upper limit on $\nu$L$_\nu$} & \multicolumn{3}{c}{3-$\sigma$ upper limit on L$_{\rm IR}$} \\
 & agr.& norm. & sec.& agr.& norm. & sec. & agr.& norm. & sec.\\
 & \multicolumn{3}{c}{mJy} &   \multicolumn{3}{c}{log$_{10}$(L/L$_\odot$)} & \multicolumn{3}{c}{log$_{10}$(L/L$_\odot$)} \\
 \hline
\endhead
\hline
\endfoot
CANDELS\_GOODSS\_12 & $<$0.164 & $<$0.241 & $<$0.236 & $<$10.47 & $<$10.63 & $<$10.63 & $<$11.34 & $<$11.51 & $<$11.50\\
CANDELS\_GOODSS\_14 & $<$0.067 & $<$0.108 & $<$0.119 & $<$10.23 & $<$10.44 & $<$10.48 & $<$11.10 & $<$11.31 & $<$11.35\\
CANDELS\_GOODSS\_21 & $<$0.066 & $<$0.141 & $<$0.145 & $<$10.22 & $<$10.55 & $<$10.57 & $<$11.10 & $<$11.43 & $<$11.44\\
CANDELS\_GOODSS\_37 & $<$0.121 & $<$0.227 & $<$0.255 & $<$10.35 & $<$10.62 & $<$10.67 & $<$11.22 & $<$11.50 & $<$11.55\\
CANDELS\_GOODSS\_38 & $<$0.088 & $<$0.136 & $<$0.178 & $<$10.35 & $<$10.54 & $<$10.65 & $<$11.23 & $<$11.41 & $<$11.53\\
CANDELS\_GOODSS\_42 & $<$0.066 & $<$0.106 & $<$0.106 & $<$10.22 & $<$10.42 & $<$10.43 & $<$11.09 & $<$11.30 & $<$11.30\\
CANDELS\_GOODSS\_47 & $<$0.086 & $<$0.157 & $<$0.167 & $<$10.34 & $<$10.60 & $<$10.63 & $<$11.21 & $<$11.48 & $<$11.50\\
CANDELS\_GOODSS\_57 & $<$0.076 & $<$0.133 & $<$0.206 & $<$10.28 & $<$10.53 & $<$10.72 & $<$11.16 & $<$11.40 & $<$11.59\\
CANDELS\_GOODSS\_75 & $<$0.074 & $<$0.097 & $<$0.092 & $<$10.28 & $<$10.39 & $<$10.37 & $<$11.15 & $<$11.27 & $<$11.25\\
CANDELS\_GOODSS\_8 & $<$0.075 & $<$0.118 & $<$0.151 & $<$10.27 & $<$10.47 & $<$10.58 & $<$11.15 & $<$11.34 & $<$11.45\\
DEIMOS\_COSMOS\_206253 & $<$0.127 & $<$0.219 & $<$0.216 & $<$10.36 & $<$10.60 & $<$10.59 & $<$11.24 & $<$11.48 & $<$11.47\\
DEIMOS\_COSMOS\_224751 & $<$0.057 & $<$0.071 & $<$0.080 & $<$10.18 & $<$10.27 & $<$10.32 & $<$11.05 & $<$11.15 & $<$11.20\\
DEIMOS\_COSMOS\_274035 & $<$0.157 & $<$0.270 & $<$0.281 & $<$10.46 & $<$10.69 & $<$10.71 & $<$11.33 & $<$11.57 & $<$11.59\\
DEIMOS\_COSMOS\_298678 & $<$0.096 & $<$0.148 & $<$0.161 & $<$10.40 & $<$10.59 & $<$10.62 & $<$11.27 & $<$11.46 & $<$11.50\\
DEIMOS\_COSMOS\_308643 & $<$0.179 & $<$0.313 & $<$0.324 & $<$10.52 & $<$10.76 & $<$10.78 & $<$11.40 & $<$11.64 & $<$11.65\\
DEIMOS\_COSMOS\_328419 & $<$0.116 & $<$0.186 & $<$0.179 & $<$10.48 & $<$10.69 & $<$10.67 & $<$11.36 & $<$11.57 & $<$11.55\\
DEIMOS\_COSMOS\_336830 & $<$0.074 & $<$0.102 & $<$0.132 & $<$10.29 & $<$10.43 & $<$10.54 & $<$11.16 & $<$11.31 & $<$11.42\\
DEIMOS\_COSMOS\_351640 & $<$0.089 & $<$0.133 & $<$0.148 & $<$10.37 & $<$10.54 & $<$10.59 & $<$11.24 & $<$11.42 & $<$11.46\\
DEIMOS\_COSMOS\_357722 & $<$0.069 & $<$0.101 & $<$0.100 & $<$10.26 & $<$10.43 & $<$10.42 & $<$11.14 & $<$11.30 & $<$11.30\\
DEIMOS\_COSMOS\_372292 & $<$0.091 & $<$0.137 & $<$0.182 & $<$10.31 & $<$10.49 & $<$10.61 & $<$11.19 & $<$11.36 & $<$11.49\\
DEIMOS\_COSMOS\_378903 & $<$0.107 & $<$0.162 & $<$0.133 & $<$10.42 & $<$10.60 & $<$10.51 & $<$11.29 & $<$11.47 & $<$11.39\\
DEIMOS\_COSMOS\_400160 & $<$0.211 & $<$0.377 & $<$0.401 & $<$10.59 & $<$10.84 & $<$10.87 & $<$11.47 & $<$11.72 & $<$11.75\\
DEIMOS\_COSMOS\_403030 & $<$0.186 & $<$0.348 & $<$0.384 & $<$10.54 & $<$10.81 & $<$10.86 & $<$11.42 & $<$11.69 & $<$11.73\\
DEIMOS\_COSMOS\_406956 & $<$0.104 & $<$0.128 & $<$0.133 & $<$10.43 & $<$10.52 & $<$10.54 & $<$11.31 & $<$11.40 & $<$11.42\\
DEIMOS\_COSMOS\_412589 & $<$0.147 & $<$0.270 & $<$0.384 & $<$10.42 & $<$10.68 & $<$10.84 & $<$11.30 & $<$11.56 & $<$11.71\\
DEIMOS\_COSMOS\_416105 & $<$0.057 & $<$0.081 & $<$0.097 & $<$10.17 & $<$10.32 & $<$10.40 & $<$11.04 & $<$11.19 & $<$11.28\\
DEIMOS\_COSMOS\_420065 & $<$0.068 & $<$0.104 & $<$0.122 & $<$10.25 & $<$10.44 & $<$10.51 & $<$11.13 & $<$11.31 & $<$11.38\\
DEIMOS\_COSMOS\_421062 & $<$0.058 & $<$0.080 & $<$0.079 & $<$10.17 & $<$10.31 & $<$10.30 & $<$11.05 & $<$11.18 & $<$11.18\\
DEIMOS\_COSMOS\_430951 & $<$0.136 & $<$0.247 & $<$0.279 & $<$10.55 & $<$10.81 & $<$10.86 & $<$11.43 & $<$11.69 & $<$11.74\\
DEIMOS\_COSMOS\_431067 & $<$0.154 & $<$0.221 & $<$0.275 & $<$10.44 & $<$10.60 & $<$10.69 & $<$11.32 & $<$11.47 & $<$11.57\\
DEIMOS\_COSMOS\_432340 & $<$0.166 & $<$0.358 & $<$0.465 & $<$10.47 & $<$10.80 & $<$10.92 & $<$11.35 & $<$11.68 & $<$11.79\\
DEIMOS\_COSMOS\_434239 & $<$0.136 & $<$0.231 & $<$0.322 & $<$10.39 & $<$10.63 & $<$10.77 & $<$11.27 & $<$11.50 & $<$11.65\\
DEIMOS\_COSMOS\_442844 & $<$0.140 & $<$0.227 & $<$0.270 & $<$10.41 & $<$10.62 & $<$10.69 & $<$11.28 & $<$11.49 & $<$11.57\\
DEIMOS\_COSMOS\_454608 & $<$0.178 & $<$0.277 & $<$0.285 & $<$10.52 & $<$10.72 & $<$10.73 & $<$11.40 & $<$11.59 & $<$11.61\\
DEIMOS\_COSMOS\_470116 & $<$0.067 & $<$0.118 & $<$0.117 & $<$10.24 & $<$10.49 & $<$10.48 & $<$11.12 & $<$11.36 & $<$11.36\\
DEIMOS\_COSMOS\_471063 & $<$0.105 & $<$0.205 & $<$0.183 & $<$10.44 & $<$10.73 & $<$10.68 & $<$11.32 & $<$11.61 & $<$11.56\\
DEIMOS\_COSMOS\_472215 & $<$0.092 & $<$0.198 & $<$0.194 & $<$10.38 & $<$10.71 & $<$10.70 & $<$11.25 & $<$11.59 & $<$11.58\\
DEIMOS\_COSMOS\_494763 & $<$0.081 & $<$0.151 & $<$0.209 & $<$10.27 & $<$10.54 & $<$10.68 & $<$11.15 & $<$11.42 & $<$11.56\\
DEIMOS\_COSMOS\_503575 & $<$0.062 & $<$0.109 & $<$0.098 & $<$10.21 & $<$10.45 & $<$10.41 & $<$11.08 & $<$11.33 & $<$11.28\\
DEIMOS\_COSMOS\_510660 & $<$0.147 & $<$0.247 & $<$0.256 & $<$10.44 & $<$10.66 & $<$10.68 & $<$11.32 & $<$11.54 & $<$11.56\\
DEIMOS\_COSMOS\_519281 & $<$0.103 & $<$0.179 & $<$0.231 & $<$10.41 & $<$10.66 & $<$10.77 & $<$11.29 & $<$11.53 & $<$11.64\\
DEIMOS\_COSMOS\_536534 & $<$0.078 & $<$0.141 & $<$0.123 & $<$10.31 & $<$10.57 & $<$10.51 & $<$11.18 & $<$11.44 & $<$11.38\\
DEIMOS\_COSMOS\_549131 & $<$0.100 & $<$0.153 & $<$0.193 & $<$10.40 & $<$10.59 & $<$10.69 & $<$11.28 & $<$11.46 & $<$11.56\\
DEIMOS\_COSMOS\_550156 & $<$0.147 & $<$0.222 & $<$0.319 & $<$10.42 & $<$10.60 & $<$10.75 & $<$11.30 & $<$11.47 & $<$11.63\\
DEIMOS\_COSMOS\_567070 & $<$0.163 & $<$0.278 & $<$0.379 & $<$10.48 & $<$10.72 & $<$10.85 & $<$11.36 & $<$11.59 & $<$11.73\\
DEIMOS\_COSMOS\_576372 & $<$0.088 & $<$0.151 & $<$0.158 & $<$10.36 & $<$10.59 & $<$10.61 & $<$11.23 & $<$11.47 & $<$11.49\\
DEIMOS\_COSMOS\_586681 & $<$0.131 & $<$0.179 & $<$0.180 & $<$10.55 & $<$10.69 & $<$10.69 & $<$11.43 & $<$11.57 & $<$11.57\\
DEIMOS\_COSMOS\_592644 & $<$0.137 & $<$0.217 & $<$0.265 & $<$10.40 & $<$10.60 & $<$10.69 & $<$11.28 & $<$11.48 & $<$11.57\\
DEIMOS\_COSMOS\_627939 & $<$0.110 & $<$0.204 & $<$0.239 & $<$10.31 & $<$10.58 & $<$10.65 & $<$11.19 & $<$11.45 & $<$11.52\\
DEIMOS\_COSMOS\_628063 & $<$0.181 & $<$0.268 & $<$0.267 & $<$10.53 & $<$10.70 & $<$10.69 & $<$11.40 & $<$11.57 & $<$11.57\\
DEIMOS\_COSMOS\_628137 & $<$0.068 & $<$0.104 & $<$0.126 & $<$10.25 & $<$10.43 & $<$10.52 & $<$11.12 & $<$11.31 & $<$11.39\\
DEIMOS\_COSMOS\_629750 & $<$0.103 & $<$0.173 & $<$0.180 & $<$10.36 & $<$10.59 & $<$10.60 & $<$11.24 & $<$11.46 & $<$11.48\\
DEIMOS\_COSMOS\_630594 & $<$0.143 & $<$0.242 & $<$0.291 & $<$10.41 & $<$10.64 & $<$10.72 & $<$11.29 & $<$11.52 & $<$11.59\\
DEIMOS\_COSMOS\_665509 & $<$0.173 & $<$0.227 & $<$0.259 & $<$10.50 & $<$10.62 & $<$10.68 & $<$11.38 & $<$11.50 & $<$11.56\\
DEIMOS\_COSMOS\_665626 & $<$0.135 & $<$0.215 & $<$0.275 & $<$10.41 & $<$10.61 & $<$10.71 & $<$11.28 & $<$11.48 & $<$11.59\\
DEIMOS\_COSMOS\_680104 & $<$0.154 & $<$0.290 & $<$0.362 & $<$10.45 & $<$10.73 & $<$10.83 & $<$11.33 & $<$11.61 & $<$11.70\\
DEIMOS\_COSMOS\_709575 & $<$0.171 & $<$0.333 & $<$0.379 & $<$10.48 & $<$10.77 & $<$10.83 & $<$11.36 & $<$11.65 & $<$11.70\\
DEIMOS\_COSMOS\_722679 & $<$0.078 & $<$0.126 & $<$0.134 & $<$10.32 & $<$10.53 & $<$10.55 & $<$11.19 & $<$11.40 & $<$11.43\\
DEIMOS\_COSMOS\_733857 & $<$0.162 & $<$0.267 & $<$0.366 & $<$10.48 & $<$10.70 & $<$10.83 & $<$11.36 & $<$11.57 & $<$11.71\\
DEIMOS\_COSMOS\_742174 & $<$0.071 & $<$0.135 & $<$0.148 & $<$10.26 & $<$10.54 & $<$10.58 & $<$11.14 & $<$11.42 & $<$11.46\\
DEIMOS\_COSMOS\_743730 & $<$0.126 & $<$0.233 & $<$0.294 & $<$10.36 & $<$10.63 & $<$10.73 & $<$11.24 & $<$11.51 & $<$11.61\\
DEIMOS\_COSMOS\_761315 & $<$0.119 & $<$0.233 & $<$0.290 & $<$10.35 & $<$10.64 & $<$10.74 & $<$11.23 & $<$11.52 & $<$11.61\\
DEIMOS\_COSMOS\_773957 & $<$0.059 & $<$0.119 & $<$0.151 & $<$10.19 & $<$10.49 & $<$10.60 & $<$11.06 & $<$11.37 & $<$11.47\\
DEIMOS\_COSMOS\_787780 & $<$0.150 & $<$0.214 & $<$0.249 & $<$10.44 & $<$10.60 & $<$10.66 & $<$11.32 & $<$11.47 & $<$11.54\\
DEIMOS\_COSMOS\_790930 & $<$0.062 & $<$0.091 & $<$0.109 & $<$10.21 & $<$10.37 & $<$10.46 & $<$11.09 & $<$11.25 & $<$11.33\\
DEIMOS\_COSMOS\_803480 & $<$0.114 & $<$0.193 & $<$0.217 & $<$10.33 & $<$10.55 & $<$10.61 & $<$11.20 & $<$11.43 & $<$11.48\\
DEIMOS\_COSMOS\_814483 & $<$0.142 & $<$0.307 & $<$0.393 & $<$10.43 & $<$10.76 & $<$10.87 & $<$11.30 & $<$11.64 & $<$11.75\\
DEIMOS\_COSMOS\_834764 & $<$0.216 & $<$0.332 & $<$0.316 & $<$10.60 & $<$10.79 & $<$10.76 & $<$11.48 & $<$11.66 & $<$11.64\\
DEIMOS\_COSMOS\_838532 & $<$0.155 & $<$0.310 & $<$0.328 & $<$10.46 & $<$10.76 & $<$10.78 & $<$11.33 & $<$11.63 & $<$11.66\\
DEIMOS\_COSMOS\_842313 & $<$0.214 & $<$9.095 & $<$9.979 & $<$10.60 & $<$12.23 & $<$12.27 & $<$11.48 & $<$13.11 & $<$13.15\\
DEIMOS\_COSMOS\_843045 & $<$0.141 & $<$0.187 & $<$0.249 & $<$10.58 & $<$10.70 & $<$10.83 & $<$11.46 & $<$11.58 & $<$11.70\\
DEIMOS\_COSMOS\_845652 & $<$0.088 & $<$0.145 & $<$0.170 & $<$10.31 & $<$10.53 & $<$10.60 & $<$11.19 & $<$11.41 & $<$11.48\\
DEIMOS\_COSMOS\_859732 & $<$0.178 & $<$0.218 & $<$0.263 & $<$10.52 & $<$10.61 & $<$10.69 & $<$11.39 & $<$11.48 & $<$11.56\\
DEIMOS\_COSMOS\_869970 & $<$0.078 & $<$0.135 & $<$0.163 & $<$10.25 & $<$10.49 & $<$10.57 & $<$11.12 & $<$11.37 & $<$11.45\\
DEIMOS\_COSMOS\_873321 & $<$0.111 & $<$0.214 & $<$0.282 & $<$10.40 & $<$10.68 & $<$10.80 & $<$11.28 & $<$11.56 & $<$11.68\\
DEIMOS\_COSMOS\_880016 & $<$0.133 & $<$0.247 & $<$0.274 & $<$10.39 & $<$10.66 & $<$10.71 & $<$11.27 & $<$11.54 & $<$11.58\\
DEIMOS\_COSMOS\_910650 & $<$0.073 & $<$0.136 & $<$0.188 & $<$10.28 & $<$10.55 & $<$10.69 & $<$11.16 & $<$11.42 & $<$11.57\\
DEIMOS\_COSMOS\_920848 & $<$0.158 & $<$0.241 & $<$0.285 & $<$10.47 & $<$10.65 & $<$10.73 & $<$11.35 & $<$11.53 & $<$11.60\\
DEIMOS\_COSMOS\_926434 & $<$0.154 & $<$0.282 & $<$0.325 & $<$10.44 & $<$10.71 & $<$10.77 & $<$11.32 & $<$11.58 & $<$11.64\\
DEIMOS\_COSMOS\_933876 & $<$0.154 & $<$0.324 & $<$0.456 & $<$10.44 & $<$10.76 & $<$10.91 & $<$11.32 & $<$11.64 & $<$11.79\\
vuds\_cosmos\_5100537582 & $<$0.130 & $<$0.267 & $<$0.336 & $<$10.38 & $<$10.70 & $<$10.80 & $<$11.26 & $<$11.57 & $<$11.67\\
vuds\_cosmos\_5100541407 & $<$0.183 & $<$0.369 & $<$0.439 & $<$10.53 & $<$10.84 & $<$10.91 & $<$11.41 & $<$11.71 & $<$11.79\\
vuds\_cosmos\_5100559223 & $<$0.165 & $<$0.339 & $<$0.391 & $<$10.49 & $<$10.80 & $<$10.86 & $<$11.37 & $<$11.68 & $<$11.74\\
vuds\_cosmos\_5101013812 & $<$0.151 & $<$0.200 & $<$0.264 & $<$10.43 & $<$10.55 & $<$10.67 & $<$11.31 & $<$11.43 & $<$11.55\\
vuds\_cosmos\_5101210235 & $<$0.115 & $<$0.164 & $<$0.226 & $<$10.33 & $<$10.49 & $<$10.63 & $<$11.21 & $<$11.37 & $<$11.50\\
vuds\_cosmos\_5101244930 & $<$0.176 & $<$0.338 & $<$0.337 & $<$10.52 & $<$10.80 & $<$10.80 & $<$11.40 & $<$11.68 & $<$11.68\\
vuds\_cosmos\_5101288969 & $<$0.050 & $<$0.090 & $<$0.103 & $<$10.12 & $<$10.37 & $<$10.43 & $<$11.00 & $<$11.25 & $<$11.31\\
vuds\_cosmos\_510148750 & $<$0.105 & $<$0.154 & $<$0.174 & $<$10.29 & $<$10.45 & $<$10.50 & $<$11.16 & $<$11.33 & $<$11.38\\
vuds\_cosmos\_510327576 & $<$0.161 & $<$0.329 & $<$0.331 & $<$10.48 & $<$10.79 & $<$10.79 & $<$11.35 & $<$11.67 & $<$11.67\\
vuds\_cosmos\_510581738 & $<$0.144 & $<$0.246 & $<$0.353 & $<$10.42 & $<$10.65 & $<$10.81 & $<$11.30 & $<$11.53 & $<$11.69\\
vuds\_cosmos\_510596653 & $<$0.135 & $<$0.285 & $<$0.320 & $<$10.40 & $<$10.73 & $<$10.78 & $<$11.28 & $<$11.60 & $<$11.65\\
vuds\_cosmos\_510605533 & $<$0.162 & $<$0.233 & $<$0.311 & $<$10.47 & $<$10.63 & $<$10.76 & $<$11.35 & $<$11.51 & $<$11.63\\
vuds\_cosmos\_510786441 & $<$0.101 & $<$0.190 & $<$0.213 & $<$10.26 & $<$10.54 & $<$10.59 & $<$11.14 & $<$11.41 & $<$11.46\\
vuds\_cosmos\_5110377875 & $<$0.161 & $<$0.311 & $<$0.428 & $<$10.48 & $<$10.76 & $<$10.90 & $<$11.35 & $<$11.64 & $<$11.78\\
vuds\_cosmos\_5131465996 & $<$0.139 & $<$0.198 & $<$0.299 & $<$10.40 & $<$10.55 & $<$10.73 & $<$11.28 & $<$11.43 & $<$11.61\\
\hline
\end{longtable}

\begin{figure*}[!h]
\centering
\includegraphics[width=18cm]{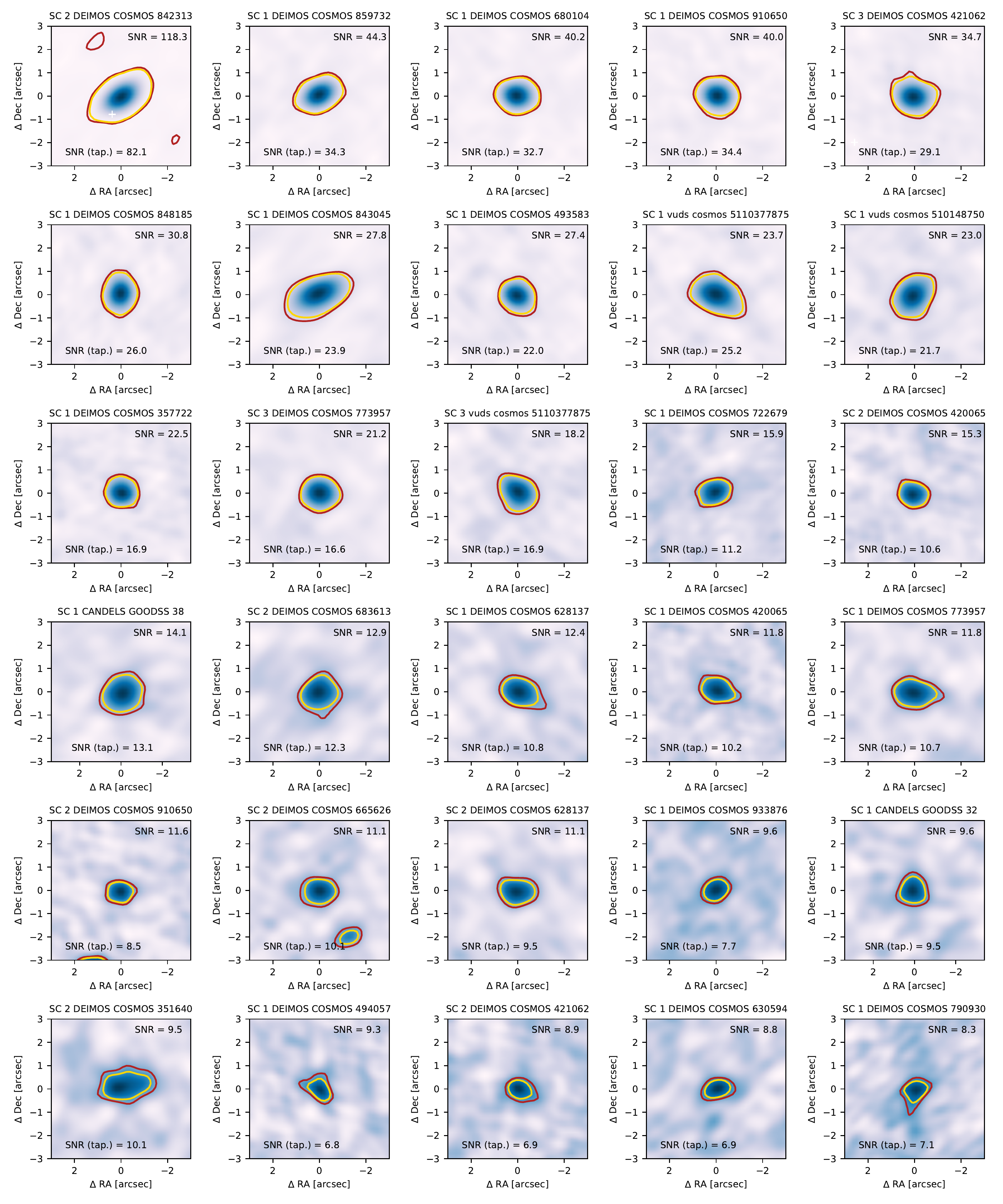}
\caption{\label{fig:cont_ser_cutouts} Cutout images of the continuum-detected nontarget sources. The red and the gold contours indicate the 3.5\,$\sigma$ and 5\,$\sigma$ levels, respectively. The white crosshair indicate the phase center. The sources are ordered by decreasing S/N.}
\end{figure*}

\begin{figure*}[!h]
\centering
\includegraphics[width=18cm]{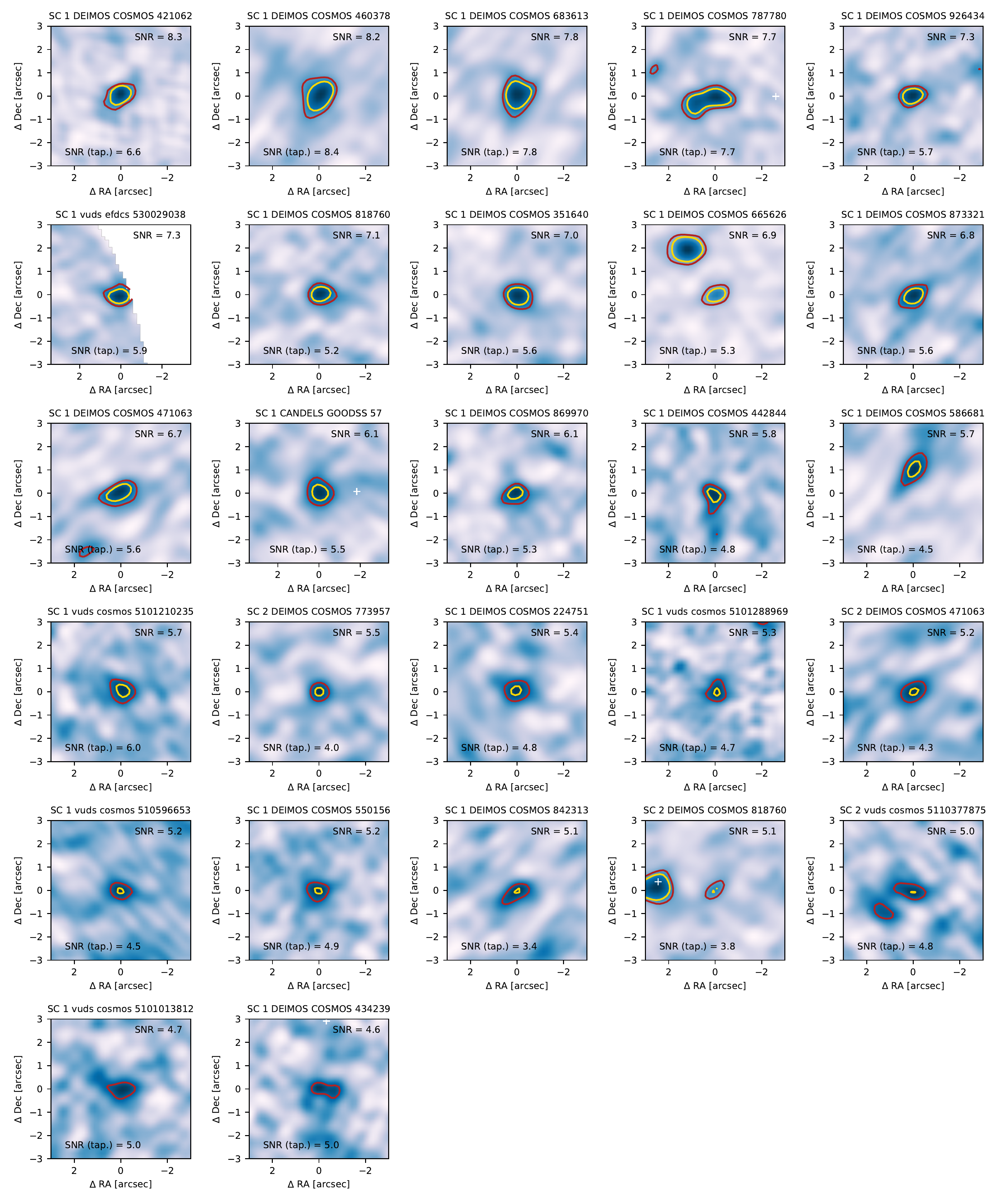}
\caption{\label{fig:cont_ser_cutouts2} Fig.\,\ref{fig:cont_ser_cutouts} continued.}
\end{figure*}

\begin{table*}[!h]
\caption{\label{tab:cont_ser_list} Characteristics of the continuum-detected nontarget sources (serendipitous detections and possible neighbors of our targets). The columns are similar to \ref{tab:cont_target_list}. The process used to deboost the flux measurements is explained in Sect.\,\ref{sect:flux_boosting}. SC\_1\_CANDELS\_GOODSS\_32 and SC\_1\_CANDELS\_GOODSS\_38 are the same source observed with the different spectral by two different but slightly overlapping pointings. We kept both in the catalog, since their observed frequencies are sufficiently different to provide useful independent information. The measurement uncertainties do not include the calibration uncertainties ($\sim$4.5\,\% on average and up to 10\,\%, see Sect.\,\ref{sect:quasars}).}
\centering
\begin{tabular}{lcccrrrr}
\hline
\hline
Name & RA & Dec & Freq. & S/N & S/N & S$_\nu$ & S$_\nu^{\rm deboost}$ \\
 & h:min:s & deg:min:s & GHz & &  (tap.) & $\mu$Jy & $\mu$Jy \\
\hline
SC\_2\_DEIMOS\_COSMOS\_842313 & 10:00:54.49 &  +02:34:36.09 & 349.1 & 118.3 & 82.1 & 8240$\pm$90 & 8240$\pm$562\\ 
SC\_1\_DEIMOS\_COSMOS\_859732 & 10:00:00.41 &  +02:36:12.37 & 348.4 & 44.3 & 34.3 & 4393$\pm$118 & 4393$\pm$318\\ 
SC\_1\_DEIMOS\_COSMOS\_680104 & 10:01:10.52 &  +02:19:52.77 & 350.2 & 40.2 & 32.7 & 3542$\pm$104 & 3542$\pm$260\\ 
SC\_1\_DEIMOS\_COSMOS\_910650 & 10:00:23.25 &  +02:40:54.91 & 290.6 & 40.0 & 34.4 & 4220$\pm$122 & 4220$\pm$309\\ 
SC\_3\_DEIMOS\_COSMOS\_421062 & 10:00:08.36 &  +01:56:06.65 & 295.7 & 34.7 & 29.1 & 763$\pm$26 & 763$\pm$58\\ 
SC\_1\_DEIMOS\_COSMOS\_848185 & 10:00:20.70 &  +02:35:20.35 & 307.6 & 30.8 & 26.0 & 5983$\pm$227 & 5983$\pm$469\\ 
SC\_1\_DEIMOS\_COSMOS\_843045 & 10:00:12.94 &  +02:34:34.94 & 283.7 & 27.8 & 23.9 & 4226$\pm$179 & 4226$\pm$341\\ 
SC\_1\_DEIMOS\_COSMOS\_493583 & 10:00:23.08 &  +02:03:05.60 & 349.4 & 27.4 & 22.0 & 1860$\pm$82 & 1860$\pm$150\\ 
SC\_1\_vuds\_cosmos\_5110377875 & 10:01:32.38 &  +02:24:23.89 & 349.4 & 23.7 & 25.2 & 3512$\pm$163 & 3492$\pm$334\\ 
SC\_1\_vuds\_cosmos\_510148750 & 09:59:22.53 &  +01:59:36.21 & 350.1 & 23.0 & 21.7 & 1428$\pm$73 & 1428$\pm$122\\ 
SC\_1\_DEIMOS\_COSMOS\_357722 & 09:59:52.36 &  +01:49:53.60 & 289.0 & 22.5 & 16.9 & 1344$\pm$71 & 1344$\pm$111\\ 
SC\_3\_DEIMOS\_COSMOS\_773957 & 10:01:09.79 &  +02:28:40.71 & 289.4 & 21.2 & 16.6 & 1123$\pm$63 & 1123$\pm$96\\ 
SC\_3\_vuds\_cosmos\_5110377875 & 10:01:32.17 &  +02:24:36.54 & 349.4 & 18.2 & 16.9 & 2216$\pm$144 & 2216$\pm$209\\ 
SC\_1\_DEIMOS\_COSMOS\_722679 & 09:59:44.78 &  +02:23:40.03 & 288.1 & 15.9 & 11.2 & 447$\pm$34 & 447$\pm$45\\ 
SC\_2\_DEIMOS\_COSMOS\_420065 & 10:01:23.86 &  +01:56:13.45 & 289.1 & 15.3 & 10.6 & 929$\pm$72 & 929$\pm$93\\ 
SC\_1\_CANDELS\_GOODSS\_38 & 03:32:16.42 &  -27:41:16.90 & 294.5 & 14.1 & 13.1 & 1022$\pm$85 & 1022$\pm$113\\ 
SC\_2\_DEIMOS\_COSMOS\_683613 & 10:00:08.97 &  +02:20:26.65 & 295.8 & 12.9 & 12.3 & 2854$\pm$266 & 2847$\pm$349\\ 
SC\_1\_DEIMOS\_COSMOS\_628137 & 09:59:55.54 &  +02:15:11.49 & 291.1 & 12.4 & 10.8 & 1784$\pm$176 & 1781$\pm$219\\ 
SC\_1\_DEIMOS\_COSMOS\_420065 & 10:01:23.86 &  +01:56:05.61 & 289.1 & 11.8 & 10.2 & 383$\pm$37 & 381$\pm$50\\ 
SC\_1\_DEIMOS\_COSMOS\_773957 & 10:01:10.73 &  +02:28:21.60 & 289.4 & 11.8 & 10.7 & 881$\pm$86 & 875$\pm$114\\ 
SC\_2\_DEIMOS\_COSMOS\_910650 & 10:00:23.18 &  +02:40:58.44 & 290.6 & 11.6 & 8.5 & 565$\pm$58 & 563$\pm$72\\ 
SC\_2\_DEIMOS\_COSMOS\_665626 & 10:01:13.91 &  +02:18:42.53 & 347.2 & 11.1 & 10.1 & 934$\pm$102 & 925$\pm$131\\ 
SC\_2\_DEIMOS\_COSMOS\_628137 & 09:59:54.67 &  +02:15:26.14 & 291.1 & 11.1 & 9.5 & 549$\pm$59 & 546$\pm$74\\ 
SC\_1\_DEIMOS\_COSMOS\_933876 & 09:59:35.98 &  +02:43:06.78 & 356.3 & 9.6 & 7.7 & 771$\pm$93 & 763$\pm$116\\ 
SC\_1\_CANDELS\_GOODSS\_32 & 03:32:16.41 &  -27:41:16.82 & 356.2 & 9.6 & 9.5 & 1637$\pm$203 & 1614$\pm$257\\ 
SC\_2\_DEIMOS\_COSMOS\_351640 & 10:01:29.51 &  +01:49:29.36 & 290.3 & 9.5 & 10.1 & 690$\pm$80 & 681$\pm$105\\ 
SC\_1\_DEIMOS\_COSMOS\_494057 & 09:58:29.19 &  +02:03:15.37 & 295.7 & 9.3 & 6.8 & 766$\pm$106 & 755$\pm$130\\ 
SC\_2\_DEIMOS\_COSMOS\_421062 & 10:00:08.33 &  +01:55:58.44 & 295.7 & 8.9 & 6.9 & 243$\pm$33 & 240$\pm$41\\ 
SC\_1\_DEIMOS\_COSMOS\_630594 & 10:00:32.77 &  +02:15:38.08 & 353.9 & 8.8 & 6.9 & 1311$\pm$180 & 1294$\pm$220\\ 
SC\_1\_DEIMOS\_COSMOS\_790930 & 10:01:28.39 &  +02:30:17.77 & 290.9 & 8.3 & 7.1 & 1144$\pm$177 & 1122$\pm$214\\ 
SC\_1\_DEIMOS\_COSMOS\_421062 & 10:00:09.09 &  +01:55:55.21 & 295.7 & 8.3 & 6.6 & 578$\pm$85 & 567$\pm$105\\ 
SC\_1\_DEIMOS\_COSMOS\_460378 & 10:00:11.82 &  +01:59:58.41 & 302.3 & 8.2 & 8.4 & 680$\pm$117 & 666$\pm$137\\ 
SC\_1\_DEIMOS\_COSMOS\_683613 & 10:00:09.24 &  +02:20:21.95 & 295.8 & 7.8 & 7.8 & 614$\pm$91 & 601$\pm$114\\ 
SC\_1\_DEIMOS\_COSMOS\_787780 & 09:59:56.83 &  +02:29:48.24 & 351.5 & 7.7 & 7.7 & 398$\pm$106 & 389$\pm$114\\ 
SC\_1\_DEIMOS\_COSMOS\_926434 & 10:00:25.79 &  +02:42:21.97 & 355.6 & 7.3 & 5.7 & 652$\pm$103 & 640$\pm$126\\ 
SC\_1\_vuds\_efdcs\_530029038 & 03:32:17.94 &  -27:52:33.19 & 356.0 & 7.3 & 5.9 & 4728$\pm$773 & 4617$\pm$944\\ 
SC\_1\_DEIMOS\_COSMOS\_818760 & 10:01:54.18 &  +02:32:24.89 & 347.2 & 7.1 & 5.2 & 2223$\pm$371 & 2168$\pm$454\\ 
SC\_1\_DEIMOS\_COSMOS\_351640 & 10:01:29.23 &  +01:49:20.54 & 290.3 & 7.0 & 5.6 & 346$\pm$58 & 341$\pm$70\\ 
SC\_1\_DEIMOS\_COSMOS\_665626 & 10:01:13.82 &  +02:18:40.60 & 347.2 & 6.9 & 5.3 & 392$\pm$87 & 384$\pm$98\\ 
SC\_1\_DEIMOS\_COSMOS\_873321 & 10:00:04.01 &  +02:37:24.82 & 315.4 & 6.8 & 5.6 & 880$\pm$164 & 857$\pm$195\\ 
SC\_1\_DEIMOS\_COSMOS\_471063 & 10:00:16.36 &  +02:01:05.54 & 289.5 & 6.7 & 5.6 & 1085$\pm$188 & 1056$\pm$230\\ 
SC\_1\_CANDELS\_GOODSS\_57 & 03:32:39.16 &  -27:52:23.00 & 296.6 & 6.1 & 5.5 & 205$\pm$39 & 199$\pm$48\\ 
SC\_1\_DEIMOS\_COSMOS\_869970 & 10:00:25.16 &  +02:37:27.53 & 311.4 & 6.1 & 5.3 & 420$\pm$85 & 406$\pm$101\\ 
SC\_1\_DEIMOS\_COSMOS\_442844 & 09:59:59.50 &  +01:58:16.74 & 352.8 & 5.8 & 4.8 & 554$\pm$105 & 535$\pm$129\\ 
SC\_1\_DEIMOS\_COSMOS\_586681 & 10:00:08.78 &  +02:11:54.12 & 283.7 & 5.7 & 4.5 & 1832$\pm$396 & 1788$\pm$466\\ 
SC\_1\_vuds\_cosmos\_5101210235 & 10:01:30.95 &  +02:21:57.33 & 347.8 & 5.7 & 6.0 & 905$\pm$181 & 871$\pm$217\\ 
SC\_2\_DEIMOS\_COSMOS\_773957 & 10:01:10.68 &  +02:28:29.30 & 289.4 & 5.5 & 4.0 & 117$\pm$33 & 114$\pm$37\\ 
SC\_1\_DEIMOS\_COSMOS\_224751 & 09:59:52.80 &  +01:37:26.82 & 289.4 & 5.4 & 4.8 & 257$\pm$54 & 249$\pm$65\\ 
SC\_1\_vuds\_cosmos\_5101288969 & 09:59:30.86 &  +02:20:01.95 & 290.3 & 5.3 & 4.7 & 224$\pm$48 & 215$\pm$58\\ 
SC\_2\_DEIMOS\_COSMOS\_471063 & 10:00:16.11 &  +02:01:06.26 & 289.5 & 5.2 & 4.3 & 488$\pm$108 & 475$\pm$129\\ 
SC\_1\_vuds\_cosmos\_510596653 & 09:59:18.53 &  +01:56:21.44 & 346.6 & 5.2 & 4.5 & 354$\pm$80 & 343$\pm$95\\ 
SC\_1\_DEIMOS\_COSMOS\_550156 & 10:01:00.33 &  +02:08:14.15 & 355.6 & 5.2 & 4.9 & 520$\pm$117 & 500$\pm$139\\ 
SC\_1\_DEIMOS\_COSMOS\_842313 & 10:00:54.30 &  +02:34:27.81 & 349.1 & 5.1 & 3.4 & 843$\pm$180 & 815$\pm$219\\ 
SC\_2\_DEIMOS\_COSMOS\_818760 & 10:01:54.69 &  +02:32:31.32 & 347.2 & 5.1 & 3.8 & 425$\pm$104 & 408$\pm$121\\ 
SC\_2\_vuds\_cosmos\_5110377875 & 10:01:31.67 &  +02:24:29.53 & 349.4 & 5.0 & 4.8 & 1644$\pm$370 & 1575$\pm$437\\ 
SC\_1\_vuds\_cosmos\_5101013812 & 10:00:50.66 &  +02:09:48.17 & 356.3 & 4.7 & 5.0 & 650$\pm$142 & 623$\pm$169\\ 
SC\_1\_DEIMOS\_COSMOS\_434239 & 10:01:17.12 &  +01:57:16.35 & 352.9 & 4.6 & 5.0 & 521$\pm$131 & 499$\pm$150\\ 
\hline
\end{tabular}
\end{table*}

\clearpage

\section{[CII] spectra, moment-0 maps, and catalogs}

\label{sect:cat_cii}

\begin{figure*}[!h]
\centering
\includegraphics[width=18cm]{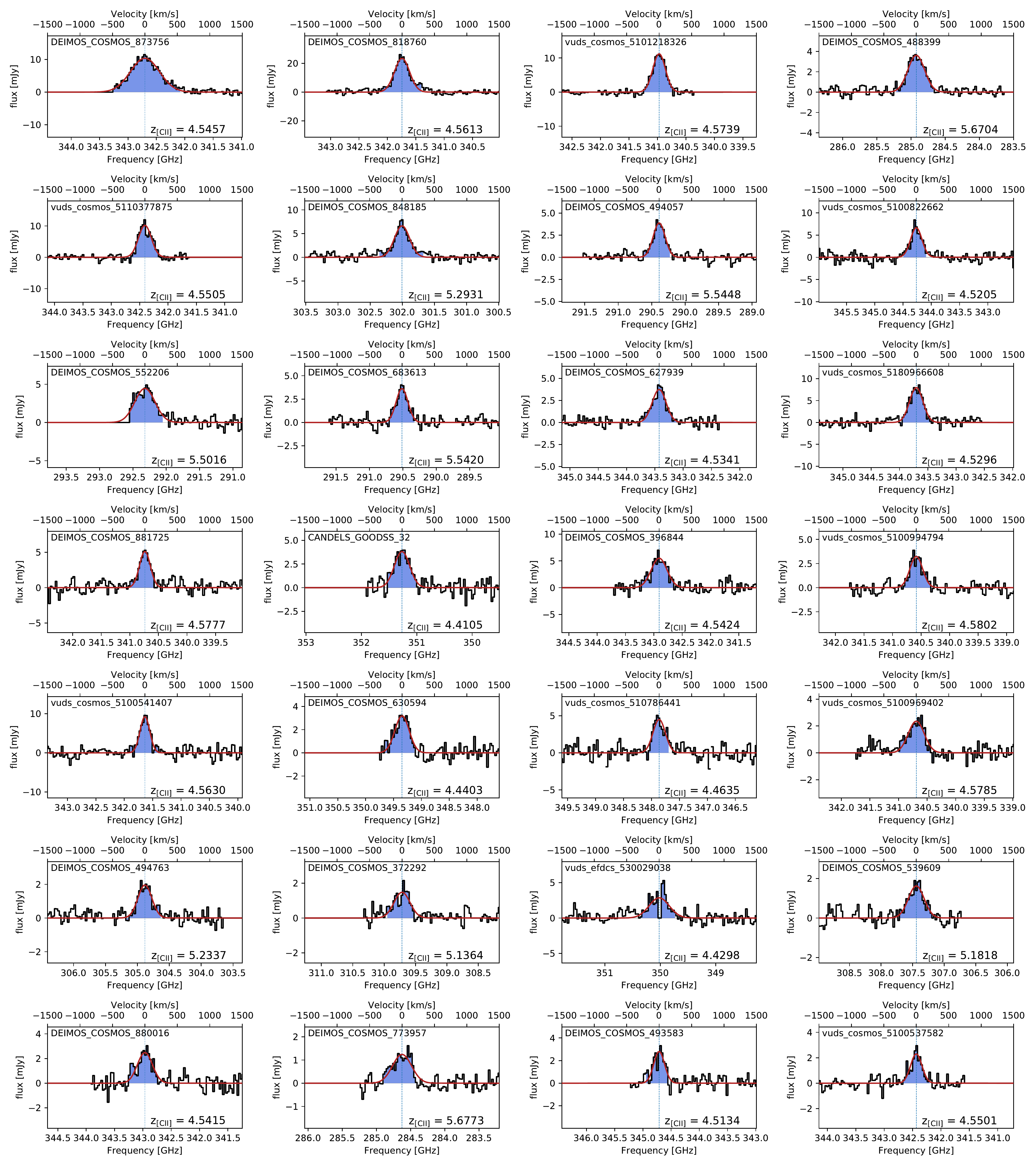}
\caption{\label{fig:spec_1} Spectra of the detected ALPINE sources (S/N$>$3.5). The blue area is the frequency range used to produce the moment-0 map.  The red solid line is the best-fit by a Gaussian. The process used to produce these spectra is described in Sect.\,\ref{sect:cii_extr}.}
\end{figure*}

\begin{figure*}[!h]
\centering
\includegraphics[width=18cm]{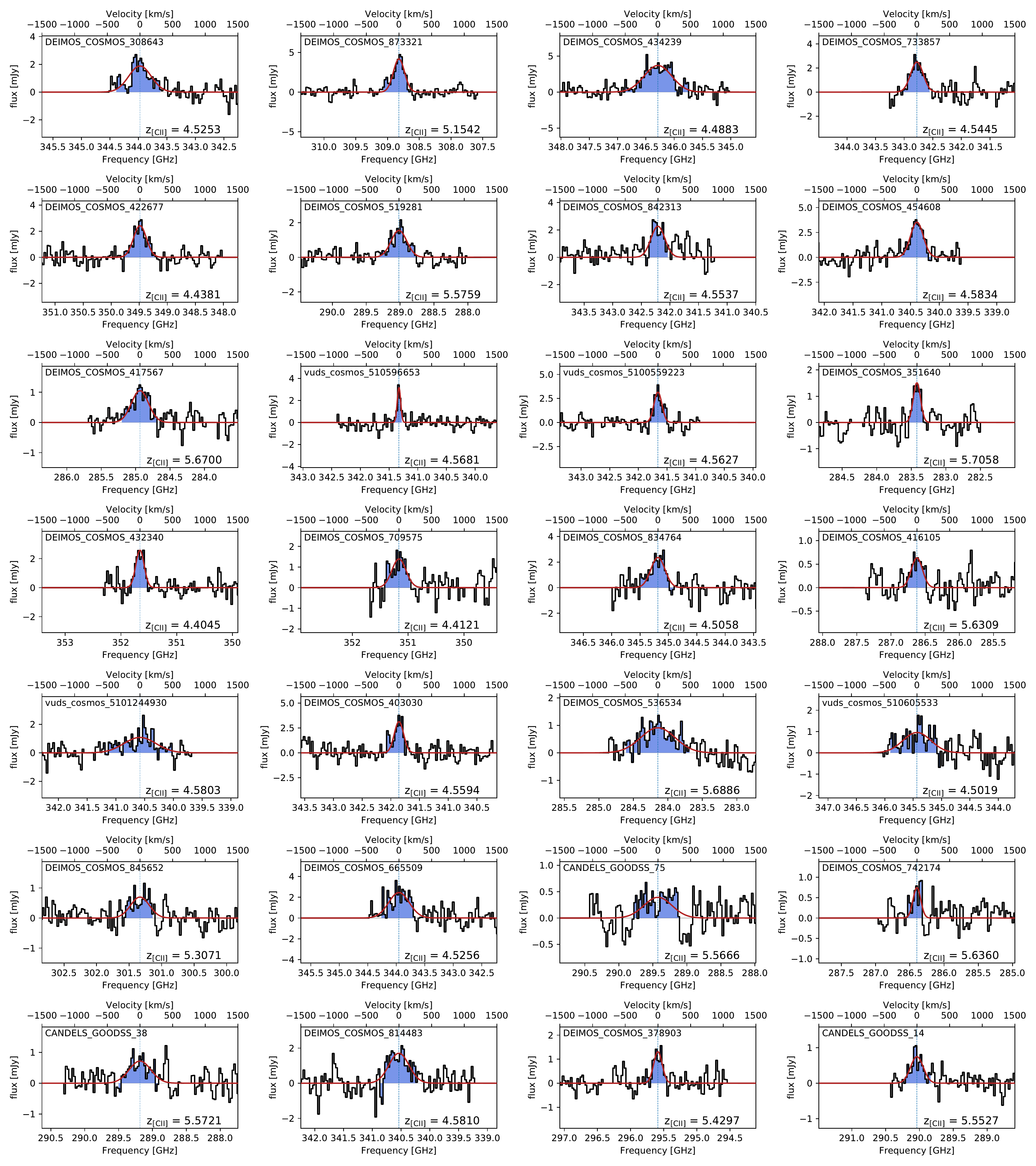}
\caption{\label{fig:spec_2} Fig.\,\ref{fig:spec_1} continued.}
\end{figure*}

\begin{figure*}[!h]
\centering
\includegraphics[width=18cm]{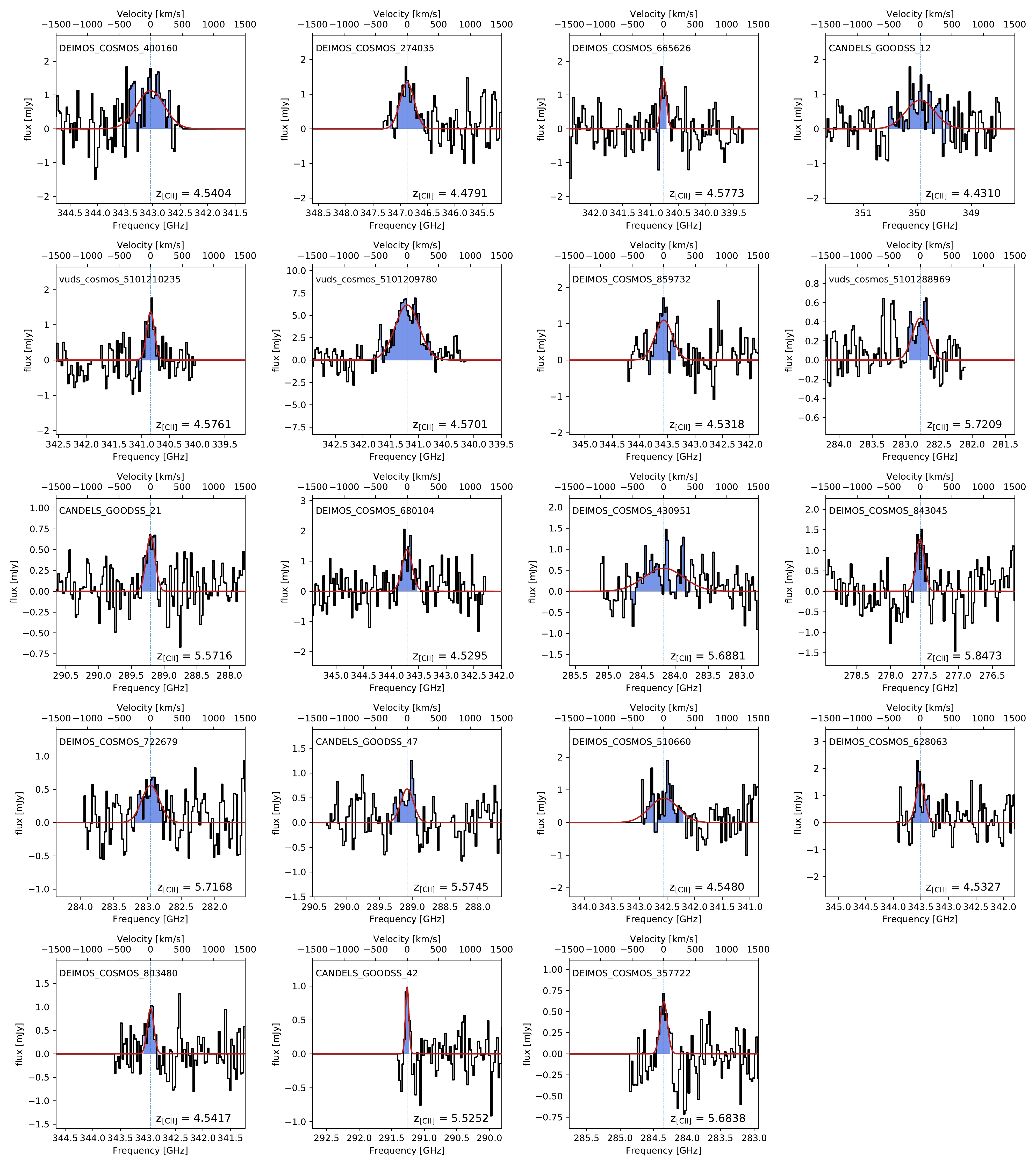}
\caption{\label{fig:spec_3} Fig.\,\ref{fig:spec_1} continued.}
\end{figure*}

\begin{figure*}[!h]
\centering
\includegraphics[width=18cm]{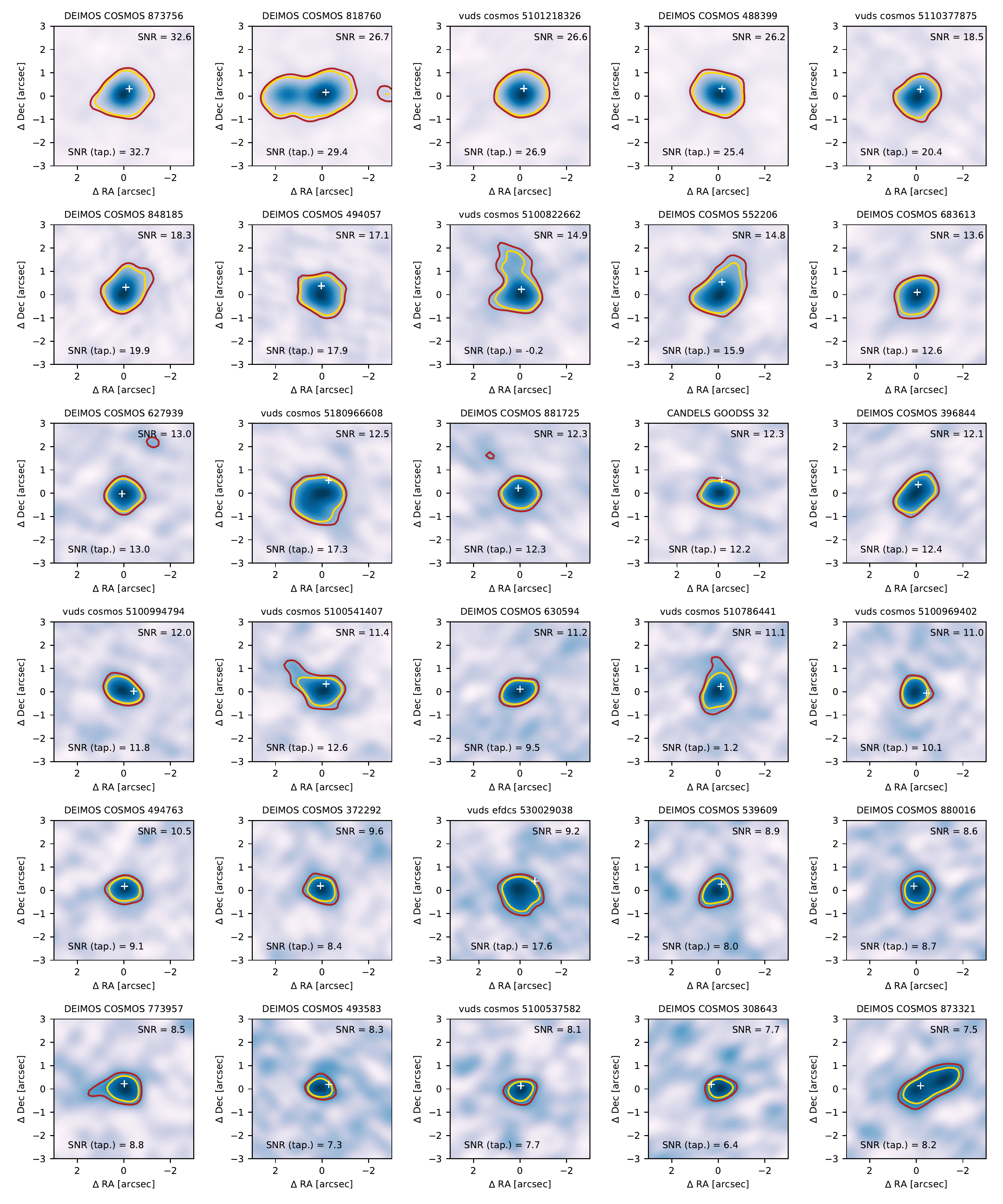}
\caption{\label{fig:target_mom0_1} Cutout images of the moment-0 maps of our detected [CII] targets. The red and the gold contours indicate the 3.5\,$\sigma$ and 5\,$\sigma$ levels, respectively. The white crosshair indicate the phase center. The sources are ordered by decreasing S/N.}
\end{figure*}

\begin{figure*}[!h]
\centering
\includegraphics[width=18cm]{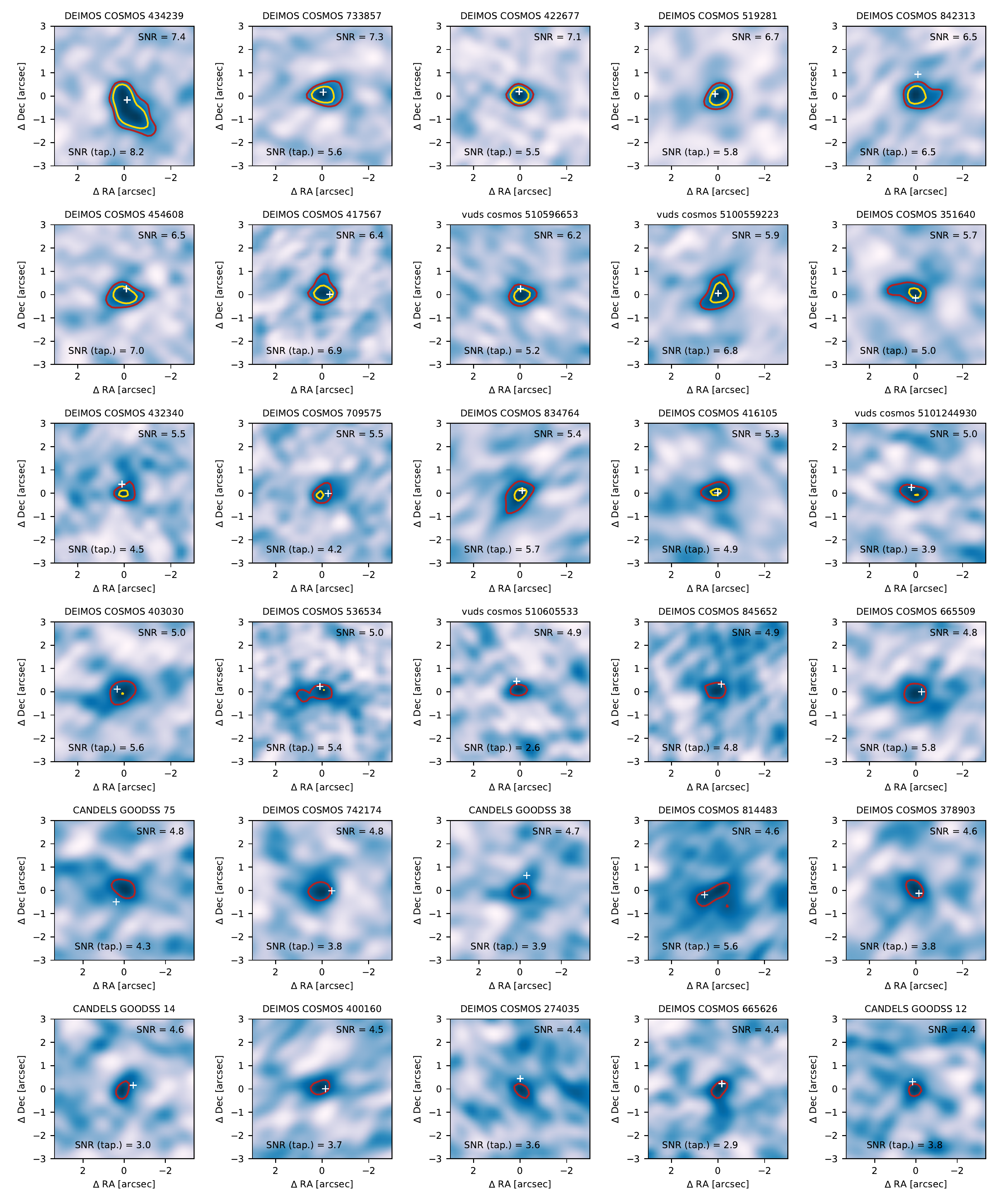}
\caption{\label{fig:target_mom0_2} Fig.\,\ref{fig:target_mom0_1} continued.}
\end{figure*}

\begin{figure*}[!h]
\centering
\includegraphics[width=18cm]{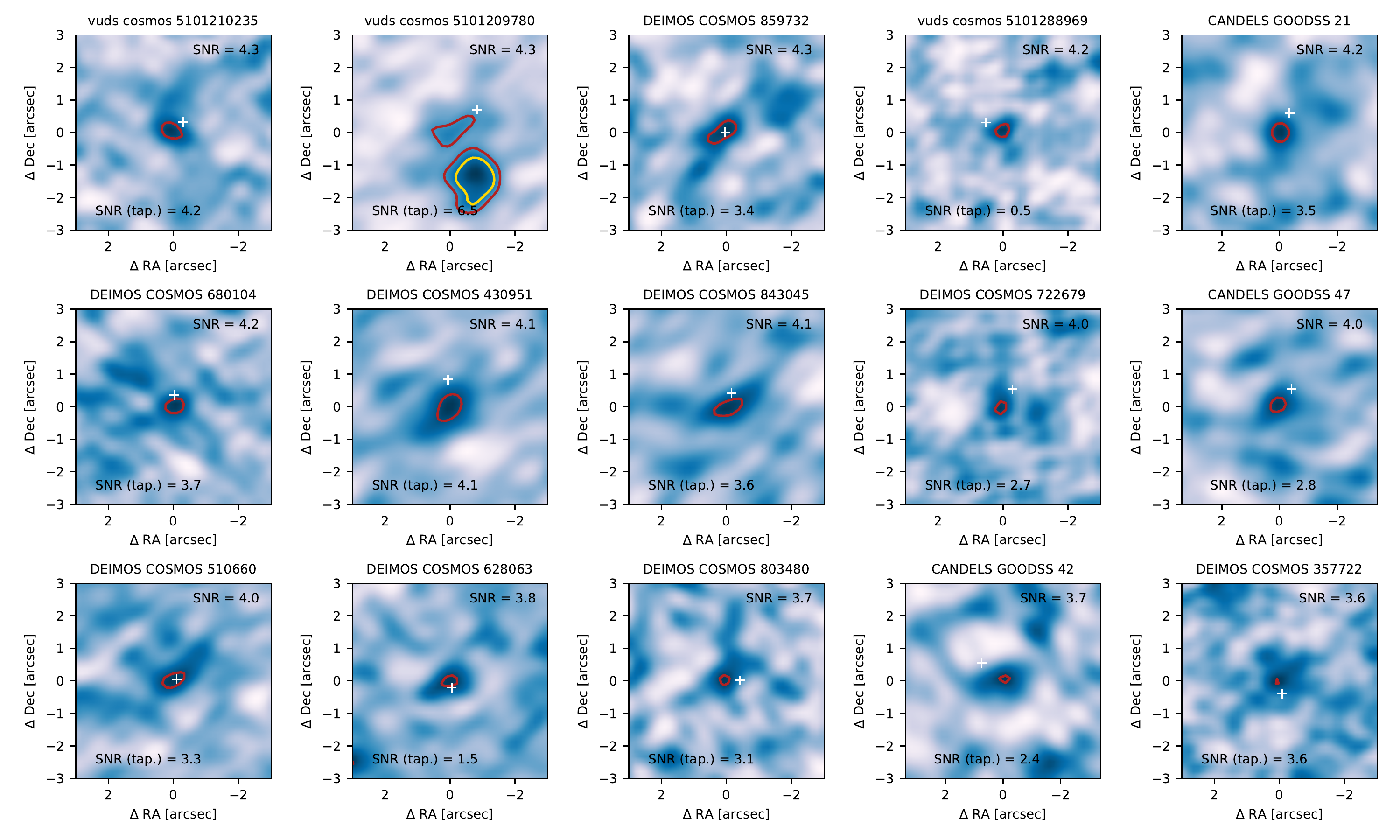}
\caption{\label{fig:target_mom0_3} Fig.\,\ref{fig:target_mom0_1} continued.}
\end{figure*}

\clearpage

\begin{longtable}{lcccccccc}
\caption{\label{tab:cii_target} List of detected [CII] target sources. The fluxes were measured in the moment-0 maps using the 2D-fit method and the luminosity is derived from this quantity. The [CII] redshift $z_{\rm [CII]}$ and the line FWHM are estimated using the centroid and the width, respectively, of the Gaussian fit of the spectrum. The typical uncertainties on the [CII] redshift are 0.0005. The measurement uncertainties do not include the calibration uncertainties ($\sim$4.5\,\% on average and up to 10\,\%, see Sect.\,\ref{sect:quasars}).}
\\\hline
\hline
Name & RA & Dec & S/N & S/N & z$_{\rm [CII]}$ & I$_{\rm [CII]}$ & FWHM & L$_{\rm [CII]}$ \\
 & h:min:s & deg:min:s & & (tap.) & & Jy\,km/s & km/s & log$_{10}$(L/L$_\odot$) \\
\hline
\endfirsthead
\caption{continued.}\\
\hline
\hline
Name & RA & Dec & S/N & S/N & z$_{\rm [CII]}$ & I$_{\rm [CII]}$ & FWHM & L$_{\rm [CII]}$ \\
 & h:min:s & deg:min:s & & (tap.) & & Jy\,km/s & km/s & log$_{10}$(L/L$_\odot$) \\
\hline
\endhead
\hline
\endfoot
DEIMOS\_COSMOS\_873756 & 10:00:02.72 &  +02:37:40.04 & 32.6 & 32.7 & 4.5457 & 5.84$\pm$0.21 & 526$\pm$13 & 9.56$\pm$0.02\\ 
DEIMOS\_COSMOS\_818760 & 10:01:54.86 &  +02:32:31.53 & 26.7 & 29.4 & 4.5613 & 6.92$\pm$0.27 & 276$\pm$13 & 9.63$\pm$0.02\\ 
vuds\_cosmos\_5101218326 & 10:01:12.50 &  +02:18:52.55 & 26.6 & 26.9 & 4.5739 & 2.92$\pm$0.13 & 229$\pm$13 & 9.26$\pm$0.02\\ 
DEIMOS\_COSMOS\_488399 & 10:03:01.15 &  +02:02:35.82 & 26.2 & 25.4 & 5.6704 & 1.24$\pm$0.06 & 303$\pm$13 & 9.03$\pm$0.02\\ 
vuds\_cosmos\_5110377875 & 10:01:32.33 &  +02:24:30.27 & 18.5 & 20.4 & 4.5505 & 2.77$\pm$0.17 & 234$\pm$13 & 9.23$\pm$0.03\\ 
DEIMOS\_COSMOS\_848185 & 10:00:21.50 &  +02:35:10.91 & 18.3 & 19.9 & 5.2931 & 2.06$\pm$0.12 & 275$\pm$13 & 9.20$\pm$0.03\\ 
DEIMOS\_COSMOS\_494057 & 09:58:28.50 &  +02:03:06.58 & 17.1 & 17.9 & 5.5448 & 0.86$\pm$0.06 & 217$\pm$13 & 8.86$\pm$0.03\\ 
vuds\_cosmos\_5100822662 & 09:58:57.91 &  +02:04:51.40 & 14.9 & 14.0 & 4.5205 & 1.28$\pm$0.11 & 208$\pm$13 & 8.90$\pm$0.04\\ 
DEIMOS\_COSMOS\_552206 & 09:58:26.78 &  +02:08:26.93 & 14.8 & 15.9 & 5.5016 & 1.84$\pm$0.14 & 365$\pm$13 & 9.18$\pm$0.03\\ 
DEIMOS\_COSMOS\_683613 & 10:00:09.42 &  +02:20:13.91 & 13.6 & 12.6 & 5.5420 & 0.95$\pm$0.08 & 216$\pm$13 & 8.90$\pm$0.04\\ 
DEIMOS\_COSMOS\_627939 & 10:01:04.86 &  +02:15:14.20 & 13.0 & 13.0 & 4.5341 & 1.17$\pm$0.10 & 252$\pm$13 & 8.86$\pm$0.04\\ 
vuds\_cosmos\_5180966608 & 10:01:37.47 &  +02:08:23.28 & 12.5 & 17.3 & 4.5296 & 2.23$\pm$0.18 & 243$\pm$13 & 9.14$\pm$0.03\\ 
DEIMOS\_COSMOS\_881725 & 10:00:13.54 &  +02:38:16.85 & 12.3 & 12.3 & 4.5777 & 1.09$\pm$0.10 & 198$\pm$13 & 8.84$\pm$0.04\\ 
CANDELS\_GOODSS\_32 & 03:32:17.00 &  -27:41:14.09 & 12.3 & 12.2 & 4.4105 & 1.38$\pm$0.14 & 279$\pm$13 & 8.91$\pm$0.04\\ 
DEIMOS\_COSMOS\_396844 & 10:00:59.64 &  +01:53:47.23 & 12.1 & 12.4 & 4.5424 & 1.86$\pm$0.17 & 287$\pm$13 & 9.06$\pm$0.04\\ 
vuds\_cosmos\_5100994794 & 10:00:41.17 &  +02:17:14.25 & 12.0 & 11.8 & 4.5802 & 0.89$\pm$0.08 & 230$\pm$13 & 8.75$\pm$0.04\\ 
vuds\_cosmos\_5100541407 & 10:01:00.92 &  +01:48:33.66 & 11.4 & 12.6 & 4.5630 & 1.90$\pm$0.19 & 177$\pm$13 & 9.07$\pm$0.04\\ 
DEIMOS\_COSMOS\_630594 & 10:00:32.61 &  +02:15:28.44 & 11.2 & 9.5 & 4.4403 & 1.04$\pm$0.10 & 260$\pm$13 & 8.79$\pm$0.04\\ 
vuds\_cosmos\_510786441 & 10:00:34.30 &  +01:59:21.14 & 11.1 & 11.3 & 4.4635 & 1.10$\pm$0.12 & 224$\pm$13 & 8.82$\pm$0.05\\ 
vuds\_cosmos\_5100969402 & 10:01:20.12 &  +02:17:01.27 & 11.0 & 10.1 & 4.5785 & 0.83$\pm$0.09 & 291$\pm$14 & 8.72$\pm$0.05\\ 
DEIMOS\_COSMOS\_494763 & 10:00:05.10 &  +02:03:12.08 & 10.5 & 9.1 & 5.2337 & 0.63$\pm$0.07 & 253$\pm$14 & 8.69$\pm$0.05\\ 
DEIMOS\_COSMOS\_372292 & 09:59:39.15 &  +01:51:28.07 & 9.6 & 8.4 & 5.1364 & 0.53$\pm$0.07 & 289$\pm$14 & 8.59$\pm$0.05\\ 
vuds\_efdcs\_530029038 & 03:32:19.05 &  -27:52:38.15 & 9.2 & 11.0 & 4.4298 & 1.17$\pm$0.13 & 367$\pm$14 & 8.84$\pm$0.05\\ 
DEIMOS\_COSMOS\_539609 & 09:59:07.27 &  +02:07:21.19 & 8.9 & 8.0 & 5.1818 & 0.65$\pm$0.09 & 287$\pm$14 & 8.69$\pm$0.06\\ 
DEIMOS\_COSMOS\_880016 & 09:59:55.16 &  +02:38:08.18 & 8.6 & 8.7 & 4.5415 & 0.89$\pm$0.12 & 274$\pm$14 & 8.74$\pm$0.06\\ 
DEIMOS\_COSMOS\_773957 & 10:01:10.05 &  +02:28:29.00 & 8.5 & 8.8 & 5.6773 & 0.48$\pm$0.06 & 344$\pm$13 & 8.62$\pm$0.06\\ 
DEIMOS\_COSMOS\_493583 & 10:00:23.36 &  +02:03:04.32 & 8.3 & 7.3 & 4.5134 & 0.70$\pm$0.10 & 198$\pm$13 & 8.64$\pm$0.06\\ 
vuds\_cosmos\_5100537582 & 10:01:33.51 &  +01:50:20.40 & 8.1 & 7.7 & 4.5501 & 0.71$\pm$0.11 & 206$\pm$15 & 8.65$\pm$0.07\\ 
DEIMOS\_COSMOS\_308643 & 10:01:26.66 &  +01:45:26.16 & 7.7 & 6.4 & 4.5253 & 0.92$\pm$0.14 & 406$\pm$40 & 8.76$\pm$0.07\\ 
DEIMOS\_COSMOS\_873321 & 10:00:04.06 &  +02:37:35.91 & 7.5 & 8.2 & 5.1542 & 1.27$\pm$0.14 & 201$\pm$13 & 8.98$\pm$0.05\\ 
DEIMOS\_COSMOS\_434239 & 10:01:17.11 &  +01:57:19.44 & 7.4 & 8.2 & 4.4883 & 2.31$\pm$0.27 & 497$\pm$13 & 9.15$\pm$0.05\\ 
DEIMOS\_COSMOS\_733857 & 10:01:19.91 &  +02:24:47.47 & 7.3 & 5.6 & 4.5445 & 0.78$\pm$0.12 & 226$\pm$13 & 8.68$\pm$0.07\\ 
DEIMOS\_COSMOS\_422677 & 10:01:59.46 &  +01:56:12.88 & 7.1 & 5.5 & 4.4381 & 0.70$\pm$0.12 & 233$\pm$14 & 8.63$\pm$0.07\\ 
DEIMOS\_COSMOS\_519281 & 09:59:00.89 &  +02:05:27.66 & 6.7 & 5.8 & 5.5759 & 0.64$\pm$0.13 & 282$\pm$14 & 8.73$\pm$0.09\\ 
DEIMOS\_COSMOS\_842313 & 10:00:54.52 &  +02:34:34.38 & 6.5 & 6.5 & 4.5537 & 0.73$\pm$0.13 & 250$\pm$15 & 8.66$\pm$0.08\\ 
DEIMOS\_COSMOS\_454608 & 10:02:43.36 &  +01:59:20.74 & 6.5 & 7.0 & 4.5834 & 1.03$\pm$0.17 & 232$\pm$15 & 8.81$\pm$0.07\\ 
DEIMOS\_COSMOS\_417567 & 10:02:04.12 &  +01:55:44.23 & 6.4 & 6.9 & 5.6700 & 0.36$\pm$0.06 & 310$\pm$16 & 8.50$\pm$0.08\\ 
vuds\_cosmos\_510596653 & 09:59:18.28 &  +01:56:17.08 & 6.2 & 5.2 & 4.5681 & 0.28$\pm$0.05 & 62$\pm$13 & 8.24$\pm$0.08\\ 
vuds\_cosmos\_5100559223 & 10:00:53.13 &  +01:51:53.50 & 5.9 & 6.8 & 4.5627 & 0.70$\pm$0.13 & 143$\pm$13 & 8.64$\pm$0.08\\ 
DEIMOS\_COSMOS\_351640 & 10:01:29.08 &  +01:49:29.67 & 5.7 & 5.0 & 5.7058 & 0.27$\pm$0.06 & 132$\pm$13 & 8.37$\pm$0.09\\ 
DEIMOS\_COSMOS\_432340 & 10:02:09.54 &  +01:57:05.53 & 5.5 & 4.5 & 4.4045 & 0.60$\pm$0.15 & 151$\pm$13 & 8.55$\pm$0.10\\ 
DEIMOS\_COSMOS\_709575 & 09:59:47.07 &  +02:22:32.95 & 5.5 & 4.2 & 4.4121 & 0.50$\pm$0.11 & 254$\pm$16 & 8.48$\pm$0.09\\ 
DEIMOS\_COSMOS\_834764 & 09:59:35.73 &  +02:34:00.55 & 5.4 & 5.7 & 4.5058 & 0.92$\pm$0.19 & 254$\pm$17 & 8.75$\pm$0.09\\ 
DEIMOS\_COSMOS\_416105 & 10:02:45.66 &  +01:55:36.02 & 5.3 & 4.9 & 5.6309 & 0.16$\pm$0.03 & 202$\pm$14 & 8.13$\pm$0.10\\ 
vuds\_cosmos\_5101244930 & 10:00:47.64 &  +02:18:02.22 & 5.0 & 3.9 & 4.5803 & 0.81$\pm$0.17 & 577$\pm$35 & 8.70$\pm$0.09\\ 
DEIMOS\_COSMOS\_403030 & 10:00:06.55 &  +01:54:21.22 & 5.0 & 5.6 & 4.5594 & 0.74$\pm$0.16 & 168$\pm$14 & 8.67$\pm$0.09\\ 
DEIMOS\_COSMOS\_536534 & 09:59:53.24 &  +02:07:05.34 & 5.0 & 5.4 & 5.6886 & 0.90$\pm$0.18 & 621$\pm$21 & 8.90$\pm$0.09\\ 
vuds\_cosmos\_510605533 & 09:59:24.59 &  +01:52:42.75 & 4.9 & 2.6 & 4.5019 & 0.43$\pm$0.11 & 504$\pm$31 & 8.42$\pm$0.11\\ 
DEIMOS\_COSMOS\_845652 & 10:00:51.60 &  +02:34:57.37 & 4.9 & 4.8 & 5.3071 & 0.45$\pm$0.12 & 339$\pm$17 & 8.55$\pm$0.11\\ 
DEIMOS\_COSMOS\_665509 & 09:58:56.45 &  +02:18:39.39 & 4.8 & 5.8 & 4.5256 & 1.38$\pm$0.31 & 372$\pm$17 & 8.93$\pm$0.10\\ 
CANDELS\_GOODSS\_75 & 03:32:32.57 &  -27:47:53.43 & 4.8 & 4.3 & 5.5666 & 0.39$\pm$0.09 & 503$\pm$15 & 8.52$\pm$0.11\\ 
DEIMOS\_COSMOS\_742174 & 10:00:39.14 &  +02:25:32.60 & 4.8 & 3.8 & 5.6360 & 0.17$\pm$0.04 & 139$\pm$14 & 8.17$\pm$0.10\\ 
CANDELS\_GOODSS\_38 & 03:32:15.91 &  -27:41:24.37 & 4.7 & 3.9 & 5.5721 & 0.31$\pm$0.09 & 392$\pm$28 & 8.42$\pm$0.12\\ 
DEIMOS\_COSMOS\_814483 & 10:01:27.12 &  +02:32:10.51 & 4.6 & 5.6 & 4.5810 & 1.27$\pm$0.29 & 360$\pm$14 & 8.90$\pm$0.10\\ 
DEIMOS\_COSMOS\_378903 & 10:01:11.42 &  +01:52:06.34 & 4.6 & 3.8 & 5.4297 & 0.22$\pm$0.06 & 155$\pm$13 & 8.26$\pm$0.12\\ 
CANDELS\_GOODSS\_14 & 03:32:18.94 &  -27:53:02.74 & 4.6 & 3.0 & 5.5527 & 0.15$\pm$0.04 & 230$\pm$16 & 8.09$\pm$0.11\\ 
DEIMOS\_COSMOS\_400160 & 10:01:04.10 &  +01:54:05.14 & 4.5 & 3.7 & 4.5404 & 0.63$\pm$0.16 & 518$\pm$19 & 8.59$\pm$0.11\\ 
DEIMOS\_COSMOS\_274035 & 09:59:32.46 &  +01:42:05.68 & 4.4 & 3.6 & 4.4791 & 0.44$\pm$0.12 & 266$\pm$19 & 8.43$\pm$0.12\\ 
DEIMOS\_COSMOS\_665626 & 10:01:14.23 &  +02:18:42.26 & 4.4 & 2.9 & 4.5773 & 0.26$\pm$0.07 & 102$\pm$13 & 8.21$\pm$0.12\\ 
CANDELS\_GOODSS\_12 & 03:32:54.01 &  -27:50:00.94 & 4.4 & 3.8 & 4.4310 & 0.84$\pm$0.25 & 541$\pm$18 & 8.70$\pm$0.13\\ 
vuds\_cosmos\_5101210235 & 10:01:31.61 &  +02:21:57.78 & 4.3 & 4.2 & 4.5761 & 0.36$\pm$0.10 & 145$\pm$13 & 8.35$\pm$0.12\\ 
vuds\_cosmos\_5101209780 & \multicolumn{8}{c}{Multi-component object, see Appendix\,\ref{sect:cii_multicomp}}\\ 
DEIMOS\_COSMOS\_859732 & 10:00:00.49 &  +02:36:19.29 & 4.3 & 3.4 & 4.5318 & 0.70$\pm$0.17 & 326$\pm$16 & 8.63$\pm$0.10\\ 
vuds\_cosmos\_5101288969 & 09:59:30.61 &  +02:19:53.45 & 4.2 & 3.5 & 5.7209 & 0.11$\pm$0.04 & 298$\pm$18 & 7.97$\pm$0.15\\ 
CANDELS\_GOODSS\_21 & 03:32:11.95 &  -27:41:57.46 & 4.2 & 3.5 & 5.5716 & 0.19$\pm$0.06 & 167$\pm$13 & 8.20$\pm$0.14\\ 
DEIMOS\_COSMOS\_680104 & 10:01:10.14 &  +02:19:56.07 & 4.2 & 3.7 & 4.5295 & 0.89$\pm$0.27 & 190$\pm$16 & 8.74$\pm$0.13\\ 
DEIMOS\_COSMOS\_430951 & 10:01:18.42 &  +01:57:02.95 & 4.1 & 4.1 & 5.6881 & 0.66$\pm$0.17 & 744$\pm$77 & 8.76$\pm$0.11\\ 
DEIMOS\_COSMOS\_843045 & 10:00:12.36 &  +02:34:43.41 & 4.1 & 3.6 & 5.8473 & 0.26$\pm$0.08 & 158$\pm$14 & 8.38$\pm$0.13\\ 
DEIMOS\_COSMOS\_722679 & 09:59:44.91 &  +02:23:45.97 & 4.0 & 2.7 & 5.7168 & 0.28$\pm$0.10 & 331$\pm$23 & 8.39$\pm$0.16\\ 
CANDELS\_GOODSS\_47 & 03:32:45.25 &  -27:49:10.17 & 4.0 & 2.8 & 5.5745 & 0.16$\pm$0.05 & 237$\pm$38 & 8.14$\pm$0.14\\ 
DEIMOS\_COSMOS\_510660 & 09:59:53.16 &  +02:04:37.78 & 4.0 & 3.3 & 4.5480 & 0.71$\pm$0.20 & 540$\pm$79 & 8.64$\pm$0.12\\ 
DEIMOS\_COSMOS\_628063 & 10:00:52.24 &  +02:15:15.77 & 3.8 & 1.5 & 4.5327 & 0.19$\pm$0.05 & 167$\pm$26 & 8.06$\pm$0.12\\ 
DEIMOS\_COSMOS\_803480 & 09:59:57.26 &  +02:31:13.01 & 3.7 & 3.1 & 4.5417 & 0.21$\pm$0.06 & 125$\pm$18 & 8.12$\pm$0.13\\ 
CANDELS\_GOODSS\_42 & 03:32:39.75 &  -27:52:58.42 & 3.7 & 2.4 & 5.5252 & 0.07$\pm$0.02 & 63$\pm$13 & 7.74$\pm$0.14\\ 
DEIMOS\_COSMOS\_357722 & 09:59:52.03 &  +01:50:06.06 & 3.6 & 3.6 & 5.6838 & 0.15$\pm$0.05 & 134$\pm$13 & 8.12$\pm$0.14\\ 
\hline
\end{longtable}

\begin{table*}[!h]
\caption{\label{tab:cii_uplims} Upper limits on the [CII] fluxes and luminosities of target sources assuming that [CII] line is in the frequency range probed by the ALPINE observations. The three methods used to compute them (aggressive, normal, and secure) are the same as in Table\,\ref{tab:uplim_cont} (method described in Sect.\,\ref{sect:uplim}), but applied to the moment-0 maps. In the absence of any external information, we recommend to use the secure upper limits. If the source is known to be point-like, the normal upper limits can be used.}
\centering
\begin{tabular}{lcccccc}
\hline
\hline
Name & \multicolumn{3}{c}{3-$\sigma$ upper limit on I$_{\rm [CII]}$} & \multicolumn{3}{c}{3-$\sigma$ upper limit on L$_{\rm [CII]}$} \\
 & aggr.& norm. & sec.& aggr.& norm. & sec. \\
 & \multicolumn{3}{c}{Jy\,km/s} &  \multicolumn{3}{c}{log$_{10}$(L/L$_\odot$)} \\
\hline
CANDELS\_GOODSS\_19 & $<$0.142 & $<$0.157 & $<$0.213 & $<$7.94 & $<$7.98 & $<$8.11\\
CANDELS\_GOODSS\_37 & $<$0.138 & $<$0.151 & $<$0.211 & $<$7.93 & $<$7.97 & $<$8.11\\
CANDELS\_GOODSS\_57 & $<$0.106 & $<$0.116 & $<$0.143 & $<$7.95 & $<$7.99 & $<$8.08\\
CANDELS\_GOODSS\_8 & $<$0.099 & $<$0.108 & $<$0.133 & $<$7.92 & $<$7.95 & $<$8.04\\
DEIMOS\_COSMOS\_206253 & $<$0.172 & $<$0.189 & $<$0.272 & $<$8.02 & $<$8.06 & $<$8.22\\
DEIMOS\_COSMOS\_224751 & $<$0.085 & $<$0.094 & $<$0.119 & $<$7.87 & $<$7.91 & $<$8.02\\
DEIMOS\_COSMOS\_298678 & $<$0.131 & $<$0.141 & $<$0.175 & $<$8.05 & $<$8.09 & $<$8.18\\
DEIMOS\_COSMOS\_328419 & $<$0.168 & $<$0.180 & $<$0.229 & $<$8.17 & $<$8.20 & $<$8.30\\
DEIMOS\_COSMOS\_336830 & $<$0.100 & $<$0.108 & $<$0.148 & $<$7.94 & $<$7.98 & $<$8.11\\
DEIMOS\_COSMOS\_406956 & $<$0.134 & $<$0.146 & $<$0.177 & $<$8.06 & $<$8.10 & $<$8.19\\
DEIMOS\_COSMOS\_412589 & $<$0.173 & $<$0.186 & $<$0.269 & $<$8.01 & $<$8.05 & $<$8.21\\
DEIMOS\_COSMOS\_420065 & $<$0.092 & $<$0.101 & $<$0.143 & $<$7.91 & $<$7.95 & $<$8.10\\
DEIMOS\_COSMOS\_421062 & $<$0.097 & $<$0.104 & $<$0.134 & $<$7.91 & $<$7.94 & $<$8.05\\
DEIMOS\_COSMOS\_431067 & $<$0.192 & $<$0.210 & $<$0.298 & $<$8.06 & $<$8.10 & $<$8.25\\
DEIMOS\_COSMOS\_442844 & $<$0.166 & $<$0.179 & $<$0.253 & $<$8.00 & $<$8.04 & $<$8.19\\
DEIMOS\_COSMOS\_460378 & $<$0.139 & $<$0.148 & $<$0.188 & $<$8.05 & $<$8.08 & $<$8.18\\
DEIMOS\_COSMOS\_470116 & $<$0.087 & $<$0.098 & $<$0.136 & $<$7.88 & $<$7.93 & $<$8.07\\
DEIMOS\_COSMOS\_471063 & $<$0.148 & $<$0.161 & $<$0.206 & $<$8.11 & $<$8.15 & $<$8.26\\
DEIMOS\_COSMOS\_472215 & $<$0.144 & $<$0.154 & $<$0.190 & $<$8.09 & $<$8.12 & $<$8.21\\
DEIMOS\_COSMOS\_503575 & $<$0.087 & $<$0.098 & $<$0.127 & $<$7.87 & $<$7.92 & $<$8.04\\
DEIMOS\_COSMOS\_549131 & $<$0.140 & $<$0.153 & $<$0.201 & $<$8.07 & $<$8.11 & $<$8.23\\
DEIMOS\_COSMOS\_550156 & $<$0.171 & $<$0.191 & $<$0.271 & $<$8.01 & $<$8.05 & $<$8.21\\
DEIMOS\_COSMOS\_567070 & $<$0.215 & $<$0.235 & $<$0.322 & $<$8.13 & $<$8.17 & $<$8.30\\
DEIMOS\_COSMOS\_576372 & $<$0.117 & $<$0.126 & $<$0.151 & $<$8.01 & $<$8.04 & $<$8.12\\
DEIMOS\_COSMOS\_586681 & $<$0.204 & $<$0.220 & $<$0.271 & $<$8.27 & $<$8.30 & $<$8.39\\
DEIMOS\_COSMOS\_592644 & $<$0.177 & $<$0.194 & $<$0.284 & $<$8.04 & $<$8.08 & $<$8.24\\
DEIMOS\_COSMOS\_628137 & $<$0.100 & $<$0.106 & $<$0.142 & $<$7.94 & $<$7.97 & $<$8.09\\
DEIMOS\_COSMOS\_629750 & $<$0.134 & $<$0.148 & $<$0.186 & $<$8.00 & $<$8.04 & $<$8.14\\
DEIMOS\_COSMOS\_743730 & $<$0.147 & $<$0.158 & $<$0.219 & $<$7.95 & $<$7.99 & $<$8.13\\
DEIMOS\_COSMOS\_761315 & $<$0.139 & $<$0.152 & $<$0.215 & $<$7.94 & $<$7.98 & $<$8.13\\
DEIMOS\_COSMOS\_787780 & $<$0.192 & $<$0.210 & $<$0.282 & $<$8.07 & $<$8.11 & $<$8.24\\
DEIMOS\_COSMOS\_790930 & $<$0.088 & $<$0.096 & $<$0.136 & $<$7.88 & $<$7.92 & $<$8.07\\
DEIMOS\_COSMOS\_838532 & $<$0.177 & $<$0.191 & $<$0.268 & $<$8.04 & $<$8.07 & $<$8.22\\
DEIMOS\_COSMOS\_869970 & $<$0.097 & $<$0.105 & $<$0.146 & $<$7.87 & $<$7.90 & $<$8.04\\
DEIMOS\_COSMOS\_910650 & $<$0.095 & $<$0.102 & $<$0.141 & $<$7.91 & $<$7.95 & $<$8.08\\
DEIMOS\_COSMOS\_920848 & $<$0.187 & $<$0.202 & $<$0.287 & $<$8.07 & $<$8.10 & $<$8.25\\
DEIMOS\_COSMOS\_926434 & $<$0.181 & $<$0.200 & $<$0.267 & $<$8.04 & $<$8.08 & $<$8.20\\
DEIMOS\_COSMOS\_933876 & $<$0.183 & $<$0.200 & $<$0.282 & $<$8.04 & $<$8.07 & $<$8.22\\
vuds\_cosmos\_5101013812 & $<$0.185 & $<$0.203 & $<$0.288 & $<$8.04 & $<$8.08 & $<$8.23\\
vuds\_cosmos\_510148750 & $<$0.132 & $<$0.143 & $<$0.190 & $<$7.91 & $<$7.94 & $<$8.07\\
vuds\_cosmos\_510327576 & $<$0.209 & $<$0.224 & $<$0.308 & $<$8.11 & $<$8.14 & $<$8.28\\
vuds\_cosmos\_510581738 & $<$0.163 & $<$0.178 & $<$0.260 & $<$8.00 & $<$8.04 & $<$8.20\\
vuds\_cosmos\_5131465996 & $<$0.162 & $<$0.174 & $<$0.251 & $<$7.99 & $<$8.02 & $<$8.18\\
\hline
\end{tabular}
\end{table*}
\twocolumn

\clearpage

\section{Sources with complex photometry}

\subsection{Continuum complexe sources}

\label{sect:DC881725}

\begin{figure}
\includegraphics[width=8.5cm]{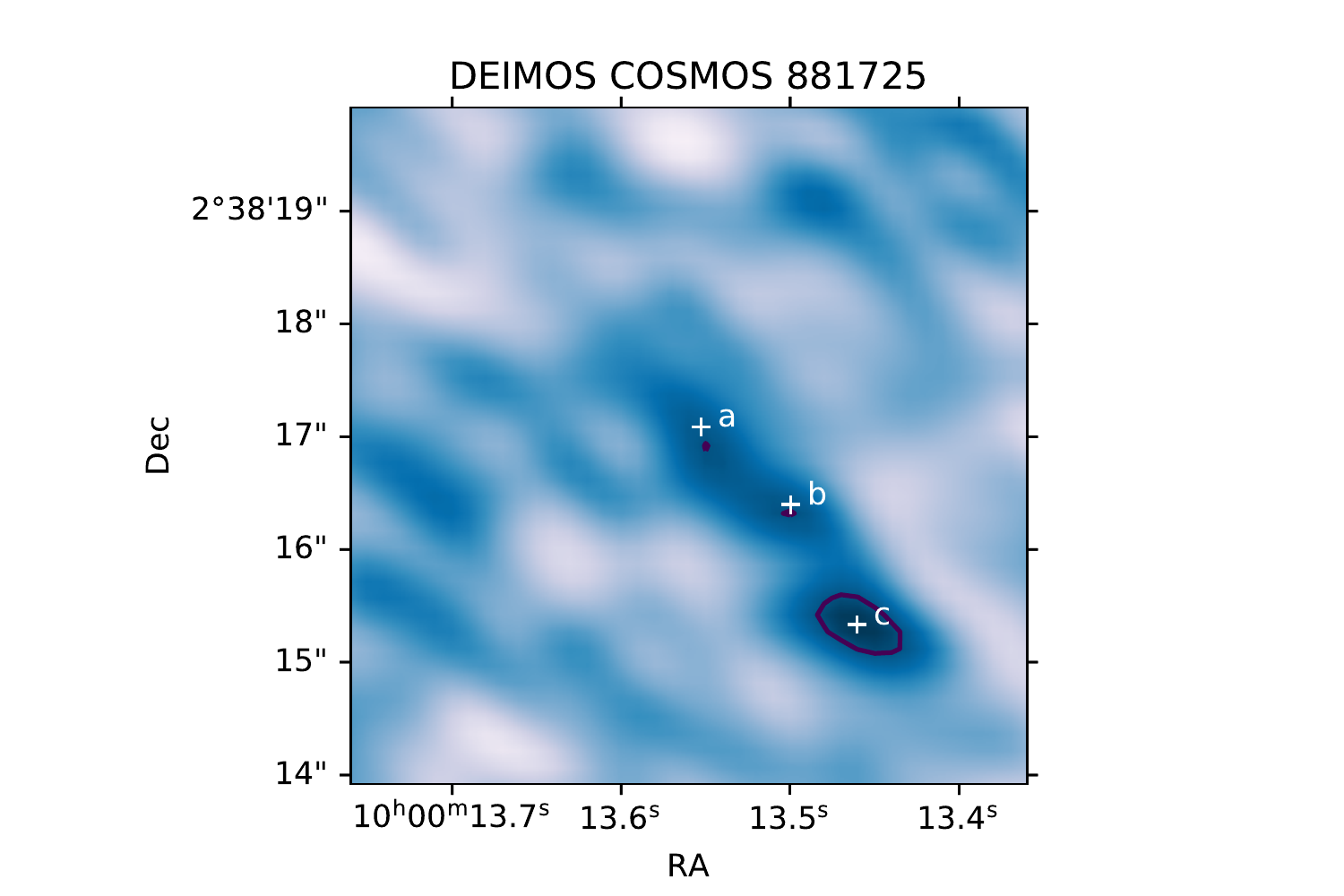}
\caption{\label{fig:DC881725} Cutout of the continuum image of DEIMOS\_COSMOS\_881725. The three components are indicated and the figure and their flux densities are listed in Table\,\ref{tab:complex_cont}.}
\end{figure}

For most of our sources, the automatic photometric measurements described in Sect.\,\ref{sect:cont_phot} are sufficient. However, a manual deblending is required for some sources with one or several neighbors. The most complex object is DEIMOS\_COSMOS\_881725, which was fit using three components simultaneously. These three components are shown in Fig.\,\ref{fig:DC881725}. Three other continuum sources have a close neighbor, which disturbs our automatic photometric procedure. They were also measured manually. All the flux densities are listed in Table\,\ref{tab:complex_cont}. As pointed out in \citet{Le_Fevre2019}, about 40\,\% of ALPINE targets are classified as merging systems (see the criteria in Le F\`evre et al.). For these systems, some deblending will be needed to properly distribute the observed line or continuum emission between the different components of the merger. This deblending of components close to the scale of the synthesized beam based on optical priors will be presented in a future paper.

\begin{table*}
\caption{\label{tab:complex_cont} Continuum photometry for multi-component target sources}
\centering
\begin{tabular}{lcccr}
\hline
\hline
Component & RA & Dec & Freq. & Flux \\
  & h:min:s & deg:min:s & GHz & $\mu$Jy \\
\hline
\multicolumn{5}{c}{DEIMOS\_COSMOS\_881725}\\
\hline
a & 10:00:13.55 & +02:38:17.08 & 347.2 & 105$\pm$55 \\
b  & 10:00:13.49 & +02:38:06.39 & 347.2 & 81$\pm$39 \\
c  & 10:00:13.46 & +02:38:15.33 & 347.2 & 163$\pm$60\\
\hline
\multicolumn{5}{c}{vuds\_cosmos\_5101209780}\\
\hline
a (target) & 10:01:33.49 & +02:22:09.81 & 348.0 & 311$\pm$112 \\ 
b (neighbor) &   10:01:33.58 & +02:22:10.72  & 348.0  & 178$\pm$78 \\
\hline
\multicolumn{5}{c}{vuds\_efdcs\_530029038}\\
\hline
a (target) & 03:32:19.02 & -027.52.38.43 & 356.0 & 125$\pm$58\\
b(neighbor) & 03:32:19.12 & -027.52.39.47 & 356.0 & 77$\pm$29 \\
\hline
\multicolumn{5}{c}{SC\_2\_vuds\_cosmos\_5110377875}\\
\hline
a (brightest)& 10:01:31.68 & +02:24:29.58 & 349.4 & 994$\pm$260 \\
b (southwest)& 10:01:31.75 & +02:24:29:02 & 349.4 & 581$\pm$210 \\
\hline
\end{tabular}
\end{table*}

\subsection{[CII] complex source}

\label{sect:cii_multicomp}

We had less problems with the deblending of [CII] target sources. Indeed, since the flux is measured in the moment-0 maps, it is very unlikely to find a CO interloper at the same exact frequency. The only possible contaminant is another physically related object at the same velocity as the ALPINE target.  For instance, DEIMOS\_COSMOS\_842313 has an extremely bright neighbor known as J1000+0234 \citep{Schinnerer2008} or AzTEC/C17 \citep{Brisbin2017}. This nearby submillimeter galaxy is identified as SC\_2\_DEIMOS\_COSMOS\_842313 in our nontarget catalog, but there is no signal at this position on the moment-0 map, since the [CII] of this source is at a different velocity.

\begin{figure}
\includegraphics[width=8.5cm]{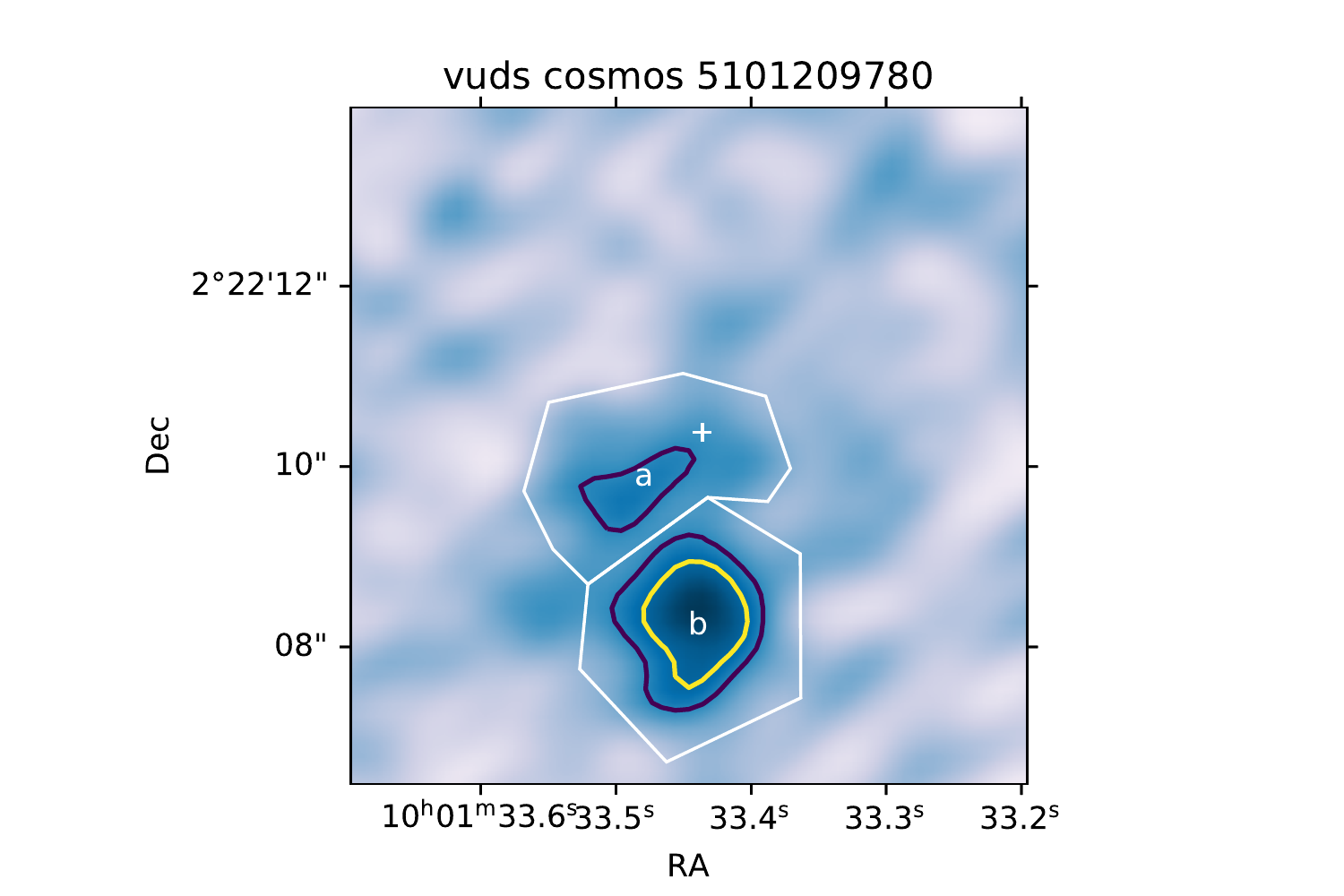}
\caption{\label{fig:vuds_cosmos_5101209780} Cutout of the moment-0 map of vuds\_cosmos\_5101209780 indicating how we defined the two components.}
\end{figure}

In the case of close mergers, our three photometric methods are compatible and tend to measure the total flux of the system. However, vuds\_cosmos\_5101209780 is a bit more problematic. The ALPINE target is faint, but it has a bright southern companion also detected in the moment-0 map. Some photometric methods include this component while other exclude it. We thus decided to perform a manual extraction after having defined manually two regions corresponding to each component (see Fig.\,\ref{fig:vuds_cosmos_5101209780}). The fluxes are provided in Table\,\ref{tab:vuds_cosmos_5101209780}.

\begin{table*}
\caption{\label{tab:vuds_cosmos_5101209780} [CII] line fluxes of the different components of vuds\_cosmos\_5101209780 (see Fig.\,\ref{fig:vuds_cosmos_5101209780}).}
\centering
\begin{tabular}{lccr}
\hline
\hline
Component & RA & Dec & Line Flux \\
  & h:min:s & deg:min:s & Jy\,km/s \\
\hline
\multicolumn{4}{c}{vuds\_cosmos\_5101209780}\\
\hline
a & 10:01:33.48 & +02:22:09.68 & 1.17$\pm$0.25\\
b & 10:01:33.43 & +02:22:08.21 & 2.02$\pm$0.28 \\
\hline
\end{tabular}
\end{table*}

\end{appendix}

\end{document}